DEVELOPMENT OF LOW CRITICAL TEMPERATURE
SUPERCONDUCTING TUNNEL JUNCTIONS FOR APPLICATION AS
PHOTON DETECTORS IN ASTRONOMY

Ph. D. committee:

Prof. dr. ir. A. Bliek, University of Twente.
Prof. dr. A. Barone, University of Naples.
Prof. dr. T. Klapwijk, TU Delft.
Dr. A. Peacock, European Space Agency.
Dr. P. Verhoeve, Aurora Technology B.V.
Prof. Dr. H. H. J. ten Kate, University of Twente.
Dr. J. Flokstra, University of Twente.
Dr. A. A. Golubov, University of Twente.
Prof. dr. H. Rogalla, University of Twente.

DEVELOPMENT OF LOW CRITICAL TEMPERATURE
SUPERCONDUCTING TUNNEL JUNCTIONS FOR APPLICATION AS
PHOTON DETECTORS IN ASTRONOMY

PROEFSCHRIFT

ter verkrijging van
de graad van doctor aan de Universiteit Twente,
op gezag van de rector magnificus,
prof. dr. F.A. van Vught,
volgens besluit van het College voor Promoties
in het openbaar te verdedigen
op woensdag 10 december 2003 om 13.15 uur

door

Guy Bernard Wilhelm Brammertz

geboren op 8 januari 1976
te Eupen (België)

Dit proefschrift is goedgekeurd door:

Promotor: Prof. dr. H. Rogalla
*Universiteit Twente*

Assistent Promotor: Dr. A.A. Golubov
*Universiteit Twente*

# Contents

# Chapter 1

# Introduction

The Superconducting Tunnel Junction used as photon detector is introduced. First, the general working principle of the junctions is described and the history of photon detection experiments performed is presented. Then two applications of the spectrometers, which are currently under development at the European Space Agency, are presented. These are on one hand the X-ray Evolving Universe Spectroscopy mission and on the other hand the optical superconducting camera used for ground based astronomy. The expression for the ideal achievable energy resolution obtainable with a Superconducting Tunnel Junction is then introduced. This leads to the main aims and the motivation of this thesis, which is the fabrication and operation of low energy gap junctions, which allow for better energy resolution performance than current higher energy gap junctions. Finally, an overview and structure of the thesis is given.

## 1.1 Background and motivation

Superconducting tunnel junctions (STJs) have been under development as photon-counting spectrometers for application in astronomy for a number of years. Since the first successful detection of 6 keV X-rays in a Sn-based STJ in 1986 [Twerenbold 86], STJs have been constantly further developed by a growing number of groups in order to exploit their potential as high resolution astronomical photon detectors.

The basic excitations of the superconducting ground state have energies of the order of one meV. This is three orders of magnitude smaller than the basic excitations in semiconductors. As a consequence, the intrinsic resolving power of superconducting detectors should be about a factor 30 better than that for the widely used semiconducting detectors. In addition, optical photons create several thousands of excitations in a superconductor, which should allow for non-dispersive photon-counting spectroscopy in the optical regime.

During the photon absorption process in a superconductor a number $Q_0$ of excitations of the superconducting ground state, called quasiparticles, are created. This number of created quasiparticles is directly proportional to the energy of the photon absorbed in the superconductor $Q_0 = E/\varepsilon$, where E is the photon energy and $\varepsilon$ is the minimum energy necessary to create one quasiparticle in the superconductor. The created quasiparticles are then read-out via tunnelling through an insulating barrier into a second superconductor. The most basic lay-out of a STJ consists of a superconducting film separated from a second superconducting film by a very thin (~1 nm) insulating barrier, which allows for quantum mechanical tunnelling of the quasiparticles. Figure 1.1 shows an example of such a junction based on Ta-Al electrodes separated by a thin Al oxide barrier. A magnetic field applied in parallel to the barrier suppresses all zero voltage Josephson currents in the junction. A small bias voltage applied between the electrodes of the detector favours the current flow into the direction indicated by the bias. Two different tunnel events of the quasiparticles are possible, the normal tunnel event and the back-tunnel event. The normal tunnel event transfers the quasiparticle as well as the charge in the direction indicated by the bias voltage. The back-tunnel event on the other hand transfers the quasiparticle against the direction indicated by the bias, whereas the charge is transferred in the direction of the bias. In this way quasiparticles can tunnel several times back and forth between the electrodes during their lifetime and create additional charge amplification [Gray 78]. The excess quasiparticle population created by the photo-absorption process decreases approximately exponentially with a timescale equal to the decay time of the pulse, because of the existence of several quasiparticle loss channels. The current pulse created by the excess quasiparticles is integrated and the total charge detected is proportional to the energy of the absorbed photon, thereby allowing for the determination of the photon energy from the integrated current pulse.

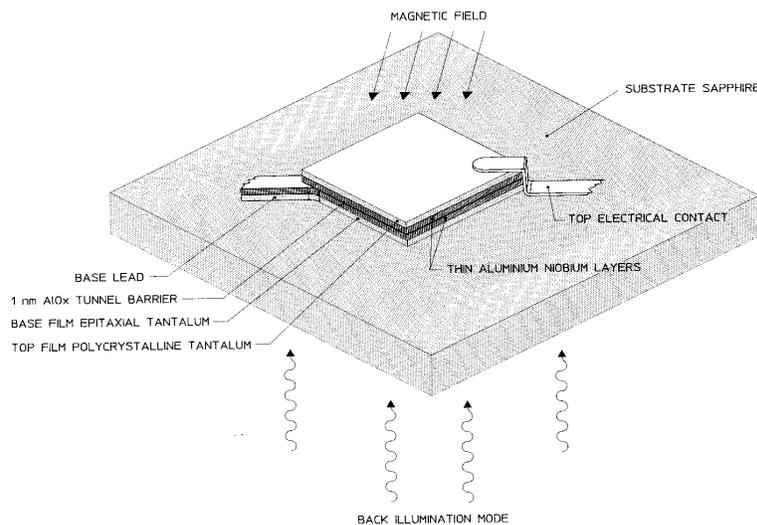

**Figure 1.1:** Schematic of a Ta-Al superconducting tunnel junction used as photon detector. The back illumination mode applies for optical photons, whereas X-ray illumination would be from the top.

The pioneering work of Twerenbold and co-workers in 1986 already showed a very promising energy resolution of 50 eV full width at half maximum (FWHM) for 6 keV x-rays absorbed in Sn junctions. Because of the better resistance to thermal cycling, the community soon switched to junctions based on Nb, with which an energy resolution of 36 eV at 6 keV was obtained by Mears and his co-workers [Mears 93b]. This was at a later stage improved to 29 eV at 6 keV [Mears 96], which is to date the best measured resolution with a Nb based STJ at 6 keV. In the mid-nineties other groups started the development of Ta based junctions, with which an energy resolution of 24 eV was obtained for 6 keV x-rays [Brammertz 01b], and Al based junctions, which have to date achieved the best 6 keV energy resolution measured with a single pixel STJ equal to 12 eV FWHM [Angloher 01]. Nevertheless, single STJs have not yet reached the predicted intrinsic resolutions at 6 keV, which are equal to 10, 7 and 3.5 eV for Nb, Ta and Al based junctions respectively. For all three junction types the measured resolution is about a factor 3-4 above the theoretical resolution. The main reason for this strong resolution degradation is generally attributed to variations in the response of the detector depending on the exact absorption position of the photon in the detector.

The detection of single optical photons with STJs was first demonstrated with Nb based junctions in 1996 [Peacock 96] and later with Ta based devices [Verhoeve 97]. Up to date the best optical result was achieved with a symmetrical Ta based junction, whose electrodes consist of a sandwich of a 100 nm thick Ta and a 65 nm thick Al layer. The lay-up of this junction is shown in Fig. 1.1. The measured energy resolution of this device is equal to 0.15 eV for 2.48 eV photons ($\lambda$ = 500 nm) [Verhoeve 02b], including a 0.09 eV broadening contribution of the read-out electronics. This corresponds to a resolving power of approximately 16 for photons with a wavelength of 500 nm. The corresponding intrinsic resolution of this detector is equal to 0.12 eV, which corresponds to the predicted optimum achievable energy resolution for this lay-up.

Because of the good spectroscopic performance over a broad energy domain with high efficiency and fast response time, STJs can be used for a large number of applications including time-of-flight mass spectroscopy [Twerenbold 96], x-ray fluorescence microanalysis [Frank 97] and time-resolved analysis of biological fluorescent samples [Fraser 03]. In this thesis the focus is on the application of STJs as photon detectors for astronomy. Two direct applications under development at the European Space Agency (ESA) are the X-ray Evolving-Universe Spectroscopy (XEUS) mission and the optical superconducting camera (S-Cam).

XEUS [Bavdaz 01, Arnauld 00, Bleeker 00, Aschenbach 01] is an ambitious project currently under study at ESA, which is aiming at probing the organisation of hot matter in the very early universe. The main scientific goal of XEUS is the analysis of the formation of the very first black holes and clusters of galaxies, when the universe was just a few percent of its current age. The photon fluxes coming from these objects are extremely low of the order of $10^{-17}$ erg cm$^{-2}$ s$^{-1}$, which requires a large collection area of the optics and excellent quantum efficiency of the detectors. In addition, the collection of high resolution spectra of these sources with medium spatial resolving power will allow the study of the physical and chemical properties of the accreting material as well as the basic properties of the black holes themselves. The current design goals of the XEUS spectrometers consist of energy resolutions of 1 and 5 eV for 1 and 8 keV x-rays respectively and an energy range extending from 50 eV to 10 keV. In addition to these resolution requirements the spectrographs should also have an inherent spatial resolution of approximately 0.6 arcsec with a field of view of 30x30 arcsec. Two detector types seem currently able to meet all of these stringent requirements, which are the STJ for the low energy range from 50 eV to 2 keV and the Transition Edge Sensor [Hoevers 02] for the energy range from 2 to 10 keV. The 10 meter diameter x-ray optics will have a focal length of 50 meters. As a consequence XEUS will consist of two separated spacecraft, one consisting of the x-ray optics and one consisting of the detector spacecraft. Figure 1.2 shows an artists impression of XEUS in its final constellation, with the two spacecrafts separated by 50 meters. As the 10 meter diameter optics is too big to be launched by any type of rocket, the final assembly of the optics will have to be made in space. The proposed tool for this assembly process is the International Space Station and XEUS will therefore be a truly international mission, whose proposed launch date is in the 2012-2015 time frame.

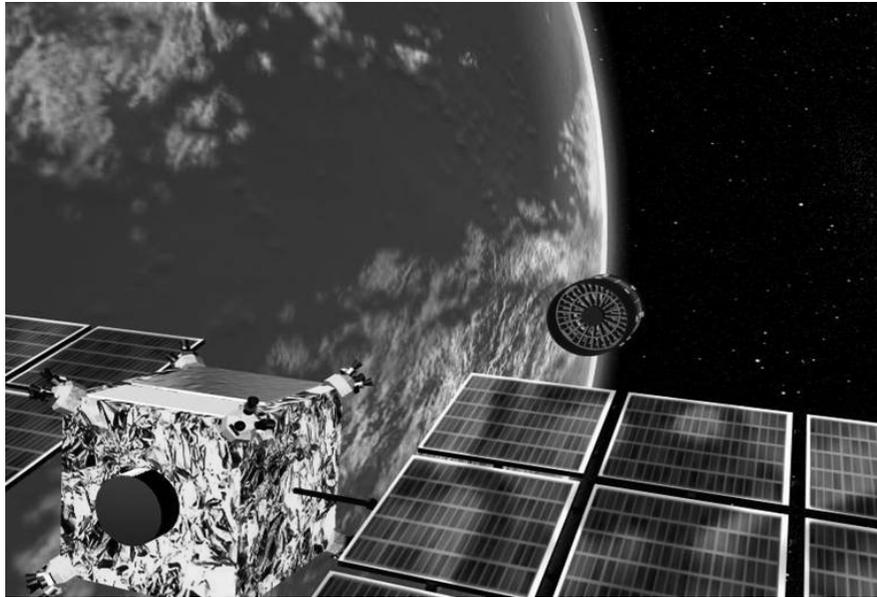

**Figure 1.2:** Artists impression of the XEUS spacecrafts in their final configuration with the mirror and detector spacecrafts separated by 50 m.

The S-Cam [Verhoeve 02b] program of the Science Payload and Advanced Concepts Office of the European Space Agency develops photon-counting spectrometers for ground-based astronomical applications in the visible domain. Up to date two different instruments have been fabricated, which were operated at the 4.2 meter William Herschel Telescope in La Palma. Both instruments consist of a 6x6 array of 30 μm side length Ta-Al based STJs, whose lay-up is similar that shown in Fig. 1.1. S-Cam 1 [Rando 00] is a technology demonstrator, which showed the potential of this new kind of spectrometer. S-Cam 1 observed the fast light pulse of the Crab pulsar, a neutron star spinning at about 30 revolutions per second, analysing simultaneously the time evolution as well as the spectral evolution of the light emitted by the source with good detection efficiency [Perryman 99]. S-Cam 2 [Verhoeve 02] is based on the same type of 6x6 Ta-based STJ array, but includes several improvements to the instrument thereby improving the resolving power ($\lambda/\Delta\lambda$ FWHM) for 500 nm photons to approximately 8, keeping the temporal resolution of the instrument equal to 5 μsec for a total photon count-rate performance per pixel of approximately 5 kHz. Three observation campaigns have been conducted with S-Cam 2 at the William Herschel Telescope in the years 1999 and 2000. The data of these campaigns, including observations of a variety of astronomical objects such as γ-ray bursts, cataclysmic variables, pulsars and dwarf novae has resulted in several publications [de Bruijne 02, Reynolds 03]. The next S-Cam instrument, which is currently in its final integration phase, will consist of a 10x12 array based on Ta-Al junctions (Fig. 1.3) and is scheduled to be operated early 2004 at the William Herschel Telescope. In addition to the larger array size the major improvements of S-Cam 3 with respect to the previous instruments are an improved resolving power, which increased from 8 to ~13 for 500 nm photons, a larger count-rate performance, which increased to ~10 kHz per pixel, as well as an increased bandpass, now ranging from 330 to 800 nm [Martin 03].

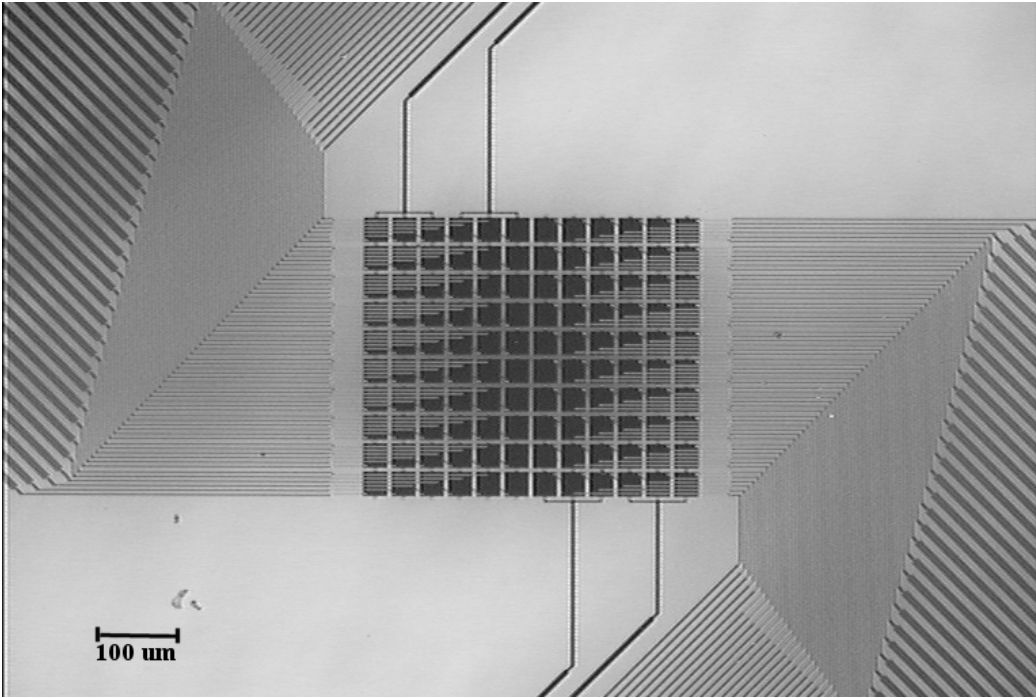

**Figure 1.3:** Optical Microscope image of the 10x12 Ta-Al based array used for the S-Cam 3 instrument.

## 1.2  Aims and objectives

The expected best achievable FWHM energy resolution of a symmetrical STJ, which allows for multiple tunnelling of the quasiparticles between the two electrodes, is given by:

$$\Delta E = 2.355\sqrt{(F + G)\varepsilon E} \ , \tag{1.1}$$

where F is the Fano factor, G is the tunnelling factor, ε is the mean energy necessary to create a quasiparticle and E is the photon energy. The factor 2.355 arises from the conversion of root mean square (RMS) resolution to full width at half maximum (FWHM) resolution.

The Fano factor F represents the broadening due to statistical variations in the initial number of quasiparticles created by the photo-absorption process in the superconductor. Its value was calculated to be approximately equal to 0.2 for Sn [Kurakado 82] and Nb [Rando 92] and in general it is accepted that this value does not depend much on the nature of the metallic superconductor.

The tunnelling factor G represents the statistical variation in the average number of tunnel events <n> a quasiparticle undergoes during its lifetime. For a symmetrical superconductor and infinite integration time of the current pulse, G was calculated to be equal to 1+1/<n> [Mears 93, Goldie 94]. For tunnel junctions with a large number of tunnels during the lifetime of the quasiparticle, the tunnel contribution G approaches one.

Both the Fano factor and the tunnelling factor are independent of the nature of the superconducting material forming the detector, not leaving a lot of space for improving the energy resolution of the detectors. On the other hand, the mean energy $\epsilon$ necessary to create a quasiparticle depends strongly on the nature of the material. In fact $\epsilon$ was calculated to be equal to $1.7\Delta_g$ [Kurakado 82, Rando 92, Kozorezov 00], where $\Delta_g$ is the minimum energy of a basic excitation of the superconducting ground state, which is called the energy gap of the superconductor. This energy gap depends strongly on the nature of the superconductor. Table 1.1 shows the energy gap and the critical temperature $T_C$ for several metallic superconductors as well as the best theoretical energy resolution achievable for three different energies of the incoming photon.

**Table 1.1:** Energy gap, critical temperature and best expected FWHM energy resolutions at three different photon energies for different superconducting materials.

| Material | Energy gap $\Delta_g$ | Critical temperature $T_C$ | $\Delta E$ at 2.48 eV | $\Delta E$ at 1 keV | $\Delta E$ at 6 keV |
| --- | --- | --- | --- | --- | --- |
| | ($\mu$eV) | (K) | (eV) | (eV) | (eV) |
| Niobium (Nb) | 1550 | 9.3 | 0.208 | 4.2 | 10.2 |
| Vanadium (V) | 820 | 5.4 | 0.15 | 3.0 | 7.5 |
| Tantalum (Ta) | 700 | 4.5 | 0.14 | 2.8 | 7 |
| Aluminium (Al) | 180 | 1.2 | 0.07 | 1.4 | 3.5 |
| Molybdenum (Mo) | 139 | 0.915 | 0.06 | 1.25 | 3.1 |
| Hafnium (Hf) | 19.4 | 0.128 | 0.023 | 0.47 | 1.15 |

The expected best energy resolution is shown graphically in Fig. 1.4 for six superconductors as a function of the incoming photon energy. The ideal resolution achievable varies as the square root of the energy gap of the superconductor. With Al, for example, which has a four times lower energy gap than Ta, an energy resolution a factor two better than for Ta STJs should be achievable. As a consequence the migration from Nb- and Ta-based STJs, as currently used by most of the groups, to lower energy gap superconductors like Al and Mo should result in an increase in energy resolution by at least a factor two. Especially for the requirements of the XEUS mission the development of lower $T_C$ STJs with an energy gap lower than 300 $\mu$eV is required in order to achieve the desired energy resolution of less than 2 eV for 1 keV photons, which will allow for the diagnostics of the profiles of the emission lines of plasma. For the S-Cam programme the development of low energy gap STJs is required in order to increase the resolving power of the detectors for 500 nm photons to values in excess of 20, because the Ta-based junctions that are used up to date approach their best possible achievable energy resolution. An increase in resolving power is achievable by reducing the energy gap of the material. This increased energy resolution comes at a certain cost of course, as the operating temperature of the junctions decreases directly proportional to the energy gap. In order to prevent any thermal excitations in the superconductor, which would be a source of additional noise, the detector has to be cooled to about one tenth of its critical temperature. Whereas Nb, Ta and V-based junctions can be operated at a temperature of 300 mK, which is the base temperature reached by most $^3$He sorption coolers, Al and Mo-based junctions have to be operated at temperatures lower than 100 mK, which is much harder to achieve.

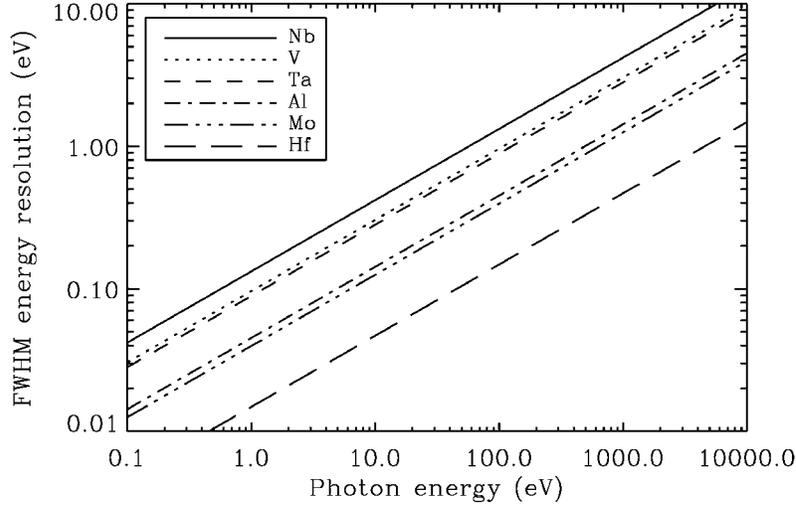

**Figure 1.4:** Expected ideal energy resolution as a function of incoming photon energy for six different superconducting materials.

The aim of this thesis is to demonstrate the fabrication and operation of high quality STJs fabricated out of alternative materials other than Ta and Nb with the main focus on lower energy gap materials in order to increase the energy resolution of the detectors. The interest lies in junctions based on V, Al and Mo, which have energy gaps of 820, 180 and 139 μeV respectively. We decided to use V as a junction fabrication process development vehicle. V has an energy gap comparable to Ta and therefore similar resolution capabilities. The possibility to operate V junctions at 300 mK allows a much faster turnaround time for testing the influence of variations in certain processing steps on the quality of the tunnel junctions. The knowledge gained can then be applied to the lower energy gap junctions, which have to be operated in an Adiabatic Demagnetisation Refrigerator (ADR) with a much longer turnaround time.

In addition to the very practical aim of fabricating and operating a functioning low energy gap STJ, another main focus is also the development of an improved model for the quasiparticle processes occurring in these very complex detectors during the electronic read-out phase of the created excitations. To date, the vast majority of the numerous developed models describing the functioning of STJs as photon detectors [Twerenbold 86, Le Grand 94, Gijsbertsen 95, Verhoeve 96, van den Berg 99, Poelaert 99] are based on the Rothwarf-Taylor balance equations approach [Rothwarf 67], whose main assumption is that all the quasiparticles stay at the gap energy during the entirety of the read-out process. The quasiparticle energy down-conversion rate on the other hand depends on the cube of the energy gap [Kaplan 76], which means that this approximation gets worse for decreasing energy gaps. Even for the larger gap junctions based on Nb and Ta, Poelaert et al. [Poelaert 99] found evidence that this condition is not fulfilled and introduced the term "balance energy", which describes the average energy of the quasiparticles during the read-out phase. In this thesis another aim is to get a better insight into the quasiparticle energy distribution during the read-out process, as this might be of increasing importance for the low energy gap junctions under study. Therefore, a model was developed, which takes into account the full energy dependence of all the processes occurring in the detectors.

## 1.3  Summary and layout

In chapter 2 the theory of STJs used as astronomical photon detectors is discussed. The main focus is given to the energy distribution of the non-equilibrium quasiparticles, which to date is a rather unknown territory and is becoming increasingly important for low energy gap junctions. First, several general properties of metallic superconductors are presented, followed by a description of the proximity effect theory, which calculates the properties of superconducting bi-layers that one can find in the electrodes of the detectors. The role of the interface parameters describing the superconductor-superconductor interface in the bi-layer is stressed and a new method for the determination of the latter is proposed. The proximity effect theory is then applied to two series of Nb-Al and Ta-Al bi-layers with varying Al thickness. Based on the results of the proximity effect theory a novel kinetic model describing the photon detection process is presented, which takes into account the full energy dependence of the excess quasiparticles. It allows the determination of the complete time-variation of the quasiparticle energy distribution. The theory is illustrated by applying it to two experimentally very well characterised Nb-Al and Ta-Al based junctions and the results are then compared to previous models based on the Rothwarf-Taylor approach.

Chapter 3 gives an overview of the experimental set-up used for testing the quality and spectral performance of the junctions. The experiments were performed in the laboratories of the Science Payloads and Advanced Concepts Office of the European Space Agency. Two different cryostats are described, which are a $^3$He sorption cooler with a base temperature of 300 mK and an Adiabatic Demagnetisation Refrigerator with a base temperature of 35 mK. The read-out electronics are described as well.

Chapter 4 is dedicated to the fabrication process of the junctions as well as the basic properties of the electrodes and insulating layers separating the latter. First, a somewhat general fabrication process is described, which is then completed by material specific details for the V-Al, Al and Mo-Al based junctions. Then the characteristics of V, Al and Mo single films are described, which form the electrodes of the detector. Finally, the quality of the insulating barrier is described for the V, Al and Mo-based junctions respectively, by analysing the current-voltage characteristics of the junctions as well as the Josephson current suppression pattern.

In chapter 5 the photon detection experiments are described. First the experimental data acquired with the V-Al based junctions that were exposed to 6 keV x-rays is presented. The model of chapter 2 is applied in order to explain the different variations of the responsivity and pulse decay time. Then the response of the Al junctions to IR to soft-UV radiation is presented, as well as the response to 6 keV x-rays. The model of chapter 2 is then applied to the Al case in order to explain interesting features in the responsivity of the different junctions. Finally, the energy resolution of the detectors is discussed.

# Chapter 2

# Theory of Superconducting Tunnel Junctions used as photon detectors for Astronomy

In this chapter the theory behind the operation of superconducting tunnel junctions used as photon detectors is presented. The theoretical developments in this chapter are mainly focused towards the determination of the quasiparticle energy distribution in the electrodes of the detector, which is important knowledge for the low energy gap junctions treated in the following chapters. The quasiparticle down-scattering rate depends roughly on the cube of the critical temperature $T_C$ and therefore the relaxation of quasiparticles towards the gap energy is much slower in the low $T_C$ devices. As a result the quasiparticles will be spread out over a very broad energy domain, having important consequences for the operation of low energy gap STJs. The results presented in this chapter include the presentation of the proximity effect theory between superconducting layers as well as recent developments in non-equilibrium quasiparticle and phonon dynamics. First the general properties of superconducting metals will be presented that have a direct influence on this work. Then the proximity effect theory between superconducting layers will be presented as well as a novel method for the determination of the parameters characterising the interface between the superconductors. Finally, based on the results of the proximity effect theory, a kinetic model of the photon detection process is presented. This kinetic model involves for the first time the full energy dependence of all relevant processes occurring in the junction and allows the determination of the complete time evolution of the quasiparticle energy distribution in both electrodes of the detector, from the moment of creation of the excess quasiparticles until the moment of disappearance of the last quasiparticle. The theory will be illustrated by applying it to two experimentally extensively characterised Nb-Al and Ta-Al based junctions.

## *2.1 Superconductivity*

At absolute zero temperature the electrons in a metal, which obey to the Pauli exclusion principle, occupy the lowest available energy levels up to a certain level, the Fermi energy $E_F$. Cooper showed in 1956 [Cooper 56] that this so-called Fermi sea of electrons is unstable against the formation of a single bound pair of electrons. He proved the existence of a bound state of electrons with opposite momentum and spin that, although composed of electrons with a kinetic energy larger than the kinetic energy of electrons at the Fermi level, has a negative total energy compared to the free electron energy at the Fermi level. Therefore, as soon as the electrons see a mutual attractive interaction potential they will start to condense into paired states until some equilibrium state is reached. This condensate of paired electrons with wave vector $\mathbf{k}$ and spin $\uparrow$ (Cooper pairs) is called the BCS ground state and it is responsible for the most obvious superconducting properties like perfect conductivity and perfect diamagnetism. By approximating the attractive interaction potential as being independent of electron momentum, Bardeen, Cooper and Schrieffer determined in 1957 the probability $v_{\mathbf{k}}^2$ that a Cooper pair state $\left(\mathbf{k}\uparrow,-\mathbf{k}\downarrow\right)$ is occupied at zero temperature [Bardeen 57]:

$$v_{\mathbf{k}}^2 = \frac{1}{2}\left(1 - \frac{\varepsilon_{\mathbf{k}}}{E_{\mathbf{k}}}\right), \text{ with } E_{\mathbf{k}} = \sqrt{\varepsilon_{\mathbf{k}}^2 + \Delta_g^2} \ . \tag{2.1}$$

Here $\varepsilon_{\mathbf{k}}$ is the energy of a free electron with momentum $\hbar\mathbf{k}$ relative to the Fermi energy

$$\varepsilon_{\mathbf{k}} = \frac{\hbar^2\mathbf{k}^2}{2m^*} - E_F , \tag{2.2}$$

and $\Delta_g$ is the energy gap of the superconductor. Fig. 2.1 shows $v_{\mathbf{k}}^2$ as a function of the free electron energy $\varepsilon_{\mathbf{k}}$. Note that even at absolute zero some of the electrons bound into Cooper pairs occupy states with a free electron energy above the Fermi level. In simple metallic superconductors, as the ones treated in this thesis, the attractive potential between electrons arises from the interaction with the crystal lattice. The qualitative idea is that an electron moving through the positively charged ion lattice distorts the latter and leaves a higher ion density behind, which in turn attracts another electron. The characteristic length

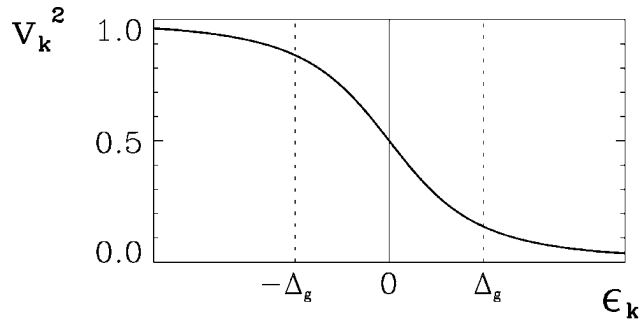

**Figure 2.1:** Occupational probability $v_{\mathbf{k}}^2$ of a Cooper pair state $\left(\mathbf{k}\uparrow,-\mathbf{k}\downarrow\right)$ as a function of free electron energy relative to the Fermi level $\varepsilon_{\mathbf{k}}$ for a BCS superconductor at zero temperature.

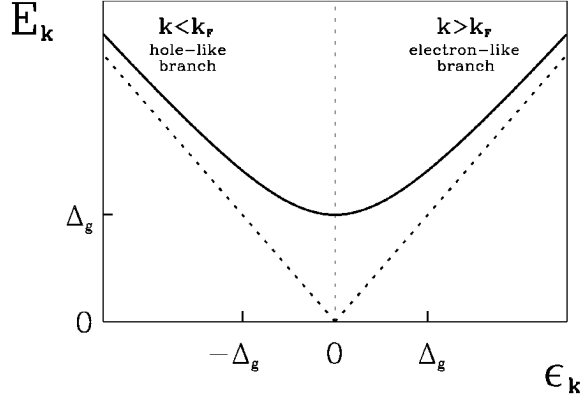

**Figure 2.2:** Quasiparticle excitation energy as a function of the free electron energy relative to the Fermi energy. Superconducting state (solid line) and normal state (dotted line).

scale of this interaction within a Cooper pair is given by the coherence length $\xi_0$:

$$\xi_0 = \frac{\hbar v_F}{\pi \Delta_g} , \qquad (2.3)$$

where $v_F$ is the Fermi velocity. This coherence length is typically much larger than the interatomic distance of the lattice, showing the strong overlap of the Cooper pairs in the superconductor.

In 1958 Bogoliubov and Valatin [Bogoliubov 58, Valatin 58] calculated the excitations of the superconducting ground state, called Bogoliubov quasiparticles (for simplicity called quasiparticles in the following). The energy of a such a quasiparticle excitation with momentum $\hbar \mathbf{k}$ is given by $E_k$, as defined in (2.1). Figure 2.2 shows the excitation energy of a quasiparticle as a function of the free electron energy relative to the Fermi energy. The figure shows that even at the Fermi surface ($\mathbf{k}=\mathbf{k_F}$) the quasiparticle excitations possess a minimum energy of $\Delta_g$, justifying the nomenclature energy gap. In addition, for every quasiparticle excitation energy above the gap energy two quasiparticle states are possible, one with $|\mathbf{k}| < |\mathbf{k_F}|$, called hole-like quasiparticle, and one with $|\mathbf{k}| > |\mathbf{k_F}|$, called electron-like quasiparticle. In fact, every excitation of the superconducting ground state with energy $E_\mathbf{k}$ is a superposition of electron- and hole-like quasiparticles, where the fractional hole-like character is given by $v_\mathbf{k}^2$, as defined in (2.1) and the fractional electron-like character is $1 - v_\mathbf{k}^2$. In order to conserve the particle number the quasiparticles are always created or annihilated in pairs by respectively breaking a Cooper pair or recombining into a Cooper pair. The BCS density of states DoS of the quasiparticles is given by:

$$\frac{DoS(E_\mathbf{k})}{2N(0)} = \begin{cases} \dfrac{E_\mathbf{k}}{\sqrt{E_\mathbf{k}^2 - \Delta_g^2}} & \left( E_\mathbf{k} > \Delta_g \right) \\ 0 & \left( E_\mathbf{k} < \Delta_g \right) \end{cases} \qquad (2.4)$$

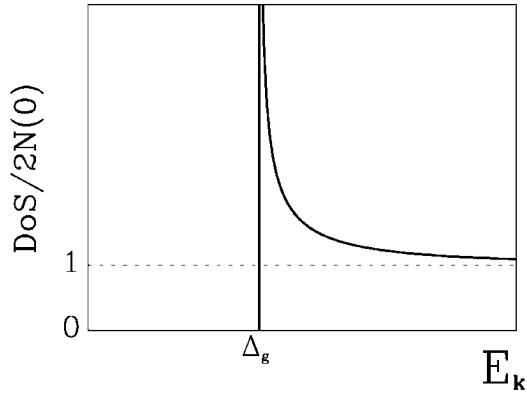

**Figure 2.3:** Quasiparticle density of states normalised to 2N(0) as a function of quasiparticle excitation energy $E_k$. Superconducting state (solid line) and normal state (dotted line).

Here $N(0)$ is the single spin electronic density of states at the Fermi energy in the normal material. Fig. 2.3 shows the density of states as a function of the quasiparticle energy. No states are available at energies below the gap energy $\Delta_g$. The dashed line represents the density of states in the normal metal. The integrated number of states in the normal metal and in the superconductor are equal.

## 2.2 The Proximity effect theory for superconducting bi-layers

Understanding the properties of the superconducting materials the detector is build from, is a basic need in order to be able to model the response of the detector. The two electrodes of the Josephson junction forming the detector consist generally of a superconducting bi-layer $S_1$-$S_2$. Here $S_1$ is the absorber material and $S_2$ is in most cases a thin Al film, which is partly oxidised in order to form the isolating layer between the electrodes. When these two superconducting materials $S_1$ and $S_2$ are put into contact, the properties of both materials are modified to within a distance of several times the coherence length $\xi$ of the bulk materials. This coherence length is generally of the order of ~100 nm for the materials used in this thesis and therefore comparable to the film thickness of the materials forming the detector. The properties of the so formed bi-layer differ considerably from the bulk properties of the materials forming the bi-layer and it does generally not show any BCS-like behaviour. Therefore, it is of utmost importance to determine the superconducting properties of the bi-layer, which will have a strong influence on all dynamic characteristics of the junction.

The physical quantities relevant for our detectors and which are affected in a proximity-coupled bi-layer are the Cooper pair potential $\Delta$, the density of states of the quasiparticles DoS and the critical temperature $T_C$. The determination of these quantities in a dirty superconductor, which have a short electronic mean free path compared to the coherence length, is possible within the framework of the Usadel equations [Usadel 70]. Usadel showed that, in the case of an almost isotropic motion of the electrons, the transport-like Eilenberger equations [Eilenberger 68] simplify into a diffusion-like equation from which all information about a dirty superconductor can be obtained.

### 2.2.1 Cooper pair potential and QP density of states

Let us consider a superconducting film $S_i$ of arbitrary thickness $d_{S_i}$ for which the dirty limit condition $l_{S_i} \leq \xi_{S_i}$ is fulfilled, where $l_{S_i}$ and $\xi_{S_i}$ are respectively the mean free path and the coherence length in the film. The x-axis is defined perpendicular to the film surface and the film is considered infinite in the y and z directions. In this case it was shown that the Usadel equations can be written in SI units as [Golubov 95]:

$$\frac{\hbar D_{S_i}}{2} \frac{\partial^2 \theta_{S_i}}{\partial x^2}(\varepsilon, x) + i\varepsilon \sin \theta_{S_i}(\varepsilon, x) + \Delta_{S_i}(x) \cos \theta_{S_i}(\varepsilon, x) = 0 , \qquad (2.5)$$

where the pair potential $\Delta_{S_i}(x)$ is determined by the self-consistency relation:

$$\Delta_{S_i}(x) \ln \frac{T}{T_{c,S_i}} + 2\pi kT \sum_{\omega_n} \left[ \frac{\Delta_{S_i}(x)}{\hbar \omega_n} - \sin \theta_{S_i}(i\hbar \omega_n, x) \right] = 0 . \qquad (2.6)$$

The function $\theta_{S_i}(\varepsilon, x)$ is a unique Green's function, which defines the QP density of states $DoS_{S_i}$ according to the relation:

$$\frac{DoS_{S_i}(\varepsilon, x)}{N_{S_i}(0)} = Re\left[ \cos \theta_{S_i}(\varepsilon, x) \right]. \qquad (2.7)$$

Here $\omega_n = (2n+1)\pi kT / \hbar$ is the Matsubara frequency, which is related to the QP energy $\varepsilon$ by the relation $\hbar \omega_n = -i\varepsilon$, $D_{S_i}$ is the normal state diffusion constant, T is the temperature and $N_{S_i}(0)$ is the electronic density of states in the normal state at the Fermi surface.

One can also define a function $ImF(\varepsilon,x)$, which is the imaginary part of the sine of $\theta_{S_i}(\varepsilon, x)$:

$$\frac{Im F_{S_i}(\varepsilon, x)}{N_{S_i}(0)} = Im\left[ \sin \theta_{S_i}(\varepsilon, x) \right]. \qquad (2.8)$$

In order to fix the ideas $S_1$ is now defined as being the higher $T_C$ film, whereas the film $S_2$ is the film with the lower $T_C$: $T_{C,S_1} > T_{C,S_2}$ .

For the bi-layer calculation it is convenient to normalise all energies ($\Delta$, $\varepsilon$ and kT) to $\pi kT_{C,S_1}$ and the distances in the x direction to $\xi^*$ of the corresponding material, where $\xi^*$ is the normalised coherence length: $\xi^*_{S_i} = \xi_{S_i} \sqrt{T_{C,S_i} / T_{C,S_1}}$ ($\xi^*_{S_1} = \xi_{S_1}$). In this way one obtains a unique energy scale in units of $\pi kT_{C,S_1}$ and a distance scale, different in every material $S_i$, in units of $\xi^*_{S_i}$. This yields the dimensionless equations:

$$\frac{\partial^2 \theta_{S_i}}{\partial x^2}(\varepsilon, x) + i\varepsilon \sin \theta_{S_i}(\varepsilon, x) + \Delta_{S_i}(x) \cos \theta_{S_i}(\varepsilon, x) = 0 , \qquad (2.9)$$

and the corresponding dimensionless self-consistency relations:

$$\Delta_{S_i}(x)\ln\frac{T}{T_{C,S_i}} + 2\frac{T}{T_{C,S_i}}\sum_{\omega_n}\left[\frac{\Delta_{S_i}(x)}{\hbar\omega_n} - \sin\theta_{S_i}(i\hbar\omega_n, x)\right] = 0. \tag{2.10}$$

For the establishment of (2.9), the following relation between the coherence length and the diffusion constant was used:

$$\xi_{S_i}^2 = \frac{\hbar D_{S_i}}{2\pi k T_{C,S_i}}. \tag{2.11}$$

Equations (2.9) and (2.10) have to be solved in both superconducting films $S_1$ and $S_2$, with the use of the appropriate boundary conditions. The origin of the coordinate system is chosen at the $S_1$-$S_2$ interface. The region with x>0 refers to the $S_1$ layer while x<0 refers to the lower gap $S_2$ film. The film thicknesses are respectively $d_{S_1}$ and $d_{S_2}$. At the free interfaces of both $S_1$ and $S_2$ layers the boundary conditions are:

$$\theta'_{S_1}(x = d_{S_1}) = 0, \tag{2.12}$$

$$\theta'_{S_2}(x = -d_{S_2}) = 0. \tag{2.13}$$

At the $S_1$-$S_2$ interface (x=0) the boundary conditions are [Kupriyanov 77]:

$$\gamma_{BN}\xi_{S_2}^*\theta'_{S_2} = \sin(\theta_{S_1} - \theta_{S_2}), \tag{2.14}$$

$$\gamma\xi_{S_2}^*\theta'_{S_2} = \xi_{S_1}^*\theta'_{S_2}, \tag{2.15}$$

where $\gamma$ and $\gamma_{BN}$ are the interface parameters describing the nature of the interface between the two materials. They are defined by:

$$\gamma = \frac{\rho_{S_1}\xi_{S_1}^*}{\rho_{S_2}\xi_{S_2}^*}, \tag{2.16}$$

$$\gamma_{BN} = \frac{R_B}{\rho_{S_2}\xi_{S_2}^*}. \tag{2.17}$$

Here $\rho_{S_1}$ and $\rho_{S_2}$ are the normal state resistivities and $R_B$ is the product of the resistance of the $S_1$-$S_2$ boundary and its interface area. $\gamma$ can be qualitatively understood as a measure of the strength of the proximity effect between the layers $S_1$ and $S_2$, whereas $\gamma_{BN}$ describes the effect of the boundary transparency between the layers.

Only in a small number of limiting cases, with limitations on the layer thickness and interface parameters $\gamma$ and $\gamma_{BN}$, can the Usadel equations be solved analytically [Gennes 64, Silwert 66, McMillan 68, Jin 89, Martinis 00, Fominov 01]. In the most general case with no limitations on film thickness and interface parameters one must solve Eqs. (2.9) and (2.10) numerically. First the differential equations (2.9) have to be solved using the bulk values of the order parameter $\Delta_0$. The solution $\theta_{S_i}(\epsilon, x)$ is then introduced into equations (2.10) in order to calculate the next iteration for the order parameter $\Delta(x)$. This procedure is repeated until convergence is achieved.

### 2.2.2 Critical temperature

In the vicinity of the critical temperature $T_C$ the Usadel equations can be linearised. In this case the unique Green's function $\theta(\varepsilon,x)$ is small and the dimensionless Usadel equations (2.9) and (2.10) take the linearised form:

$$\frac{\partial^2 \theta_{S_i}}{\partial x^2}(\varepsilon, x) + i\varepsilon \theta_{S_i}(\varepsilon, x) + \Delta_{S_i}(x) = 0 \text{ , and} \tag{2.18}$$

$$\Delta_{S_i}(x) \ln \frac{T}{T_{c,S_i}} + 2 \frac{T}{T_{C,S_i}} \sum_{\omega_n} \left[ \frac{\Delta_{S_i}(x)}{\hbar \omega_n} - \theta_{S_i}(i\hbar\omega_n, x) \right] = 0 \text{ .} \tag{2.19}$$

The boundary conditions (2.12) – (2.15) remain the same.
The critical temperature $T_C$ of the bi-layer is defined as the maximum temperature for which a non-trivial solution for the order parameter $\Delta_{S_i}(x)$ exists. Again, analytical solutions are only possible under certain limitations on layer thicknesses and interface parameters. In the general case the Usadel equations have to be solved numerically. First, (2.18) and (2.19) are solved for a temperature exactly in the middle of the temperature interval $(0, T_{C,S_i})$: $T_{C,S_i}/2$. If the order parameter $\Delta_{S_i}(x)$ converges to a non-trivial solution, the critical temperature $T_C$ of the bi-layer lies in the temperature interval $(T_{C,S_i}/2, T_{C,S_i})$. If the order parameter converges to the trivial solution $\Delta_{S_i}(x) = 0$, the critical temperature of the bi-layer lies in the interval $(0, T_{C,S_i}/2)$. The calculation then has to be repeated with the temperature in the middle of the corresponding interval. After several iterations the critical temperature $T_C$ of the bi-layer is known to within a good accuracy.

### 2.2.3 Interface parameters

The crucial point, when calculating the solution of the Usadel equations, is the correct determination of the interface parameters $\gamma$ and $\gamma_{BN}$. Recently, several methods have been used in order to determine the interface parameters for a certain bi-layer. Golubov et al. and Poelaert et al. [Golubov 94, Poelaert 99, Brammertz 01c] measure experimentally the energy gap and the critical current of a Josephson junction formed out of two bi-layers as a function of temperature. This data is then compared to simulations with different combinations of interface parameters in order to yield the requested pair of interface parameters for the bi-layer. Zehnder et al. [Zehnder 99] calculate $\gamma$, using Eq. (2.16), from resistivity and coherence length data obtained experimentally. The parameter $\gamma_{BN}$ is obtained from comparison of experimental IV-curves with simulated curves.

### *2.2.3.1 Interface parameter determination*

Here a method is applied, which consists of the comparison of the experimental critical temperature and low-temperature energy gap of the bi-layer with simulated values [Brammertz 01a and 02b]. First, the $T_C$ and $\Delta_g$ of the bi-layer have to be determined experimentally. The critical temperature is easily determined from a simple resistor made out of the corresponding bi-layer. In order to determine the energy gap an SIS, SNS or SIN junction has to be fabricated with the corresponding bi-layer as the electrode(s). Then simulations with different combinations of interface parameters $\gamma$ and $\gamma_{BN}$ have to be made. The requested pair of interface parameters $\gamma$ and $\gamma_{BN}$ is the combination for which both the $T_C$ and the $\Delta_g$ agree with the experimental values. Usually, the determination can be done to within a good accuracy, as the region in $(\gamma, \gamma_{BN})$-space for which the $T_C$ agrees with the experimental value is almost orthogonal to the region for which the $\Delta_g$ agrees with the experimental energy gap.

### *2.2.3.2 Film thickness dependence*

From the definition of the interface parameters (2.16) and (2.17) one can isolate the parts, which are independent of the thickness $d_{S_1}$ and $d_{S_2}$ of the two films. Replacing the coherence length in the films by the dirty limit expression:

$$\xi = \sqrt{\xi_0 l/3} \ , \qquad (2.20)$$

where $\xi_0$ is the coherence length in the bulk material and $l$ is the mean free path in the film, yields:

$$\gamma = C_\gamma \sqrt{\frac{l_{S_2}}{l_{S_1}}} \ , \text{ with } C_\gamma = \frac{\rho_{S_1} l_{S_1}}{\rho_{S_2} l_{S_2}} \sqrt{\frac{\xi_{0,S_1} T_{C,S_1}}{\xi_{0,S_2} T_{C,S_2}}} \ , \qquad (2.21)$$

$$\gamma_{BN} = C_{\gamma_{BN}} \sqrt{l_{S_2}} \ , \text{ with } C_{\gamma_{BN}} = \frac{R_B}{\rho_{S_2} l_{S_2}} \sqrt{\frac{3 T_{C,S_1}}{\xi_{0,S_2} T_{C,S_2}}} \ . \qquad (2.22)$$

Here, the quantities $C_\gamma$ and $C_{\gamma_{BN}}$ are independent of the thickness of the $S_1$ and $S_2$ films, because $\rho l$ is a material constant. The critical temperature is considered to be independent of the film thickness. This is a good approximation, as only for very thin films a minor thickness dependence can be seen [Cooper 62]. The constant $C_\gamma$ depends only on the nature of the two materials involved, whereas $C_{\gamma_{BN}}$ also depends on the quality of the interface between the two films. For bi-layers deposited under similar conditions and having only different film thickness, the same values of $C_\gamma$ and $C_{\gamma_{BN}}$ can be assumed.

The dependence of the interface parameters on the film thickness can then be determined by substituting the film thickness dependence of the mean free path into (2.21) and (2.22). For complete electron scattering at the film surfaces, the following equation for the mean free path $l$ as a function of film thickness $d$ holds [Movshovitz 90]:

$$l(d) = l_0 + l_0^2 / d \left\{ \frac{3}{8} \left[ E_3(d/l_0) - E_5(d/l_0) \right] - \frac{3}{8} \right\}, \qquad (2.23)$$

where the exponential integrals are defined by $E_n(x) = \int_1^\infty t^{-n}e^{-xt}dt$ and $l_0$ is the mean free path in the bulk material.

### 2.2.3.3 Theoretical determination of $C_\gamma$ and the interface transmissivity $T^*$

Replacing the resistivity by:

$$\rho^{-1} = e^2 DN(0),$$ (2.24)

and the normal state diffusion constant by:

$$D = \frac{1}{3}v_F l,$$ (2.25)

using (2.3), one can rewrite the interface constant $C_\gamma$ (2.21) as:

$$C_\gamma = \sqrt{\frac{v_{F,S_2}}{v_{F,S_1}}} \frac{N_{S_2}(0)}{N_{S_1}(0)}.$$ (2.26)

Replacing the corresponding quantities in equation (2.21) by their values in the respective materials, yields a theoretical estimate of the interface constant $C_\gamma$. Here, the Fermi velocity $v_F$ can be easily obtained from the BCS relation of the coherence length [Tinkham 96]:

$$v_F = \frac{\xi_0 \pi \Delta_g(0)}{\hbar}.$$ (2.27)

On the other hand the parameter $\gamma_{BN}$ can be re-written as [Kupriyanov 88]:

$$\gamma_{BN} = \frac{2l_{S_2}}{3\xi_{S_2}^*} \frac{1 - T^*}{\left(m_{S_1}^* v_{F,S_1} / m_{S_2}^* v_{F,S_2}\right)^2 T^*},$$ (2.28)

where $T^*$ is the charge carrier transmission coefficient, which gives the transmission probability of a quantum-mechanical particle through an interface of two metals with different Fermi velocities. Using Eqs. (2.15) and (2.17) and solving (2.28) with respect to $T^*$, one can write:

$$\frac{1}{T^*} = \sqrt{\frac{T_{C,S_2}}{T_{C,S_1}}} \left(\frac{m_{S_1}^* v_{F,S_1}}{m_{S_2}^* v_{F,S_2}}\right)^2 \frac{\sqrt{3\xi_{0,S_2}} C_{\gamma_{BN}}}{2} + 1,$$ (2.29)

independent of the film thickness.

On the other hand $T^*$ is defined in the free-electron model as:

$$T^* = \frac{4 v_{F,S_1} v_{F,S_2}}{\left( v_{F,S_1} + v_{F,S_2} \right)^2 + 4 U_0^2},$$  (2.30)

where $U_0$ is the height of a $\delta$-potential barrier at the $S_1$-$S_2$ interface.

### 2.2.4  Application to Ta-Al and Nb-Al bi-layers

In order to illustrate the proximity effect theory it will now be applied to two series of high quality Ta-Al and Nb-Al bi-layers. These two series consist of a 100 nm thick Ta or respectively Nb layer grown on a sapphire substrate, which is covered with an Al film, whose thickness varies between 5 and 265 nm. Table 2.1 gives the values of all material parameters needed for the proximity effect calculations. The $T_C$, $\Delta_g(0)$ and $l_0$ were obtained from thin film measurements in the laboratories of the European Space Agency. The Fermi velocities were obtained by applying Eq. (2.3) and N(0) was taken from [Gladstone 69].

**Table 2.1:** Nb, Ta and Al film parameters.

|  | $T_C$ (K) | $\Delta_g(0)$ ($\mu$eV) | $l_0$ (nm) | $\xi_0$ (nm) | $v_F$ ($10^6$m/sec) | N(0) ($10^{21}$/(eV cm$^3$)) |
|---|---|---|---|---|---|---|
| Nb | 9.3 | 1550 | 125 | 38 | 0.280 | 31.7 |
| Ta | 4.5 | 700 | 90 | 90 | 0.3 | 40.8 |
| Al | 1.2 | 180 | 52 | 1600 | 1.37 | 12.2 |

The first step is the experimental determination of the interface constants $C_\gamma$ and $C_{\gamma_{BN}}$. For this purpose the $T_C$ and energy gap of a bi-layer with an Al thickness of 30 nm was measured experimentally. Then a series of calculations using different values for the two interface parameters was performed. In the case of the Nb-Al bi-layer, 100 calculations were made with 10 values for $\gamma$ ranging from 0.5 to 1.0 and 10 values for $\gamma_{BN}$ ranging from 2.5 to 3.5. For the Ta-Al bi-layer, 10 values for $\gamma$ ranging from 0.2 to 0.4 and 10 values for $\gamma_{BN}$ ranging from 3.0 to 5.0 were chosen. For every $\gamma$-$\gamma_{BN}$ combination the critical temperature and the energy gap at 300 mK were calculated. Then an interpolation of the calculated points is made in order to yield continuous surface plots of $T_C$ and $\Delta_g$ as a function of $\gamma$ and $\gamma_{BN}$. The intersections of these surfaces with the planes corresponding to the experimental values of $T_C$ and $\Delta_g$ yield two lines in the $\gamma$-$\gamma_{BN}$ plane, which intersect at the $\gamma$, $\gamma_{BN}$ combination corresponding to the deposited bi-layer. Fig. 2.4 shows the intersections of the calculated surfaces of $T_C$ and $\Delta_g$ with the planes corresponding to the experimental values.

The two lines in ($\gamma$, $\gamma_{BN}$)-space, corresponding to an agreement of calculations with experiment of respectively the $T_C$ and the $\Delta_g$, are near to being orthogonal, therefore making an accurate determination of the interface parameters possible. The pair of interface parameters corresponding to the bi-layer with 30 nm of Al are respectively $\gamma = 0.324$ and $\gamma_{BN} = 3.631$ for Ta-Al and $\gamma = 0.829$ and $\gamma_{BN} = 3.454$ for Nb-Al. From these values one can determine the interface constants by applying (2.21) and (2.22), using the values from table 2.1 and (2.23). $C_\gamma = 0.481$ and $C_{\gamma_{BN}} = 0.675$ nm$^{-\frac{1}{2}}$ is obtained for the Ta-Al bi-layer and $C_\gamma = 1.372$ and $C_{\gamma_{BN}} = 0.642$ nm$^{-\frac{1}{2}}$ is obtained for the Nb-Al bi-layer.

This experimentally derived value of $C_\gamma$ can be compared with the alternative definitions of section 2.2.3.3. Applying (2.26) with the values from table 2.1 yields $C_\gamma = 0.639$ for Ta-

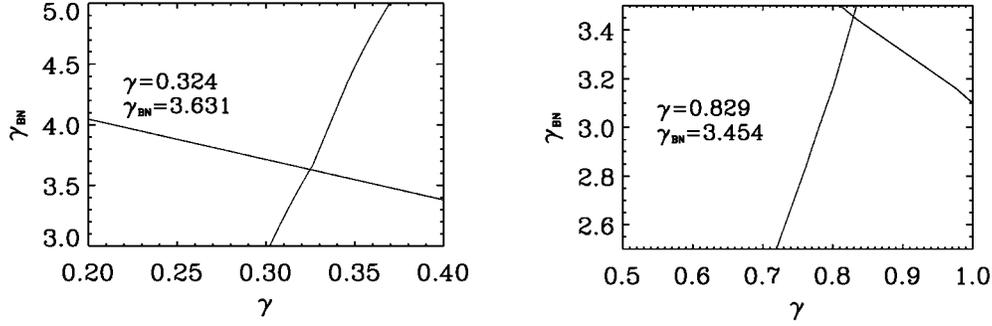

**Figure 2.4**: Intersections of the calculated surfaces of $T_C$ and $\Delta_g$ as a function of interface parameters with the planes corresponding to the experimental values of the Ta-Al bi-layer **(left)** and the Nb-Al bi-layer **(right)** with 30 nm of Al. The two lines intersect in the point corresponding to the $\gamma, \gamma_{BN}$ combination of the respective bi-layer.

Al bi-layers and $C_\gamma = 0.851$ for Nb-Al bi-layers. Even though the values do not agree completely with the experimentally derived values, they are still reasonably close. One can also determine the interface transmission using (2.29) and compare it to the definition given by (2.30) in order to determine the value of the $\delta$-potential at the $S_1$-$S_2$ interface. The transmission probability determined from the experimental value of $C_{\gamma_{BN}}$ and (2.29) is $T^* = 0.633$ for the Ta-Al bi-layers and $T^* = 0.75$ for the Nb-Al bi-layers. For this calculation the ratio of effective masses $m^*_{S_1} / m^*_{S_2}$ was considered to be equal to 1. The theoretical transmission probability calculated from the Fermi velocity mismatch using (2.30) without taking the interface potential into account is $T^* = 0.589$ for the Ta-Al bi-layer and $T^* = 0.563$ for the Nb-Al bi-layer. In both cases the experimentally derived transmission is higher than the theoretical value without taking any interface potential barrier into account. It can therefore be concluded that in our case the influence of the $\delta$-potential at the $S_1$-$S_2$ interface is of minor importance and that the transmission probability can be deduced from the simple Fermi velocity mismatch. It is of the order of 0.6 for both material combinations. All the values derived in the preceding paragraph are summarised in table 2.2.

**Table 2.2:** Comparison of experimentally and theoretically derived values of the interface constants $C_\gamma$ and $C_{\gamma_{BN}}$ and the interface transmission $T^*$.

|  | $C_\gamma$ | | $C_{\gamma_{BN}}$ (nm$^{-\frac{1}{2}}$) | $T^*$ | |
|---|---|---|---|---|---|
|  | Theory | Experiment | Experiment | Theory | Experiment |
| Ta-Al | 0.639 | 0.481 | 0.675 | 0.589 | 0.633 |
| Nb-Al | 0.851 | 1.372 | 0.642 | 0.563 | 0.75 |

$C_\gamma$ and $C_{\gamma_{BN}}$ can be considered independent of the film thickness of the Nb, Ta and Al films. Therefore it is possible to predict the $T_C$ and $\Delta_g$ of any bi-layer deposited under the same conditions as the bi-layer for which the determination of the interface constants was performed, independent of the thickness of the different films. The corresponding interface and input parameters for the calculations are found by using (2.20) to (2.23). Fig. 2.5 shows the variation of the interface parameters for the same bi-layers. Fig. 2.6 shows how the experimental and calculated $T_C$ and $\Delta_g$ vary as the thickness of the Al film is varied. The agreement between experiment and calculation is good over the whole range of Al thickness. The interface parameters vary according to (2.21) – (2.23). The variation is not a simple square root dependence, but a somewhat more complicated relationship because of the variation of the mean free path with Al thickness. Note that the values for $\gamma$ are in very good agreement with the values found by Zehnder et al. [Zehnder 99] for Nb-Al bi-layers over the whole Al thickness range. On the other hand the values found for $\gamma_{BN}$ differ considerably from the results by Zehnder et al. Whereas Zehnder et al. find a linear dependence varying from zero to a value of $\gamma_{BN} = 10$ for an Al thickness of 100 nm, a non-linear relationship with a value of $\gamma_{BN}$ equal to 4 for an Al thickness of 100 nm was found in this work.

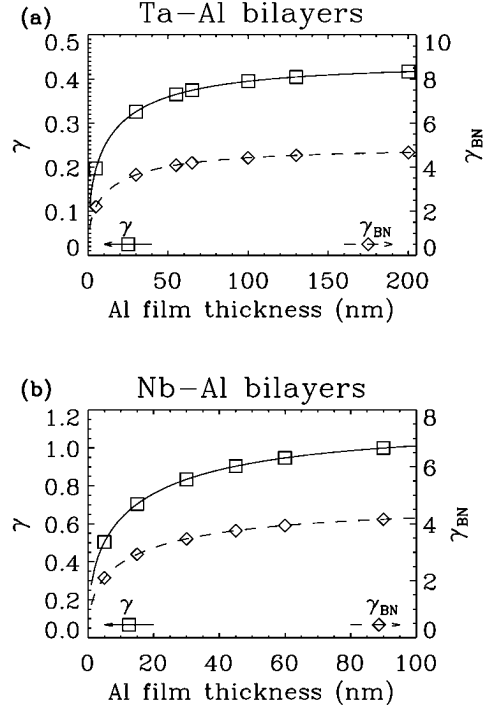

**Figure 2.5:** Interface parameters $\gamma$ (solid line, left scale) and $\gamma_{BN}$ (dashed line, right scale) as a function of Al film thickness for **(a)** Ta-Al and **(b)** Nb-Al bi-layers. Squares ($\gamma$, left scale) and diamonds ($\gamma_{BN}$, right scale) indicate the points for which calculations of the energy gap and $T_C$ were made.

The calculated variation of the order parameter $\Delta(x)$ with position in the bi-layer is shown in Fig. 2.7 (a) for a Ta-Al bi-layer with 100 nm of Ta and 55 nm of Al (left) and a Nb-Al bi-layer with 100 nm of Nb and 120 nm of Al (right). Also the variation of the density of states $DoS(\varepsilon,x)$ (b) and the function $ImF(\varepsilon,x)$ defined by (2.8) (c) are shown as a function of energy for the same Ta-Al (left) and Nb-Al (right) bi-layer systems. The variations of the density of states and ImF are shown at four different positions in the bi-layer: at the free interface and at the material interface in both superconductors.

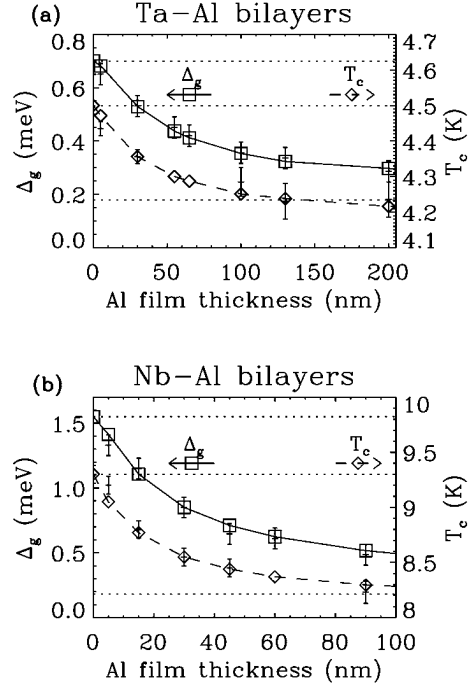

**Figure 2.6:** Energy gap at 300 mK and $T_C$ as a function of Al film thickness for **(a)** Ta/Al and **(b)** Nb-Al bi-layers. The Ta and Nb film thickness has a constant value of 100 nm. Squares ($\Delta_g$, left scale) and diamonds ($T_C$, right scale) represent the calculated values from the model. The solid ($\Delta_g$, left scale) and dashed ($T_C$, right scale) lines are a guide to the eye between the calculated points. Crosses with error bars represent the corresponding experimental values. The dotted lines represent the bulk energy gap of Nb, Ta, Al and the bulk $T_C$ of Nb and Ta.

The behaviour of the two material combinations is similar. The pair potential is discontinuous at the $S_1$-$S_2$ interface and varies between the two bulk values. The discontinuity is a measure of the interface resistance represented by $\gamma_{BN}$. The energy gap is constant throughout the bi-layer with a value in between the two bulk values. The reason for this is the non-local nature of the Cooper pairs and the fact that the film thicknesses are not large compared to the coherence lengths in the films. In the high gap material $S_1$ the density of states peaks at a value equal to the bulk energy gap $\Delta_{g,S_1}(0)$. Below this energy the number of states is reduced, but nevertheless enough states are still present at these energies. In the Al the density of states peaks at an energy slightly above the gap in the bi-layer. At an energy $\Delta_{g,S_1}(0)$ a small second peak can be observed induced by the proximity of the high gap material. At higher energies the density of states converges to the normal state value. At the interface between the two materials the density of states is discontinuous. This discontinuity induces Andreev reflections at the interface, thereby creating some extent of QP trapping in the low gap material. The function ImF has a similar behaviour as the density of states, except that it converges towards zero for large energies.

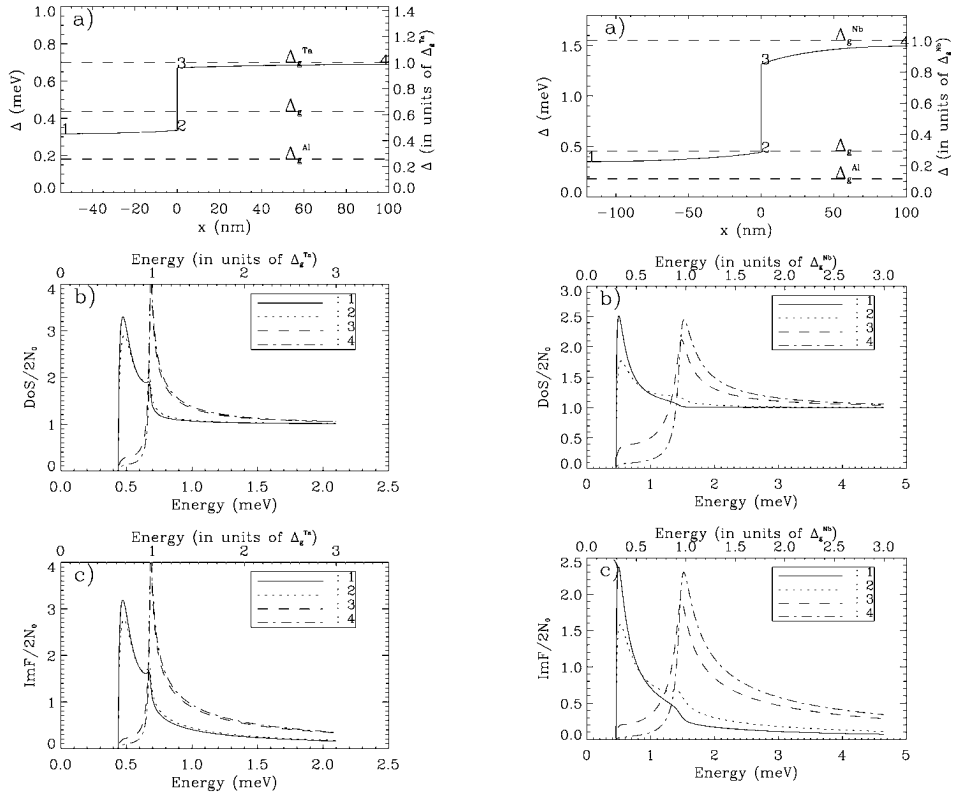

**Figure 2.7:** Pair potential as a function of position in the bi-layer **(a)**. Density of states **(b)** and ImF **(c)** as a function of quasiparticle energy for the 100 nm - 55 nm Ta-Al **(left)** and 100 nm – 120 nm Nb-Al **(right)** bi-layers. **(a)** The upper dashed line is the bulk energy gap of the high gap material. The lower dashed line is the bulk energy gap of Al. The intermediate dashed line is the energy gap of the bi-layer, as determined from (b). The points 1, 2, 3 and 4 correspond to the four positions in the bi-layer for which the density of states is given in (b). **(b)** and **(c)** The densities of states and ImF are represented for both materials at the free interface and at the $S_1$-$S_2$ interface. The points 1 to 4 in (a) indicate the positions in the bi-layer for which the energy variations of the density of states and ImF are given.

## 2.3  Kinetics of the quasiparticle energy distribution in tunnel junctions used as photon detectors

In the following a major advance in the treatment of quasiparticle dynamics is described, which is essential for modelling the latest generation of low gap, multi-tunnelling STJs designed to operate at mK temperatures.

Previously the response of a biased STJ to the absorption of a photon, creating non-equilibrium quasiparticles, has commonly been modelled within the framework of the Rothwarf-Taylor balance equations [Rothwarf 67]. The main assumption of this model is that during the initial down-conversion process quasiparticles relax very rapidly to the superconducting edge.  Further stages of charge transfer, loss and recombination are evaluated under the assumption that all active quasiparticles reside at the superconducting edge and hence that all have this same energy.  However, even in experiments involving large gap STJs based on Nb or Ta, evidence was found that the mean energy of the quasiparticles lies above the superconducting edge, and that the energy distribution remains relatively broad during the whole current integration time [Poelaert 99, Poelaert 98]. Poelaert and co-workers developed a model that was still based on the consideration that all the quasiparticles occupy the same energy level. As opposed to the classical Rothwarf-Taylor approach, this energy level was in their case not necessarily the gap energy anymore, but an energy level above the gap energy, which they call "balance energy". The different rates of the processes in the superconductors forming the electrodes of the junctions were then calculated for quasiparticles being at the balance energy, the balance energy itself being a fitting parameter of the model. For the lower gap Al- or Mo-based multi-tunnelling STJs developed in the framework of this thesis, the relaxation times of excited quasiparticles are greatly increased to the point where it is impossible to describe the experimental results with an over-simplified mono-energetic model. In order to describe the phenomena appearing in these low-gap superconductors the complete energy dependence of the quasiparticles will have to be taken into account.

In this section the first description of a STJ photon detection model that includes the full energy dependence of tunnelling, relaxation and loss processes is given. The model is presented for the most general type of STJ, one in which the two electrodes are not BCS-type superconductors but are proximised. Such electrodes have properties intermediate between the properties of the two superconductors forming the electrodes, and cannot be accurately described by the simpler BCS relationships. However the BCS forms can easily be retrieved from the expressions given. The kinetic equations for the quasiparticle numbers as a function of energy in both electrodes are derived, which can be solved in order to obtain the complete quasiparticle energy distributions as a function of time. In order to illustrate the model, it will be applied to two experimentally extensively tested junctions, one based on Nb-Al electrodes with 100 nm of Nb and 120 nm of Al and the other one based on Ta-Al electrodes with 100nm of Ta and 55 nm of Al. These junctions were already extensively studied by Poelaert et al. [Poelaert 99]. The application of the model to these junctions allows for the comparison with the previous model based on a mono-energetic approach.

### 2.3.1  Characteristic rates

In order to determine the quasiparticle energy distribution the energy, position and time dependent kinetic equations for the quasiparticle numbers in both electrodes has to be solved [Chang 86]. We are interested in modelling the evolution of the quasiparticle distribution starting from the moment when the generated quasiparticles fill homogeneously the whole volume of the electrode and hence the lateral gradients in the kinetic equations are neglected. Nevertheless, the position dependence comes from the fact that the detector is not homogeneous in the direction perpendicular to the barrier. The time it takes for a quasiparticle to traverse the tickness of the electrode (~1psec) is much faster than any of the quasiparticle processes occurring in the junction. For this reason the kinetic equations can be averaged over the position perpendicular to the barrier. This removes the position dependence in the final expression of the energy dependent kinetic equations.

Many processes occur in the electrodes of STJ detectors, for which characteristic rates must be calculated. Figure 2.8 shows a semiconductor representation of all processes included in the model.

In order to be able to calculate the characteristic rates of all the processes shown in Fig. 2.8, one needs to know several basic characteristics of the bi-layer forming the electrodes of the detector. These characteristics are: the normalized energy and position dependent density of states in the electrode i (i = 1,2) $DoS_i(x,\varepsilon)$, the function $ImF_i(x,\varepsilon)$, and the position dependent order parameter $\Delta_i(x)$. All three quantities can be calculated with the proximity effect theory elaborated in section 2.2 (see also Fig. 2.7). Here, x is the direction perpendicular to the barrier and $\varepsilon$ is the quasiparticle energy. At low enough temperatures (typically $T<T_C/10$) all three quantities are independent of the temperature of the superconductor.

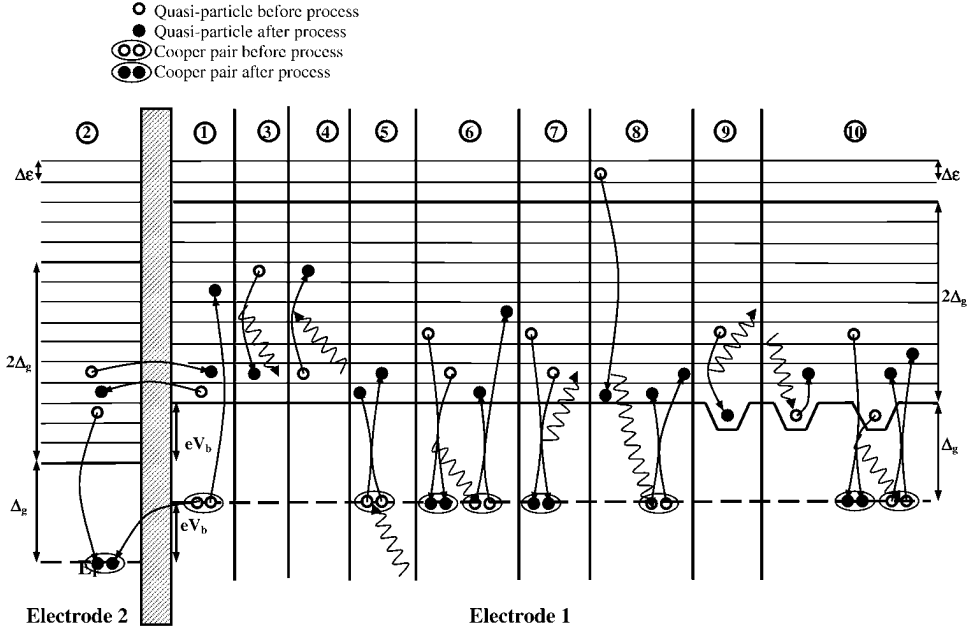

**Figure 2.8:** Schematical semiconductor representation of all processes included in our kinetic equations model. **1:** Tunnelling and back-tunnelling **2:** Cancellation tunnelling. The cancellation back-tunnelling is not shown for simplicity. It can be found by reversing the arrows on the back-tunnelling schematic of 1. **3:** Electron-phonon scattering with emission of a phonon (quasiparticle relaxation). **4:** Electron-phonon scattering with absorption of a phonon (quasiparticle excitation). **5:** Cooper pair breaking. **6:** Quasiparticle recombination with energy exchange. **7:** Quasiparticle recombination with phonon loss. **8:** Quasiparticle multiplication. **9:** Quasiparticle trapping by relaxation. **10:** Quasiparticle de-trapping by phonon absorption and by recombination with an untrapped quasiparticle.

Of course, the model is also valid for homogeneous junctions, for which the electrodes obey BCS relationships. In this case the results of the proximity effect theory have to be replaced by their BCS counterparts in all the equations:

$$DoS(x,\varepsilon) \rightarrow \frac{\varepsilon}{\sqrt{\varepsilon^2 - \Delta_g^2}} \qquad (2.31)$$

$$ImF(x,\varepsilon) \rightarrow \frac{\Delta_g}{\sqrt{\Delta_g^2 - \varepsilon^2}} \qquad (2.32)$$

$$\Delta(x) \rightarrow \Delta_g \qquad (2.33)$$

In the BCS case the position dependence disappears.

Using the results of the proximity effect theory one can then calculate the energy dependent characteristic rates of all processes occurring in biased STJs (Fig. 2.8). Since the kinetic equation for quasiparticles in superconductors is non-linear due to the presence of recombination collision integrals, it cannot in general be analysed analytically for the case of significant deviations from the equilibrium state. Therefore, a numerical approach is used. For this purpose the energy domain is divided into $N_{en}$ energy intervals of an arbitrary width $\delta\varepsilon$, which can be made as small as one prefers, with the cost of increased calculation time. A range of the energy domain from $\Delta_g$ to $4\Delta_g$ typically divided into 30 intervals was chosen, a choice which is usually a good compromise between acceptable calculation time and sufficient accuracy.

In the following the characteristic rates of the different processes will be calculated. All rates will be given for electrode 1, but are of course also valid for electrode 2 by interchanging the indices 1 and 2.

**Table 2.3**: Parameters used for the characteristic times calculations.

| Symbol | Name | Unit | Ta | Al | Nb |
|--------|------|------|-----|-----|-----|
| $R_nA$ | Normal resistivity of junction | $\mu\Omega\ cm^2$ | | $2.2 \pm 0.2$ | |
| | | | | | $2.35 \pm 0.2$ |
| $T_C$ | Critical temperature | K | 4.5 | 1.2 | 9.4 |
| $\Delta_g$ | Energy gap | $\mu eV$ | 700 | 180 | 1550 |
| $N_0$ | Single spin normal state density of states at Fermi energy | $10^{27}$ states $eV^{-1}\ m^{-3}$ | 40.8 | 12.2 | 31.7 |
| $\alpha^2$ | Average square of the electron-phonon interaction matrix element | meV | 1.38 | 1.92 | 4.6 |
| N | Ion number density | $10^{28}\ m^{-3}$ | 5.57 | 6.032 | 5.57 |
| $\tau_0$ | Electron-phonon interaction characteristic time | nsec | 1.78 | 440 | 0.149 |
| T | Temperature | K | | 0.3 | |

### 2.3.1.1   Forward tunnelling

The forward (i. e. with energy gain), tunnelling rate from an energy $\varepsilon_\alpha$ in electrode 1 to an energy $\varepsilon_\alpha+eV_b$ in electrode 2 is given by [Golubov 94, van den Berg 99, de Korte 92] :

$$\Gamma_{tun}\left(\varepsilon_\alpha \rightarrow \varepsilon_\alpha + eV_b\right) = \frac{1}{4eR_nA} \frac{DoS_1\left(0,\varepsilon_\alpha\right)DoS_2\left(0,\varepsilon_\alpha + eV_b\right)}{\int\limits_{electr1} N_0(x)DoS_1\left(x,\varepsilon_\alpha\right)dx}, \qquad (2.34)$$

where $N_0$ is the single spin density of states at the Fermi energy in the normal state (Its dependence on x is to indicate that the material is not the same throughout the electrode), $R_n$ is the normal resistance of the junction, A is the area of the junction and $V_b$ is the positive potential difference between electrodes 1 and 2. Values of all material parameters appearing in (2.34)-(2.51) are summarized in table 2.3 for Al, Ta and Nb.

The notation $(\varepsilon_\alpha\rightarrow\varepsilon_\beta)$ indicates that during the process the quasiparticle changes its energy from $\varepsilon_\alpha$ at the beginning of the process to $\varepsilon_\beta$ at the end of the process. This notation will be the same throughout the chapter.

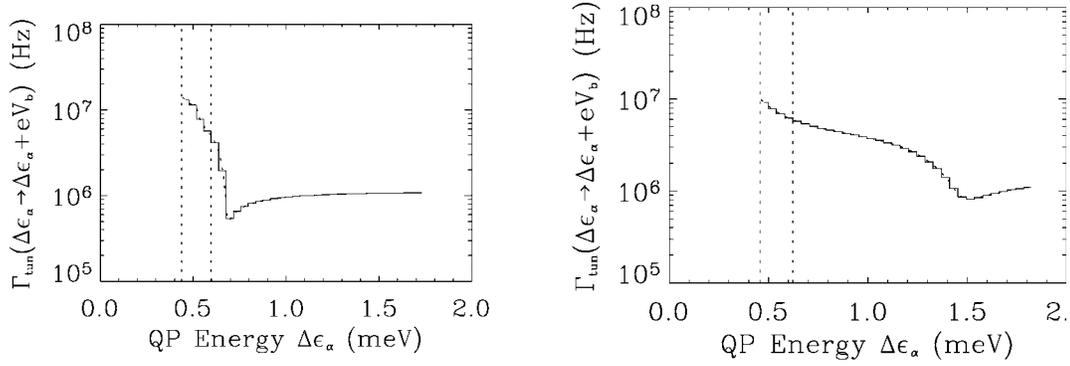

**Figure 2.9:** Direct tunnel rate as a function of quasiparticle energy for the 100 nm − 55 nm Ta-Al (left) and the 100 nm − 120 nm Nb-Al (right) junctions. The solid line gives the tunnel rate as a function of energy intervals $\Gamma_{tun}(\Delta\varepsilon_\alpha)$, whereas the dashed line shows the tunnel rate as a function of quasiparticle energy $\Gamma_{tun}(\varepsilon_\alpha)$. Note that the two lines essentially coincide. The first vertical dotted line indicates the energy gap, whereas the second vertical dotted line indicates the bias energy above the gap.

One can now determine the mean forward tunnelling rate in the energy interval $\Delta\varepsilon_\alpha$, where $\Delta\varepsilon_\alpha$ is the interval $[\varepsilon_\alpha\text{-}\delta\varepsilon/2,\varepsilon_\alpha\text{+}\delta\varepsilon/2]$, by integrating $\Gamma_{tun}(\varepsilon_\alpha\rightarrow\varepsilon_\alpha\text{+eV}_b)$ over $\varepsilon$ in $\Delta\varepsilon_\alpha$ and dividing by $\delta\varepsilon$:

$$\Gamma_{tun}\left(\Delta\varepsilon_\alpha \rightarrow \Delta\varepsilon_\alpha + eV_b\right) = \frac{\int\limits_{\Delta\varepsilon_\alpha}\Gamma_{tun}\left(\varepsilon \rightarrow \varepsilon + eV_b\right)d\varepsilon}{\delta\varepsilon} \qquad . \qquad (2.35)$$

Figure 2.9 shows the tunnel rate for the 100 nm − 55 nm Ta-Al (left) and 100 nm − 120 nm Nb-Al (right) junctions as a function of energy intervals for a bias voltage of 160 μV. The particularities of the density of states are reflected in the tunnel rates, with the minimum tunnel rate being at the respective bulk energy gap of the Ta or Nb. The tunnelling rate is independent of temperature, under the condition that the density of states is independent of temperature (typically $T<T_C/10$). One can see that a non-essential technical limitation of the model is that it only works for bias energies $eV_b$ which are an integer multiple of the energy interval $\Delta\varepsilon$, because otherwise the quasiparticles have to be distributed over two energy intervals, which creates numerical errors.

### 2.3.1.2   Cancellation tunnelling

When the quasiparticles have an energy $eV_b$ above the gap energy, they can tunnel against the direction indicated by the bias voltage. During these tunnel processes the quasiparticles will lose an energy $eV_b$ and will create a current in the direction opposite to the forward tunnel currents. For these cancellation tunnel events the following equation holds [Golubov 94, van den Berg 99, de Korte 92]:

$$\Gamma_{can}\left(\varepsilon_\alpha \rightarrow \varepsilon_\alpha - eV_b\right) = \frac{1}{4eR_nA}\frac{DoS_1\left(0,\varepsilon_\alpha\right)DoS_2\left(0,\varepsilon_\alpha - eV_b\right)}{\int\limits_{electr1}N_0(x)DoS_1\left(x,\varepsilon_\alpha\right)dx}, \qquad (2.36)$$

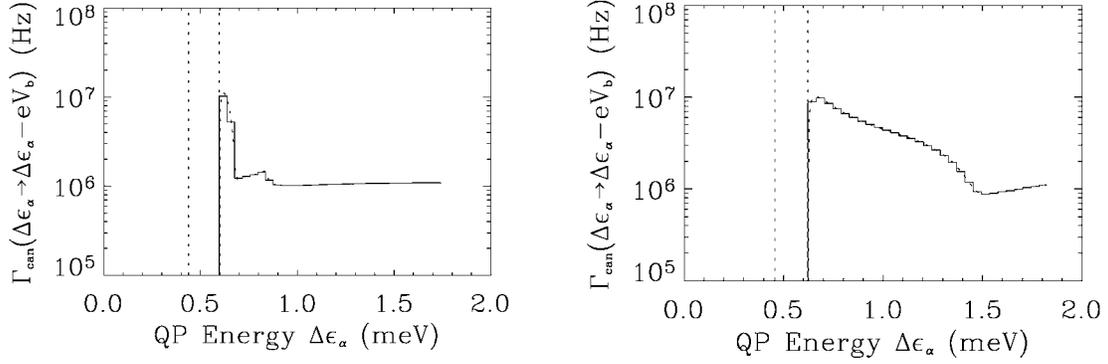

**Figure 2.10:** Cancellation tunnel rate as a function of quasiparticle energy for the 100 nm – 55 nm Ta-Al (left) and the 100 nm – 120 nm Nb-Al (right) junctions. The solid line gives the tunnel rate as a function of energy intervals $\Gamma_{can}(\Delta\varepsilon_\alpha)$, whereas the dashed line shows the tunnel rate as a function of quasiparticle energy $\Gamma_{can}(\varepsilon_\alpha)$. The first vertical dotted line indicates the energy gap, whereas the second vertical dotted line indicates the bias energy above the gap.

Note that the cancellation rate is zero for quasiparticle energies lower than $\Delta_g + eV_b$, simply because no states are available at the corresponding energies in electrode 2.

The cancellation rate in the energy interval $\Delta\varepsilon_\alpha$ can be written similarly to (5):

$$\Gamma_{can}\left(\Delta\varepsilon_\alpha \to \Delta\varepsilon_\alpha - eV_b\right) = \frac{\int\limits_{\Delta\varepsilon_\alpha}\Gamma_{can}\left(\varepsilon \to \varepsilon - eV_b\right)d\varepsilon}{\delta\varepsilon}. \tag{2.37}$$

Figure 2.10 shows the cancellation tunnel rate for the 100 nm – 55 nm Ta-Al (left) and 100 nm – 120 nm Nb-Al (right) junctions as a function of energy intervals for a bias voltage of 160 µV.

### 2.3.1.3 Rate for electron-phonon scattering with emission of a phonon (Relaxation)

A quasiparticle can scatter from energy $\varepsilon_\alpha$ to a lower energy $\varepsilon_\beta$ by emitting a phonon of energy $\varepsilon_\alpha$-$\varepsilon_\beta$. The mean rate for the relaxation of a quasiparticle of energy $\varepsilon_\alpha$ at the position x in the electrode to the energy interval $\Delta\varepsilon_\beta$, by emission of a phonon of energy $\varepsilon_\alpha$-$\varepsilon_\beta$ is given by [Golubov 94, Kaplan 76]:

$$\Gamma_{emi}\left(x,\varepsilon_\alpha \to \Delta\varepsilon_\beta\right) = \frac{1}{\tau_0(x)\left[kT_C(x)\right]^3}\int\limits_{\varepsilon_\alpha-\varepsilon_\beta-\delta\varepsilon/2}^{\varepsilon_\alpha-\varepsilon_\beta+\delta\varepsilon/2}\Omega^2\left[DoS_1\left(x,\varepsilon_\alpha-\Omega\right) - \frac{\Delta_1(x)}{\varepsilon_\alpha}ImF_1\left(x,\varepsilon_\alpha-\Omega\right)\right]\left[1+n(\Omega)\right]d\Omega \tag{2.38}$$

where $\tau_0$ is a material constant defined in Ref. [Kaplan 76]. $T_C$ is the bulk critical temperature of the material and $n(\Omega)$ is the phonon distribution function, which is in most cases much smaller than unity.

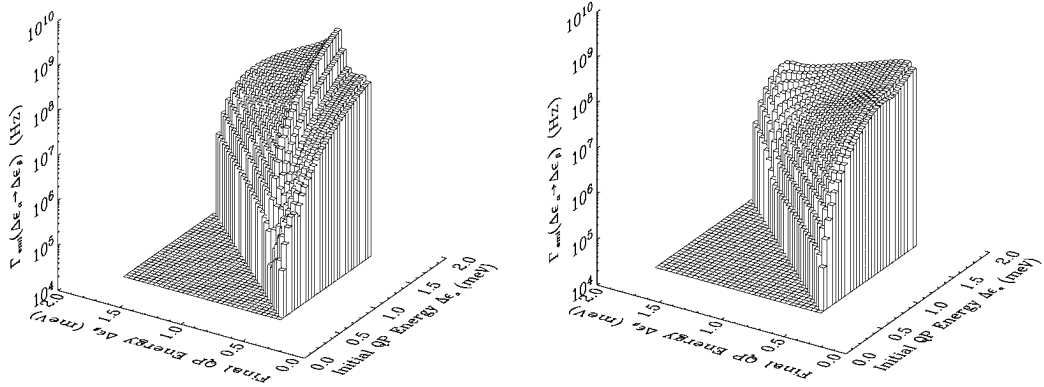

**Figure 2.11:** Rate for electron-phonon scattering with emission of a phonon, $\Gamma_{emi}(\Delta\varepsilon_\alpha \rightarrow \Delta\varepsilon_\beta)$, as a function of initial and final quasiparticle energy for the 100 nm − 55 nm Ta-Al (left) and the 100 nm − 120 nm Nb-Al (right) junctions. Note that above the diagonal all rates are zero, because the quasiparticle cannot go to a higher energy after phonon emission. The phonon emission rate is proportional to the emitted phonon energy cubed.

This phonon emission rate still depends on the position in the bi-layer x. As already stated previously, transport over the vertical direction is much faster than the typical time constant of phonon emission. Therefore, one can average the phonon emission rate over the vertical position in the bi-layer x:

$$\Gamma_{emi}\left(\varepsilon_\alpha \rightarrow \Delta\varepsilon_\beta\right) = \frac{\int\limits_{electr1} N_0(x)DoS_1(x,\varepsilon_\alpha)\Gamma_{emi}\left(x,\varepsilon_\alpha \rightarrow \Delta\varepsilon_\beta\right)dx}{\int\limits_{electr1} N_0(x)DoS_1(x,\varepsilon_\alpha)dx}. \qquad (2.39)$$

Finally, one can average the phonon emission rate from an energy $\varepsilon_\alpha$ into the energy interval $\Delta\varepsilon_\beta$ over the energy interval $[\varepsilon_\alpha\text{-}\delta\varepsilon/2,\ \varepsilon_\alpha\text{+}\delta\varepsilon/2]$, in the same way as was done for the tunnel and the cancellation rate:

$$\Gamma_{emi}\left(\Delta\varepsilon_\alpha \rightarrow \Delta\varepsilon_\beta\right) = \frac{\int\limits_{\Delta\varepsilon_\alpha}\Gamma_{emi}\left(\varepsilon_\alpha \rightarrow \Delta\varepsilon_\beta\right)d\varepsilon_\alpha}{\delta\varepsilon}. \qquad (2.40)$$

Figure 2.11 shows the phonon emission rate as a function of the initial energy of the quasiparticle $\Delta\varepsilon_\alpha$ and as a function of the quasiparticle energy after phonon emission $\Delta\varepsilon_\beta$ for the 100 nm − 55 nm Ta-Al (left) and 100 nm − 120 nm Nb-Al (right) junctions.

### 2.3.1.4 Electron-phonon scattering with absorption of a phonon (Excitation)

A quasiparticle can scatter from an energy $\varepsilon_\alpha$ to a higher energy $\varepsilon_\beta$ by absorbing a phonon of energy $\varepsilon_\beta\text{-}\varepsilon_\alpha$. The mean rate for the excitation of a quasiparticle of energy $\varepsilon_\alpha$ at the position x in the electrode to the energy interval $\Delta\varepsilon_\beta$, by absorption of a phonon of energy $\varepsilon_\beta\text{-}\varepsilon_\alpha$ is given by [Golubov 94, Kaplan 76]:

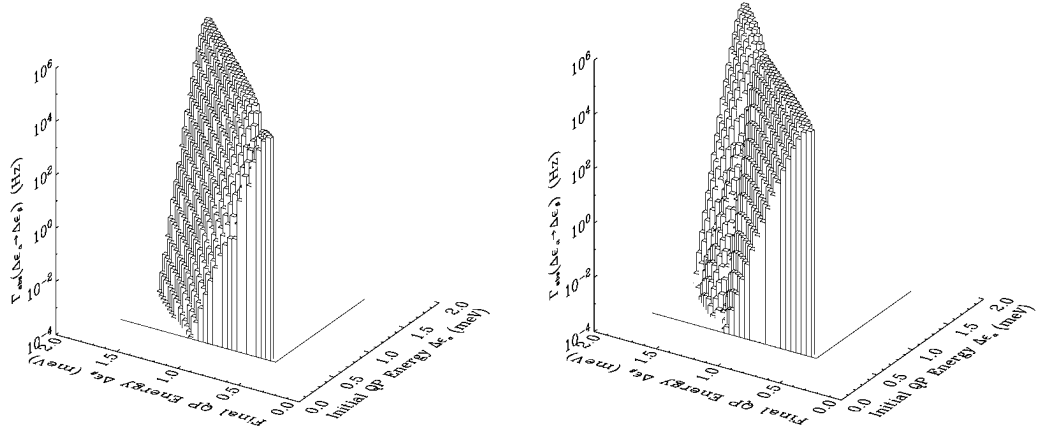

**Figure 2.12:** Rate for electron-phonon scattering with absorption of a phonon, $\Gamma_{abs}(\Delta\varepsilon_\alpha \rightarrow \Delta\varepsilon_\beta)$, as a function of initial and final quasiparticle energy for the 100 nm – 55 nm Ta-Al (left) and the 100 nm – 120 nm Nb-Al (right) junctions. Below the diagonal all rates are zero, because the quasiparticle cannot go to a lower energy after phonon absorption. The phonon absorption rate decreases exponentially with increasing phonon energy, because of the exponential dependency of the Bose distribution function.

$$\Gamma_{abs}\left(x, \varepsilon_\alpha \rightarrow \Delta\varepsilon_\beta\right) = \frac{1}{\tau_0(x)\left[kT_C(x)\right]^3} \int_{\varepsilon_\beta - \varepsilon_\alpha - \delta\varepsilon/2}^{\varepsilon_\beta - \varepsilon_\alpha + \delta\varepsilon/2} \Omega^2 \left[DoS_1(x, \varepsilon_\alpha + \Omega) - \frac{\Delta_1(x)}{\varepsilon_\alpha} Im F_1(x, \varepsilon_\alpha + \Omega)\right] n(\Omega) d\Omega$$

(2.41)

The next two steps are the same as for the previous section. The position independent phonon absorption rate is determined by:

$$\Gamma_{abs}\left(\varepsilon_\alpha \rightarrow \Delta\varepsilon_\beta\right) = \frac{\int\limits_{electr1} N_0(x) DoS_1(x, \varepsilon_\alpha) \Gamma_{abs}(x, \varepsilon_\alpha \rightarrow \Delta\varepsilon_\beta) dx}{\int\limits_{electr1} N_0(x) DoS_1(x, \varepsilon_\alpha) dx}.$$

(2.42)

And finally $\Gamma_{abs}(\varepsilon_\alpha \rightarrow \Delta\varepsilon_\beta)$ is averaged over the energy interval [$\varepsilon_\alpha$-$\delta\varepsilon/2$, $\varepsilon_\alpha$+$\delta\varepsilon/2$]:

$$\Gamma_{abs}\left(\Delta\varepsilon_\alpha \rightarrow \Delta\varepsilon_\beta\right) = \frac{\int\limits_{\varepsilon_\alpha - \delta\varepsilon/2}^{\varepsilon_\alpha + \delta\varepsilon/2} \Gamma_{abs}\left(\varepsilon_\alpha \rightarrow \Delta\varepsilon_\beta\right) d\varepsilon_\alpha}{\delta\varepsilon}$$

(2.43)

For the situation when the phonon distribution is not disturbed and remains in equilibrium $n(\Omega)$ is the Planck distribution and the described rate is the scattering rate with absorption of a thermal phonon. The phonon absorption rate in this case is strongly temperature dependent. The excess phonons created by the photon absorption and quasiparticle relaxation in the biased STJ should of course be considered as well. For simplicity, in this work these contributions have not been taken into account and only thermal phonons are taken into account.

Figure 2.12 shows the thermal phonon absorption rate as a function of the initial energy of the quasiparticle $\Delta\varepsilon_\alpha$ and as a function of the quasiparticle energy after phonon absorption $\Delta\varepsilon_\beta$ for the 100 nm − 55 nm Ta-Al (left) and 100 nm − 120 nm Nb-Al (right) junctions at a temperature of 300mK.

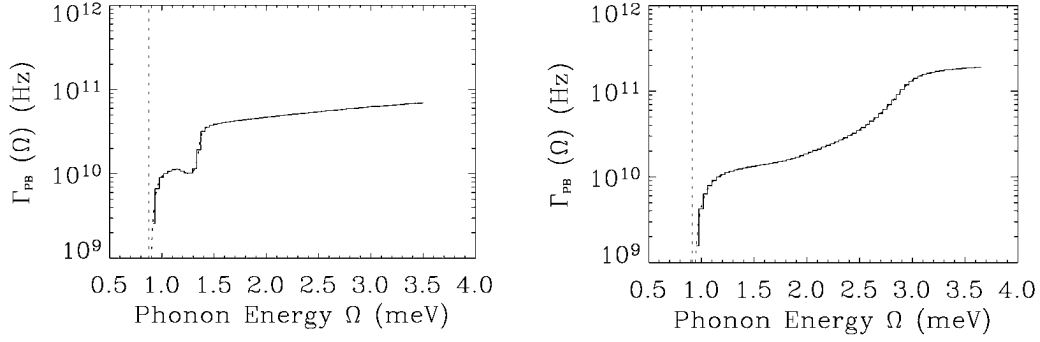

**Figure 2.13:** Cooper pair breaking rate, $\Gamma_{PB}(\Omega)$, as a function of phonon energy $\Omega$ for the 100 nm − 55 nm Ta-Al (left) and the 100 nm − 120 nm Nb-Al (right) bi-layers. The vertical dashed line represents twice the gap energy of the superconducting electrode.

### 2.3.1.5 Cooper pair breaking

The rate at which a phonon of energy $\Omega > 2\Delta_g$ breaks a Cooper pair into two quasiparticles is given by [Poelaert 99, Kaplan 76]:

$$\Gamma_{PB}(x,\Omega) = \frac{4\pi N_0(x)\alpha^2(x)}{\hbar N(x)}\int\limits_{\Delta_g}^{\Omega-\Delta_g}[DoS_1(x,\varepsilon')DoS_1(x,\Omega-\varepsilon') + Im\,F_1(x,\varepsilon')Im\,F_1(x,\Omega-\varepsilon')]d\varepsilon',$$

(2.44)

where $\alpha^2$ is the square of the matrix element of the electron-phonon interaction and N is the ion number density of the material. In general $\alpha^2$ depends on energy. Average values can be found in Ref. [Kaplan 76]. It is recalled that the position dependences for $N_0$, $\alpha^2$ and N come from the fact that the electrode material is not homogeneous across the junction. Therefore the value is different depending on the nature of the material at position x.

The Cooper pair breaking rate can be averaged over the position in the bi-layer x and over the energy interval $\Delta\varepsilon_\alpha$ in the same way as it was done before, to yield the Cooper pair breaking rate $\Gamma_{PB}(\Omega)$ as a function of phonon energy averaged over the energy intervals $\Delta\varepsilon_\alpha$. Figure 2.13 shows the Cooper pair breaking rate as a function of phonon energy for the 100 nm − 55 nm Ta-Al (left) and 100 nm − 120 nm Nb-Al (right) junctions.

### 2.3.1.6 Recombination-mediated energy exchange in the quasiparticle system

Let us consider the following sequence of events: A quasiparticle from the energy interval $\Delta\varepsilon_\alpha$ recombines with a quasiparticle from the energy interval $\Delta\varepsilon_\beta$, thereby releasing a phonon of energy $\varepsilon_\alpha+\varepsilon_\beta$. This phonon then breaks a Cooper pair into two quasiparticles, one being released into the energy interval $\Delta\varepsilon_\gamma$ and the other into the interval $\Delta\varepsilon_{\alpha+\beta-\gamma}$, which corresponds to the energy interval around the energy $\varepsilon_\alpha+\varepsilon_\beta-\varepsilon_\gamma$. This sequence annihilates two quasiparticles from the intervals $\Delta\varepsilon_\alpha$ and $\Delta\varepsilon_\beta$, and creates two quasiparticles in the intervals $\Delta\varepsilon_\gamma$ and $\Delta\varepsilon_{\alpha+\beta-\gamma}$. The rate for this sequence of events is given by [Kozorezov 03a]:

$$\Gamma_{\text{ree}}\left(x,\left(\Delta\varepsilon_\alpha,\Delta\varepsilon_\beta\right)\rightarrow\left(\Delta\varepsilon_\gamma,\Delta\varepsilon_{\alpha+\beta-\gamma}\right)\right)=\frac{\left(\varepsilon_\alpha+\varepsilon_\beta\right)^2\delta\varepsilon}{2N_0(x)\tau_0(x)\left(k_BT_C(x)\right)^3V}\frac{4N_0(x)\pi\alpha^2(x)}{\hbar N(x)}\left[\frac{1}{\Gamma_{\text{esc}}+\Gamma_{\text{PB}}\left(x,\Delta\varepsilon_{\alpha+\beta}\right)}\right]\cdot$$

$$\left[\text{DoS}_1\left(x,\Delta\varepsilon_\alpha\right)\cdot\text{DoS}_1\left(x,\Delta\varepsilon_\beta\right)+\text{Im}F_1\left(x,\Delta\varepsilon_\alpha\right)\cdot\text{Im}F_1\left(x,\Delta\varepsilon_\beta\right)\right]\cdot$$
$$\left[\text{DoS}_1\left(x,\Delta\varepsilon_\gamma\right)\cdot\text{DoS}_1\left(x,\Delta\varepsilon_{\alpha+\beta-\gamma}\right)+\text{Im}F_1\left(x,\Delta\varepsilon_\gamma\right)\cdot\text{Im}F_1\left(x,\Delta\varepsilon_{\alpha+\beta-\gamma}\right)\right]$$

$$(2.45)$$

where $\Gamma_{\text{esc}}$ is the phonon escape rate out of the film, V is the volume of the electrode and $\text{DoS}(x,\Delta\varepsilon)$ is the average density of states in the interval $\Delta\varepsilon$ and at the position x:

$$\text{DoS}(x,\Delta\varepsilon_\alpha)=\frac{1}{\delta\varepsilon}\int_{\Delta\varepsilon_\alpha}\text{DoS}(x,\varepsilon)d\varepsilon.\qquad(2.46)$$

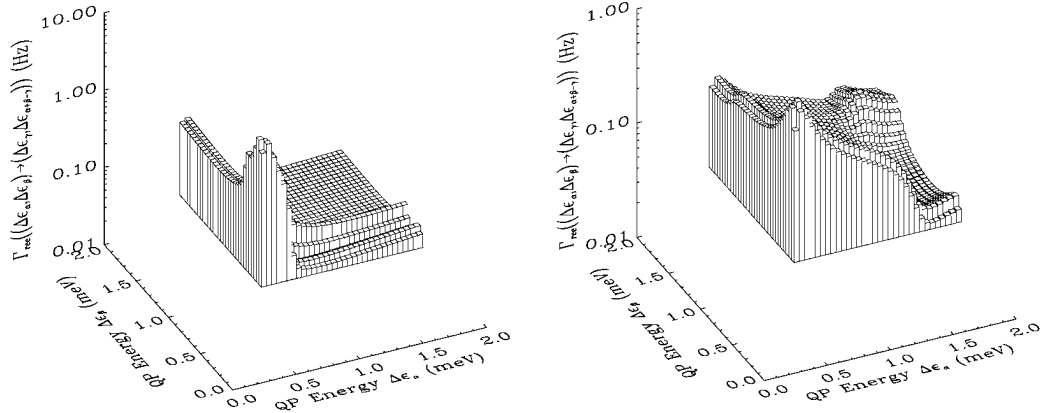

**Figure 2.14 :** Rate for the quasiparticle recombination with energy exchange process, $\Gamma_{\text{ree}}\left(x,\left(\Delta\varepsilon_\alpha,\Delta\varepsilon_\beta\right)\rightarrow\left(\Delta\varepsilon_\gamma,\Delta\varepsilon_{\alpha+\beta-\gamma}\right)\right)$, as a function of the initial quasiparticle energies $\Delta\varepsilon_\alpha$ and $\Delta\varepsilon_\beta$, and for the particular final energy $\Delta\varepsilon_\gamma$ equal to the gap energy of the electrode. The energy exchange rate is shown for the 100 nm – 55 nm Ta-Al bi-layer (left) and for the 100 nm – 120 nm Nb-Al bi-layer (right).

The same holds for $\text{Im}F(x,\Delta\varepsilon_\alpha)$.

This rate can then be averaged over the position x in the bi-layer. The rate depends on three independent indices $\alpha$, $\beta$ and $\gamma$, determining the energies of the two initial and one of the two final quasiparticles, the energy of the other final quasiparticle being fixed by the energy conservation law.

The phonon-mediated process of energy exchange in the electronic system of a superconductor, which was discussed above, may be treated exactly like an electron-electron collision process due to Coulomb interaction. One only needs to disregard the short-lived intermediate pair-breaking phonon emitted in the initial collision of the pair of quasiparticles. Whilst the Coulomb interaction is important for establishing the equilibrium within the quasiparticle system at relatively large quasiparticle densities either near $T_C$ or at quasiparticle densities comparable to that of the condensate, the phonon mediated process considered above is by far the most important equilibration mechanism at small and moderate quasiparticle densities. Figure 2.14 shows the recombination mediated energy exchange rate in the quasiparticle system as a function of the initial quasiparticle energies $\Delta\varepsilon_\alpha$ and $\Delta\varepsilon_\beta$, and for a final energy $\Delta\varepsilon_\gamma$ equal to the gap energy of the electrode. The energy exchange rate is shown for the Ta-Al bi-layer (left) and for the Nb-Al bi-layer (right).

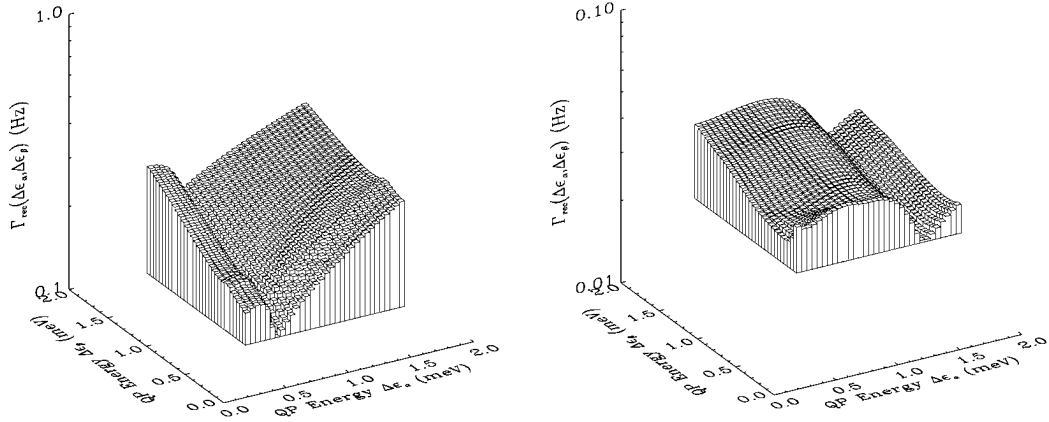

**Figure 2.15:** Rate for quasiparticle recombination with subsequent phonon loss, $\Gamma_{rec}\left(x, \left(\Delta\varepsilon_\alpha, \Delta\varepsilon_\beta\right)\right)$, as a function of the initial quasiparticle energies $\Delta\varepsilon_\alpha$ and $\Delta\varepsilon_\beta$. The recombination rate is shown for the 100 nm – 55 nm Ta-Al bi-layer (left) and for the 100 nm - 120 nm Nb-Al bi-layer (right).

### 2.3.1.7  *Quasiparticle recombination*

In the situation that the phonon released by a recombination process does not break another Cooper pair, but is lost into the substrate, the two quasiparticles are effectively lost from the system. This process is described by the following rate [Kozorezov 03a]:

$$\Gamma_{rec}\left(x, \left(\Delta\varepsilon_\alpha, \Delta\varepsilon_\beta\right)\right) = \frac{\left(\varepsilon_\alpha + \varepsilon_\beta\right)^2}{2N_0(x)\tau_0(x)(k_B T_C(x))^3 V}\left[\frac{\Gamma_{esc}}{\Gamma_{esc} + \Gamma_{PB}\left(x, \Delta\varepsilon_{\alpha+\beta}\right)}\right]\left[\frac{DoS_1\left(x, \Delta\varepsilon_\beta\right) + \dfrac{\Delta_1(x)}{\varepsilon_\beta}\mathrm{Im}\,F_1\left(x, \Delta\varepsilon_\beta\right)}{DoS_1\left(x, \Delta\varepsilon_\beta\right)}\right]$$

$$(2.47)$$

Again, the rate can be averaged over the position in the bi-layer x to yield the characteristic rate for quasiparticle recombination with phonon loss, depending on two indices $\alpha$ and $\beta$ representing the energy of the two initial quasiparticles. A practically important case is when the recombination is bottlenecked due to fast phonon re-absorption. In this case the recombination mediated energy exchange term is large and the probabilities of the two processes add up to unity. Thus for that case the energy exchange is a factor $\Gamma_{PB}/\Gamma_{esc} \gg 1$ faster than recombination.

Figure 2.15 shows the recombination rate as a function of the two initial quasiparticle energies for the 100 nm – 55 nm Ta-Al (left) and 100 nm – 120 nm Nb-Al (right) junctions. The junction side length was taken equal to 20 μm.

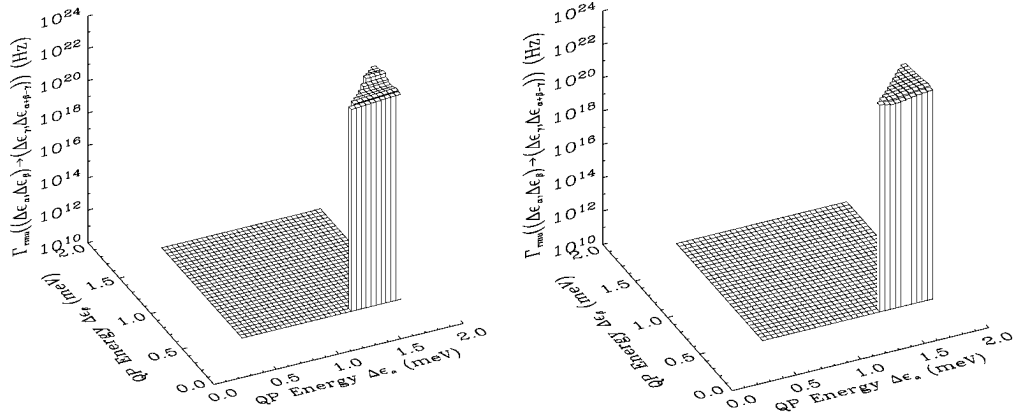

**Figure 2.16:** Rate for the quasiparticle multiplication process as a function of the initial ($\Delta\varepsilon_\alpha$) and final ($\Delta\varepsilon_\beta$) energy of the first quasiparticle for the 100 nm − 55 nm Ta-Al junction (left) and the 100 nm − 120 nm Nb-Al junction. For the figure, the energy of one of the generated quasiparticles was fixed to being equal to the gap energy $\Delta_g$. The energy of the second generated quasiparticle is fixed by the energy conservation law.

### 2.3.1.8 Quasiparticle multiplication

If a quasiparticle has an energy $\varepsilon_\alpha$ that is larger than $3\Delta_g$, it can relax to an energy $\varepsilon_\beta$ close to the gap energy and thereby release a phonon that has an energy larger than $2\Delta_g$. This phonon has enough energy to break a Cooper pair and create two additional quasiparticles in the energy intervals $\Delta\varepsilon_\gamma$ and $\Delta\varepsilon_{\alpha-\beta-\gamma}$. The following rate describes the quasiparticle multiplication process [Kozorezov 03a]:

$$\Gamma_{\text{rmu}}\left(x,\left(\Delta\varepsilon_\alpha,\Delta\varepsilon_\beta\right)\rightarrow\left(\Delta\varepsilon_\gamma,\Delta\varepsilon_{\alpha-\beta-\gamma}\right)\right)=\frac{8\pi N_0(x)\alpha^2(x)}{\hbar N(x)}\frac{(\varepsilon_\alpha-\varepsilon_\beta)^2\delta\varepsilon^2}{\tau_0(x)[k_B T_C(x)]^3}\frac{1}{\Gamma_{\text{esc}}+\Gamma_{\text{PB}}\left(x,\Delta\varepsilon_{\alpha-\beta}\right)}\cdot$$

$$\left[\frac{\text{DoS}_l\left(x,\Delta\varepsilon_\alpha\right)\cdot\text{DoS}_l\left(x,\Delta\varepsilon_\beta\right)+\text{Im}\,F_l\left(x,\Delta\varepsilon_\alpha\right)\cdot\text{Im}\,F_l\left(x,\Delta\varepsilon_\beta\right)}{\text{DoS}_l\left(x,\Delta\varepsilon_\alpha\right)}\right]\cdot$$
$$\left[\text{DoS}_l\left(x,\Delta\varepsilon_\gamma\right)\cdot\text{DoS}_l\left(x,\Delta\varepsilon_{\alpha-\beta-\gamma}\right)+\text{Im}\,F_l\left(x,\Delta\varepsilon_\gamma\right)\cdot\text{Im}\,F_l\left(x,\Delta\varepsilon_{\alpha-\beta-\gamma}\right)\right]$$

$$(2.48)$$

Again, the rate can be averaged over the position in the bi-layer x.

Fig. 2.16 shows the rate for the quasiparticle multiplication process as a function of the initial ($\Delta\varepsilon_\alpha$) and final ($\Delta\varepsilon_\beta$) energy of the first quasiparticle for the 100 nm − 55 nm Ta-Al junction (left) and the 100 nm − 120 nm Nb-Al junction. For the figure, the energy of one of the generated quasiparticles was fixed to being equal to the gap energy $\Delta_g$. The energy of the second generated quasiparticle is fixed by the energy conservation law. The multiplication rate is only non-zero for initial quasiparticle energies $\Delta\varepsilon_\alpha$ larger than twice the gap energy and for a final energy of the first quasiparticle $\Delta\varepsilon_\beta$, which lies at least twice the gap energy lower. Only in this case the released phonon has enough energy to break another Cooper pair. The multiplication rate is very high compared to the recombination and energy exchange rates, but the final number of multiplication processes is strongly reduced because of the very low quasiparticle densities at energies higher than three times the gap energy.

### 2.3.1.9 Quasiparticle trapping

Some regions in the superconducting film can have a local energy gap, which is lower than the energy gap of the superconductor surrounding it [Poelaert 99a]. Possible reasons for a locally reduced gap are dislocations, single magnetic impurity atoms or their clusters giving discrete or continuous states inside the superconducting gap or small normal metal inclusions in the superconducting film. A quasiparticle, which is close to such a region of reduced energy gap, can emit a phonon and scatter down to the lower energy gap region, where it is restrained from diffusing any further. Such a quasiparticle is trapped in the region of lower energy gap and is therefore effectively lost from the tunnelling system. One can approximate the trapping rate of a quasiparticle in the energy interval $\Delta\varepsilon_\alpha$ as:

$$\Gamma_{trap}\left(\Delta\varepsilon_\alpha\right) = C_{trap}\Gamma_{emi}\left(\Delta\varepsilon_\alpha + d_{trap} \rightarrow \Delta_g\right), \qquad (2.49)$$

where $d_{trap}$ is the trap depth, $C_{trap}$ is the trapping probability and $\Gamma_{emi}(\Delta\varepsilon_\alpha + d_{trap} \rightarrow \Delta_g)$ is the rate of quasiparticle scattering into the trap with emission of a phonon of energy $\Delta\varepsilon_\alpha - \Delta_g + d_{trap}$.

### 2.3.1.10 Quasiparticle de-trapping

Phonon absorption can free a trapped electron out of the region of reduced energy gap and make it available to the tunnel system again. Similarly to the previous paragraph, one can write the de-trapping rate by phonon absorption of a trapped quasiparticle into the energy interval $\Delta\varepsilon_\alpha$ as:

$$\Gamma_{deabs}\left(\Delta\varepsilon_\alpha\right) = \Gamma_{abs}\left(\Delta_g \rightarrow \Delta\varepsilon_\alpha + d_{trap}\right). \qquad (2.50)$$

A second possibility for a quasiparticle to escape from the trap is via recombination with another quasiparticle from the electrode, having an energy $\Delta\varepsilon_\alpha$. Depending on the energy $\Delta\varepsilon_\alpha$ of this free quasiparticle the trapped quasiparticle is either completely lost from the electrode or freed from the trap. If the energy of the free quasiparticle is larger than $\Delta_g + d_{trap}$, the released phonon has enough energy to directly break a Cooper pair. The two quasiparticles created by this pair breaking process have an energy larger than the energy gap and can diffuse away from the trap:

$$\Gamma_{deree}\left(\Delta\varepsilon_\alpha, \Delta\varepsilon_\beta\right) = \Gamma_{ree}\left((\Delta_g, \Delta\varepsilon_\alpha) \rightarrow (\Delta\varepsilon_\beta, \Delta\varepsilon_{\Delta_g + \alpha - \beta})\right), \qquad \Delta\varepsilon_\alpha > \Delta_g + d_{trap}. \qquad (2.51)$$

At low temperatures and small non-equilibrium phonon densities this mechanism may become the only significant de-trapping process.

### 2.3.1.11 Quasiparticle loss

The quasiparticle loss rate $\Gamma_{loss}$ is modelled as being independent of the quasiparticle energy. It includes all other losses than the ones by trapping or recombination. An example of these direct losses is diffusion of the quasiparticles out of the junction area through the leads. In order to prevent quasiparticles from leaving the junction area the contacts to the base and top film are fabricated out of a higher $T_C$ material than the electrodes. In an ideal case these additional losses would be negligible compared to the trapping and recombination losses.

### 2.3.2 Energy dependent balance equations

By regrouping all the terms calculated in the previous section, one can write the energy dependent balance equation of the QP number in the energy interval $\Delta\varepsilon_\alpha$ in the first electrode, $N_1(\Delta\varepsilon_\alpha)$:

$$
\begin{aligned}
\frac{dN_1(\Delta\varepsilon_\alpha)}{dt} &= \Gamma_{tun,2}\left(\Delta\varepsilon_\alpha - eV_b \rightarrow \Delta\varepsilon_\alpha\right) \cdot N_2\left(\Delta\varepsilon_\alpha - eV_b\right) - \Gamma_{tun,1}\left(\Delta\varepsilon_\alpha \rightarrow \Delta\varepsilon_\alpha + eV_b\right) \cdot N_1(\Delta\varepsilon_\alpha) \\
&+ \Gamma_{can,2}\left(\Delta\varepsilon_\alpha + eV_b \rightarrow \Delta\varepsilon_\alpha\right) \cdot N_2\left(\Delta\varepsilon_\alpha + eV_b\right) - \Gamma_{can,1}\left(\Delta\varepsilon_\alpha \rightarrow \Delta\varepsilon_\alpha - eV_b\right) \cdot N_1(\Delta\varepsilon_\alpha) \\
&+ \sum_\beta \Gamma_{emi,1}\left(\Delta\varepsilon_\beta \rightarrow \Delta\varepsilon_\alpha\right) \cdot N_1(\Delta\varepsilon_\beta) - \sum_\beta \Gamma_{emi,1}\left(\Delta\varepsilon_\alpha \rightarrow \Delta\varepsilon_\beta\right) \cdot N_1(\Delta\varepsilon_\alpha) \\
&+ \sum_\beta \Gamma_{abs,1}\left(\Delta\varepsilon_\beta \rightarrow \Delta\varepsilon_\alpha\right) \cdot N_1(\Delta\varepsilon_\beta) - \sum_\beta \Gamma_{abs,1}\left(\Delta\varepsilon_\alpha \rightarrow \Delta\varepsilon_\beta\right) \cdot N_1(\Delta\varepsilon_\alpha) \\
&- \sum_\beta \sum_\gamma \Gamma_{ree,1}\left(\left(\Delta\varepsilon_\alpha, \Delta\varepsilon_\beta\right) \rightarrow \left(\Delta\varepsilon_\gamma, \Delta\varepsilon_{\alpha+\beta-\gamma}\right)\right) \cdot \left[\frac{N_1(\Delta\varepsilon_\alpha) \cdot N_1(\Delta\varepsilon_\beta)}{DoS_1(\Delta\varepsilon_\alpha) \cdot DoS_1(\Delta\varepsilon_\beta)} - \frac{N_1(\Delta\varepsilon_\gamma) \cdot N_1(\Delta\varepsilon_{\alpha+\beta-\gamma})}{DoS_1(\Delta\varepsilon_\gamma) \cdot DoS_1(\Delta\varepsilon_{\alpha+\beta-\gamma})}\right] \\
&- \sum_\beta \Gamma_{rec,1}\left(\Delta\varepsilon_\alpha, \Delta\varepsilon_\beta\right) \cdot \left[N_1(\Delta\varepsilon_\alpha) \cdot N_1(\Delta\varepsilon_\beta) - N_{th,1}(\Delta\varepsilon_\alpha) \cdot N_{th,1}(\Delta\varepsilon_\beta)\right] \\
&+ \sum_\beta \sum_\gamma \Gamma_{rmu,1}\left(x, \left(\Delta\varepsilon_\beta, \Delta\varepsilon_\alpha\right) \rightarrow \left(\Delta\varepsilon_\gamma, \Delta\varepsilon_{\alpha+\beta-\gamma}\right)\right) \cdot N_1(\Delta\varepsilon_\beta) \\
&+ 2\sum_\beta \sum_\gamma \Gamma_{rmu,1}\left(x, \left(\Delta\varepsilon_\beta, \Delta\varepsilon_\gamma\right) \rightarrow \left(\Delta\varepsilon_\alpha, \Delta\varepsilon_{\beta+\gamma-\alpha}\right)\right) \cdot N_1(\Delta\varepsilon_\beta) \\
&- \Gamma_{trap,1}(\Delta\varepsilon_\alpha)\left[n_1^{traps} - N_1^t\right] \cdot N_1(\Delta\varepsilon_\alpha) + \Gamma_{det\,rap,1}(\Delta\varepsilon_\alpha) \cdot N_1^t \\
&+ \sum_\beta \Gamma_{derec,1}\left(\Delta\varepsilon_\alpha, \Delta\varepsilon_\beta\right) \cdot \frac{N_1^t \cdot N_1(\Delta\varepsilon_\alpha)}{DoS_1(\Delta_g) \cdot DoS_1(\Delta\varepsilon_\alpha)} \\
&- \Gamma_{loss,1} \cdot N_1(\Delta\varepsilon_\alpha)
\end{aligned}
$$

$$(2.52)$$

where $N_{th,1}(\Delta\varepsilon_\alpha)$ is the number of thermal quasiparticles in the energy interval $\Delta\varepsilon_\alpha$ of electrode 1, $N_1^t$ is the number of trapped quasiparticles in electrode 1, $n_1^{traps}$ is the number of available traps in electrode 1 and $DoS(\Delta\varepsilon_\alpha)$ is the density of states in the energy interval $\Delta\varepsilon_\alpha$, averaged over the position in the bi-layer.

Of course one has one equation per energy interval $\Delta\varepsilon_\alpha$. This system of equations in electrode 1 has to be completed by the equation giving the variation of the number of trapped quasiparticles in electrode 1, which can be written as:

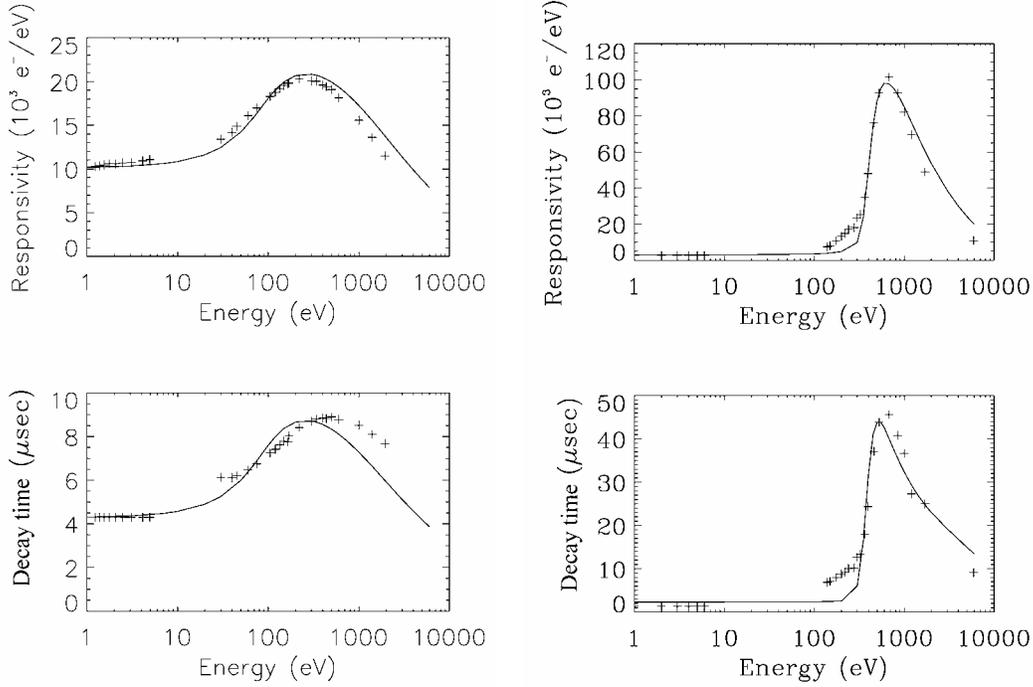

**Figure 2.17:** Responsivity and decay time of the Ta-Al based (left) and the Nb-Al based (right) 20 μm side length junctions at a bias voltage of 180 μV as a function of incoming photon energy. The crosses represent the experimental data, whereas the solid line represents the calculated fit to the data with the parameters from table 2.4.

$$\frac{dN_1^t}{dt} = \sum_\alpha \left[ \Gamma_{trap,1}(\Delta\varepsilon_\alpha) \cdot \left[ n_1^{traps} - N_1^t \right] \cdot N_1(\Delta\varepsilon_\alpha) - \Gamma_{det\,trap,1}(\Delta\varepsilon_\alpha) \cdot N_1^t \right]$$

$$- \sum_\alpha \sum_\beta \Gamma_{deree,1}(\Delta\varepsilon_\alpha, \Delta\varepsilon_\beta) \cdot \frac{N_1^t \cdot N_1(\Delta\varepsilon_\alpha)}{DoS_1(\Delta_g) \cdot DoS_1(\Delta\varepsilon_\alpha)}. \qquad (2.53)$$

A similar set of equations can be written for electrode 2 by interchanging the indices 1 and 2. If there are $N_{en}$ energy intervals in one electrode, one ends up with a system of $2N_{en}+2$ coupled, non-linear, first order differential equations, which has to be solved numerically. The numerical method used is a simple Euler iterations scheme, where the step size is varied from $10^{-11}$ seconds at the beginning of the pulse to approximately $10^{-8}$ sec at the end of the pulse. The variation of the step size is linked to the variation of the number of quasiparticles in the different energy intervals.

The initial conditions are found by remarking that the model is started after the second stage of the electronic down-conversion process, as defined in [Kozorezov 01]. At this point all the phonons in the system have an energy lower than twice the gap energy of the superconducting absorber and cannot break Cooper pairs anymore. The number of quasiparticles $Q_0$ created by the first two stages of the down-conversion process was calculated to be equal to [Kozorezov 01]:

$$Q_0 = \frac{E}{1.7\Delta_g}, \qquad (2.54)$$

where E is the photon energy.

At t = 0, this number of quasiparticles is put in the highest energy interval below $3\Delta_g$ of the absorbing electrode. The exact energy distribution at t = 0 is of no importance, as the excess quasiparticles very quickly (~0.1 μsec) find a "quasi-equilibrium" distribution via tunnelling and relaxation, totally independent of the initial distribution of the quasiparticles (see next section).

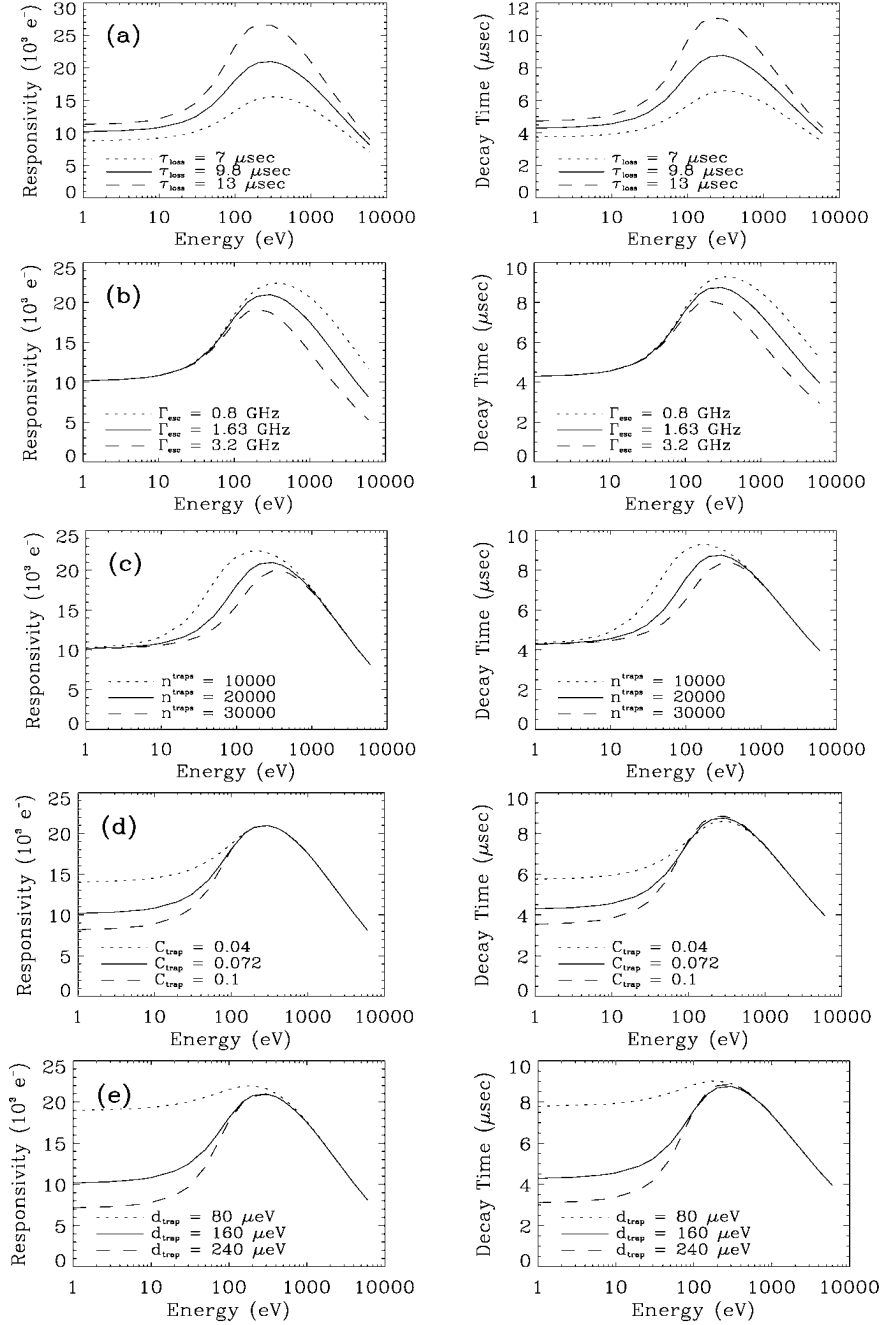

**Figure 2.18:** Variation of the responsivity and decay time curves as the five input parameters of the model are independently varied: (a) quasiparticle loss time $\tau_{loss}$, (b) phonon escape rate $\Gamma_{esc}$, (c) number of traps in electrode $n^{traps}$, (d) Trapping probability $C_{trap}$ and (e) trap depth $d_{trap}$. The simulations were made for the case of the 20 μm side-length Ta-Al based junction.

### 2.3.3   Results of simulations

Simulations made for the two symmetrical Ta-Al and Nb-Al STJs with 20 μm side length are described in the following in order to illustrate the model. For the Ta-Al bi-layer the thickness of the Ta is equal to 100 nm and the Al thickness equals 55 nm, whereas for the Nb-Al based junction the Nb thickness equals 100 nm and the Al film thickness is equal to 120 nm. Both junctions have an energy gap which is close to 450 μeV. The junctions were tested extensively at 350 mK in a portable ³He sorption cryostat at the BESSY synchrotron radiation facility from 30 to 2000 eV, in the optical regime from 1 to 5 eV and with an ⁵⁵Fe radiation source around 6keV radiation. The variation of the responsivity (charge output per unit of incoming photon energy) and the pulse decay time, which is equal to the rise time of the output of the charge sensitive pre-amplifier, were measured as a function of incoming photon energy at a bias voltage of 180 μeV. The experimental data, which was taken from [Poelaert 99], is shown in figure 2.17 for the Ta-Al (left) and Nb-Al (right) based junction.

The model has five unknown fitting parameters, which are: the energy independent quasiparticle loss rate $\Gamma_{loss}$, the energy independent phonon escape rate from the electrode $\Gamma_{esc}$, the trapping probability $C_{trap}$, the number of available traps in the electrode $n^{traps}$ and the trap depth $d_{trap}$. All of the fitting parameters have a profound physical interpretation and all of them, except for $C_{trap}$ and $d_{trap}$, have a single, very specific region of influence on the curves in figure 2.17, which allows them to be determined with certainty:

(a) The quasiparticle loss time $\tau_{loss}=1/\Gamma_{loss}$ reflects the rise-time of the signal, when all other loss channels, like losses by trapping and losses by recombination, are negligible. Therefore, it determines the height of the maximum of the curve, when all the trapping states are saturated and filled with quasiparticles and quasiparticle recombination has not yet set in. In case the losses by trapping or recombination are dominant over the whole energy range this fitting parameter can be safely neglected. Figure 2.18(a) shows the variation of the calculated responsivity and pulse decay time curves as the quasiparticle loss time is varied.

(b) The phonon escape rate out of the electrode $\Gamma_{esc}$ determines the losses by quasiparticle recombination that only become dominant at high quasiparticle densities in the electrodes. It therefore only influences the steepness of the negative slope of the curve in the high photon energy range. The direct determination of the phonon escape time through the usage of average phonon transmission values through material interfaces, as derived by [Kaplan 79] for different material combinations, is difficult and not very accurate in case of our very thin junctions. As the junctions have characteristic length scales that are of the order of the phonon wavelength the different reflections and transmissions within the junctions are difficult to account for and the phonon loss rate will depend on the exact microscopic structure of the edges of the junctions or the roughness of the interfaces. In practice it was found that it is necessary to keep the phonon loss rate as a free parameter. Figure 2.18(b) shows the variation of the calculated responsivity and pulse decay time curves as the phonon escape rate is varied.

(c) The number of available traps in the electrode $n^{traps}$ determines the onset of the positive slope part of the curve, when the quasiparticle traps start to be saturated with quasiparticles. The more available traps there are, the more the onset of the positive slope is shifted towards high photon energies, because it takes more quasiparticles to fill the available traps. Figure 2.18(c) shows the variation of the calculated responsivity and pulse decay time curves as the number of traps in the electrode is varied.

(d) The trapping probability $C_{trap}$ influences the speed at which a quasiparticle is trapped. This parameter mainly determines the decay time of the pulse in the low energy part of the curve, by setting the speed at which a small number of quasiparticles is lost into a large amount of traps. This region of the curve is linear, as the traps are far from being saturated. Figure 2.18(d) shows the variation of the calculated responsivity and pulse decay time curves as the trapping probability is varied.

(e) The trap depth $d_{trap}$ has the same influence on the trapping speed as the trapping probability $C_{trap}$. As these two parameters both act in the same direction, it is fairly difficult to determine these two characteristics in an absolute manner just with the data of figure 2.17. Nevertheless, the trap depth can be derived from other experiments, like from responsivity and rise-time measurements as a function of temperature [Kozorezov 00]. Figure 2.18(e) shows the variation of the calculated responsivity and pulse decay time curves as the trap depth is varied.

The preceding five parameters were varied in order to find a fit to the experimental data of Fig. 2.17. The calculated curves are also shown in the figure. Table 2.4 shows the set of parameters associated to the 20 µm Nb-Al and Ta-Al junctions, which were determined to fit the results of the model to the experimental data. For both junctions the fit to the responsivity is very good, whereas the fit to the decay time of the pulse shows certain discrepancies especially in the high energy domain, where non-linear quasiparticle recombination becomes important. This effect is probably due to the experimental determination of the decay time, which assumes a perfectly exponential pulse (see section 3.2.2), whereas, especially in the high energy region, the pulse is not purely exponential. There exists therefore an uncertainty in the experimental determination of the decay time of the pulse.

**Table 2.4:** Fitting parameters of the model for the Ta-Al based and Nb-Al based 20 µm side length junctions.

| Symbol | Name | Unit | Ta-Al | Nb-Al |
|--------|------|------|-------|-------|
| $\tau_{loss}=1/\Gamma_{loss}$ | Quasiparticle loss time | µsec | 9.8 | 100 |
| $\Gamma_{esc}$ | Phonon escape rate | Hz | $1.6\ 10^9$ | $5\ 10^9$ |
| $C_{trap}$ | Trapping probability | / | 0.072 | 0.22 |
| $n^{traps}$ | Number of traps in electrode | / | 20 000 | 185 000 |
| $d_{trap}$ | Trap depth | µeV | 160 | 240 |

The quasiparticle loss time in the Ta-Al based junction is a factor 10 lower than the loss time in the Nb-Al junction. The reason for this is magnetic flux trapping in the Ta-Al junction during the photon detection experiments. The trapped flux quantum acts as a region with zero gap and represents therefore a strong loss channel for the quasiparticles. On the other hand the Nb-Al junction possesses a considerably larger number of traps, as well as a higher trapping probability and larger trap depth.

Let us now discuss the main scope of the model, the quasiparticle energy distribution and its variation with time.

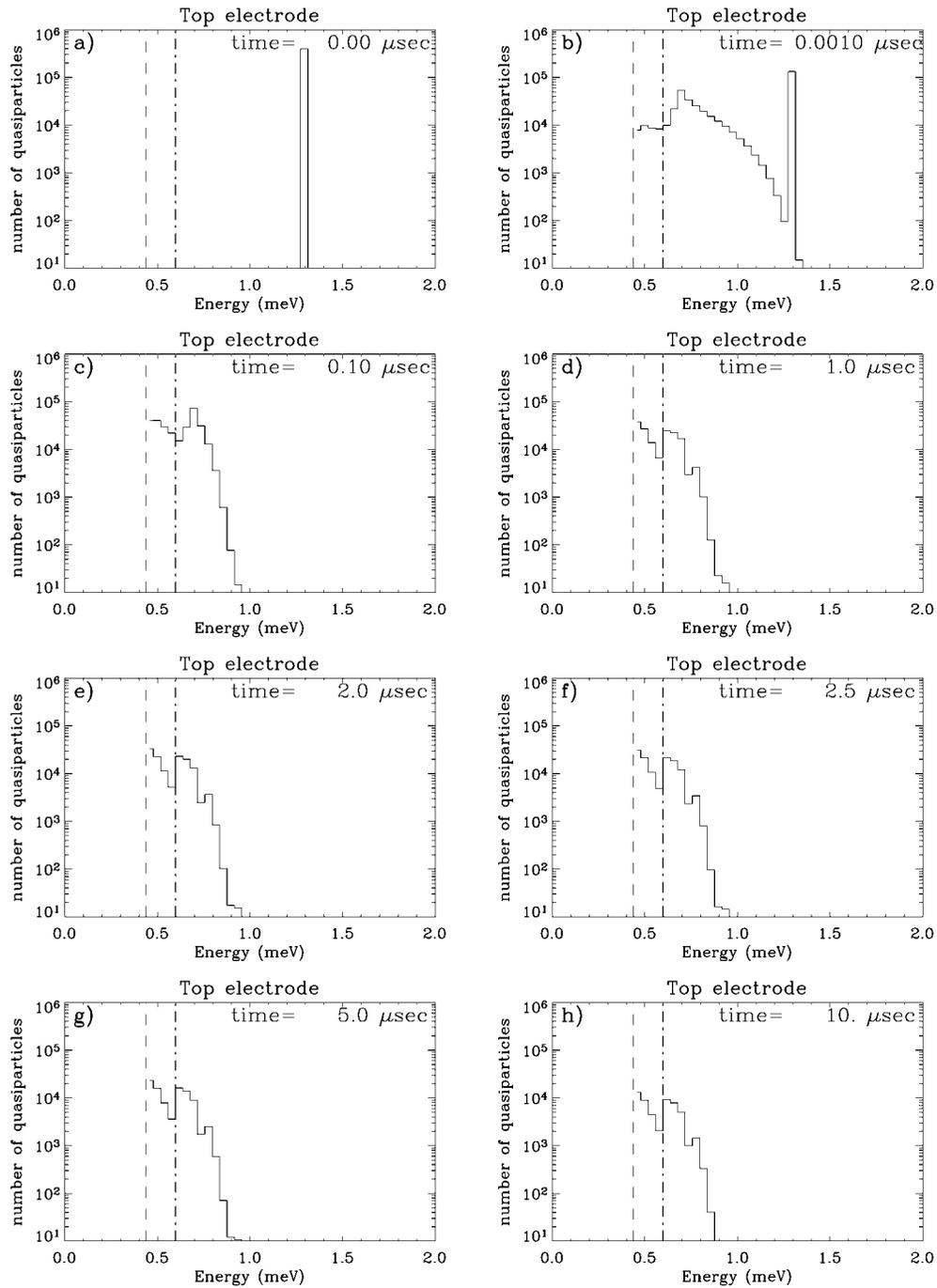

**Figure 2.19:** Quasiparticle energy distribution at eight different instants after absorption of a 300 eV photon in the top electrode of the Ta-Al based junction. In every graph the dashed vertical line represents the energy gap of the superconducting electrode of the junction, whereas the dashed-dotted vertical line represents the bias energy $eV_b$ above the energy gap.

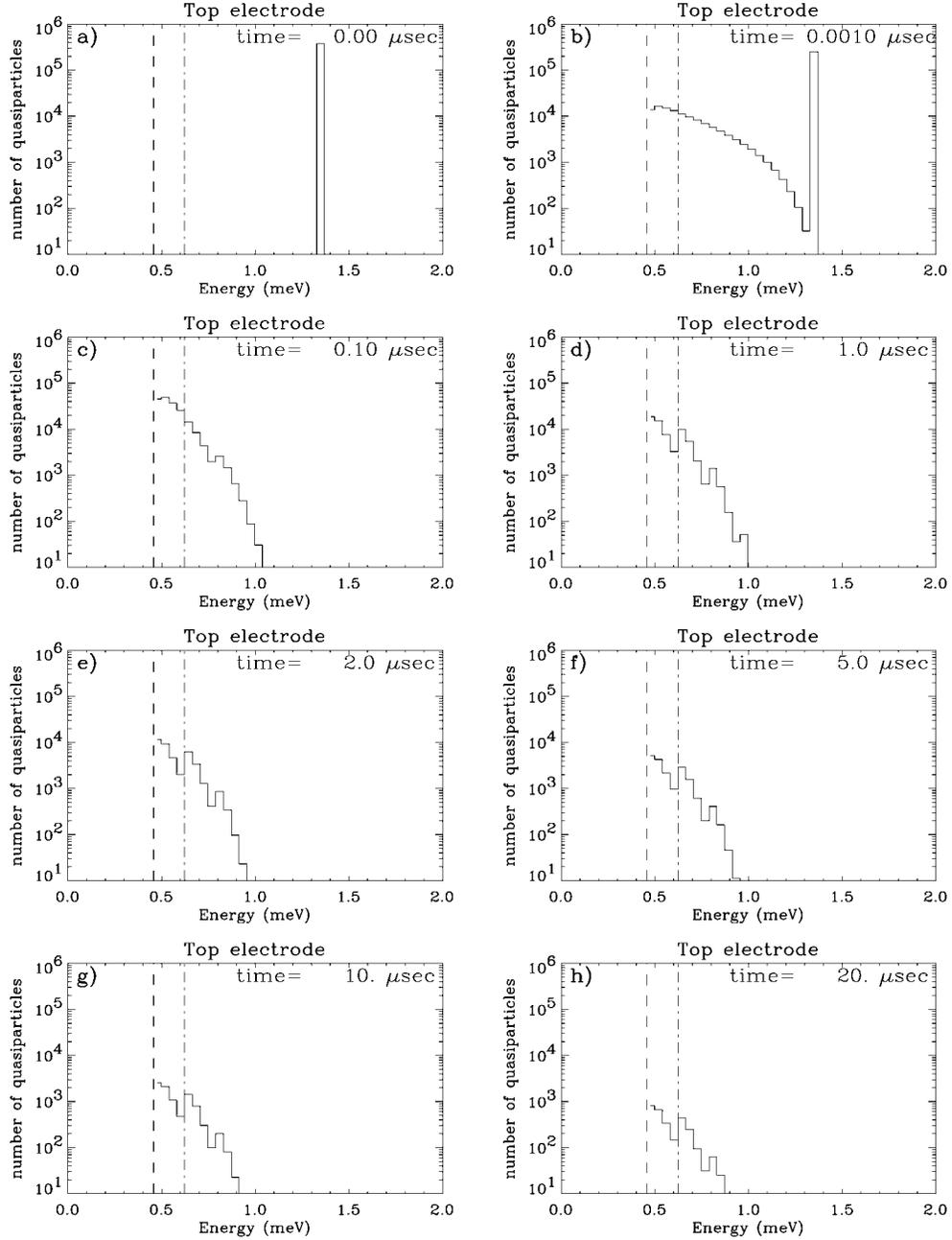

**Figure 2.20:** Quasiparticle energy distribution at eight different instants after absorption of a 300 eV photon in the top electrode of the Nb-Al based junction. In every graph the dashed vertical line represents the energy gap of the superconducting electrode of the junction, whereas the dashed-dotted vertical line represents the bias energy $eV_b$ above the energy gap.

### 2.3.3.1  Convergence to a "quasi-equilibrium" distribution

After 0.1 to 0.5 µsec the quasiparticle distribution converges from its initial state to a "quasi-equilibrium" distribution. This distribution is called to be in "quasi-equilibrium" in the sense that the normalized energy distribution of the quasiparticles stays constant, whereas the total number of quasiparticles in the electrodes diminishes, because of the different quasiparticle loss channels. This is illustrated in Figs. 2.19 and 2.20, where the quasiparticle energy distribution is shown at different instants of time in the absorbing electrode of the Ta-Al and Nb-Al junctions after the absorption of a 300 eV photon.

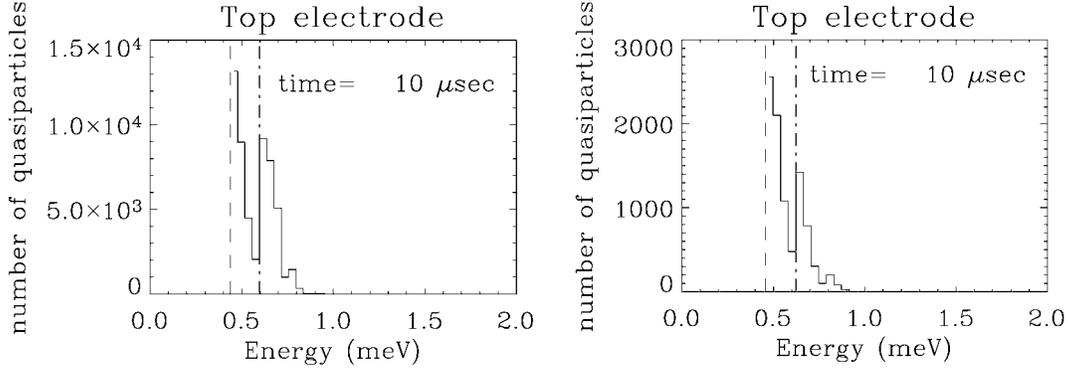

**Figure 2.21:** Quasiparticle energy distribution at t=10 μsec after the absorption of a 300 eV photon in the top electrode of a 20 μm side length Ta-Al (left) and Nb-Al (right) based junction. The dashed vertical line represents the energy gap of the superconducting electrode of the junction, whereas the dashed-dotted vertical line represents the bias energy $eV_b$ above the energy gap. The step-like structure of the quasiparticle distribution can be clearly identified.

At the moment of photon absorption (t = 0) all the quasiparticles are in an elevated energy level of the absorbing electrode. An exact knowledge of the initial conditions is not necessary, as, through a series of phonon scattering and tunnel events, the quasiparticle population converges within a fraction of a microsecond towards a stable configuration. One can see on the figure that for times t>1 μsec, the shape of the distribution does not vary anymore. Only the absolute number of quasiparticles decreases with time because of the different quasiparticle loss channels. This "quasi-equilibrium distribution" shows a step-like structure, caused by the discrete energy gain (loss) of $eV_b$ during a tunnel(cancellation-tunnel) process. Fig. 2.21 shows the quasiparticle energy distribution in the Ta-Al (left) and Nb-Al (right) junction at t=10 μsec on a linear scale. In this figure the step-like structure can be seen more clearly. The energy difference between two consecutive steps is $eV_b$. Clearly, even for these rather high gap Ta-Al and Nb-Al based junctions, the quasiparticles are spread out over a broad energy domain. They do not reside at the gap energy as it is assumed in the Rothwarf-Taylor approach. The observed distribution is similar to the results of Kozorezov et al. [Kozorezov 03], who already observed the step-like structure of the quasiparticle energy distribution for BCS-like junctions in thermal equilibrium. In their case this particular structure was responsible for current steps in the IV-curves of the junctions. Segall et al [Segall 99] , who developed a similar, but less complete model for the special case of BCS-like junctions, do not mention the step-like nature of the quasiparticle distribution. They compare the quasiparticle energy distribution to a thermal distribution with an effective temperature $T_{eff}$ that is higher than the bath temperature. The junctions for which they make their computations have relatively long tunnel times of the order of 2 μsec, and they only give the quasiparticle distribution at a time t = 1 μsec, which is smaller than the tunnel time. This is why the steps are not yet visible in their graphs, but should build up for times larger than the tunnel time. Therefore, their results are not in contradiction to the findings in this thesis.

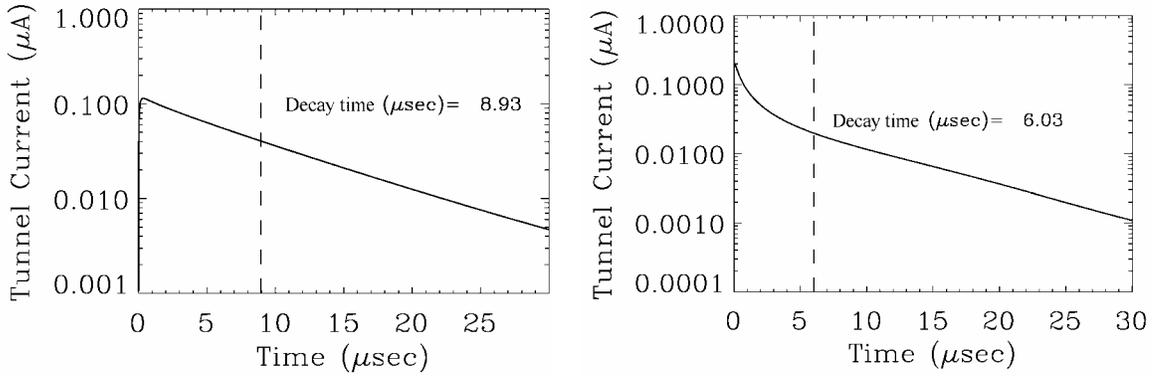

**Figure 2.22:** Tunnel current pulse on a logarithmic scale as a function of the time after the absorption of a 300 eV photon in the top electrode of the Ta-Al (left) and the Nb-Al (right) junction. The applied bias voltage is 180 μeV. The vertical dashed line represents the decay time of the pulse.

### 2.3.3.2 Tunnel Current

The tunnel current is given by the sum of the tunnel terms minus the sum of the cancellation terms. After reaching a maximum within a fraction of a microsecond, the tunnel current decays mainly exponentially. Deviations from the simple exponential decay are observed in case of high losses by recombination and in case of trap saturation during the pulse. Figure 2.22 shows the tunnel current pulse from the Ta-Al and Nb-Al based junctions under discussion after the absorption of a 300 eV photon on a logarithmic scale. In the case of the Nb-Al junction one can clearly identify that the pulse does not decay purely exponentially at the beginning of the pulse because of fast quasiparticle trapping. In the case of the Ta-Al junction the decay is mainly exponential because the number of traps is much smaller than the available quasiparticles. Fig. 2.23 shows the integrated current pulse, representing the charge output of the junction. In order to determine the charge output accurately, the pulse has to be calculated up to several times the decay time. In order to shorten the calculations, the last part of the current pulse is fitted with an exponentially decaying curve of the form:

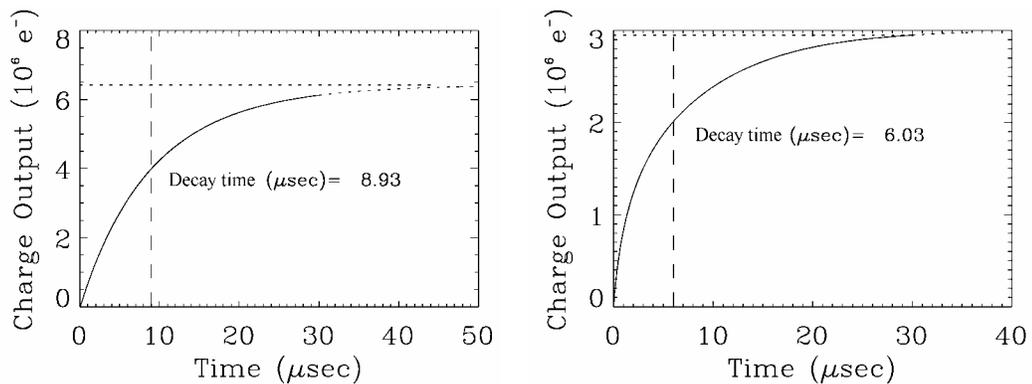

**Figure 2.23** Integrated current pulse as a function of time after the absorption of a 300 eV photon in the top electrode of the Ta-Al (left) and Nb-Al (right) based junction. The applied bias voltage is 180 μeV. The vertical dashed line represents the rise time of the integrated pulse and the horizontal dashed line represents the total charge output of the junction.

$$I(t) = I_m \exp(-t/\tau_D). \tag{2.55}$$

The calculation can then be stopped after a time slightly larger than the decay time of the pulse. The total charge output can then determined from the combined calculated and fitted curves. The decay time is determined as the time when the integrated current pulse reaches 63 % of its final value.

### 2.3.3.3   Average values and tunnel to cancellation ratio

Knowing the quasiparticle energy distribution during a current pulse, one can now determine the average quasiparticle energy and the average characteristic tunnel times according to:

$$\langle \Gamma \rangle = \frac{\sum_i N(\Delta \varepsilon_i) \Gamma(\Delta \varepsilon_i)}{\sum_i N(\Delta \varepsilon_i)}, \tag{2.56}$$

where $\Gamma(\Delta \varepsilon_i)$ is a characteristic parameter in the energy interval $\Delta \varepsilon_i$ and $N(\Delta \varepsilon_i)$ is the number of quasiparticles in the interval $\Delta \varepsilon_i$. From these values the tunnel to cancellation ratio can be derived, which characterizes the fraction of charge output lost due to the cancellation tunnel events

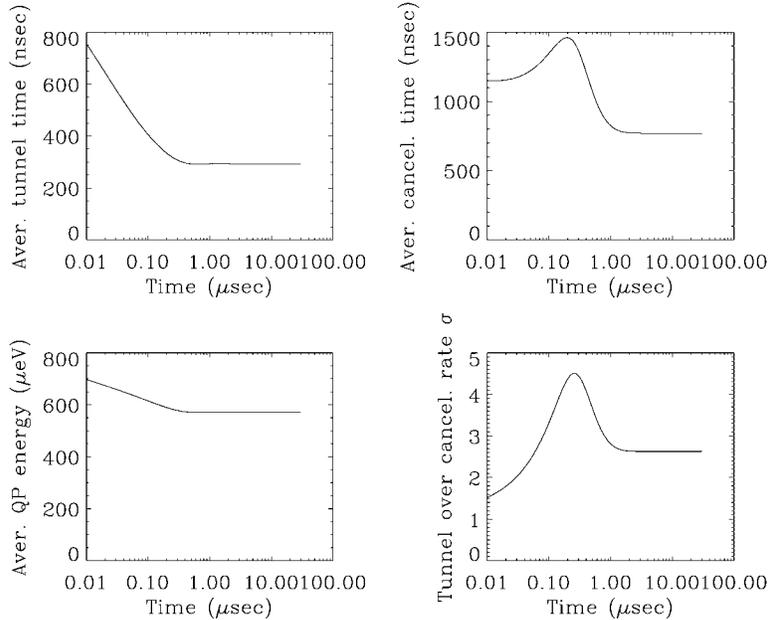

**Figure 2.24:** Average tunnel time, average cancellation tunnel time, average quasiparticle energy and tunnel to cancellation ratio as a function of time after the absorption of a 300 eV photon in the top electrode of the Ta-Al junction. The applied bias voltage is 180 μeV. All four variables converge after a microsecond. The time scale is exponential in order to be able to see the very rapid convergence towards the "quasi-equilibrium" values.

$$\sigma = \frac{\langle \Gamma_{tun} \rangle}{\langle \Gamma_{can} \rangle}. \tag{2.57}$$

The knowledge of this ratio, which cannot be determined experimentally and is important in order to derive for example the cancellation noise [Segall 99], allows the determination of the mean number of tunnel $\langle n_{tun} \rangle$ and cancellation tunnel $\langle n_{can} \rangle$ events per quasiparticle as well as the total number of tunnels per quasiparticle $\langle n \rangle$ from the charge amplification factor $\bar{n} = Q/Q_0$, which can be derived from experiment.

$$\langle n_{tun} \rangle = \frac{\sigma}{\sigma - 1} \frac{Q}{Q_0}, \tag{2.58}$$

$$\langle n_{can} \rangle = \frac{1}{\sigma - 1} \frac{Q}{Q_0}, \tag{2.59}$$

$$\langle n \rangle = \langle n_{tun} \rangle + \langle n_{can} \rangle = \frac{\sigma + 1}{\sigma - 1} \frac{Q}{Q_0}, \tag{2.60}$$

where $Q_0$ is the initial number of quasiparticles created in the electrode and $Q$ is the measured charge output.

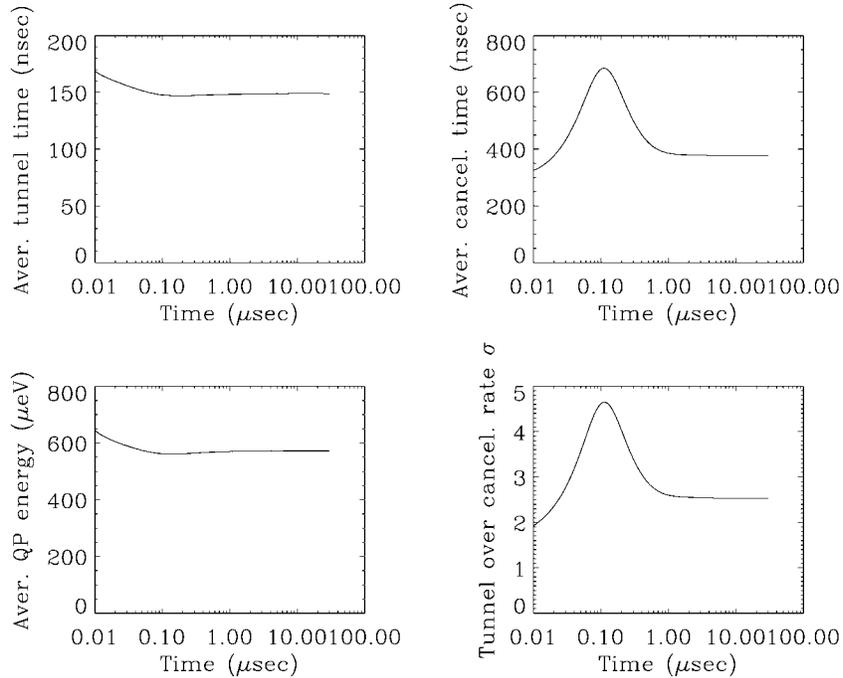

**Figure 2.25:** Average tunnel time, average cancellation tunnel time, average quasiparticle energy and tunnel to cancellation ratio as a function of time after the absorption of a 300 eV photon in the top electrode of the Nb-Al junction. The applied bias voltage is 180 $\mu$eV. All four variables converge after a microsecond. The time scale is exponential in order to be able to see the very rapid convergence towards the "quasi-equilibrium" values.

Figures 2.24 and 2.25 show how the average quasiparticle energy, the average tunnel time, the average cancellation tunnel time and the tunnel to cancellation ratio vary during the current pulse for the Ta-Al and the Nb-Al junction respectively. As the quasiparticle energy distribution converges towards the quasi-equilibrium distribution, the four characteristics also converge. In the quasi-equilibrium distribution, the quasiparticles have in both cases an average energy of 570 μeV, which is considerably higher than the gap energy of the electrode, equal to approximately 450 μeV. This feature was already predicted qualitatively by A. Poelaert et al. [Poelaert 99] by taking advantage of a non energy dependent kinetic model. Taking the "balance energy" as an additional fitting parameter and making the very gross approximation that all quasiparticles have energy equal to the balance energy, he derived balance energies equal to 650 and 500 μeV for the Ta-Al and the Nb-Al junction respectively. In order to derive these values Poelaert et al. fitted the very same experimental data. Here, of course, the average quasiparticle energy is not a free fitting parameter, but is derived naturally from the energy distribution of the quasiparticles. Because of the large average energy of the quasiparticles, one can thus talk of a quasiparticle heating effect, caused by the energy gain due to tunnelling. The exact value of the average quasiparticle energy will depend strongly on the tunnelling time, the bias voltage and the energy gap of the superconductors forming the electrodes. Another interesting feature is that the tunnel to cancellation ratio is approximately equal to 2.5 in both material combinations. This means that 30 % of the quasiparticles actually transfer a charge in the wrong direction when tunnelling and thereby reducing the measured signal at the output of the detector. For junctions with a lower energy gap the tunnel to cancellation ratio will progressively decrease towards one, having a considerable effect on the measured responsivity and the energy resolution of the detector.

## 2.4 Conclusions

Within the framework of the proximity effect theory a new method for determining the interface parameters of a superconductor-superconductor interface was presented. The method is based on the experimental determination of the energy gap and the critical temperature of the bi-layer, from which the two parameters $\gamma$ and $\gamma_{BN}$ characterising the interface can be deduced. Expressions for the interface parameters as a function of the two film thicknesses of the bi-layer are given, from which two interface constants $C_\gamma$ and $C_{\gamma_{BN}}$ can be derived, which are independent of the film thickness. These interface constants were determined from measurements of the energy gap and the critical temperature for a Ta-Al bi-layer with 100 nm of Ta and 30 nm of Al and a Nb-Al bi-layer with a 100 nm thick Nb film and a 30 nm thick Al film. From the knowledge of the interface constants the variation of the interface parameters as a function of Al film thickness could be calculated and the energy gap and critical temperatures of bi-layers with varying Al thickness could be predicted. The simulations were compared to experimental results on a series of Ta-Al and Nb-Al bi-layers with 100 nm thick Ta and Nb films and an Al film thickness varying between 5 and 265 nm. The experimental data corresponds very well to the values predicted with the theory.

Based on the results of the proximity effect theory a new kinetic model for the photon detection process in an STJ was presented, which takes the full energy dependence of all the quasiparticle processes occurring in the junctions into account. The model allows the calculation of the full temporal variation of the non-equilibrium quasiparticle energy distribution in the electrodes of the junction during the photon detection process. The model has five unknown fitting parameters, which all have a profound physical interpretation. These parameters are the quasiparticle loss time, the phonon escape time out of the junction, the number of available local quasiparticle trapping states and the depth of the latter, as well as the trapping probability. In order to illustrate the theory it was applied to two different junctions, one based on a Ta-Al bi-layer with 100 nm of Ta and 55 nm of Al and the other based on a Nb-Al bi-layer with 100 nm of Nb and 120 nm of Al. For both junctions a fit to experimental data showing the responsivity and decay time of the detector pulses as a function of incoming photon energy was determined by varying the different fitting parameters. It was shown that it is possible to obtain a good knowledge of the number of traps, the quasiparticle loss time and the phonon escape time with just these two experimental curves, as these parameters influence different parts of the curves. In order to determine the quasiparticle trap depth and trapping probability independently, temperature-dependent data of the responsivity has to be available.

The calculations on the Nb and Ta-based junctions show that the non-equilibrium quasiparticle energy distribution created during the photon absorption process converges within a fraction of a microsecond towards a "quasi-equilibrium" distribution. This distribution is called to be in "quasi-equilibrium" in the sense that the normalised distribution of the quasi-particles does not change with time. Only the absolute number of quasiparticles decreases in the electrodes because of the different loss channels. The quasi-equilibrium distribution itself shows a step-like structure, with maxima separated by the bias energy, which are created because of the discrete energy gain of the quasiparticles during a tunnel event. The average quasiparticle energy can be calculated and it is shown that it lies above the gap energy of the bi-layer forming the electrodes. The average energy of the quasiparticles depends strongly on the tunnelling time, the applied bias voltage and the energy gap of the junction. The knowledge of the quasiparticle energy distribution during the current pulse also allows the determination of the tunnel to cancellation ratio, which is important knowledge for the calculation of the different resolution broadening factors.

# Chapter 3

# Experimental set-up

Almost all the experimental measurements, which are presented in this thesis, were performed in the laboratories of the Science Payloads and Advanced Concepts Office within the Research and Scientific Support Department of the European Space Agency (ESA). Situated in Noordwijk, the Netherlands, the infrastructure of the European Space Research and Technology Center (ESTEC) includes different kind of cryogenic apparatus, able to reach the very low temperatures needed for the successful operation of the junctions. Within the framework of this thesis two different cryostats have been extensively used, a $^3$He sorption cooler, able to achieve a base temperature of 300mK and an Adiabatic Demagnetisation Refrigerator (ADR), which has a base temperature of 35 mK. In the following the cryostats will be presented, as well as the electronic set-up used for reading out the junctions either in voltage sweep mode for IV-curve acquisitions or in stable DC biasing mode for spectral applications.

## *3.1 Cryostats*

STJs need to be operated at temperatures lower than about a tenth of their critical temperature, in order to freeze out all thermal excitations of the Cooper pair bath. Nb has a $T_C$ of about 9.3 K. Therefore, for Nb based STJs a pumped $^4$He bath with a base temperature of about 1.2 K is already sufficient for a thermal excitation free operation of the devices. Ta has a critical temperature of about 4.5K and therefore needs a colder environment than the Nb based junctions. Here, a cryostat based on a pumped $^3$He reservoir is useful, which has a base temperature of about 300mK. The same is true for V based junctions, which have a $T_C$ of 5.4 K. For junctions with an even lower critical temperature, like junctions based on Al (1.2K) and Mo (0.915K), even colder environments have to be created. In order to operate these junctions, an Adiabatic Demagnetisation Refrigerator (ADR) with a base temperature of about 35mK [White 02] is used.

### 3.1.1 $^3$He sorption cooler

The $^3$He cryostat is cooled by means of a closed $^3$He reservoir. First, a small reservoir of liquid $^4$He (1K pot) is pumped out with a rotary pump, which cools the small $^4$He reservoir to a temperature of about 1.5K. In the part of the $^3$He reservoir, which is in close contact with the 1.5K environment, the $^3$He gas condensates and aided by gravity falls to the bottom of the reservoir. When most of the gaseous $^3$He is condensed, a piece of charcoal located inside the $^3$He reservoir is cooled from 40K to about 4K. At temperatures below 20K the adsorption capabilities of the charcoal become important and the reservoir of $^3$He is effectively pumped out. This lowers the temperature of the liquid $^3$He in the bottom of the reservoir to about 300mK and as a consequence lowers the temperature of the sample space, which is physically attached to the bottom of the $^3$He reservoir. If the $^4$He in the pumped 1K pot is constantly refilled, the hold time of the system is limited by the evaporation rate of the liquid $^3$He at the bottom of the cold finger. Of course the hold time will depend strongly on the thermal load on to the cold finger, but typically for STJ operation, the temperature is stable at about 300mK for 8 to 12 hours. The two $^3$He cryostats in the laboratories of the Research and Space Science Department are top loading Heliox coolers fabricated by Oxford Instruments [OI]. The whole cryostat is magnetically shielded from the earth's magnetic field by a double mu-metal shield. Inside the cryostat a superconducting magnet immersed in the liquid $^4$He bath provides the magnetic field parallel to the junction's insulating layer, which suppresses the Josephson currents of STJs. Figure 3.1(a) shows a schematic of the $^3$He sorption cooler, whereas Fig. 3.1(b) shows a picture of the cold part of the insert.

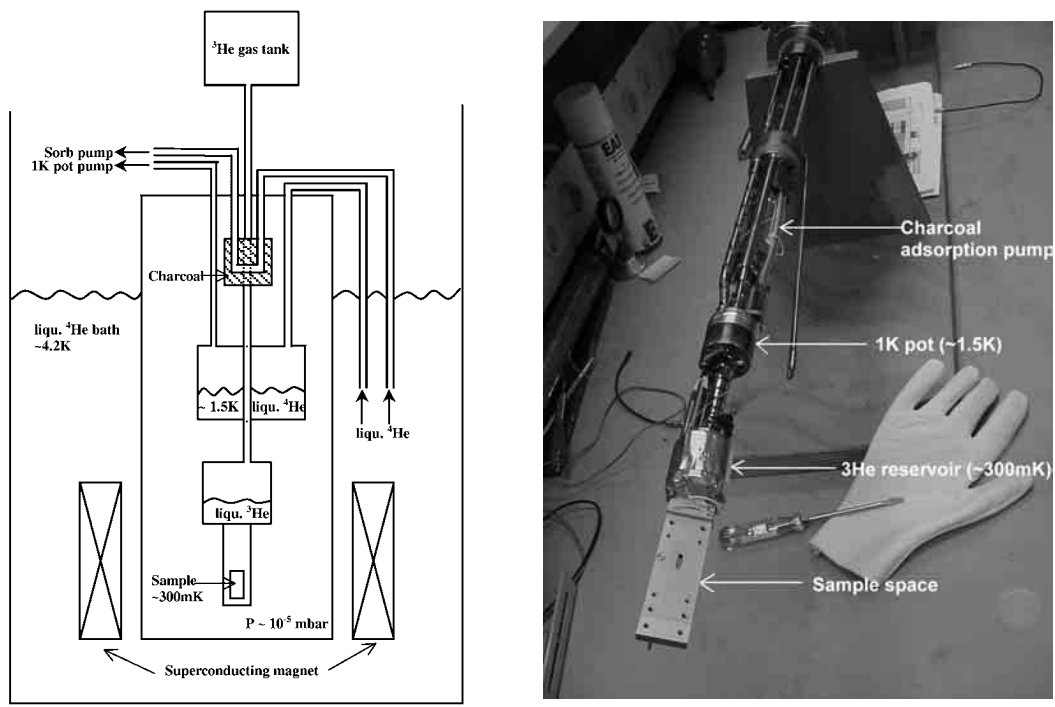

**Figure 3.1: (a)** Schematic of the $^3$He sorption cooler. **(b)** Photograph of the $^3$He sorption cooler insert.

### 3.1.2 Adiabatic Demagnetisation Refrigerator (ADR)

The ADR is cooled through the disorganisation of the dipoles of a previously ordered paramagnetic material. When a strong magnetic field is applied to a paramagnetic material at low temperature, the randomly distributed dipoles of the material will tend to align with the external magnetic field and the entropy of the paramagnetic material is reduced considerably. This organisation of the dipoles generates heat, which has to be fed into a heat sink via a heat switch. After full magnetisation of the paramagnetic material and thermalisation to the temperature of the heat sink, the heat switch is opened and the paramagnetic material as well as the sample space, which is physically attached to it, are thermally isolated from the environment. Then, the external magnetic field is adiabatically reduced, which causes the ordered dipoles of the paramagnetic material to relax and to extract entropy from the crystalline structure of the paramagnetic material as well as the sample space and, as a consequence, reduces their temperature.

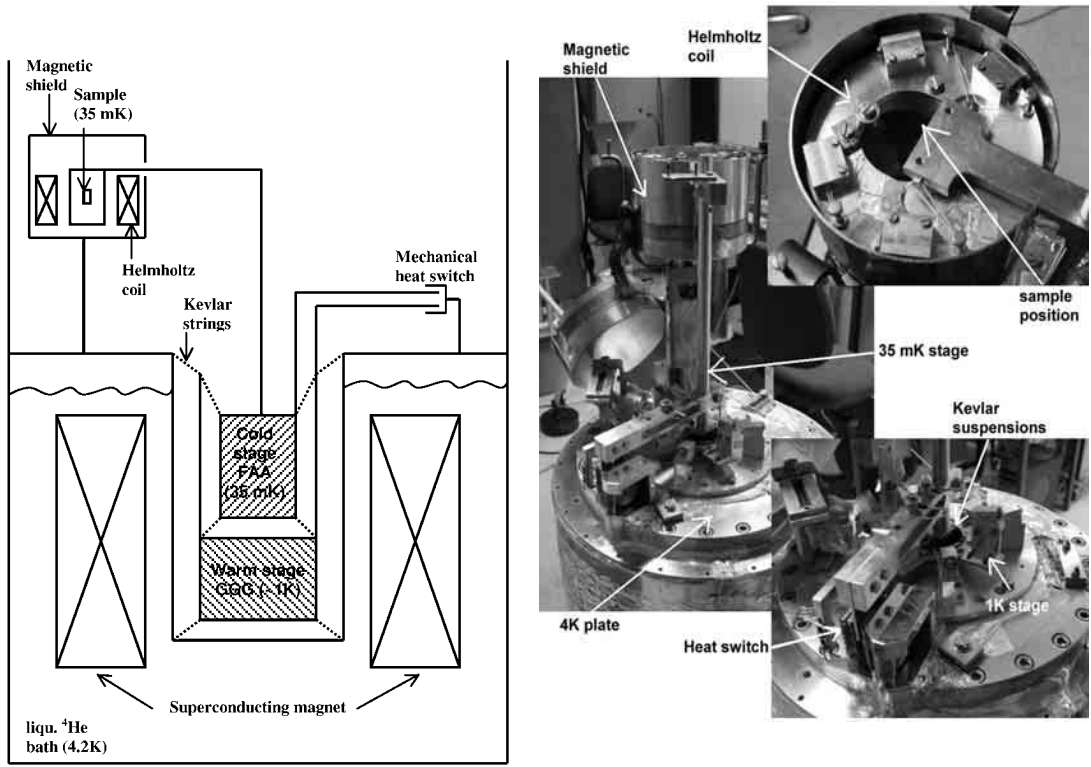

**Figure 3.2: (a)** schematic representation of the two stage ADR system. **(b)** photograph of the ADR. The big picture shows the system without the vacuum cover. The picture in the top corner shows a close-up of the sample area, whereas the lower corner picture shows a close up of the 4K plate with the heat switch. The magnet and the two paramagnetic salt pills are hidden below the 4K plate.

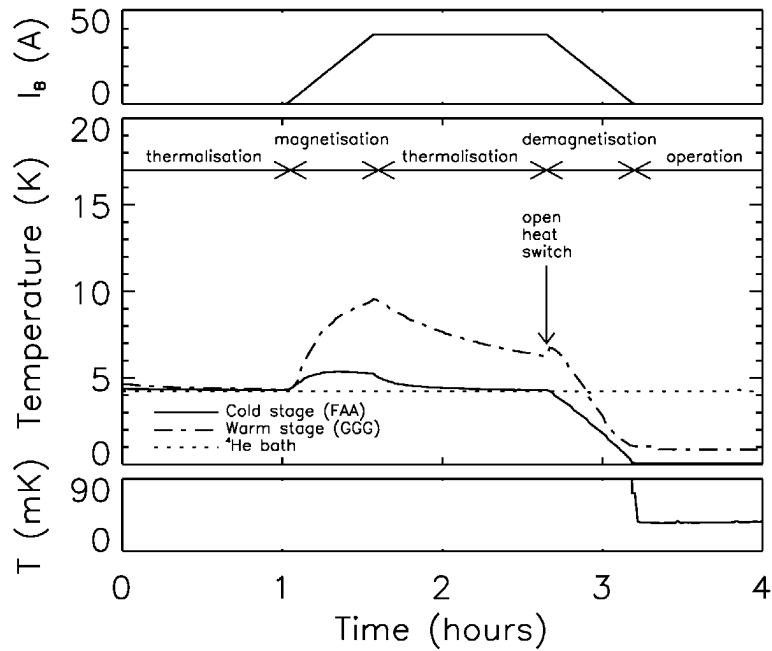

**Figure 3.3:** Cool down cycle of the ADR. The middle plot shows the variation of the temperatures of the cold stage (FAA), the warm stage (GGG) and the $^4$He bath as a function of time. The lower plot gives a close-up of the very low temperature region (0 - 90 mK), whereas the upper plot shows the variation of the current in the superconducting magnet.

The ADR system, which was used within the framework of this thesis, is a two stage ADR acquired from Vericold Technologies [Vericold]. A schematic of the two stage ADR system is shown in Fig. 3.2(a), whereas Figure 3.2(b) shows a photograph of the system used.

The ADR consists of two concentric paramagnetic salt pills. The high temperature stage uses as a paramagnetic material Gadolinium Gallium Garnet (GGG) and has a base temperature of ~1K. Its function is to reduce the heat flow from the $^4$He bath, which is at a temperature of 4.2K, to the cold stage comprising the sample area. The cold stage uses Ferric Ammonium Alum (FAA) as a paramagnetic material and has a base temperature of ~35mK. Both paramagnetic pills are suspended via a set of 14 very resistant Kevlar strings to the 4.2K environment, which serves as a heat bath. This suspension system with the high thermal impedance Kevlar strings thermally isolates the two pills from the environment. Both pills can be directly thermally connected to the bath via a mechanical switch, which can be opened and closed using a stepper motor. The system of pills is surrounded by a single superconducting magnet immersed in the $^4$He bath. This magnet is able to produce a magnetic field of up to 7 Tesla at its centre and magnetises both stages simultaneously. The sample space, which is physically connected to the low temperature paramagnetic material, is shielded from the magnetic fields created by the magnet via a small mu-metal box. This shielding reduces the stray field in the sample area to less than a Gauss during the complete cool-down procedure. Inside the magnetic shielding a small superconducting Helmholtz coil creates the magnetic field parallel to the junction, which is necessary for the Josephson current suppression.

Figure 3.3 shows a complete cool-down cycle of the ADR. First the heat switch is closed, which thermally connects both the cold and the warm stage paramagnetic pills to the 4.2K bath. After complete thermalisation of the two salt pills to the 4.2K environment, the magnetic field is slowly ramped up to ~7 Tesla, which creates heat in the paramagnetic materials and warms them up. This heat is removed from the paramagnetic pills via the heat switch to the bath. After 30 minutes the maximum magnetic field is reached and the system is left in this state for about an hour in order for both pills to thermalise with the ${}^4$He bath again. Now the mechanical heat switch is opened and the two paramagnetic materials are thermally isolated from the bath. The magnetic field is now slowly removed. The whole "adiabatic" demagnetisation of the magnet takes about 30 minutes. The removal of the magnetic field causes the paramagnetic materials to cool down to their respective base temperatures, which is reached as the magnetic field is reduced to zero. Note that there is a possibility to control the temperature, as the application of a magnetic field will cause the dipoles of the paramagnetic materials to align with the field again, which causes them to release heat and to warm up the crystalline structure to a temperature, which is directly related to the value of the applied magnetic field. In our case absolute temperature stability is not crucial and the temperature is not controlled by means of a magnetic field. The temperature just slowly drifts up, as a small heat flux from the warm surroundings warms the cold paramagnetic materials up. Typically, the temperature of the cold stage naturally drifts from 35 mK to 100mK in about 10 to 12 hours.

### 3.2 Electronics

The junctions tested in this thesis are single pixel junctions, which means that every single junction has a separate contact to the top electrode of the junction and is read out individually. In general, the contact to the base film is in common between all the junctions on a chip and set to ground. Therefore, a chip comprising n junctions needs n+1 read-out wires going from the chip at low temperatures to the read-out electronics, which in our case are at room temperature. In all cases the first part of the wiring going from the cold stage to the bath at 4.2K is made out of superconducting Nb-Ti wires, in order to reduce the heat load on the cold stage. For the part of the wiring going from the bath to room temperature several options are possible, depending on the application. If one wants to reduce the resistance of the wiring, as necessary for X-ray photon detection and IV-curve tracing, Cu wires are best suited. In case of optical photon detection, the resistance of the wiring is of no importance and the heat load on the bath can be minimised by choosing a low thermal conductance material, like Mn.

Two different room temperature electronic set-ups are possible. IV-curve tracing is a very effective junction diagnostic technique, whereas a pulse height analyser set-up is necessary for photon detection experiments with the STJs [Knoll 00].

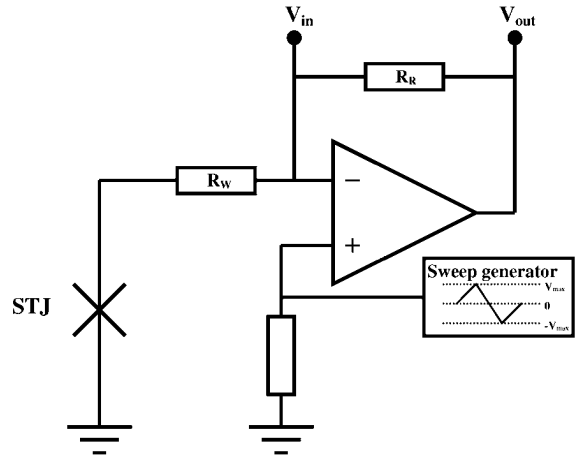

**Figure 3.4:** Schematic of the IV-curve tracer set-up.

### 3.2.1 IV-curve tracer

IV-curve tracing is a good tool for accessing junction characteristics like sub-gap currents and normal resistance of the junction or sum gap of the two superconducting electrodes. Two IV-curve sweep modes are in general possible, the current sweep and the voltage sweep mode. In our set-up the voltage sweep mode is used, as it gives information about the sub-gap currents, even if the zero-voltage Josephson current or the Fiske resonance current steps are not completely suppressed. The layout of the IV-curve tracer is schematically shown in Fig. 3.4. The voltage $V_{in}$ applied to the top electrode of the junction is stepwise increased from zero to a maximum value $V_{max}$. At every voltage step the output voltage $V_{out}$ is read and stored together with $V_{in}$. Then the voltage is decreased stepwise towards zero and continues into the negative voltage region to perform the same kind of cycle in the negative voltage domain. For every read-out step the current passing through the junction $I_{STJ}$ and the voltage across the junction $V_{STJ}$ are given by:

$$I_{STJ} = \frac{V_{out} - V_{in}}{R_R} \tag{3.1}$$

$$V_{STJ} = V_{in} - I_{STJ}R_W, \tag{3.2}$$

where $R_w$ is the resistance of the wiring going from the room temperature electronics to the junction at low temperature and $R_R$ is the range resistor. Three different range resistors are available (100$\Omega$, 10k$\Omega$ and 1M$\Omega$), depending on the maximum current which has to be supplied to the junction.

### 3.2.2 Photo-pulse analyser

The room temperature pulse height analyser consists of a charge sensitive pre-amplifier followed by a pulse shaping stage peak detector and AD converter. The schematic of the standard set-up is shown in Fig. 3.5.

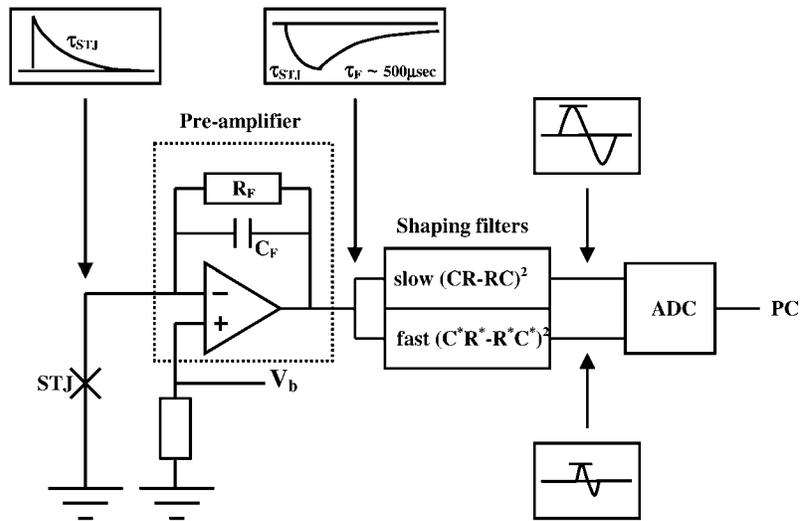

**Figure 3.5**: schematics of the read-out electronics

The signal created by the detector is fed to the input of the pre-amplifier, where it is integrated to yield a total charge output. A complete range of pre-amplifiers is available, which all have a RC time constant of the order of 500 μsec. Depending on the application, the feedback capacitance takes values between 1pF (optical photon detection) and 1nF (X-ray photon detection). At the output of the pre-amplifier one has a signal with a fast rise time, corresponding to the decay time of the detector pulse, and a slow exponential decay, corresponding to the RC time constant of the integrator in the pre-amplifier.

This signal is then simultaneously fed into two semi-Gaussian bipolar CR-RC-CR-RC shaping filters, which limit the noise bandwidth and the pulse duration. The first differentiation stage performs the pole zero cancellation, which prevents any undershoot in the signal after the first differentiation-integration stage. At the outputs of the shaping filters one now has a bipolar signal with a maximum quickly followed by a minimum. The maxima of these signals are then measured by means of an analogue to digital converter. Both values are transmitted via a fibre link to a PC, which stores the values for every pulse.

One of the two shaping stages used has a slow RC time constant and one presents a faster time constant, defining the slow and fast channel. The two different channels perform the charge output and rise time determination. The slow channel time constant is chosen so that the maximum of the filter output signal is representative of the fully integrated pulse. On the other hand, the fast channel time constant is chosen in order for the maximum of the filter output signal to be representative of the first, fast rise part of the pre-amplifier pulse. As a consequence, the output of the slow channel presents a measure of the total charge output of the detector, whereas the ratio of the output of the slow channel to the output of the fast channel represents a measure of the decay time of the detector pulse. Three different sets of shaping filters were used for spectral analysis during this thesis. The first set has a slow channel centre frequency of 5 kHz and a fast channel centre frequency of 33 kHz. This set is used for optical photon detection, for which the detector signals typically present decay times longer than 10 μsec. A second set is used for the much faster X-ray detection experiments, where the decay time of the signals is typically of the order of a microsecond. The slow channel of this shaping filter has a centre frequency of 17 kHz, whereas the fast channel centre frequency is of the order of 328 kHz. A third set of shaping filters was used, which has centre frequencies of respectively 16 and 37 kHz.

For every combination of pre-amplifier and shaping stage a calibration was made using a waveform synthesizer. Signals with an instantaneous rise-time and an exponential decay were fed into the preamplifier. For every combination of pre-amplifier and shaping stage the full-scale charge of the slow channel was measured as a function of decay time of the input signal. The ratio of the slow channel output to the fast channel output as a function of decay time of the input signal was determined as well. In practice, the signals produced by the junctions do not have an instantaneous rise time and a perfect exponential decay. Therefore, there exists a certain uncertainty in the determined values of the decay times of the detector pulses.

## 3.3  Photon sources

STJs are operational as photon detectors for photon energies ranging from the near-IR to X-ray energies. In order to get information on the performance in the different energy domains several different photon sources have to be employed. In this thesis three different photon sources were utilised for testing the junctions.

### 3.3.1  Near-IR and optical light source

For near-IR and optical measurements a Xenon lamp is used as a light source in combination with a double grating monochromator, able to produce photons having a wavelength varying from 250 to 1000 nm. Via an optical fibre the output of the monochromator is coupled to the detector space. The illumination of the junctions is made through the back of the chip via the sapphire substrate.

### 3.3.2  $^{55}$Fe X-ray source

For X-ray photon detection experiments a small $^{55}$Fe source is used for illumination of the junctions. The $^{55}$Fe decays via capture of an orbital electron to $^{55}$Mn. The vacancy created in one of the inner shells, most often the K shell, is subsequently filled, accompanied by the emission of the characteristic X-rays. The decay of $^{55}$Fe emits predominantly X-rays from the Mn-$K_\alpha$ series with an energy of ~5.9 keV. The radioactive source is placed directly above the sample at a distance of about 5 mm, so that ideally about 50 to 100 events are detected per second in the junction.

### 3.3.3  Synchrotron radiation

Some of the experimental results in this thesis involve photon counting experiments in the energy domain between optical and 6 keV X-ray energies. These experiments were made at the BESSY II synchrotron radiation facility in Berlin. The SX-700 plane grating monochromator in the laboratories of the Physikalisch Technische Bundesanstalt (PTB) covers the region from 30 to 2000 eV with very high energy resolving power. The junctions were operated in a small portable $^3$He cryostat with a base temperature of 330 mK. Through a window in the cryostat the detectors can be directly coupled to the beamline. A small Al window of 100nm thickness installed in the aperture of the window prevents the detection of stray IR photons from the warm beamline environment.

# **Chapter 4**

# **Junction fabrication and properties**

The quality of the tunnel junction is of major importance for the successful operation of STJs as photon detectors. Mainly the insulating barrier separating the two superconducting electrodes needs to fulfil several stringent conditions. The insulating layer needs to be continuous and very homogeneous over the whole area of the detector, in order to reduce leakage currents and spatial variations, which would degrade the energy resolution of the detector. On the other hand the insulating layer also needs to be extremely thin, in order to increase the probability of quasiparticle tunnelling from one electrode to the other. These two competing properties of the insulating layer make the fabrication of junctions with high responsivity a very delicate task. All the vanadium, aluminium and molybdenum-based junctions in this thesis were fabricated by Cambridge MicroFab Ltd to specifications provided by the STJ research team at the European Space Agency. In the following the detailed fabrication steps used for the production of the junctions are presented as well as the properties of the materials forming the electrodes. Then, the quality of the insulating barrier between the two electrodes of the junctions is inspected by analysing the current-voltage characteristics of the junctions.

## *4.1 Junction fabrication*

In this section the fabrication of three different types of tunnel junctions for photon detection is described. The first type is based on an electrode consisting of a vanadium (V)-Al bi-layer, the second is a junction based on pure Al and the third type is based on a molybdenum (Mo)-Al bi-layer. The exact fabrication procedure for the junctions depends on the nature of the material used. Nevertheless, a general fabrication procedure can be sketched and is described in Fig. 4.1.

**(a)** The junctions are deposited on a polished r-plane (1-102) sapphire substrate acquired from Kyocera Industrial Ceramic Corporation [Kyocera]. The wafers are first cleaned and then loaded into a four-station UHV deposition system. One of the stations is used for the ion beam miller, whereas the other three stations have very pure (99.99% purity) V, Mo, Al, Nb, Ta or Si targets loaded, depending on the type of junction to be deposited. The system is pumped down to a residual base pressure of $\sim 10^{-10}$ mbar. Then, typically 100 nm of the absorbing, usually higher $T_C$ material are deposited by DC-sputtering onto the sapphire wafer. The temperature of the wafer is kept at the optimum deposition temperature with respect to flatness and epitaxy of the deposited film. This optimum temperature depends of course on the nature of the material deposited. The available temperatures at which the films can be deposited range from liquid nitrogen temperature to 800° C.

**(b)** Without breaking vacuum the wafer is moved to the next deposition station. On top of the ideally very flat and epitaxial base film is then sputtered an Al film with a thickness typically ranging from 5 to 100 nm, depending on the application and the flatness of the underlying film. Al is used because of its good wetting characteristics and the very favourable properties of the aluminium oxides (AlOx), which allow the growth of the extremely thin and transparent tunnel barrier. The deposition temperature of this Al film is usually -120° C, a deposition temperature for which the Al film shows the best flatness. In general this film is not epitaxial, but forms vertical crystals with a lateral diameter of ~50nm.

**(c)** The top surface of this ideally flat Al layer is then oxidised in an atmosphere of Ar and $O_2$. In order to obtain a different insulator layer thickness for different tunnel barrier characteristics, the wafer temperature, partial $O_2$ pressure and the oxidation time can be varied. At the end of the oxidation procedure a very thin (~1nm) and homogeneous AlOx film should have grown on top of the Al layer. From this moment on the temperature of the wafer should not exceed anymore 120° C, as otherwise the characteristics of the tunnel barrier will degrade due to atom migration.

**(d)** After pump-out of the deposition system to base pressure, another Al film is deposited on top of the AlOx insulating barrier. The temperature of the substrate is kept stable at about -120° C.

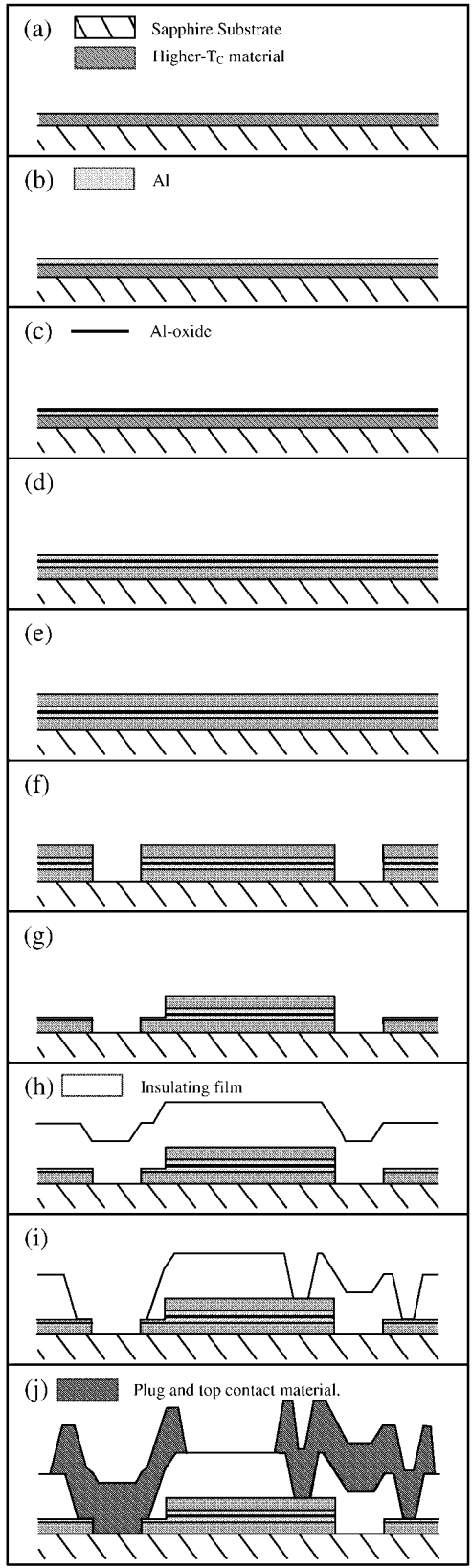

- (a) Sapphire Substrate / Higher-$T_C$ material — • DC-sputtering of base, higher-$T_C$ material film.

- (b) Al — • DC-sputtering of base Al film.

- (c) Al-oxide — • Oxidation of the base Al film.

- (d) — • DC-sputtering of top Al film.

- (e) — • DC-sputtering of top, higher-$T_C$ material film.

- (f) — • Base etch (Definition of STJs and leads).

- (g) — • Mesa etch (Remove top layers and barriers from leads).

- (h) Insulating film — • Deposition of insulating layer.

- (i) — • Fabrication of vias through the insulating layer.

- (j) Plug and top contact material. — • Deposition of plugs and top contacts via lift-off procedure.

**Figure 4.1**: Fabrication sequence of the V-Al, Al and Mo-Al superconducting tunnel junctions.

**(e)** Still without breaking the vacuum another higher $T_C$ material layer is deposited at room temperature on top of the Al film. Then, the wafer is removed from the deposition system and a natural oxide layer of 2-10 nm thickness, depending on the nature of the material, forms on top of the multi-layer.

**(f)** UV lithography is used for patterning the resist for the following steps. First, a Shipley S1813 resist [Shipley] is spin-coated onto the multi-layer, soft-baked, developed and finally hard-baked. Then, the base etch is performed, which defines the geometry of the junctions. The base etch goes through the complete multi-layer. The etch profile should be as vertical as possible, in order to avoid energy gap variations at the edges of the junctions, but in practice the edge profile will always have non-vertical slopes and steps [den Hartog 01]. Depending on the material of the multi-layer, several methods can be used for performing the base etch. Some processes used by Cambridge MicroFab are: anodisation, ion beam milling, wet etching or plasma etching with SF6 plasma. Sometimes several of the preceding methods have to be used on the same multi-layer, because of strong etch selectivity for some material combinations. The different methods used with the different materials will be explained in more detail in the next section.

**(g)** The base etch is followed by the mesa etch, which only etches through the top electrode and is stopped as soon as the etch went through the aluminium oxide layer. This step is necessary for making a contact to the base film. In the same way as for the base etch, several processes can be utilised, depending on the materials to be etched, including anodisation, ion beam milling, wet etching and plasma etching. The important point for this step is the exact determination of the end point, in order to leave the base electrode as far as possible untouched.

**(h)** The complete wafer is then electrically isolated by depositing an approximately 300 nm thick layer of insulator on top of the processed multi-layer. Several choices of materials were tested including a photosensitive epoxy called SU-8 [MicroChem], a homemade epoxy based on hardener and resin dissolved in photoresist solvent, and a reactively sputtered silicon oxide. All materials have good insulating properties. The difficulty lies in making the openings through the dielectric for the contacts and plugs.

**(i)** In order to make the contacts to the base and top electrodes, vias have to be created through the dielectric film. Depending on the nature of the dielectric, different methods are utilised. The vias through the SU-8 resist are photo-patterned, whereas $SF_6$ plasma etching is used for making the openings through the homemade epoxy and the silicon oxide. Reliable and complete opening of these small, typically 1.5 μm wide vias is very important, as even a very thin residual layer of dielectric will result in open-circuit junctions or junctions with a very large series resistance, which prohibits any low-noise operation.

**(j)** Prior to top contact and plug deposition, the free surfaces of the electrodes are first cleaned with the ion beam miller in order to get rid of eventual impurities and natural oxide films on top of the metals. Finally, as a last processing step the top contacts and plugs are deposited at room temperature in order to make a contact from the top and base electrodes to the corresponding wiring. These top contacts and plugs are made out of a higher $T_C$ material than the electrodes, in order to prevent out-diffusion of the quasiparticles out of the electrodes into the leads. As top contact and plug material Nb and Ta were used. The plugs and top contacts are patterned through a two resist lift-off procedure.

In the following the processing particularities of every material combination treated in this thesis is described. For every lay-up the process route that gives the best results with respect to leakage performance is described.

### 4.1.1  V-Al technology.

First 100 nm of base V are deposited at a substrate temperature of 650° C. Then 30 nm of Al are sputtered at a temperature of –120° C followed by oxidation of the Al film. The multi-layer is then completed by the deposition of 30 nm of Al at -120° C followed by the deposition of 100 nm of V at room temperature. The main specific change in the process route concerning the processing of the V-Al multi-layer is that the mesa etch is carried out before the base etch. The mesa etch, which removes only the top electrode and should stop right after penetration of the AlOx layer, consists of two subsequent operations. First, the top V is removed by anodic electro-dissolution, followed by removal of the top Al until the barrier is penetrated.

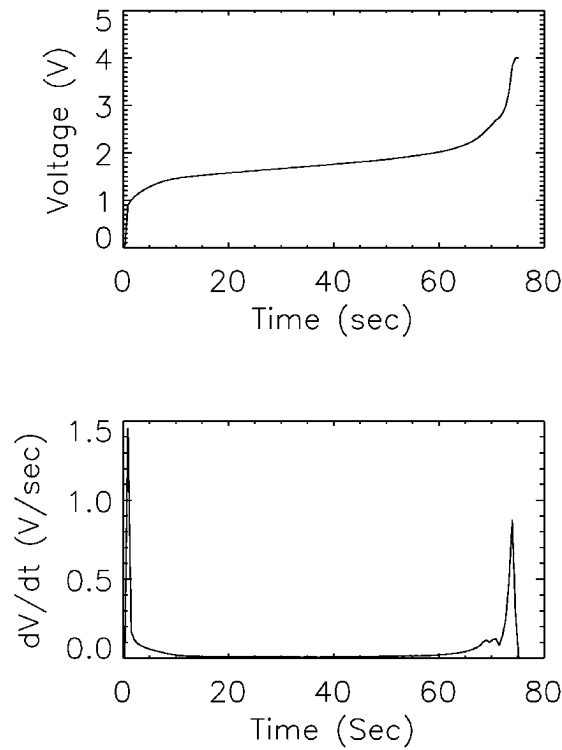

**Figure 4.2: (a)** Variation of the voltage difference between anode and cathode during the electrolytic dissolution of a 100 nm thick V film. **(b)** Variation of the anodisation speed. The steep rise of the speed indicates that the complete V film is oxidised.

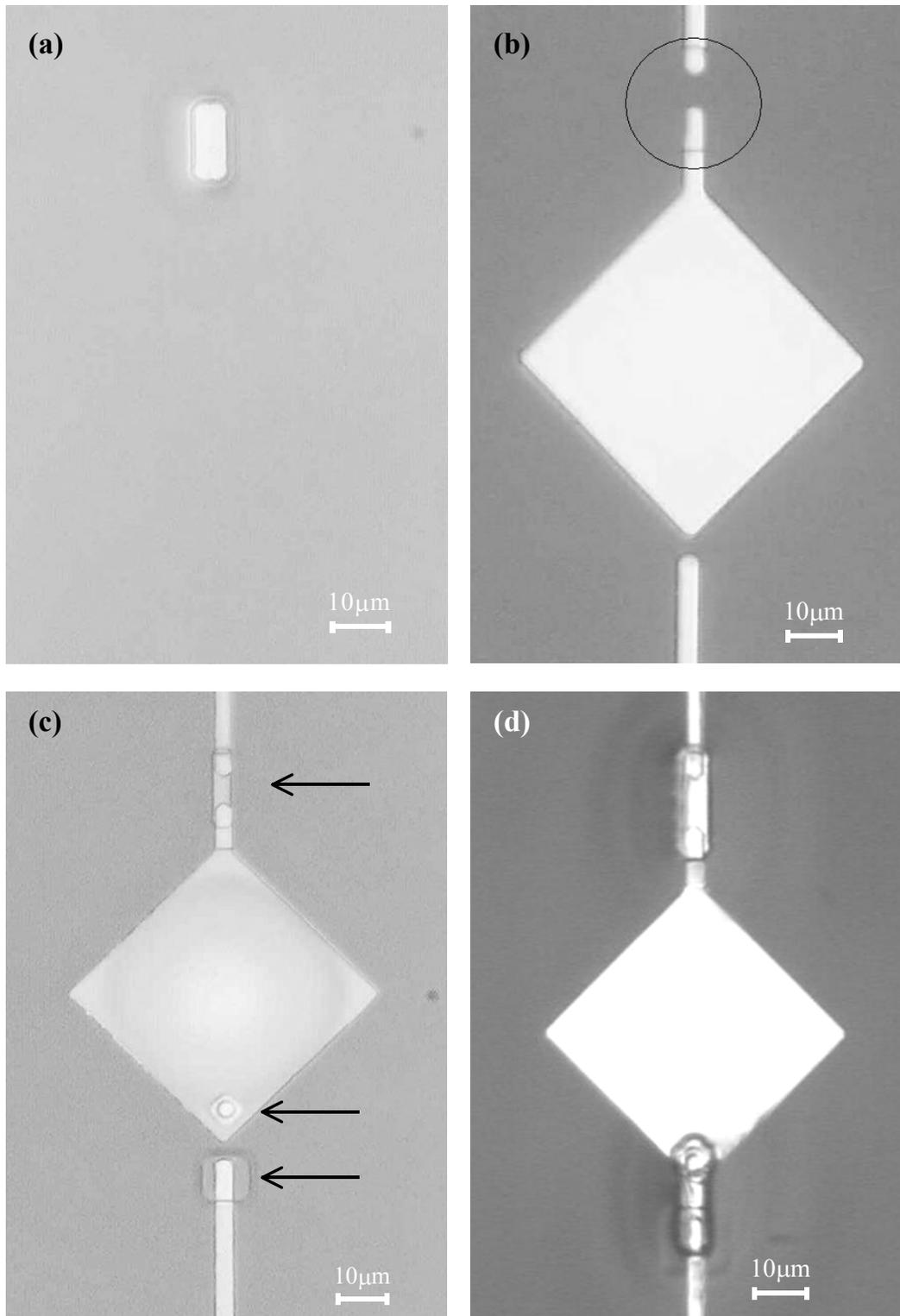

**Figure 4.3:** Optical microscope image of a V-Al multi-layer: **(a)** After mesa etch. **(b)** After base etch. The circle highlights the mesa etched region. **(c)** After opening of the vias through the SU-8 dielectric. The arrows indicate the positions of the three vias through the dielectric. **(d)** After plug and top contact deposition.

For the anodic electro-dissolution, the multi-layer is patterned with photoresist in a way that only the V that has to be oxidised is exposed to an electrolytic solution. The anode of an electrical circuit providing a constant current is electrically connected to the multi-layer and the cathode is immerged into the solution. The following reaction then takes place at the anode:

$$2V + 5H_2O \rightarrow V_2O_5 + 10H^+ + 10e^-. \tag{4.1}$$

Figure 4.2 shows how the voltage difference between anode and cathode varies during the dissolution process. The increasing voltage reflects the increasing thickness of oxidised V. The oxidation speed (dV/dt) is shown as well. As soon as all of the V is transformed into $V_2O_5$, the oxidation speed varies, as the speed with which the O atoms move through the film depends on the nature of the film. The current is then removed and the electro-dissolution process stopped. The top Al film is then wet-etched by immersion of the wafer into a phosphoric acid solution. Figure 4.3 (a) shows an optical microscope image of a V-Al multi-layer after the mesa etch process. The resist for the mesa operation is then removed and replaced by the pattern for the base etch operation. The base etch is made by immersion of the wafer into a phosphoric acid solution until the complete multi-layer is etched away. The end point is determined by visual inspection of the wafer. Figure 4.3 (b) shows an optical microscope image of a V-Al multi-layer after the base etch process. The geometry of the junction as well as the top and base film leads can now be recognised. The circle highlights the previously mesa etched region. The wafer is now spin-coated with a photosensitive epoxy called SU8. The thickness of the insulating film is close to 350 nm. The vias in the insulating layer are photo-patterned. Figure 4.3 (c) shows an optical microscope image of a junction after photo-patterning of the vias through the dielectric. The three arrows indicate the vias through the dielectric, one for the base lead plug and two for the top contact. As a final step Nb for the top contacts and plugs is deposited and patterned with a two-layer resist lift-off procedure. First, a 1.3 μm thick layer of Shipley S1813 resist is spin-coated and flood exposed with UV radiation. On top of this is spin-coated a 1.5 μm thick layer of Shipley SPR220 resist, which is normally exposed to UV radiation. When the resists are developed, the developer will dissolve the lower resist film faster than the upper film, which undercuts material from the lower resist layer. The arising resist pattern is shown in Fig. 4.4. When 400 nm of contact material are then sputtered onto this resist pattern, the material that goes through the opening will not make a contact with the material that condenses on top of the resist. When the resist is dissolved, the material on top of the resist can be nicely separated from the wafer. In practice however the resist pattern is not as perfect as shown on the graph, because the upper resist cannot be correctly hardened due to the upper temperature limit of 120° C, needed for leaving the insulating AlOx barrier intact. Therefore, the sidewalls do not show the nice roof structure as shown in Fig. 4.4, but have rather vertical sidewalls. When the contact material is then sputtered onto this two-layer resist, the material that will go through the opening is still in contact with the material that condensed on top of the resist via material that condensed on the vertical sidewalls of the resist. When the resist is then dissolved the contact material that is lifted off the wafer exerts a force onto the material in the opening, which gives rise to the edge pattern of the contact material, as shown in Fig. 4.5. This figure shows a scanning electron microscopy (SEM) picture of a Nb top contact patterned by the described lift-off procedure. The edges are curled upwards because of the force exerted by the Nb that was lifted off the wafer via the material deposited on the vertical sidewalls of the resist. Figure 4.3 (d) shows the completed junction after deposition of the Nb contact material.

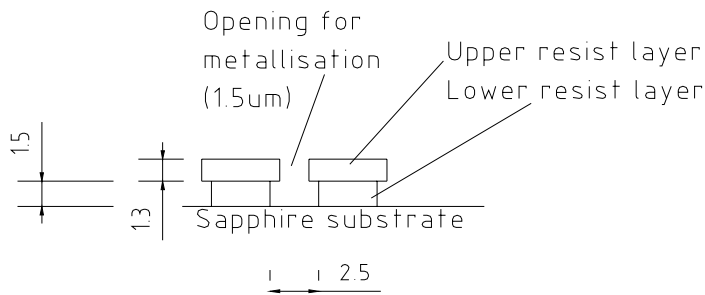

**Figure 4.4:** Ideal resist pattern for the two-layer lift-off procedure.

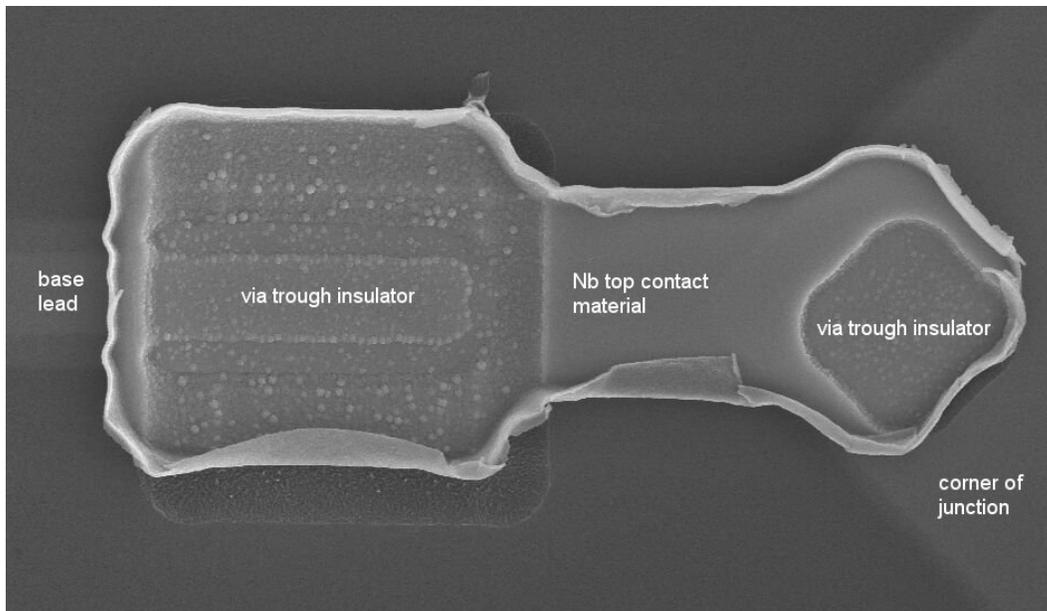

**Figure 4.5:** SEM picture of a Nb top contact patterned with the two-layer resist lift-off procedure.

### 4.1.2  Al technology.

100 nm of Al are DC-sputtered onto the R-plane sapphire substrate at a wafer temperature of -120° C. This layer is then partly oxidised in order to form the insulating barrier. On top of this oxide barrier is then sputtered another 50nm of Al. The wafer is taken out of the deposition system and resist is spin-coated and photo-patterned for the base etch step. The base etch is done by immersion of the wafer into a phosphoric acid solution until the whole Al film is etched through. The end point is reached after a few minutes and determined by visual inspection of the wafer. Now the base etch resist is applied for the mesa etch step. The mesa etch is performed with a neutralised ion beam miller. The Ar ions that are accelerated by means of an acceleration potential etch away the Al film at a constant etch rate of several nm per minute. When approximately 60nm of Al are etched away, enough to go through the complete top electrode and the barrier, the ion beam milling is stopped. The wafer is then covered by a 300 nm thick film of reactively sputtered silicon oxide (SiOx), which acts as the insulator for the top contacts. The SiOx is reactively sputtered from a high purity Si target in an $O_2$ environment. The vias through the SiOx are created by plasma etching with a sulphur hexafluoride ($SF_6$) plasma. Finally, the top contacts are patterned with the two-layer resist lift-off procedure in the same way as for the V-Al based devices.

### 4.1.3  Mo-Al technology.

50 nm of Mo are deposited on the R-plane sapphire substrate at a temperature of 800° C. On top of this 15 nm of Al are DC-sputtered at a temperature of –120° C. Then, the Al is partly oxidised. On top of this are deposited another 15 nm of Al and 50 nm of Mo. The mesa etch is a two stage process. First the top Mo is removed with an $SF_6$ plasma and then the Al is etched with the usual phosphoric acid wet etch solution. The base etch is made by ion beam milling the sample for several minutes. Then the wafer is covered by SU8. The vias through the SU8 are created with an $O_2$ plasma etch. Finally the Nb top contacts and plugs are deposited and patterned with the usual two-layer lift-off process.

## *4.2  Material characteristics and single film qualities*

Several properties of the materials used for fabricating the electrodes of the detectors are of utmost importance for the performance of STJs as photon detectors. For the best performance all possible quasiparticle loss channels have to be eliminated from the electrodes. The main loss channel for quasiparticles in the superconducting electrodes is quasiparticle trapping into regions that present a lower energy gap than the surrounding material [Poelaert 99a]. Such regions can be normal metal inclusions present because of impurity atoms or metallic oxide inclusions, as well as localised states below the gap formed by dislocations and magnetic impurities. It should therefore be taken care of that none of the materials forming the electrodes of the detector create natural oxides that are in the metallic state. Also, the lowest impurity and dislocation level of the materials should be achieved. In the following the main properties of the three materials that were used as electrode materials in this thesis are presented as well as the quality of the single films sputtered in the laboratories of Cambridge MicroFab Ltd.

### 4.2.1  Vanadium

Vanadium forms a body-centred cubic lattice with a lattice parameter a = 3.03Å [Kittel 96]. When deposited on sapphire the V (001) plane forms an angle of ~3° with the R-plane of the substrate [Gutsche 95]. A series of epitaxial films of V were deposited at 550° C for which the film thickness was varied from 13 to 385nm. For all samples the resistivity was measured at room temperature and at a temperature of 10K. From these values the residual resistance ratio (RRR) can be deduced, which is defined by:

$$\text{RRR} = \frac{\rho_{300}}{\rho_{10}} = \frac{l_{10}}{l_{300}},\qquad(4.2)$$

where $\rho_{300}$, $\rho_{10}$, $l_{300}$ and $l_{10}$ are respectively the resistivities and mean free paths at 300 K and 10 K. At 300 K the resistivity is completely limited by electron-phonon scattering. Therefore, $\rho_{300}$ is a material characteristic, which is independent of the quality and geometry of the sample. In the same way the mean free path at 300 K is limited by electron-phonon interactions and does not depend on the purity or thickness of the sample. In this work a value $l_{300} = 2.8$ nm is adopted, which is an average of 4 different values found in the literature, which range from 1.8 to 4.9 nm [Gutsche 94, Radebaugh 76, Reale 74, Tsai 81]. On the other hand, at 10 K the electron-phonon scattering contribution to the resistivity and the mean free path is negligible and the resistivity and mean free path are completely defined by scattering of the electrons with impurities, dislocations or sample boundaries. The RRR is thus a measure of the mean free path at 10 K of the sample, which can simply be found by:

$$l_{10} = \text{RRR} \cdot l_{300}.\qquad(4.3)$$

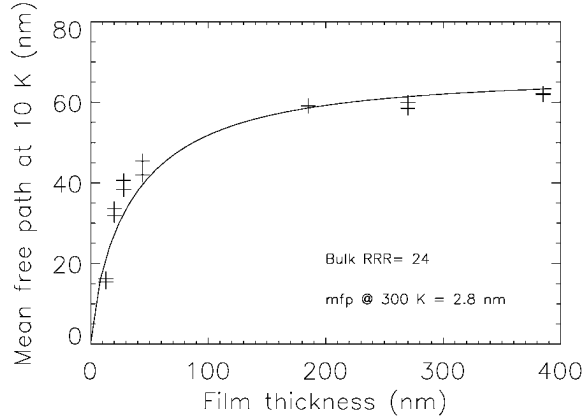

**Figure 4.6:** Mean free path at 10 K in V thin films with thickness varied between 10 and 400 nm. Experimental data (+) and fit to the data using (4.4) (solid line).

For films with a sample thickness smaller than the mean free path, the latter is limited by scattering at the boundaries of the film. The variation of the mean free path as a function of film thickness d is given by [Movshovitz 90]:

$$l(d) = l_0 + l_0^2 / d \left\{ \frac{3}{2} \left[ E_3(d/l_0) - E_5(d/l_0) \right] - \frac{3}{8} \right\},\qquad(4.4)$$

where the exponential integrals are defined by $E_n(x) = \int^\infty t^{-n} e^{-xt} dt$ and $l_0$ is the mean free path in the bulk material. Figure 4.6 shows the mean free path of the different V samples as a function of the thickness of the films. A fit to the experimental data using (4.4) is shown as well in the figure. A value for the bulk RRR equal to 24 gave the best fit to the data. From this the mean free path at 10 K in a 100 nm thick V film is deduced, which is equal to 52 nm.

V naturally reacts with $O_2$ to form $V_2O_5$ [Cotton 99, Rao 98]:

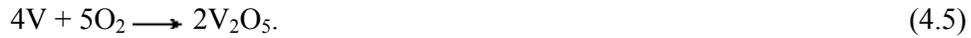

$$4V + 5O_2 \longrightarrow 2V_2O_5. \tag{4.5}$$

Therefore, a natural oxide film forms on top of the V layer. By resistivity measurements this thickness of the natural oxide film was determined to be ~7 nm [Strade 99]. $V_2O_5$ is a semiconductor with a bandgap of ~2eV [Eyert 98]. This vanadium pentoxide is contaminated by some of the other numerous vanadium oxide forms. V is able to combine with O in 2-, 3-, 4- and 5-valence states and forms a series of oxides out of which at least eight undergo metal to insulator transitions. $VO_2$ and $V_2O_3$ for example show a transition from metal at high temperature to insulator at low temperature at the respective transition temperatures of 340 and 150 K [Chudnovskiy 02]. VO on the other hand is metallic down to liquid helium temperature.

### 4.2.2 Aluminium

Al forms a face-centred cubic lattice with a lattice parameter a = 4.05 Å [Kittel 96]. As the aluminium films are deposited at liquid nitrogen temperature in order to optimise the flatness of the film, the sputtered Al does not form an epitaxial single crystal but rather a polycrystalline film. The RRR of such a 100 nm thick polycrystalline film is equal to 10 and probably limited by the interfaces of the different crystals composing the film. If one adopts a value for the mean free path at room temperature, which is equal to 41 nm [Reale 73], it can be concluded that the mean free path at 10 K in the 100 nm thick polycrystalline film is equal to 410 nm.

Al naturally reacts with $O_2$ to form $Al_2O_3$:

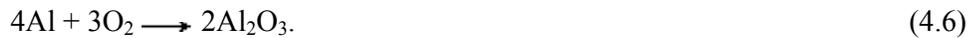

$$4Al + 3O_2 \longrightarrow 2Al_2O_3. \tag{4.6}$$

In this way a thin film of insulator forms on top of the Al film, of which the thickness was determined to be ~3nm [Peacock 00]. The $Al_2O_3$ is thermodynamically very stable and has very good dielectric properties down to liquid He temperatures.

### 4.2.3 Molybdenum

Mo forms a body-centred cubic lattice with a lattice parameter a = 3.15 Å [Kittel 96]. A series of single epitaxial Mo films were deposited on a sapphire substrate at a deposition temperature of 800° C, with film thickness ranging from 11 to 400 nm. Figure 4.7 shows the mean free path at 10 K of the different samples as determined from the measured RRR and the value of the mean free path of Mo at 300 K, which is equal to 16.1 nm [Reale 73]. The mean free path of all samples, including the thickest samples, is clearly limited by boundary scattering at the two surfaces of the films. The solid line on the figure shows a fit to the data using equation (4.4) and a value for the mean free path in the bulk material equal to 5.3μm. For a 100 nm thick film the mean free path at 10 K is approximately equal to 600 nm. At room temperature Mo does not react with air, but at elevated temperatures above 500° C the non-metallic trioxide, $MoO_3$, is formed:

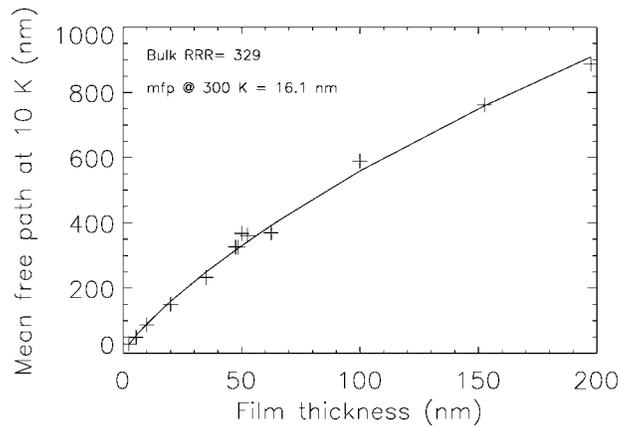

**Figure 4.7:** Mean free path at 10 K in Mo thin films with thickness varied between 10 and 400 nm. Experimental data (+) and fit to the data using (4.4) (solid line).

$$2Mo + 3O_2 \longrightarrow 2MoO_3. \tag{4.7}$$

This insulating molybdenum trioxide is contaminated by some of the other numerous molybdenum oxide forms. Mo is able to combine with O in 2-, 3-, 4-, 5- and 6-valence states and forms a series of oxides out of which at least $Mo_2O_3$ and $MoO_2$ are metallic with good conduction properties.

### 4.3 Multi-layer and AlOx barrier characteristics

The most important characteristic of a tunnel junction is the quality of its insulating barrier through which the quasiparticles will tunnel. The barrier needs to fulfil several stringent and contradictory conditions. First of all it needs to be as thin as possible, in order to allow fast tunnelling of the quasiparticles through the barrier. The tunnel time of a quasiparticle is directly proportional to the resistance of the insulating barrier (see section 2.3.1.1 for more information) and this tunnel time has to be minimised in order to achieve the highest possible collected charge in the pre-amplifier. On the other hand the barrier needs to be continuous and pinhole-free in order to reduce the leakage currents through the insulating barrier. To achieve an acceptable electronic noise level for optical photon detection experiments, leakage currents as low as several tens of picoampere have to be reached. In addition to these two conditions the insulating barrier needs to be as homogeneous as possible, in order to reduce variations of the charge output over the area of the detector.

In the following the characteristics of the multilayer and in particular the structure of the base films on which the AlOx barrier is grown are presented. Then the quality of the barrier is quantified by means of measured IV curves of the different devices and by analysing the Josephson current suppression pattern, which gives important information about the homogeneity of the tunnel current distribution over the area of the junction.

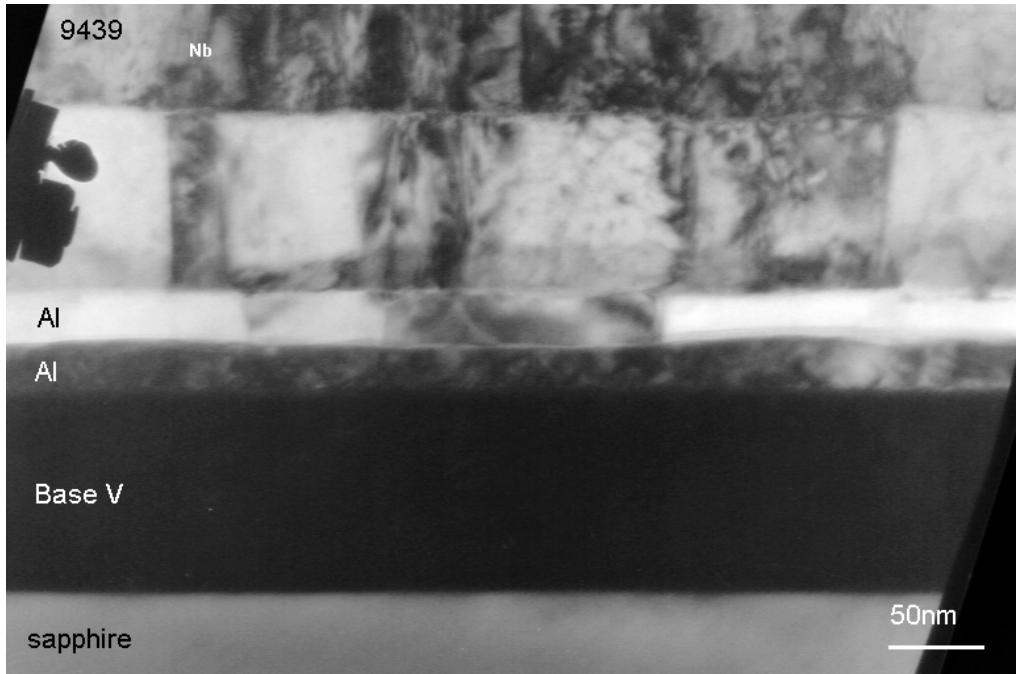

**Figure 4.8:** Transmission electron microscopy image of a V-Al based multi-layer.

### 4.3.1   V-Al based junctions

Figure 4.8 shows a transmission electron microscopy (TEM) image of a V-Al based multi-layer. The 100nm thick base V film is clearly epitaxial. Whether the Al film of the base electrode is epitaxial as well cannot be determined with certainty from this picture, but it can be seen that the flatness of the film is good. The thickness of the Al film varies between 20 nm and the nominal 30 nm. The thin, approximately 1 nm thin, AlOx insulating barrier can be seen on parts of the picture. The Al film deposited on top of the insulating barrier is polycrystalline with columnar crystal grains. The crystal size is of the order of 50 to 100 nm. The top V film is also polycrystalline with columnar grains showing a horizontal diameter of approximately 30-100 nm. The upper polycrystalline film is the Nb contact material.

#### 4.3.1.1   IV-curves

Figure 4.9 shows IV-curves of two symmetrical V-Al based junctions with a 100 nm thick V film and a 25 nm thick Al film. Figure 4.9 (a) shows the complete IV-curve of a 20 μm side-length junction. When the bias energy applied to the junction exceeds the sum of the gaps of the two electrodes, superconductivity is lost and the junction switches into the normal resistive behavior. For this reason the almost vertical transition between the superconducting junction behavior and the normal resistive behavior reveals the sum of the energy gaps of the two V-Al electrodes, which is equal to 1073 μeV. This value is determined from the intersection of the almost vertical dashed line, which is a fit to the gap region of the IV-curve, with the voltage axis.  As the lay-up of the junction is symmetrical we will assume the same gap in both top and base electrodes and we therefore conclude that the energy gap $\Delta_g$ in the V-Al bi-layer is equal to 536 μeV. The proximity effect between the superconducting V and Al films is responsible for this intermediate gap value, which lies in between the two values of the gap in the bulk materials respectively equal to 820 and 180 μeV (see section 2.2). In practice a small deviation between the energy gaps of top and base electrode exists because of the difference in film quality between the two electrodes. Nevertheless, this difference is very small and only of secondary importance. The normal state resistance of the junction can be determined on this figure as well. As superconductivity is lost in the junction because of the large energy furnished by the bias voltage, the normal resistive behavior of the metal-insulator-metal junction can be observed. The dashed line through the origin of Fig. 4.9(a) is a fit to the normal resistance $R_n$, which is equal to 0.29 Ω for this 20 μm side length junction. This corresponds to a normal resistivity $\rho_{nn}$ of 1.16 μΩ cm². This value is lower than for comparable Nb-Al and Ta-Al junctions, for which normal resistivity values of 2-10 μΩ cm² [Poelaert 99, Rando 92, Monaco 92] and ~2.5 μΩ cm² [Verhoeve 02] respectively are generally reported.

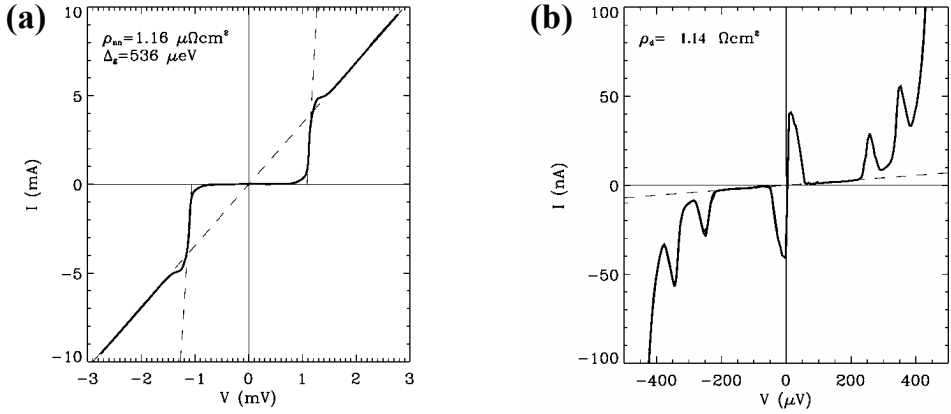

**Figure 4.9:** Current-voltage characteristics of V-Al based junctions with 100 nm of V and 25 nm of Al. **(a)** Complete IV-curve of a 20 μm side-length junction. The dashed lines represent fits to the energy gap $\Delta_g$ and the normal resistance $R_n$ of the junction. **(b)** IV-curve of the sub-gap region of a 40 μm side length junction. The dashed line is a fit to the dynamical resistance $R_d$ of the junction. The three higher current excursions in the sub-gap regime are the Josephson current at zero bias voltage and two Fiske modes at bias voltages of respectively 250 and 340 μeV

Figure 4.9(b) shows the IV-curve of a 40 μm side-length junction in the sub-gap domain, which is the region where the bias voltage energy is lower than the gap energy. This sub-gap region is the preferred biasing region for photon detection experiments, because of the low equilibrium currents generated by the junction in this domain. The three higher current excursions in the sub-gap regime are the Josephson current at zero bias voltage and two Fiske resonances at bias voltages of respectively 250 and 340 μeV. In order to quantify the quality of a junction with respect to leakage currents, the dynamical resistance $R_d$ is defined, which is the resistance of the junction in the sub-gap regime. In Fig. 4.9(b) the dashed line is a fit to the sub-gap currents and represents the dynamical resistance. Its value is equal to ~15 kΩ for the 40 μm junction, which corresponds to a dynamical resistivity of 1.14 Ω cm². From this one can deduce the quality factor of the junction defined as $Q_f = \rho_d/\rho_{nn}$ equal to $10^6$ for the V-Al based junctions. The quality factor of the V-Al junctions are comparable to the quality factors of the best Nb-Al and Ta-Al based junctions, which show a $Q_f$ of respectively $10^6$ [Poelaert 99] and $10^7$ [Verhoeve 02]. Note that with both the Nb-Al and Ta-Al based junctions successful

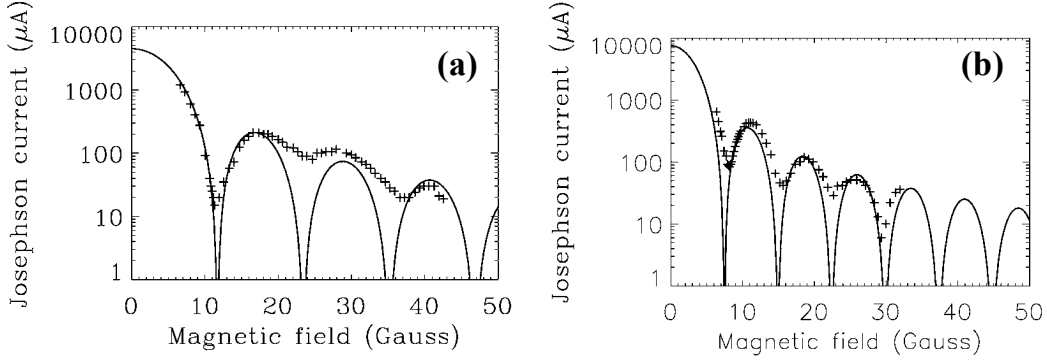

**Figure 4.10:** Josephson current as a function of applied parallel magnetic field for V-Al based junctions with 100 nm of V and 25 nm of Al. **(a)** 25µm side–length junction from a 6x6 pixel array **(b)** 40µm side-length single pixel junction.

optical photon detection experiments were performed, which underlines the very high quality of our V-Al junctions from a leakage current point of view.

### 4.3.1.2 Josephson current suppression

A good way of quantifying the uniformity of the insulating barrier separating the two electrodes is the examination of the Josephson current suppression pattern as a function of the applied parallel magnetic field. For a completely uniform current distribution over the area of the junction and for our geometrical set-up, which is a square junction with the magnetic field applied with an inclination of 45 degrees with respect to the side of the junction, the Josephson current dependence on parallel magnetic field is given by [Peterson 91]:

$$I_c = I_m \left| \frac{1 - \cos \pi h}{\pi^2 h^2} \right|, \text{with h=B/B}_0 \text{ and } B_0 = \frac{\phi_0}{\sqrt{2} L d_{eff}}, \quad (4.8)$$

where B is the applied magnetic field, $\phi_0$ is the magnetic flux quantum, L is the side length of the junction and $d_{eff}$ is the effective thickness of the junction. The effective thickness is the depth of penetration of the magnetic field into the electrodes on both sides of the insulator. For a junction with an electrode thickness larger than the London penetration depth $\lambda_L$, it can be approximated by:

$$d_{eff} \cong t + \lambda_1 + \lambda_2, \quad (4.9)$$

where $\lambda_i$ is the London penetration depth in top and base film and t is the thickness of the insulating barrier. For the case where the penetration depth is of the order of the film thickness the effective thickness is given by:

$$d_{eff} = t + \lambda_1 \tanh \frac{d_1}{2\lambda_1} + \lambda_2 \tanh \frac{d_2}{2\lambda_2}, \quad (4.10)$$

where $d_1$ and $d_2$ are the thickness of the top and the base film.

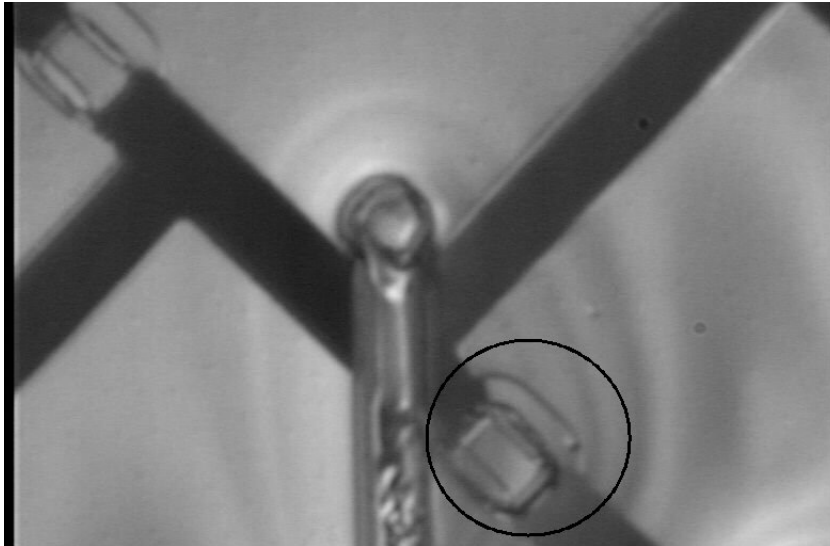

**Figure 4.11:** Optical microscope image of pixels within a 6x6 pixel V-Al array. The circle highlights the Nb plug area, which makes a contact between the base films of adjacent pixels. One can clearly identify that the mesa etch goes well into the junction area.

The crosses in Fig. 4.10 show experimental data of the variation of the Josephson current as the applied parallel magnetic field is varied. Data for a 25 μm side-length junction within a 6x6 array was acquired as well as for a single pixel 40 μm side-length junction. No data could be acquired for magnetic fields lower than 5 Gauss, because the large current densities break down superconductivity in the 2 to 3 μm wide leads. The solid lines in the figure show a fit to the data using (4.8). For both junctions the best fit was obtained for an effective thickness $d_{eff}$ equal to 102 nm. This corresponds to a London penetration depth $\lambda_L = 50$ nm on both sides of the tunnel barrier. The fit for the 40 μm single pixel junction is fairly good and shows that the tunnel current is rather uniform over the area of the junction. In case of the 25 μm side length junction the fit is much worse. This is due to the geometry of the pixels in the array, which are not perfectly square pixels anymore. In fact the mesa etch made for the Nb base film plugs, which connect adjacent pixels, goes well into the square junction area (Fig 4.11). Therefore, the junction is not a perfect square anymore and (4.8) is not strictly applicable. Nevertheless, the maxima can still be identified at the positions corresponding to an effective thickness equal to 102 nm. For photon detection experiments a magnetic field of about 100 to 200 Gauss is generally applied parallel to the junction area in order to suppress the Josephson current in V-Al based junctions to values lower than 10 nA. Note that the critical field in V is 1408 Gauss [Vonsovsky 82] and therefore the application of a parallel field of 100 to 200 Gauss does not considerably suppress the superconducting state in V.

### 4.3.2  Al based junctions

The quality of the pure Al based junctions with a base electrode thickness of 100nm and a top electrode, which is 50 nm thick, will now be presented. Both Al films are polycrystalline with a RRR of ~10.

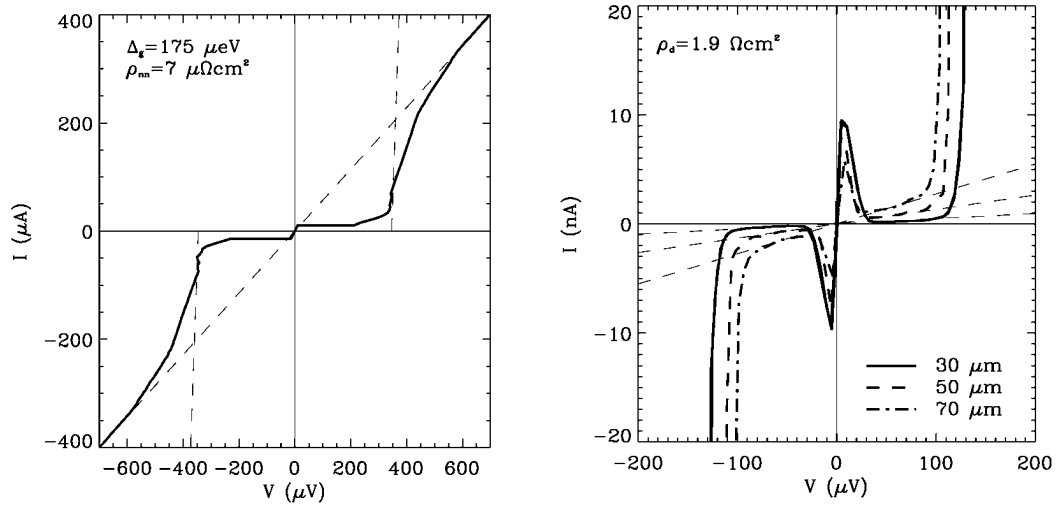

**Figure 4.12:** IV-curves of Al based junctions taken at a temperature of 40 mK. The applied parallel magnetic field is of the order of 30 Gauss. **(a)** Full IV-curve of a 20 μm side-length STJ. The dashed lines are fits to the sum-gap of the two electrodes and the normal resistance of the junction. **(b)** IV-curves of the sub-gap region of three junctions with side-lengths of 30, 50 and 70 μm. The thin dashed lines are fits to the dynamical resistances in the sub-gap regime.

### 4.3.2.1    IV-curves

Figure 4.12(a) shows the complete IV-curve of a 20 μm side length junction taken at 40 mK. A parallel magnetic field of approximately 30 Gauss was applied parallel to the junction in order to suppress the zero voltage Josephson current to an acceptable level. The top and base leads of this junction are 6 μm wide in order to prevent breakdown in the leads at high current densities. On the figure one can identify the sum-gap of the electrodes equal to 350 μeV. The almost vertical dashed line is a fit to the sum-gap region of the curve. Assuming a symmetrical lay-up, one deduces an energy gap equal to 175 μeV in both electrodes, which is reasonably close to the bulk energy gap of Al equal to 180 μeV. The normal state resistance $R_n$ of the junction can also be determined from the figure, and is equal to 1.75 Ω for this 20 μm side length junction. This gives a normal state resistivity $\rho_{nn}$ equal to 7 μΩ cm$^2$. One can compare this value to the results of the Yale University and the University of Munich group, who also develop high quality Al junctions for application as photon detectors. The Al junctions fabricated by the Yale group show a normal resistivity $\rho_{nn}$ equal to 10 μΩ cm$^2$ [Wilson 01] and the junctions of the Munich group show $\rho_{nn}$=100 μΩ cm$^2$ [Angloher 00]. The devices from the Munich group have therefore a considerably thicker insulating barrier, whereas the devices from the Yale group have insulating barriers, which have a thickness comparable to our barriers.

Figure 4.12(b) shows the sub-gap regimes of three different junctions with side lengths of respectively 30, 50 and 70 μm. The IV-curves were acquired at a temperature of 40 mK. A magnetic field of approximately 30 Gauss was applied in parallel to the junction in order to suppress the Josephson current below 10 nA. The dashed lines are fits to the dynamical resistances $R_d$ in the sub-gap domain, which is the region in which the junctions are biased for photon detection experiments. The sub-gap currents in the operational bias voltage regime (~30-100 μV) are proportional to the area of the junction and are equal to 260 fA per μm$^2$ of junction area at a bias voltage of 50 μV. This shows

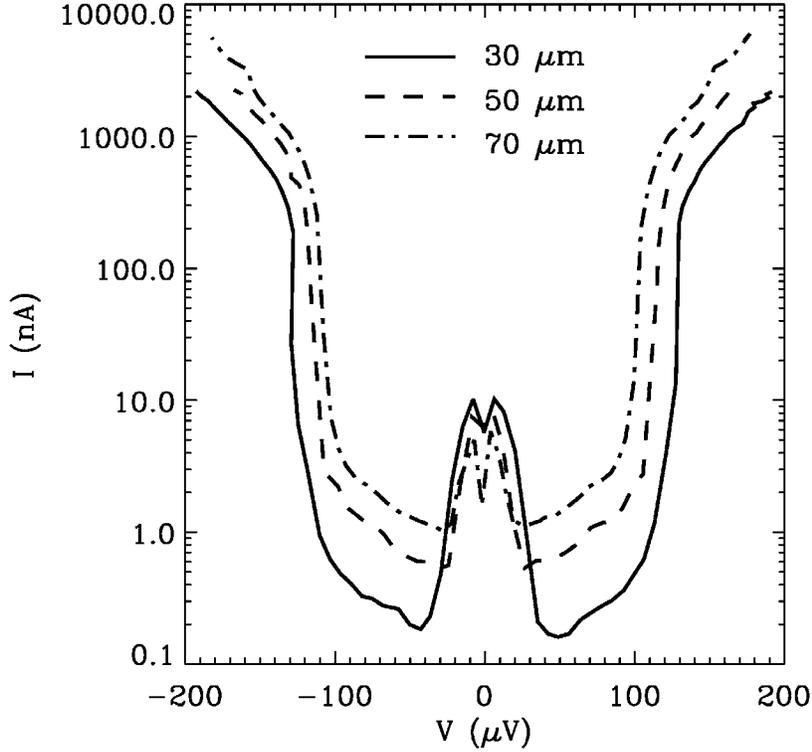

**Figure 4.13:** IV-curves of the 30, 50 and 70 μm side-length junctions on a logarithmic scale.

that the residual currents in the bias domain are leakage currents arising from very small pinholes in the insulating barrier distributed homogeneously over the area of the junction. The dynamical resistivity $\rho_d$ of the Al based junctions can be derived and is equal to 1.9 Ω cm$^2$. This yields a quality factor Q = $\rho_d/\rho_{nn}$ = 2.7 10$^5$. One can compare this to the dynamical resistance observed for the Al devices of the Munich group, which is equal to 5 Ω cm$^2$ [Angloher 00], yielding a quality factor Q equal to 5 10$^4$, a value 5 times lower than for the junctions presented in this thesis. Unfortunately, the Al devices of the Yale group have not been measured at temperatures lower than 220 mK. Therefore, their sub-gap currents in the bias region are limited by thermal currents giving a dynamical resistance in the bias area of approximately $\rho_d$ = 0.03 Ω cm$^2$. This yields a quality factor for the devices of the Yale group equal to 3 10$^3$.

Another interesting feature is the sudden current rise in the IV-curves at a bias voltage of approximately 100 μV. As can be seen in Fig. 4.12(b), the sub-gap currents rise dramatically at bias voltages of 125, 110 and 100 μV for the 30, 50 and 70 μm side-length junctions respectively. In Fig 4.13, which shows the same IV-curves on a logarithmic scale, it can be seen that these current steps increase the sub-gap currents by almost three orders of magnitude. Well known mechanisms [Wolf 85], such as Fiske resonances, multi-particle tunneling, self-coupling of Josephson radiation and multiple Andreev reflections, which are known to introduce a certain structure in the sub-gap currents of tunnel junctions, cannot explain this very drastic current step at a bias voltage that depends on the size of the junction. The current steps observed in these high quality, low $T_C$ and low loss junctions arise because of the very special non-equilibrium state, which forms due to the interplay of energy gain of the quasiparticles caused by sequential tunneling and energy loss due to down-scattering (see also section 2.3.3.1). During its lifetime a quasiparticle undergoes a large number of tunnel and back-tunnel processes.

During every single tunnel or back-tunnel process the quasiparticle gains an energy $eV_b$. In addition, quasi-particle down-scattering to the gap energy is slow in low $T_C$ junctions because of the cubic dependence of the characteristic electron phonon scattering rate $\tau_0^{-1}$ on the $T_C$ of the material [Kaplan 76] (see also section 2.3.1.3). As a consequence, the quasi-particles in a biased low $T_C$ junction form a non-equilibrium energy distribution with quasiparticles at energies high above the gap energy. Note that it takes a certain amount of time for a quasiparticle to access the highest energy states, as it has to go through a cycle of subsequent tunnel, back-tunnel and down-scattering events. Therefore, in junctions with slow down-scattering, the highest energy state at which quasiparticles reside is limited by the loss time of the quasiparticles, which determines the maximum time available to a quasiparticle to go through the tunnel, back-tunnel and down-scattering cycle. Now, if the quasiparticles that reside at this highest energy level possess an energy that lies $2\Delta_g$ above the energy gap of the material, they will release a phonon of energy $2\Delta_g$, when relaxing down to the gap. This phonon can then break a Cooper pair and create two more quasiparticles. This process is called quasiparticle multiplication and is described in section 2.3.1.8. These newly formed quasiparticles are then free themselves to undergo the tunnel, back-tunnel and down-scattering cycle. As a consequence, the number of quasiparticles in the electrode will be greatly enhanced and the tunnel current will rise sharply when the bias voltage is reached, which will lift the first quasiparticles into the active region. The fact that this threshold bias voltage is different for the three Al junctions of different size comes from the different loss times in the three junctions. As will be shown in Chapter 5, the quasiparticle loss times are respectively equal to 20, 40 and 90 μsec in the 30, 50 and 70 μm junctions. Less time is available for the tunnel, back-tunnel and down-scattering cycle in the smaller junctions, which is the reason why the energy gained per tunnel event has to be larger in order for the quasiparticles to reach the active region within the loss time. Therefore, the threshold bias voltage at which the current step occurs is higher for the junctions with the faster losses. The structure of the IV-curves as shown in Fig. 4.13 can be successfully simulated with a model with characteristics similar to the kinetic equation model developed in section 2.3, but adapted to the case of a stationary regime. More details about the sub-gap structure in IV-curves of low $T_C$ and low loss junctions can be found in [Kozorezov 03a, Kozorezov 03b].

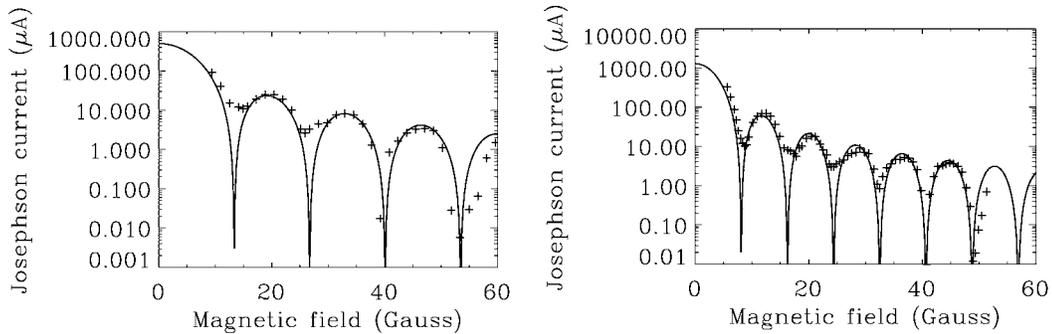

**Figure 4.14:** Josephson current suppression as a function of applied parallel magnetic field for **(a)** a 30 μm side length and **(b)** a 50 μm side length Al junction. Crosses represent measured values, whereas the solid lines are fits to the experimental data. The temperature is equal to 40 mK.

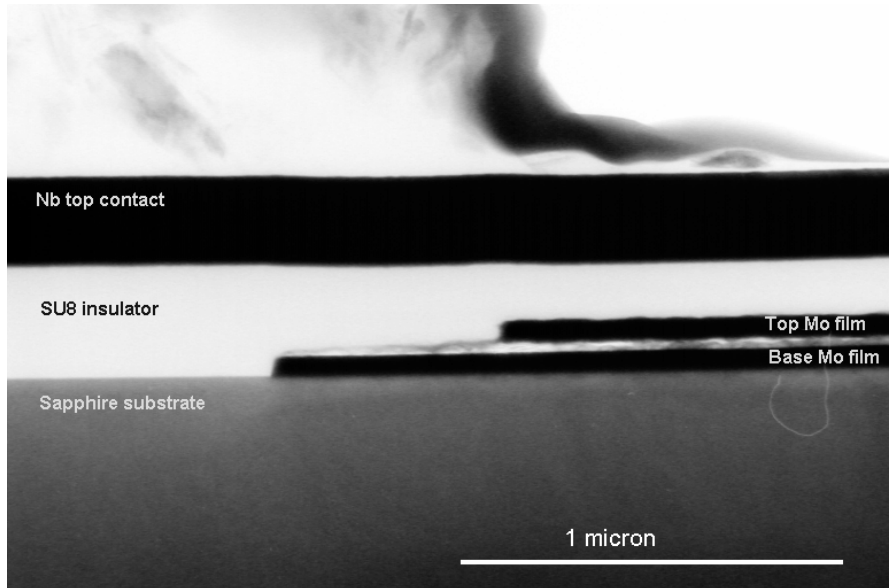

**Figure 4.15:** TEM image of a Mo-Al based multi-layer.

#### 4.3.2.2    Josephson current suppression

Josephson current suppression is expected to be more critical in pure Al based junctions, as the critical field of Al is equal to 105 Gauss [Vonsovsky 82]. In order to avoid trapping of a magnetic flux quantum within the junction, which favors normal electron tunneling through the insulating barrier, the magnetic field applied in parallel to the junction should not exceed 50 Gauss. The variation of the Josephson current with the parallel magnetic field applied to the junction is shown in Fig. 4.14 for a 30 and a 50 μm side-length junction. The crosses represent the experimental data points, whereas the solid lines represent a fit using (4.8). For both fits the maximum critical current density $J_m$ was chosen equal to 0.52 μA/μm$^2$ and the effective thickness $d_{eff}$ was chosen equal to 73 nm. This yields a London penetration depth $\lambda_L$ on both sides of the barrier equal to 36 nm. In both cases the Josephson suppression scheme is very regular, showing the good uniformity of the insulating barrier. Also the minima at around 50 Gauss are very pronounced, allowing the suppression of the Josephson current to values below 100nA, which is required for the stable biasing of the junction.

### 4.3.3   Mo-Al based junctions

Figure 4.15 shows a TEM picture of a symmetrical Mo-Al based multilayer with a 50 nm thick film of Mo and a 15 nm thick Al film. The edge of the junction can be clearly seen and shows a step-like structure. The top Mo film is etched 600 nm further than the base Mo film and the Al film in between does not show a vertical edge profile. This step structure can also be identified on the optical microscope image of the junction in Fig. 4.16.

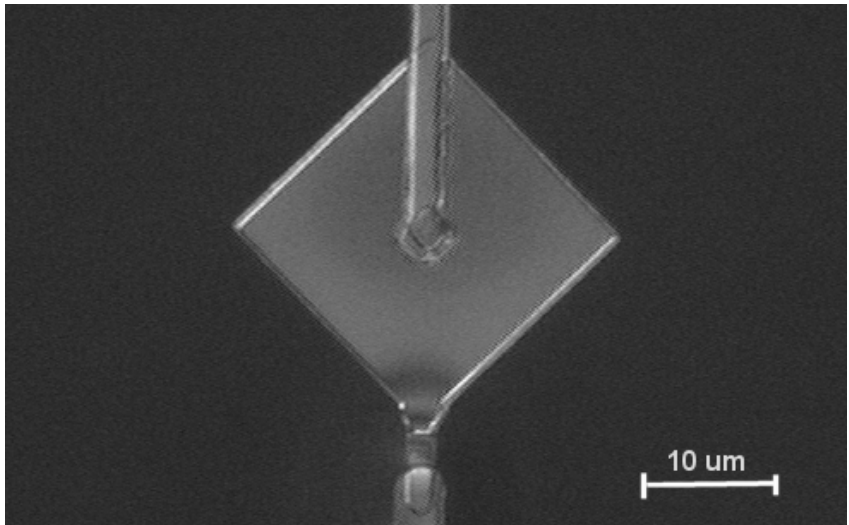

**Figure 4.16:** Optical microscope image of a 20 μm side length Mo-Al based junction. The step-like edge profile can be clearly identified.

### 4.3.3.1   IV-curves

This inhomogeneous edge structure of the Al film does of course destroy the uniformity of the very thin insulating AlOx layer. As a consequence the IV-curves of these Mo-Al junctions show very large leakage currents, which show a dependence on the perimeter length of the junction. As can be seen in Fig. 4.16 the leakage currents are approximately equal to 1.25 μA per μm of perimeter length. Figure 4.17 shows the IV-curves of two Mo-Al junctions taken at a temperature of 300 mK. The sizes of the two junctions are respectively 30 and 70 μm. The leakage currents are much higher than the expected currents due to thermal quasiparticle tunnelling, which should be of the order of one μA at 300 mK assuming a $T_C$ of 0.9 K. At a bias voltage of approximately 40 μV the leakage currents through the insulating layer even exceed the critical current density of the 3 μm wide superconducting leads, which breaks down superconductivity in these parts of the structure. This suddenly adds a considerable resistance to the circuit and prevents the measurement of the normal resistance and the energy gap of the junctions.

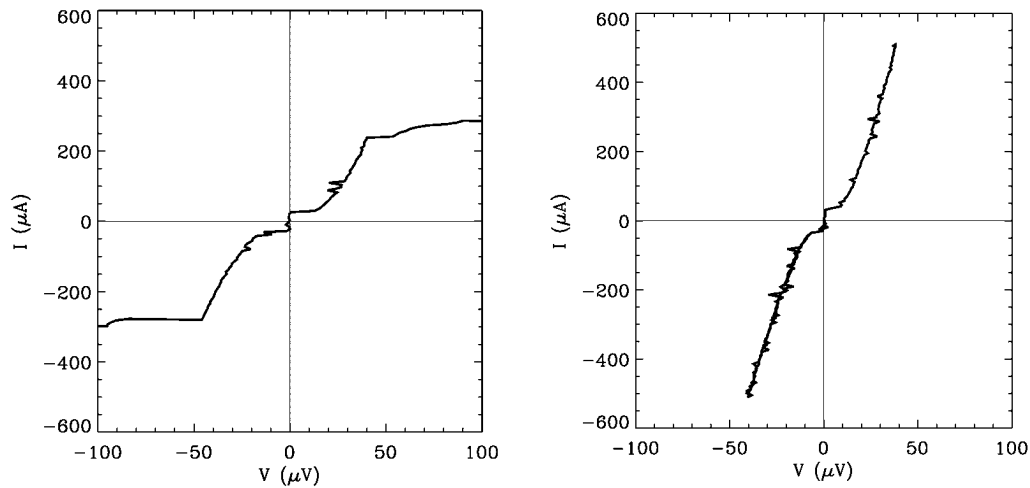

**Figure 4.17:** IV-curve of a 20 μm side length (a) and a 50 μm side length (b) Mo-Al based junction. The IV-curve was acquired at a temperature of 300 mK.

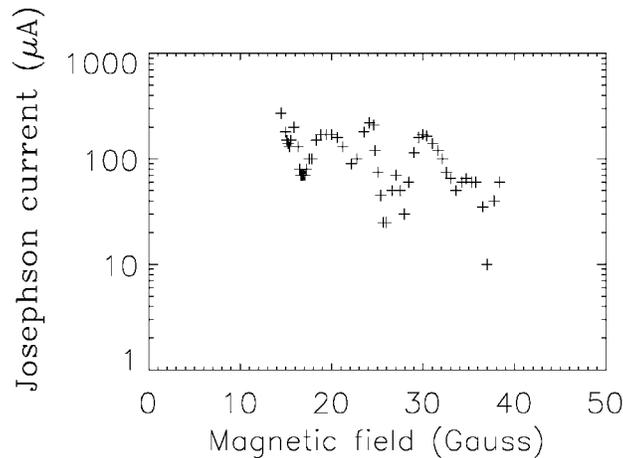

**Figure 4.18:** Josephson current as a function of applied parallel magnetic field for a 20 μm side length Mo-Al junction.

### 4.3.3.2 Josephson current suppression

The Josephson current suppression pattern as a function of applied parallel magnetic field also shows that the current distribution over the area of the junction is very inhomogeneous (Fig. 4.18). The pattern is in no way comparable to the pattern described by equation (4.8). No significant decay of the maxima within the pattern is visible, which is very typical for a junction with a localised source of leakage currents.

In order to get rid of these high leakage currents localized at the edge of the junctions, the edge profile has to be improved. In order to achieve this, the base etch procedure will have to be modified. Such an approach is currently under development.

## 4.4 Conclusions

The junction fabrication procedure details were described for the V-Al, Al and Mo-Al based STJs.

The fabricated V-Al junctions are of good quality. The 100 nm base V film of the junction is epitaxial and has a RRR of approximately 18. The 100 nm thick V film of the top electrode is polycrystalline with columnar crystal grains with diameters varying between 50 and 100 nm. Inspection of the IV-curves of the V-based devices reveal a normal resistivity of the insulating barrier $\rho_{nn}$ equal to 1.16 μΩ cm$^2$ and a dynamical resistivity in the bias domain $\rho_d$ equal to 1.14 Ω cm$^2$, which corresponds to a quality factor $Q_f = \rho_d/\rho_{nn}$ of the order of $10^6$. This value is comparable to the best Ta and Nb-based junctions fabricated up to date, with which single optical photon counting was achieved. The Josephson current suppression pattern is very regular, showing the good uniformity of the insulating barrier. A fit to the experimental data revealed a London penetration depth at both sides of the barrier equal to 50 nm.

The fabricated Al junctions possess a 100 nm thick polycrystalline base film covered by a 50 nm polycrystalline top electrode. The base Al electrode has a RRR of approximately 10. Inspection of the IV-curves reveal a normal resistivity of the insulating barrier equal to 7 μΩ cm$^2$ and a dynamical resistivity equal to 1.9 Ω cm$^2$, which corresponds to a quality factor of 2.7 $10^5$. The sub-gap currents in the bias domain are directly proportional to the area of the insulating barrier and equal to 260 fA per $\mu m^2$ of junction area at a bias voltage of 50 $\mu V$. The IV-curves of the Al junctions show a very strong current step over three to four orders of magnitude in current at a bias voltage of approximately 100 $\mu V$. This strong current step is typical for low energy gap and low loss junctions and cannot be explained by any of the existing well known mechanism that introduce current steps in the sub-gap regime. The strong current step is caused by a quasiparticle generation mechanism, called quasiparticle multiplication, caused by pair breaking of $2\Delta_g$ phonons released by quasiparticles with energy larger than $3\Delta_g$ relaxing down to the energy gap. For bias voltages below the current step the energy gained due to subsequent tunnel events is not sufficient to lift the quasiparticles above the $3\Delta_g$ threshold energy. As soon as the bias energy is large enough, the first quasiparticles will reach the threshold energy and start multiplying. The quasiparticle population will then grow until losses by recombination outnumber the gains by multiplication. This causes the very large increase in tunnel currents observed for bias voltages above the current step level. The Josephson current suppression pattern is very regular, showing the uniformity of the tunnel barrier. A fit to the experimental data reveals a London penetration depth on both sides of the barrier equal to 36 nm.

The fabricated Mo-Al based junctions show a strong step-like structure in the edge profile. The top Mo film is etched 600 nm further than the base Mo film. The Al film in between does not show a vertical edge profile, causing very strong leakage currents at the edges of the device. The leakage currents of all the Mo-Al based junctions are proportional to the length of the perimeter of the junction equal to 1.25 $\mu A/\mu m$. The Josephson current suppression pattern is very irregular and shows no real decay of the maxima of the curve, which is typical for junctions with a localised source of leakage currents. Further work on the base etch procedure has to be performed in order to reduce the leakage currents in the Mo-Al based junctions.

# Chapter 5

# Photon detection experiments

.

In this chapter the main point of interest of this thesis is discussed, which is the operation of the fabricated tunnel junctions as near IR to X-ray photon detectors. The junctions described in chapter 4 were operated at low temperature and exposed to electro-magnetic radiation. The ultimate goal is of course to obtain the best possible energy resolution and at the same time to conserve an acceptable response time of the system with good detection efficiency. First, the experimental data acquired with V-Al junctions, which were exposed to 6 keV X-rays from a radioactive $^{55}$Fe source, is presented. The model presented in chapter 2 is then applied in order to analyse the different variations of the responsivity and the decay time of the pulses. The energy resolution is discussed as well. Then the response of Al junctions to near IR to soft UV radiation is presented, as well as the response to 6 keV X-rays. The model of chapter 2 is again applied in order to analyse the data and explain interesting features. Then the energy resolution of the junctions is discussed.

## *5.1 V-Al based junctions*

The V-Al junctions presented in the previous chapter were operated in a 300mK [3]He sorption cooler and exposed to 6 keV X-ray radiation from a [55]Fe radioactive source. The devices tested as photon detectors are symmetrical junctions with a 100 nm thick V film covered by an approximately 25 nm thick Al film. The devices are single pixel junctions with side lengths of respectively 7, 10, 20 and 30 μm. In addition to these single pixel junctions a 6x6 pixel array (Fig. 5.1) was also fabricated and tested as photon detector. The devices composing this array have a side length of 25 μm and a slightly different geometry compared to the single pixel devices, as the all the base film electrodes of the array are interconnected via Nb base film plugs. All the interconnected base electrodes are then connected to a single return wire, which is connected to ground. As a consequence there are two Nb plugs in every pixel of the array (See also Fig. 5.1) as opposed to a single base film plug in the single pixel structures.

Prior to presenting the experimental results, the different characteristic times of the junctions, which are necessary in order to simulate the experimental data with the model, will be calculated.

### 5.1.1  Proximity effect theory applied to V-Al

For a V-Al junction with a 100 nm thick V film and a 12 nm thick Al film the energy gap and critical temperature were measured and found to be equal to 653 μeV and 5.28K respectively. From these two measured values one can determine the interface parameters $\gamma$ and $\gamma_{BN}$ for this particular lay-up. The knowledge of these values allows the determination of the interface constants $C_{\gamma}$ and $C_{\gamma_{BN}}$, which are independent of the film thickness. More details can be found in section 2.2.3. A series of simulations was made for the 100 nm-12nm lay-up with $\gamma$ ranging from 0.2 to 1.2 and $\gamma_{BN}$ ranging from 1 to 8. The parameters used for these simulations can be found in table 5.1.

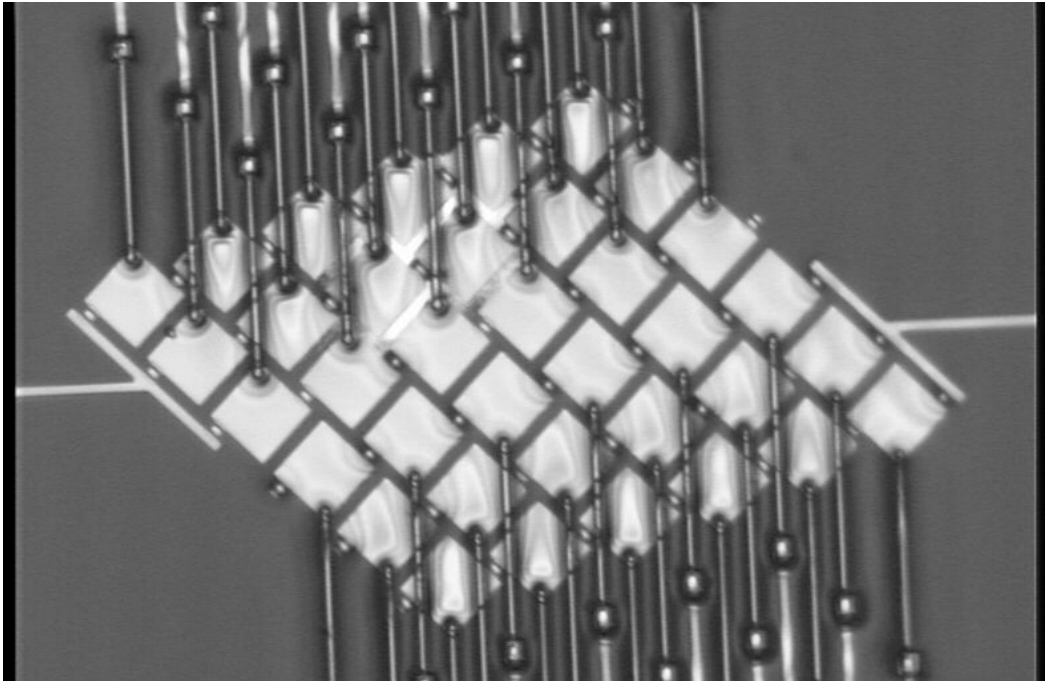

**Figure 5.1:** 6x6 array of V-Al based STJs.

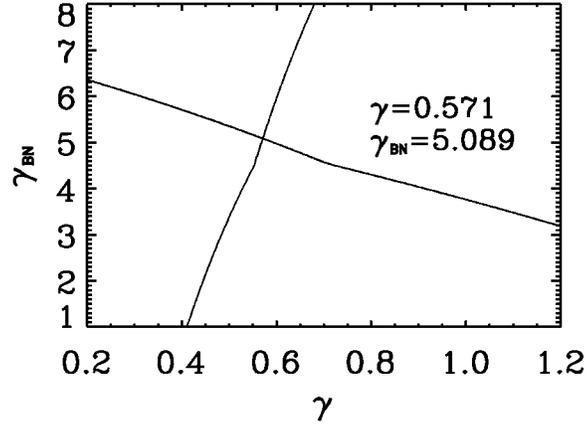

**Figure 5.2:** Intersections of the calculated surfaces of respectively $T_C$ and $\Delta_g$ as a function of interface parameters with the planes corresponding to the experimental values of the V-Al bi-layer with 100nm of V and 12 nm of Al. The intersection of the two lines determines the interface parameters $\gamma$ and $\gamma_{BN}$ for this bilayer.

**Table 5.1:** V and Al film parameters.

|   | $T_C$ (K) | $\Delta_g(0)$ ($\mu$eV) | $l_0$ (nm) | $\xi_0$ (nm) | $v_F$ ($10^6$m/sec) | $N(0)$ ($10^{21}$/(eV cm$^3$)) |
|---|---|---|---|---|---|---|
| V | 5.4 | 820 | 67 | 45 | 0.176 | 38.1 |
| Al | 1.2 | 180 | 52 | 1600 | 1.37 | 12.2 |

Figure 5.2 shows the two lines in the $\gamma$-$\gamma_{BN}$ space, for which respectively the calculated energy gap and critical temperature agree with the experimental values. The intersection of these two lines is therefore the single combination of interface parameters for the lay-up under consideration. The derived interface parameters are $\gamma$=0.571 and $\gamma_{BN}$=5.089. From these values, using the material parameters from table 5.1 and equations (2.21)-(2.22), one can deduce the interface constants, which are equal to $C_\gamma = 0.959$ and $C_{\gamma_{BN}} = 1.193$. One can then calculate the order parameter, the density of states and the imaginary part of the sine of the Green function for the lay-up with 100 nm of V and 25 nm of Al, for which the photon detection experiments will be presented. The results are shown in Fig. 5.3, and are very similar to the results for the Nb-Al and Ta-Al bi-layers presented in chapter 2. The energy gap is continuous throughout the bi-layer with a gap value in between the bulk energy gaps in V and Al. The calculated energy gap of the bi-layer is equal to 517 $\mu$eV. This value is very close to the value determined experimentally from the IV-curves presented in the previous chapter, which is equal to 536 $\mu$eV. In the V layer a maximum appears in the density of states at the bulk gap energy in V. At energies below the bulk gap energy the number of available states is greatly reduced. In the Al, on the other hand, the number of states presents a maximum at the energy gap of the bi-layer. A smaller local maximum can be observed at the bulk energy gap of V, because of the proximity of the V layer. These features in the density of states will have strong implications on all the characteristic quasiparticle rates in the bi-layer.

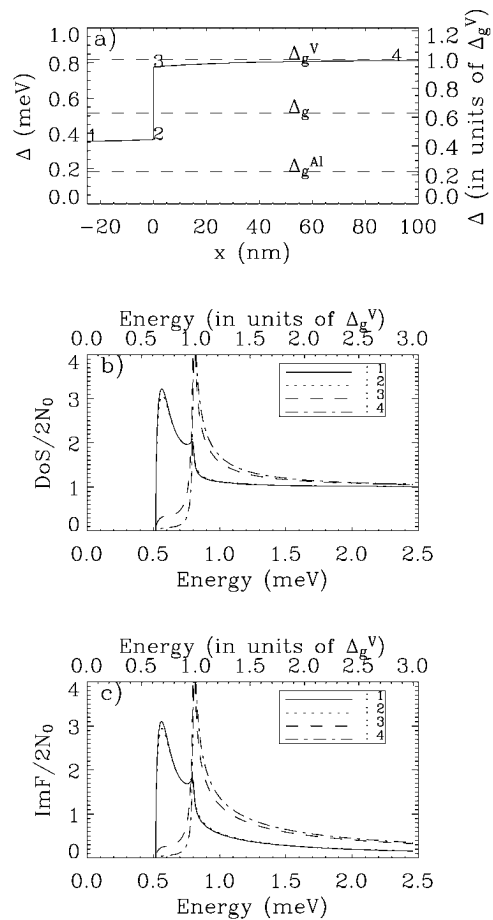

**Figure 5.3: (a)** Pair potential $\Delta$ for a V-Al bi-layer with 100 nm of Nb and 25 nm of Al. The upper dashed line is the bulk energy gap of V. The lower dashed line is the bulk energy gap of Al. The intermediate dashed line is the energy gap of the bi-layer, as determined from the density of states. The points 1, 2, 3 and 4 correspond to the four positions in the bi-layer for which the density of states is given in (b). **(b)** Density of states DoS for a V-Al bi-layer with 100 nm of V and 25 nm of Al. The densities of states are represented for both materials at the free interfaces and at the V-Al interface. The points 1 to 4 in (a) indicate the positions in the bi-layer for which the densities of states are given. **(c)** Imaginary part of the Green's function ImF for the same bi-layer. The imaginary part of the Green's function ImF is given at the same four positions in the bi-layer as for the density of states.

### 5.1.2 Quasiparticle characteristic rates in V-Al

With the results of the proximity effect theory one can calculate the quasiparticle characteristic rates in the electrodes of the junction. The parameters used for the calculations are found in table 5.2. The results of the characteristic rate calculations are summarised in Fig. 5.4.

The most interesting feature is that the tunnel rate shows a strong maximum for quasiparticles at the gap energy of the bi-layer and a minimum for quasiparticles at the bulk gap energy of V. The reason for this lies in the relative number of states available at both sides of the tunnel barrier and in the bulk of the two materials composing the electrodes. The maximum at the gap energy arises from the maximum of the density of states at this energy in the Al at the position of the tunnel barrier. The minimum at the bulk gap energy in V is the effect of the raised density of states at these energies in the V film, away from the barrier.

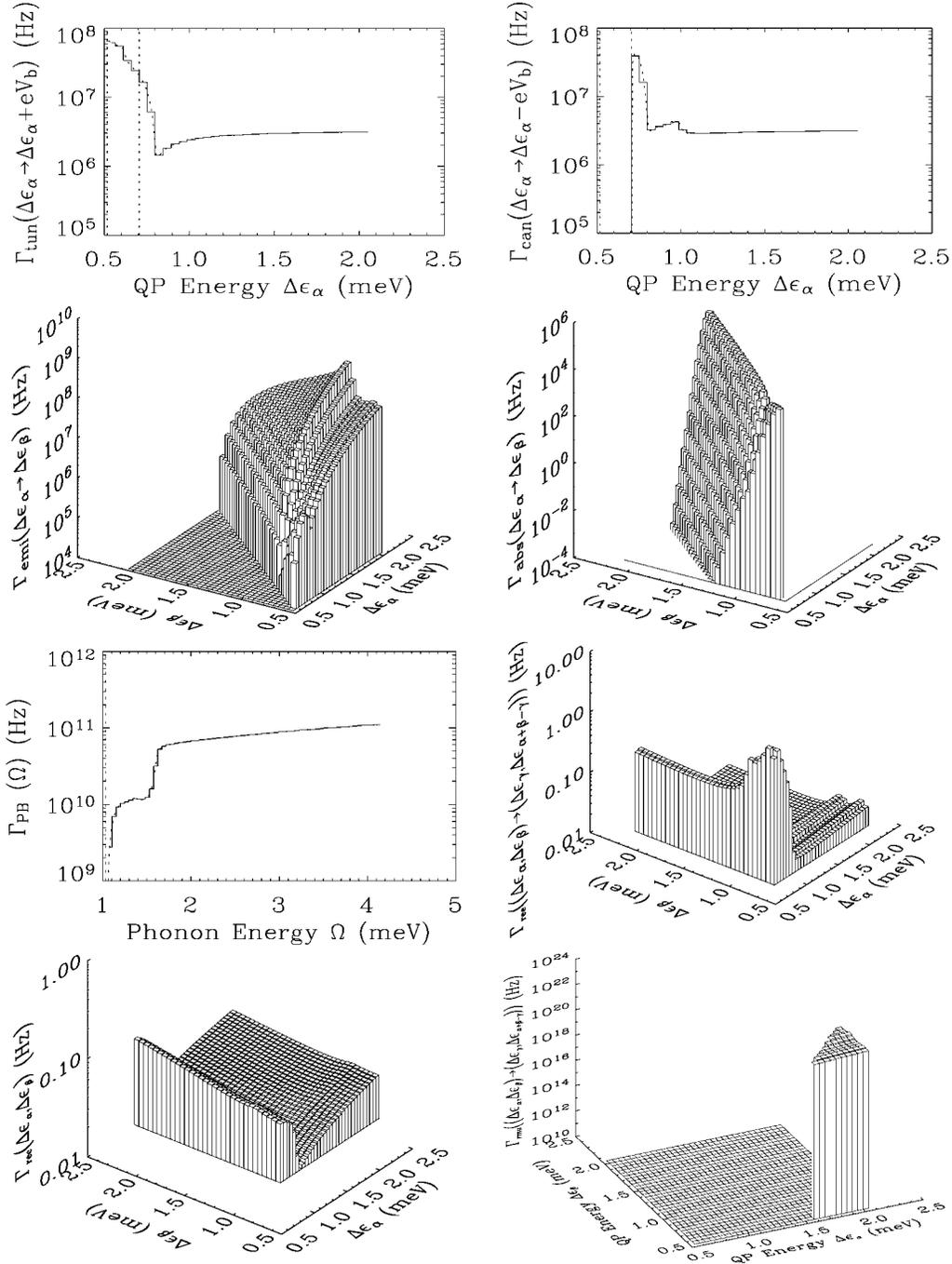

**Figure 5.4:** Characteristic quasiparticle rates in a V-Al junction with electrodes composed of 100 nm thick V and 25 nm thick Al. **(a)** Tunnel rate as a function of quasiparticle energy ($V_b$=190μV). **(b)** Cancellation tunnel rate as a function of quasiparticle energy ($V_b$=190μV). **(c)** Rate for electron-phonon scattering with emission of a phonon as a function of initial and final quasiparticle energy. **(d)** Rate for electron-phonon scattering with absorption of a thermal phonon as a function of initial and final quasiparticle energy (T=300mK). **(e)** Cooper pair breaking rate as a function of phonon energy. **(f)** Rate for quasiparticle recombination with subsequent pair breaking as a function of the initial quasiparticle energies $\Delta\varepsilon_\alpha$ and $\Delta\varepsilon_\beta$, and for the particular final energy $\Delta\varepsilon_\gamma$ equal to the gap energy of the electrode (junction size = 20 μm). **(g)** Rate for quasiparticle recombination with subsequent phonon loss as a function of the initial quasiparticle energies (size = 20 μm). **(h)** Rate for the quasiparticle multiplication process as a function of the initial ($\Delta\varepsilon_\alpha$) and final ($\Delta\varepsilon_\beta$) energy of the first quasiparticle. The energy of one of the generated quasiparticles $\Delta\varepsilon_\gamma$ is fixed to being equal to the gap energy $\Delta_g$. (size = 20 μm).

The cancellation rate is zero for energies below the bias energy level for the obvious reason that no states are available at the other side of the barrier to tunnel into. The maximum cancellation rate occurs at energies right above the bias energy, because of the larger density of states in the Al near to the barrier. For quasiparticles, which have energy larger than approximately $3\Delta_g$, the direct tunnel rate and the cancellation rate are the same. As a consequence the effective tunnel current for these quasiparticles is zero.

**Table 5.2**: Material parameters used for the characteristic times calculations in V-Al junctions.

| Symbol | Name | Unit | V | Al |
|---|---|---|---|---|
| $R_nA$ | Normal resistivity of junction | $\mu\Omega\ \mathrm{cm}^2$ | $1.16 \pm 0.2$ | |
| $T_C$ | Critical temperature | K | 5.4 | 1.2 |
| $\Delta_g$ | Energy gap | $\mu$eV | 820 | 180 |
| $N_0$ | Single spin normal state density of states at Fermi energy | $10^{27}$ states $\mathrm{eV}^{-1}\ \mathrm{m}^{-3}$ | 38.1 | 12.2 |
| $\alpha^2$ | Average square of the electron-phonon interaction matrix element | meV | 2.5 | 1.92 |
| N | Ion number density | $10^{28}\ \mathrm{m}^{-3}$ | 7.25 | 6.032 |
| $\tau_0$ | Electron-phonon interaction characteristic time | nsec | 3.75 | 440 |
| T | Temperature | K | 0.3 | |

### 5.1.3  6 keV soft X-ray photon detection experiments in V-Al junctions

The different V-Al based junctions were tested in the $^3$He sorption cooler at a temperature of 300 mK. The $^{55}$Fe radioactive source was located approximately 3-5 mm away from the sample in order to have a total count-rate of approximately 100 photon absorption events per second, which corresponds to approximately 5 counts per second in the electrodes of the detector. A parallel magnetic field of the order of 200 Gauss was applied in order to suppress the Josephson currents and eventual Fiske modes in the junctions, which would create excess electronic noise unwanted for our application.

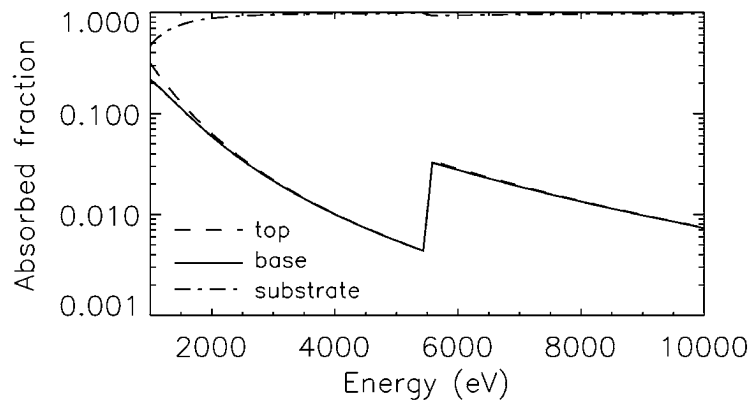

**Figure 5.5:** Fraction of X-ray photons absorbed in the top and base electrodes of a V-Al based junction and in the sapphire substrate directly underneath the junction. Both electrodes of the STJs are composed of a 100nm thick V film. The absorption in the Al film is neglected.

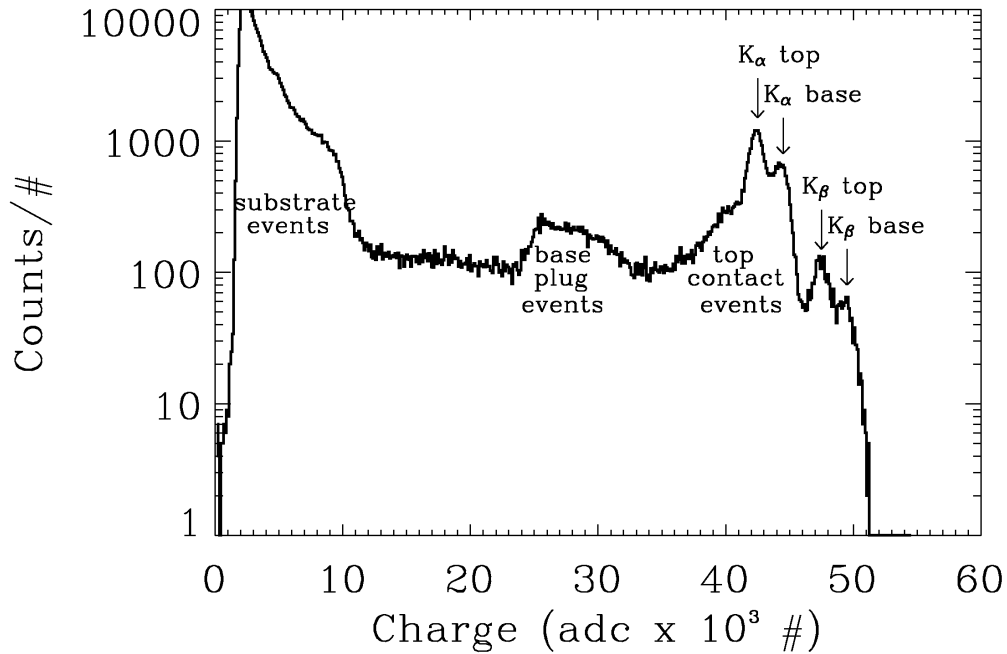

**Figure 5.6:** [55]Fe spectrum acquired with the 10 μm side length V-Al based junction with 25 nm of Al. The applied bias voltage is equal to 260 μV and the applied parallel magnetic field is 210 Gauss.

The radioactive [55]Fe source emits X-rays from the Mn-$K_\alpha$ (5895 eV) and the Mn-$K_\beta$ (6490 eV) emission lines with probabilities of respectively 90% and 10%.

Figure 5.5 shows the absorption efficiencies of the 100 nm thick top and base V films for 1 to 10 keV X-rays [Henke 93, LBNL]. The absorption in the Al is neglected. At 6 keV the absorption probability in both films is approximately 3%, which shows that the majority (~94%) of the X-rays emitted by the radioactive source are absorbed in the sapphire substrate. In the 350 nm thick Nb forming the top contact and the base lead plug the absorption efficiency is equal to 10%.

A typical [55]Fe spectrum acquired with a 10 μm junction biased at 260 μV is shown in Fig. 5.6. As expected, most events are absorbed in the substrate, which is coupled to the detector through phonons and gives rise to the typical substrate absorption structure in the low charge output part of the spectrum. The events absorbed in the Nb of the base lead plug give rise to some structure in between the substrate events and the events absorbed in the detector material. The same is true for the events absorbed in the Nb top contact except that the charge output for those events is slightly higher because of the better coupling to the detector. Quasiparticles absorbed in the base lead plug first have to travel through part of the lead before reaching the junction area, whereas the Nb top contact is in direct contact with the top electrode. The two peaks corresponding to the 5.9keV $K_\alpha$ events absorbed in top and base film respectively can be clearly discerned, as well as the two peaks corresponding to the 6.49keV photons from the $K_\beta$ emission line.

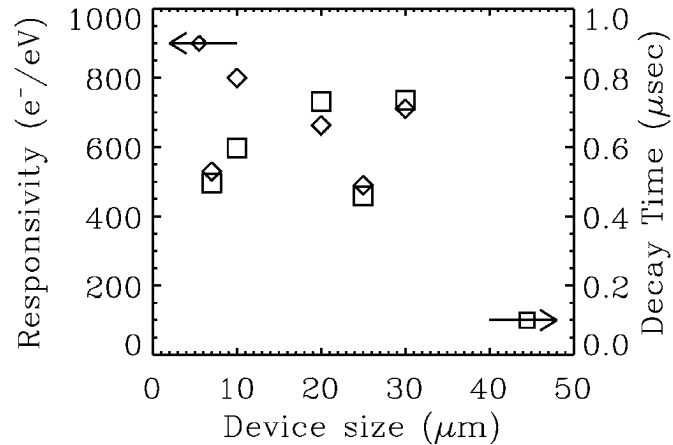

**Figure 5.7:** Responsivity and decay time in the base film of the V-Al STJs for the detection of 5.9 keV X-ray photons. The applied bias voltage is for all junctions equal to 200 μV and the parallel magnetic field of the order of 200 Gauss.

### 5.1.3.1   Responsivity and pulse decay time

From the analysis of the slow and fast channel spectra (see section 3.2.2) one can determine the responsivity, which is the charge output per eV of incoming photon energy, and the decay time of the signal pulses.

#### 5.1.3.1.1   Size dependence

Fig. 5.7 shows the responsivity and decay time for the absorption of 5.9 keV X-rays in the tested V-Al devices with an applied bias voltage of 200 μV as a function of the size of the devices. The responsivities of all the devices are of the order of 600 electrons per eV of photon energy absorbed. This responsivity is extremely low, considering that 1130 quasiparticles are created in the electrodes of the V-Al junction per eV of incoming photon energy. This responsivity corresponds to a charge amplification factor of approximately 0.5, which means that only half of the quasiparticles created in the junction tunnel once across the insulating barrier before they are lost. Both the responsivity and decay time do not depend on the size of the device. The variation around the mean values of about 600 e⁻/eV and 0.6 μsec is rather random.

#### 5.1.3.1.2   Energy dependence

From the spectrum of the 10 μm side length device shown in Fig. 5.6 one can deduce the responsivity and pulse decay time of both the $K_\alpha$ and the $K_\beta$ emission lines of the $^{55}$Fe radioactive source. Therefore, the responsivity and decay time for two different photon energies are known. In Fig. 5.8 these four experimental points are plotted as a function of the photon energy. A fit was made to these data points and to the temperature dependent data, presented in the next section, with the model presented in chapter 2. The result is shown along with the experimental points in the figure.

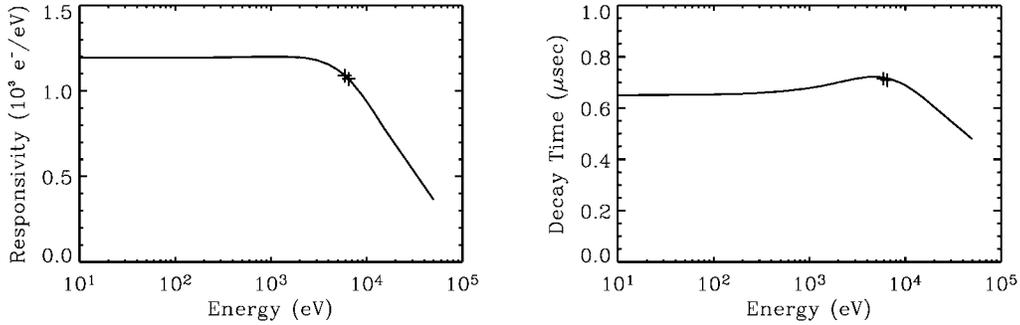

**Figure 5.8:** Responsivity and pulse decay time of the 10 μm V-Al junction as a function of photon energy ($V_b$=260 μeV). The crosses are the experimental points for the Mn-$K_\alpha$ (5.9keV) and $K_\beta$ (6.49keV) emission lines. The line is a fit to the data with the model from chapter 2. The values used for the free parameters of the model are shown in table 5.3.

### 5.1.3.1.3   Temperature dependence

Naturally, the four experimental points in the energy dependent plot are not enough information to assure a reliable determination of the five free parameters of the model, which are the quasiparticle loss time, the phonon escape time, the number of available traps, the trapping probability and the trap depth. In order to get a more accurate determination of the free parameters, experimental data of the responsivity at 6 keV as a function of device temperature was acquired as well and fitted in parallel with the energy dependent data. The experimental data taken with the 10 μm junction with an applied bias voltage of 250 μV as well as the result of the model fit are shown in Fig. 5.9. The set of free parameters chosen for both the simulations of the energy dependent and the temperature dependent data is shown in table 5.3.

The very large amount of traps in V-Al junctions as well as the large depth of the latter is striking. The 900 000 trapping states present in the V-Al electrode is a large number compared to the respective number of trapped states in Nb-Al and Ta-Al junctions (185 000 and 20 000 respectively). Also the depth of the traps equal to 330 μeV is larger than the corresponding values in Nb-Al and Ta-Al (240 and 160 μeV respectively). This strong trapping has as a consequence that the responsivity curve as a function of photon energy does not show a pronounced maximum anymore. As opposed to Ta-Al and Nb-Al junctions, the available trapping states never become saturated and therefore the responsivity curve remains flat. For photon energies larger than 3 keV quasiparticle self-recombination sets in and the responsivity starts to drop. The strong temperature dependence on the other hand arises from the interaction of the trapped quasiparticles with the thermal phonon bath. As the number of thermal phonons increases with increasing temperature, more trapped quasiparticles can be freed by phonon absorption from the traps. As a consequence de-trapping is more efficient with increasing substrate temperature and the responsivity increases accordingly. If the density of thermal quasiparticles in the electrodes becomes too important, recombination with thermal quasiparticles sets in and the responsivity starts decaying [Kozorezov 01].

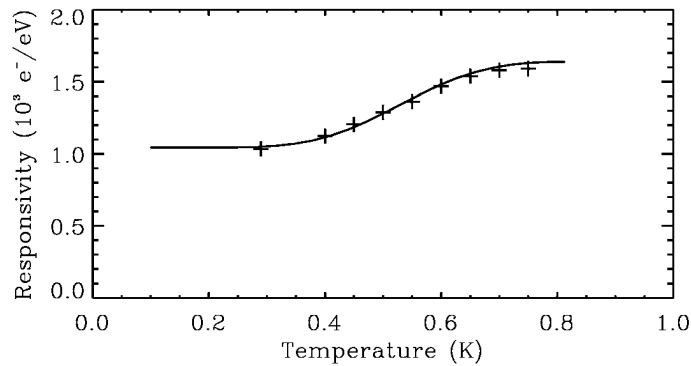

**Figure 5.9:** Responsivity in the base film for the absorption of 5.9 keV X-rays of the 10 μm side length junction as a function of substrate temperature. The crosses represent the experimental data, whereas the line is a fit to the data made with the model using the parameters from table 5.3. The applied bias voltage is equal to 250 μV and the magnetic field is 206 Gauss.

**Table 5.3:** Fitting parameters of the model for the V-Al based 10 μm side length junction.

| Symbol | Name | Unit | V-Al |
|---|---|---|---|
| $\tau_{loss}=1/\Gamma_{loss}$ | Quasiparticle loss time | μsec | 1.15 |
| $\Gamma_{esc}$ | Phonon escape rate | Hz | $1.8 \times 10^9$ |
| $C_{trap}$ | Trapping probability | / | 0.25 |
| $n^{traps}$ | Number of traps in electrode | / | 900 000 |
| $d_{trap}$ | Trap depth | μeV | 330 |

The reason for the enormous amount of traps in the V-based junctions is most probably related to the strong reactivity of V with oxygen and the metallic nature of some of the V oxides. As the multi-layer is removed from the deposition system, the upper V surface is exposed to the normal atmosphere and a natural film of oxides forms on top of the multi-layer. In addition, after the base etch processing step the outer edge of the junction also gets exposed to the atmosphere and natural V oxides form at this edge as well. If some of these oxides are metallic and in contact with the superconductor, they will, through the proximity effect with the Cooper pairs from the superconductor, form a small, very localised region of suppressed energy gap. If quasiparticles from the superconductor now scatter into these lower energy states, they cannot diffuse into other regions and are effectively trapped at this position.

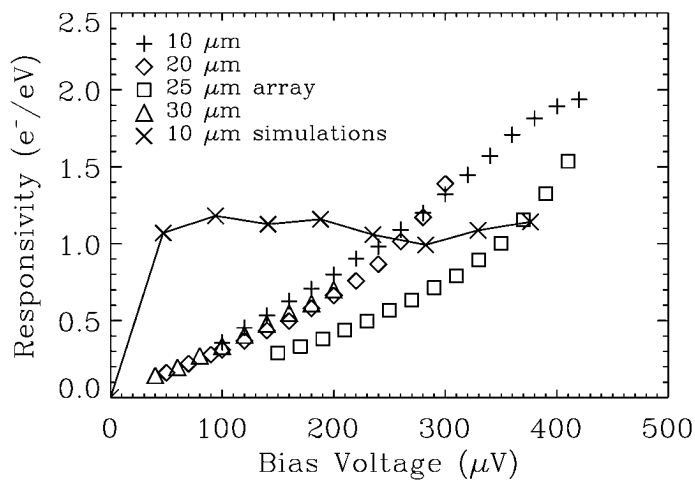

**Figure 5.10:** Responsivity in the base film of the V-Al based junctions with 25 nm of Al as a function of the applied bias voltage. Experimental data for the 10 (crosses), 20 (diamonds), 25 (squares) and 30 (triangles) μm side-length junctions is shown, as well as values calculated with the model for the 10 μm side-length device. The model parameters from table 5.3 were used. The applied magnetic field is of the order of 200 Gauss.

### 5.1.3.1.4    Bias voltage dependence

Figure 5.10 shows the variation of the base electrode responsivity for the absorption of 5.9 keV photons as the bias voltage is varied between 0 and 400 µV. The experimental points for the 10, 20, 25 and 30 µm side length junctions are shown along with the result of calculations made with the model and the parameters from table 5.3. A large discrepancy can be seen between the experimental data and the simulations. The experimental responsivities for all the devices show a very strong increase with increasing bias voltage. On the other hand, the simulated curve for the 10 µm device remains rather flat with some structure caused by the different quasiparticle energy distributions induced by the different bias voltages. The very strong increase of responsivity with bias voltage is a rather uncommon feature, which is not generally observed in Ta and Nb-based junctions [Hettl 97, Katagiri 97, Verhoeve 02b]. For these junctions the responsivity increases to a small maximum at relatively low bias voltages and then stays rather constant over the rest of the bias domain, similar to the result of the simulation. In order to try to explain the rapid rise of the responsivity in V-Al junctions one has to take a deeper look into the variation of the quasiparticle energy distribution as a function of the bias voltage.

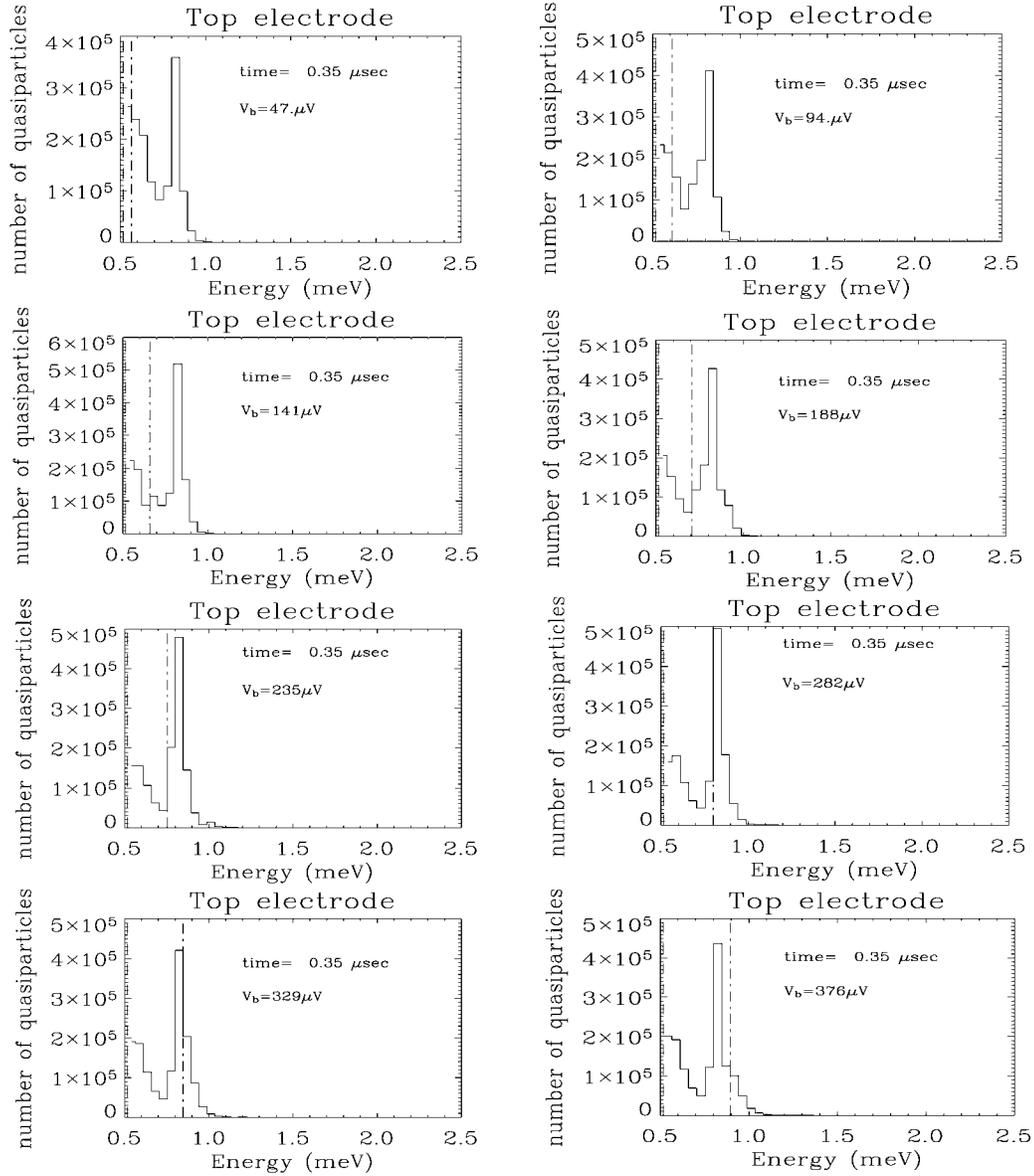

**Figure 5.11:** Quasiparticle energy distribution 0.35 µsec after the absorption of a 5.9 keV X-ray in the electrode of a 10 µm V-Al junction for eight different bias voltages applied to the junction. The dashed line represents the energy gap of the junction whereas the dashed dotted line represents the bias energy above the gap.

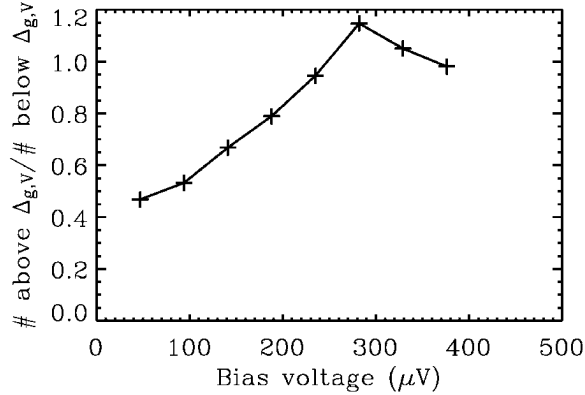

**Figure 5.12:** Ratio of the number of quasiparticles above the energy gap in bulk V to the number of particles below the energy gap in bulk V as a function of applied bias voltage.

Figure 5.11 shows the simulated quasiparticle energy distributions in the electrodes of the detector 0.35 microseconds after the absorption of a 6 keV photon for the eight different bias voltage settings for which simulations were made. Note that simulations can only be made for bias energies, which are an integer multiple of the energy interval $\delta\varepsilon$, because otherwise quasiparticles would, after tunnelling, end up in between energy intervals, which would induce considerable numerical errors. Talking about a "quasi-equilibrium" distribution the same as for the Nb and Ta-based devices described in chapter 2 is not appropriate for the V-based junctions under consideration, as the formation of the quasi-stationary distribution takes about half a microsecond, which is approximately equal to the lifetime of the quasiparticles in the electrodes. Therefore, the stable distribution is not reached until the moment when most quasiparticles are already lost and the signal pulse is already strongly reduced. Nonetheless, interesting features can be seen in the energy distributions of the quasiparticles. It can be observed that, due to fast tunnelling for quasiparticles at energies below the bulk gap in V and slow tunnelling for quasiparticles at energy levels above the bulk gap in V, inversion of the population of quasiparticles is obtained. Most quasiparticles actually reside at an energy equal to the bulk gap energy in V, whereas the lower levels are not as populated as would be expected from a purely thermal distribution. It can also be seen in the figure that the fraction of quasiparticles residing above the bulk gap energy of V increases as the bias voltage increases, because of stronger energy gain due to tunnelling. Figure 5.12 shows the ratio of the number of quasiparticles residing at an energy above 820 $\mu$eV to the number of quasiparticles below that energy level. The increase is almost linear with bias voltage up to a bias energy of approximately 300 $\mu$V.

A possible scenario could be that the excess phonons of energy $\Delta_g^V$-$\Delta_g$ provoke the relaxation of the quasiparticles residing at the bulk V energy gap by stimulated emission of another phonon of energy $\Delta_g^V$-$\Delta_g$. Thereby, the population of phonons with energy $\Delta_g^V$-$\Delta_g$ is increased and the effect gets even stronger. The stimulated emission would be larger for large quasiparticle densities at the bulk V energy gap, thus for larger bias energies. Therefore, quasiparticle relaxation from $\Delta_g^V$ to $\Delta_g$ would be stronger for larger bias voltages. On the other hand, quasiparticles at the energy gap of the bi-layer $\Delta_g$ have a very fast tunnel time and this would thus increase the measured signal. Stimulated phonon emission is not included in the model and therefore this effect can at present not be modelled.

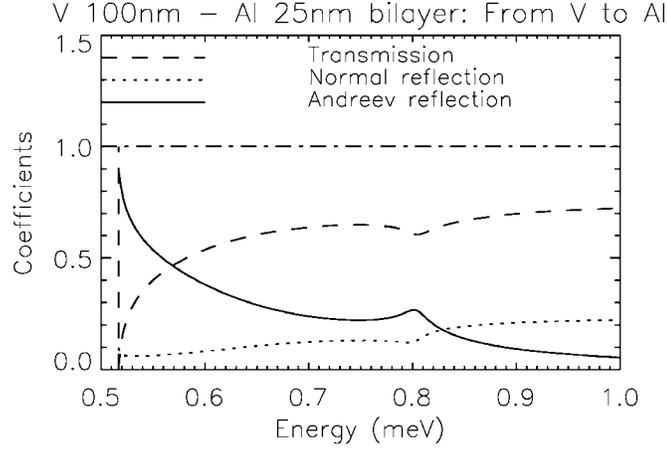

**Figure 5.13:** Probabilities of Andreev reflection A(ε), normal reflection B(ε) and transmission T(ε) for a quasiparticle of energy ε transferred from the V layer to the Al layer.

Andreev reflections at the V-Al interface could also play a role in increasing the charge output for larger bias voltages. Due to the discontinuity in the density of states at the V-Al interface (Fig. 5.3b), the quasiparticles undergo normal reflections as well as Andreev reflections when hitting the interface. Only a fraction of the incident quasiparticles actually passes the interface between the films. For a quasiparticle of energy ε hitting the interface, the probabilities of undergoing an Andreev reflection A(ε), normal reflection B(ε) or a transmission T(ε) are respectively given by [Aminov 96]:

$$A(\varepsilon) = \frac{\left|\sin\theta(\varepsilon, 0+)\right|^2}{\left|1 + 2Z^2 + \cos\theta(\varepsilon, 0+)\right|^2}, \tag{5.1}$$

$$B(\varepsilon) = \frac{4Z^2\left(1 + Z^2\right)}{\left|1 + 2Z^2 + \cos\theta(\varepsilon, 0+)\right|^2}, \tag{5.2}$$

$$T(\varepsilon) = \frac{2\left(1 + 2Z^2\right)\mathrm{Re}\left[\cos\theta(\varepsilon, 0+)\right]}{\left|1 + 2Z^2 + \cos\theta(\varepsilon, 0+)\right|^2}, \tag{5.3}$$

where θ(ε,x) is the unique Green's function calculated with the proximity effect theory (Section 2.2.1), the position 0+ is the position right across the interface and the Z factor is related to the normal state transmission coefficient $T^*$ via $1 + 2Z^2 = 1/T^*$. The normal state transmission coefficient can be calculated using equation (2.29), where the ratio of effective masses is chosen equal to one. Using the material parameters from table 5.1 and the determined values for the interface constant $C_{\gamma_{BN}}$ one finds $T^* = 0.756$. Now, knowing the results of the proximity effect theory for the 100 nm V – 25 nm Al bi-layer, one can calculate the three probabilities as a function of the quasiparticle energy. The result is shown in Fig. 5.13. The figure shows that Andreev reflections are quite strong close to the gap energy, which means that the quasiparticles at low energies are confined in the V layer, where the losses due to trapping are strongest and they are prevented from tunneling. For larger bias voltages the lower states are less populated, thereby making this effect smaller and increasing the charge output.

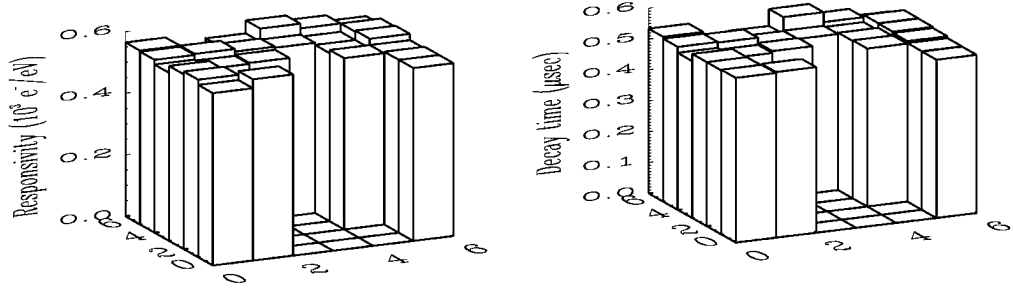

**Figure 5.14:** Responsivity and decay time of the signal pulses of a 6x6 pixel V-Al array for the absoption of 6 keV X-rays in the base electrode. The applied bias voltage is 250 µV and the parallel magnetic field equal to 208 Gauss.

Neither Andreev reflection, nor stimulated emission are included in the model. Therefore the effects of both mechanisms on the charge output could not be measured with our model. For future developments it should be considered to include the effects into the model and simulate the effect of the bias voltage on the responsivity of the devices.

### 5.1.3.1.5    Variations over an array

As mentioned in the introduction of the section, a 6x6 pixel array of V-Al junctions was fabricated as well. The size of an individual junction from the array is 25x25 µm². The base electrodes of the junctions in the array are interconnected via bridges and then connected to a common ground wire. In order to prevent diffusion of quasiparticles from one pixel into another, a Nb base plug is present in the bridge connecting neighbouring pixels. The top electrodes are connected individually to the read out electronics. The responsivity of all the pixels was determined for the absorption of 6 keV photons in the detectors. Figure 5.14 shows the uniformity of the responsivity of the different pixels in the array. The applied bias voltage is 250 µV and the applied parallel magnetic field equal to 208 Gauss. Unfortunately, several pixels in the array have interconnected base electrodes due to incompletely etched base V material (see Fig. 5.**1**). These pixels are represented by the zero responsivity junctions in the figure.

### 5.1.3.2    Energy resolution

The full width at half maximum (FWHM) energy resolution for a STJ is given by:

$$\Delta E = 2.355\sqrt{\sigma_{el}^2 + 1.7E\Delta_g(F + G + H) + \alpha E^2} \, , \qquad (5.4)$$

where $\sigma_{el}^2$ is the electronic noise contribution, E is the photon energy, F is the Fano factor, G is the tunneling factor, H is the cancellation factor and $\alpha$ is the spatial broadening factor.

In the following the different contributions to the noise are explained in more detail:

- $\sigma_{el}^2$ is the electronic noise at the output of the charge sensitive preamplifier, consisting of the JFET's series and parallel noise. It can easily be measured by introduction of a constant pulser signal. The electronic noise contribution is independent of the photon energy and depends, in the V-Al case, mainly on the dynamical resistance in the bias domain.

- The Fano factor F represents the resolution broadening due to statistical variations in the number of quasiparticles created during the photon absorption process. This broadening has a square-root dependence on the photon energy. The factor F was calculated to be equal to 0.2 in Sn [Kurakado 82] and Nb [Rando 92] and it is generally accepted that this value is the same in other materials.

- The tunneling factor G represents the resolution broadening due to statistical variations in the average number of tunnel events a quasiparticle undergoes. This contribution is non-zero only for STJs that allow back-tunneling of the quasiparticles. It has a square root dependence on the energy of the absorbed photons. For symmetrical STJs with infinite integration time of the signal pulse the contribution was calculated to be equal to G=1+1/<n>, where <n> is the average number of tunnel events per quasiparticle [Mears 93, Goldie 94]. As the signal to noise ratio of this noise contribution changes during the current pulse, the contribution can be minimized by choosing an optimum integration time of the pulse [Hiller 01, Verhoeve 02a]. In the V-Al case the decay times of the current pulses are too short to allow for serious minimisation of the factor. As the average number of tunnels in the measured V-Al devices is approximately equal to 0.5 independent of device size, one can derive G ~ 3 for V-Al.

- The cancellation factor H represents the resolution broadening because of statistical variations in the ratio of direct tunnel events, which add charge to the output signal, to cancellation tunnel events, which remove charge from the final signal. This broadening term has a square-root dependence on the photon energy E. The contribution was calculated to be equal to [Segall 99]:

$$H = \frac{4\sigma}{<n>(\sigma-1)^2},\qquad\qquad(5.5)$$

where $\sigma$ is the average of the ratio of direct tunnel events <n$_{tun}$> per quasiparticle to cancellation tunnel events <n$_{can}$> per quasiparticle, $\sigma = \frac{<n_{tun}>}{<n_{can}>}$, and <n> is the average total number of tunnel events per quasiparticle <n>= <n$_{can}$>+<n$_{tun}$>. Our simulations of photon detection experiments in V-Al devices revealed $\sigma = 10$, which gives a cancellation factor H approximately equal to one.

- The spatial broadening factor $\alpha$ represents the signal broadening because of spatial variations of the responsivity over the area of the device. Depending on the absorption position of the photon in the junction the generated output signal can present considerable variations because of the localised nature of certain loss sources [den Hartog 02]. The spatial broadening term has a linear dependence on the photon energy and is therefore the dominating factor for large photon energies.

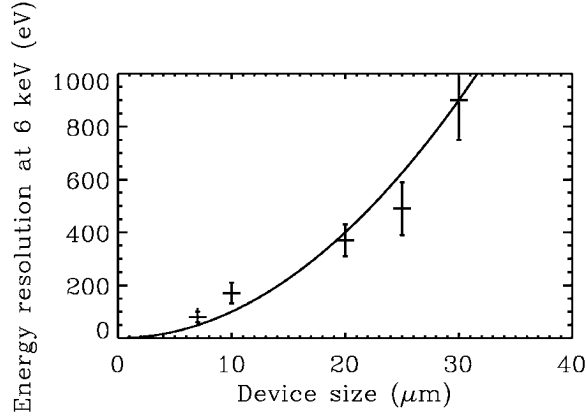

**Figure 5.15:** Measured energy resolution of the V-Al devices as a function of device side-length. The crosses with error bars represent the experimental points. The solid line represents the curve y = $x^2$.

Figure 5.15 shows the variation of the FWHM energy resolution as measured in the different spectra as a function of the side-lengths of the V-Al junctions. The energy resolution is extremely bad and varies approximately as the square of the side length of the junctions. Table 5.4 shows the different contributions to the resolution broadening at 5.9 keV as expected from the different terms in (5.4), as well as the derived spatial broadening factor $\alpha$ as a function of the side length of the detector.

**Table 5.4:** Different calculated and measured contributions to the resolution broadening of V-Al STJs at 5.9 keV as a function of the device size and derived spatial broadening factor $\alpha$.

| Device size | Measured Total FWHM | Measured electronic noise | Calculated Fano noise | Calculated tunnel noise | Calculated cancellation noise | Derived spatial broadening | Derived $\alpha$ |
|---|---|---|---|---|---|---|---|
| (µm) | (eV) | (eV) | (eV) | (eV) | (eV) | (eV) | / |
| 7 | 80 | 17 | 2.4 | 9.5 | 5.5 | 77 | $3 \cdot 10^{-5}$ |
| 10 | 170 | 50 | 2.4 | 9.5 | 5.5 | 162 | $1.4 \cdot 10^{-4}$ |
| 20 | 370 | 102 | 2.4 | 9.5 | 5.5 | 355 | $6.5 \cdot 10^{-4}$ |
| 25 | 490 | 150 | 2.4 | 9.5 | 5.5 | 466 | $1.1 \cdot 10^{-3}$ |
| 30 | 900 | 236 | 2.4 | 9.5 | 5.5 | 868 | $3.9 \cdot 10^{-3}$ |

The spatial broadening is clearly the single most important contribution to the resolution broadening. The spatial broadening factor $\alpha$ varies between $3 \cdot 10^{-5}$ for the smallest 7 µm junction and $3.9 \cdot 10^{-3}$ for the largest 30 µm junction. Note that in Ta-Al based junctions, which show a total energy resolution of 24 eV at 5.9 keV, values for $\alpha$ of the order of $10^{-6}$ have already been achieved [Brammertz 01b]. The fact that the spatial broadening depends on the square of the device size shows that the responsivity curve is very inhomogeneous over the area of the detector and that the losses are stronger at the edges. The quasiparticles created by a photo-absorption process in the middle of the detector will only reach the stronger loss centers at the edges of the detector after a certain time $\tau_D$, which depends on the diffusion speed in the electrode. During this time $\tau_D$ the losses will be much smaller than for the quasiparticles created by an absorption process close to the edge.

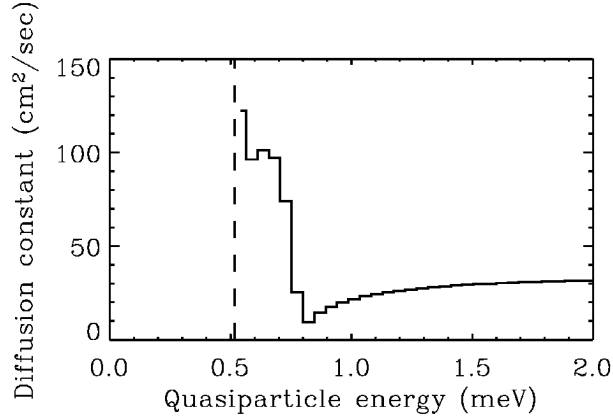

**Figure 5.16:** Diffusion constant in the superconducting V-Al bi-layer with 100nm of V and 25 nm of Al as a function of the quasiparticle energy. The vertical dashed line is the energy gap in the bi-layer.

Let us now try to estimate this diffusion speed in V-Al and the time it takes to fill the electrode of the detector homogeneously with quasiparticles as a function of the side-length of the detector. In the normal state the diffusion constant in a metal is given by (2.25). On the other hand, in a superconductor the quasiparticles move according to the group velocity:

$$v_g(\varepsilon) = \frac{v_F}{\text{DoS}(\varepsilon)}, \tag{5.6}$$

where $v_F$ and $\text{DoS}(\varepsilon)$ are the well known Fermi velocity and density of states in the superconductor. Replacing the Fermi velocity by the group velocity in (2.25) and using (4.2), one finds the expression for the diffusion constant in a superconductor:

$$D(\varepsilon) = \frac{1}{3} v_g(\varepsilon) \text{RRRl}_{300}. \tag{5.7}$$

As our electrode is formed of a V-Al bi-layer, the diffusion constant will depend on the position x in the bi-layer. One can find the average diffusion constant $D(\varepsilon)$ in the electrodes, by averaging $D(x,\varepsilon)$ over the number of states $2N_0(x) \text{DoS}(x,\varepsilon)$ in the bi-layer, which was calculated with the proximity effect theory:

$$D(\varepsilon) = \frac{\int_{\text{electr.}} N_0(x) v_F(x) \text{RRR}(x) l_{300}(x) dx}{3 \int_{\text{electr.}} N_0(x) \text{DoS}(x,\varepsilon) dx}. \tag{5.8}$$

Using $v_F$ and $N_0$ from table 5.1, the mean free paths at 300 K of respectively 2.8 and 41 nm and the RRR of the 100 nm V and 25 nm thick Al films of respectively 18 and 3, one finds the diffusion constant as a function of quasiparticle energy in the V-Al bi-layer. The result of the theoretical estimate of the diffusion constant is shown in Fig. 5.16. The diffusion is fastest for the quasiparticles close to the gap energy, as for those energies most states are available in the Al film, which presents fast diffusion. At the bulk gap in V the diffusion constant is minimum and of the order of 10 cm$^2$/sec.

Using the quasiparticle energy distributions from our simulations (Fig. 5.11), one can now calculate the average diffusion constant in the V-Al bi-layer according to (2.56). For bias voltages between 100 and 400 µV, the average theoretical diffusion constant in the superconducting state is approximately equal to 50 cm$^2$/sec. Experimentally it has been found though, that the diffusion in superconductors is much slower than the theoretical estimates. In Ta the experimental diffusion constant is found to be a factor 4 to 7 slower than the theoretical estimate [Friedrich 97, Nussbaumer 00], whereas in Nb the diffusion coefficient is found to be a factor 6 to 12 smaller than the theoretical estimate [den Hartog 02]. The discrepancy between the theoretical estimates and the experimental values is suspected to arise from subsequent quasiparticle trapping and de-trapping events during the diffusion process, which would confine the quasiparticles in the trap for the amount of time it takes to de-trap them again. The larger discrepancies seen for Nb films compared to Ta films could then be explained by the larger trap density in Nb as compared to Ta (see chapter 2). Because in V the trap density was found to be even larger than in Nb, an experimental diffusion constant reduced by a factor 10-15 as compared to the theoretical estimate is rather likely in our V-Al electrodes. Nevertheless, a direct link between quasiparticle trapping and slow diffusion has not been established yet. Experiments involving the determination of the diffusion constant as a function of temperature might give more certainty as to the origin of the slow diffusion. A higher temperature effectively increases the de-trapping rate due to phonon absorption and would, in the trapping/de-trapping model, also increase the diffusion speed of the particles.

Therefore a diffusion constant of the order of 4 cm$^2$/sec is assumed in our electrodes. After a time $\tau_D$ the quasiparticles have on average diffused a length $l_{diff} = \sqrt{D\tau_D}$ . One can now calculate the time after which the quasiparticles created by a photo-absorption process in the middle of the square junction will on average reach the corner of the junction. This time is a very simple approximation of the time necessary to fill the complete electrode with quasiparticles. In table 5.5 the results are shown as a function of the device size for a diffusion constant equal to 4 and 50 cm$^2$/sec respectively.

**Table 5.5:** Approximate time necessary to fill the junction homogeneously with quasiparticles as a function of device side length for two different values of the diffusion constant.

| Diffusion constant (cm$^2$/sec) | Time necessary for homogeneous junction filling (µsec) | | | | |
|---|---|---|---|---|---|
| | 7 µm | 10µm | 20 µm | 25 µm | 30 µm |
| 4 | 0.06 | 0.12 | 0.5 | 0.75 | 1.1 |
| 50 | 0.005 | 0.01 | 0.04 | 0.06 | 0.09 |

It is clear from table 5.5 that the diffusion constant in our electrodes cannot be as large as 50 cm$^2$/sec. For such a large value of the diffusion constant even the 30 µm side-length electrode would have a homogeneous quasiparticles distribution after only a tenth of a microsecond. Therefore, for most of the current pulse the quasiparticles would be homogeneously distributed over the area of the junction and spatial inhomogeneities should not be as strong as observed. For a diffusion constant value of 4 cm$^2$/sec the situation looks much more plausible. In the 7 µm junction the quasiparticles are distributed homogeneously over the electrode after only 60 nsec, which explains the relatively good energy resolution for those devices. For the larger devices the fraction of time for which the losses depend strongly on the absorption position gradually increases and for the 30 µm device the quasiparticle distribution never reaches a homogeneous distribution during the current pulse. This explains qualitatively the dependence of the energy resolution on the device size. A more thorough analysis should include the diffusion equation of quasiparticles and spatially distributed sources of quasiparticle loss channels and then analyse the effect on the energy resolution. This could unfortunately not be incorporated into the model because of calculation time limitations. A good model, which includes the mentioned features, but does not include the energy distribution of the quasiparticles, was developed by Kozorezov et al. [Kozorezov 02].

## *5.2 Al based junctions*

The fabricated Al junctions presented in the previous chapter were operated as photon detectors in an ADR at temperatures between 35 and 100mK. The junctions were illuminated with monochromatic near-IR to soft-UV radiation (1-5 eV) as well as with 6 keV X-rays from a $^{55}$Fe radioactive source. The tested devices are square single pixel junctions with side-lengths of respectively 10, 30, 50 and 70 μm. The base electrode consists of a 100 nm thick Al film covered by a 50 nm thick top electrode.

Prior to presenting the photon detection experiments, first the characteristic rates of the quasiparticle processes in the electrodes of the junctions have to be calculated.

### 5.2.1 Quasiparticle characteristic rates in Al

Naturally, the proximity effect model does not need to be applied in this case, as the electrodes of the junctions are not sandwiches of two superconducting materials anymore. In all the equations from section 2.3.1 one can therefore replace the results of the proximity effect theory by their BCS counterparts:

$$\text{DoS}(x,\varepsilon) \rightarrow \frac{\varepsilon}{\sqrt{\varepsilon^2 - \Delta_g^{Al\,2}}}, \tag{5.9}$$

$$\text{ImF}(x,\varepsilon) \rightarrow \frac{\Delta_g^{Al}}{\sqrt{\Delta_g^{Al\,2} - \varepsilon^2}}, \tag{5.10}$$

$$\Delta(x) \rightarrow \Delta_g^{Al}. \tag{5.11}$$

The parameters used for the calculation of the different characteristic rates can be found in table 5.6.

Before proceeding with presenting the results for the characteristic times in the electrodes of the junction, a further remark has to be made. The Al junctions under consideration are not symmetrical, because of the difference in thickness between the top and base electrodes. On the other hand, the model was only implemented numerically for symmetrical lay-ups. Note that this limitation applies of course only to the numerical code written and not to the model presented in chapter 2, which is also valid for asymmetric devices. Therefore, an approximation will have to be made and the junction will be treated as an equivalent symmetrical junction with an electrode thickness intermediate between the thickness of the top and base electrodes. The condition for being able to perform this "electrode thickness averaging approximation" is that the quasiparticles tunnel very often back and forth between the two electrodes, before they finally are lost. In this way the properties of the base and top electrodes average out. Of course, a model valid for asymmetrical junctions would be more appropriate here, but until this model is available we will have to work with this approximation and consider the changes to the stationary state of the quasiparticle energy distribution of minor importance. Therefore the characteristic rates in an Al electrode with a thickness of 75 nm will be calculated. The results of the calculations are shown in Fig. 5.17.

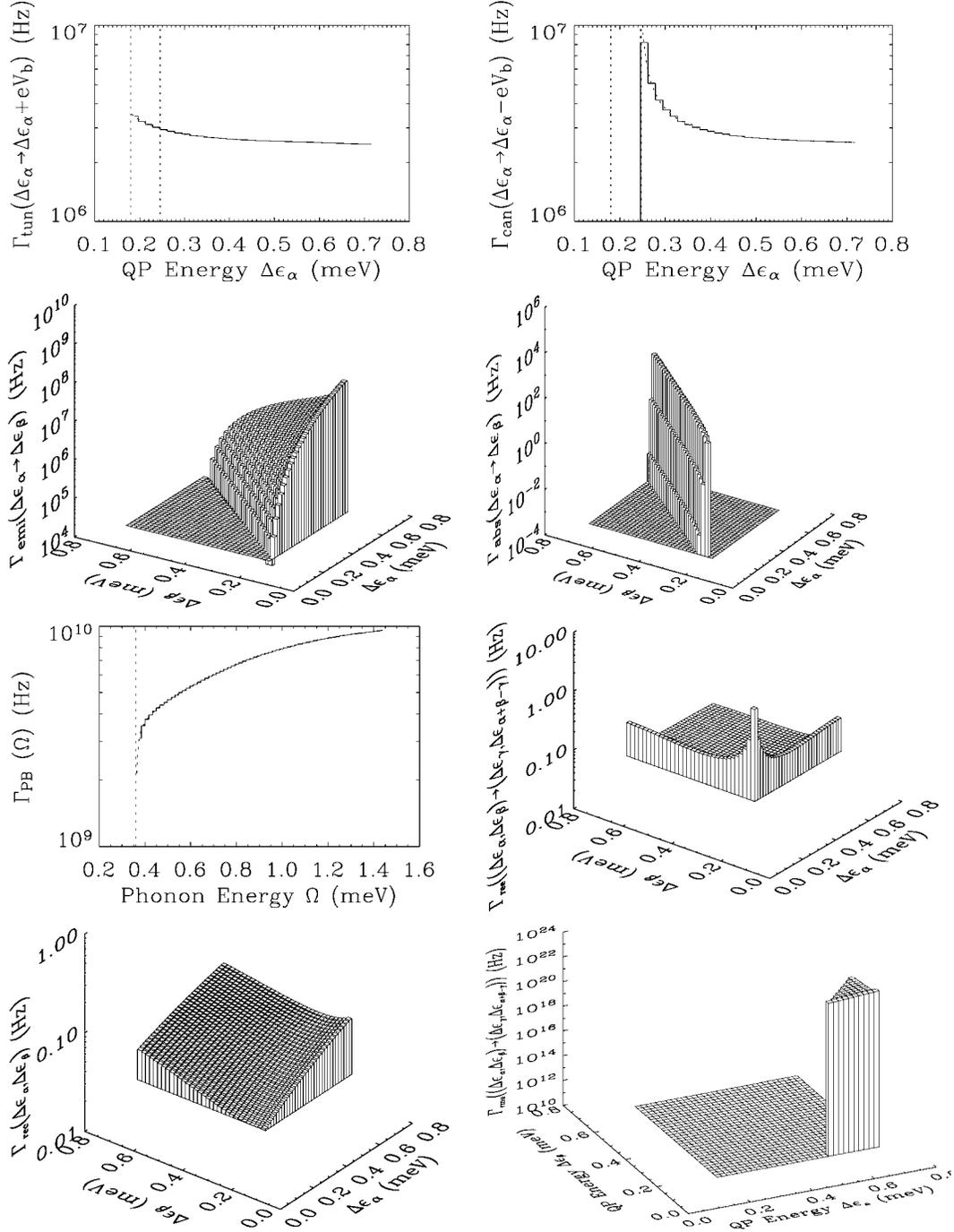

**Figure 5.17**: Characteristic quasiparticle rates in our Al junctions. **(a)** Tunnel rate as a function of quasiparticle energy ($V_b$ = 75 µV). **(b)** Cancellation tunnel rate as a function of quasiparticle energy ($V_b$ = 75 µV). **(c)** Rate for electron-phonon scattering with emission of a phonon as a function of initial and final quasiparticle energy. **(d)** Rate for electron-phonon scattering with absorption of a thermal phonon as a function of initial and final quasiparticle energy (T=40mK). **(e)** Cooper pair breaking rate as a function of phonon energy. **(f)** Rate for quasiparticle recombination with subsequent pair breaking as a function of the initial quasiparticle energies $\Delta\epsilon_\alpha$ and $\Delta\epsilon_\beta$, and for the particular final energy $\Delta\epsilon_\gamma$ equal to the gap energy of the electrode (junction size = 30 µm). **(g)** Rate for quasiparticle recombination with subsequent phonon loss as a function of the initial quasiparticle energies (size = 30 µm). **(h)** Rate for the quasiparticle multiplication process as a function of the initial ($\Delta\epsilon_\alpha$) and final ($\Delta\epsilon_\beta$) energy of the first quasiparticle. The energy of one of the generated quasiparticles $\Delta\epsilon_\gamma$ is fixed to being equal to the gap energy $\Delta_g$ (size = 30 µm).

**Table 5.6**: Material parameters used for the characteristic times calculations in Al junctions.

| Symbol | Name | Unit | Al |
|--------|------|------|-----|
| $R_n A$ | Normal resistivity of junction | $\mu\Omega\,cm^2$ | $7 \pm 0.5$ |
| $T_C$ | Critical temperature | K | 1.2 |
| $\Delta_g$ | Energy gap | $\mu eV$ | 180 |
| $N_0$ | Single spin normal state density of states at Fermi energy | $10^{27}$ states $eV^{-1}\,m^{-3}$ | 12.2 |
| $\alpha^2$ | Average square of the electron-phonon interaction matrix element | meV | 1.92 |
| N | Ion number density | $10^{28}\,m^{-3}$ | 6.032 |
| $\tau_0$ | Electron-phonon interaction characteristic time | nsec | 440 |
| T | Temperature | K | 0.04 |

Of course, for pure Al junctions all features caused by the second proximity layer disappear. The only feature left is the singularity at the gap energy. The tunnel rate at a typical bias voltage of 75 μV is rather constant in Al junctions and for our junctions of the order of 3 MHz. The cancellation rate on the other hand presents a strong maximum just above the bias energy, because of the strong singularity in the density of states at the gap energy. The phonon absorption process at 40 mK is slow because of the small number of thermal phonons at this low temperature. This implies that de-trapping by phonon re-absorption is going to be very ineffective.

### 5.2.2 Near-IR to soft-UV photon detection experiments in Al junctions

The different Al junctions were tested in the Adiabatic Demagnetisation Refrigerator. The photons from the monochromatic light source with energies ranging from one to approximately five eV were coupled to the detectors at ~40 mK via an optical fibre. The illumination was made through the transparent sapphire substrate on to the base electrode of the detector. The sapphire-Al interface is very reflective in the energy domain under consideration. Figure 5.18 shows the absorption efficiency of a 100 nm thick Al film on a 500 μm thick sapphire substrate for incoming photon wavelengths ranging from 100 to 1800 nm [Palik 85]. For photons of wavelength varying between 300 and 1000 nm the efficiency roughly varies between 10 and 20%, the incomplete absorption efficiency in the thin Al film being completely due to reflections at the sapphire-Al interface. The intensity of the Xenon light source was adjusted in order to have about 100 counts per second in the base electrodes of the different detectors.

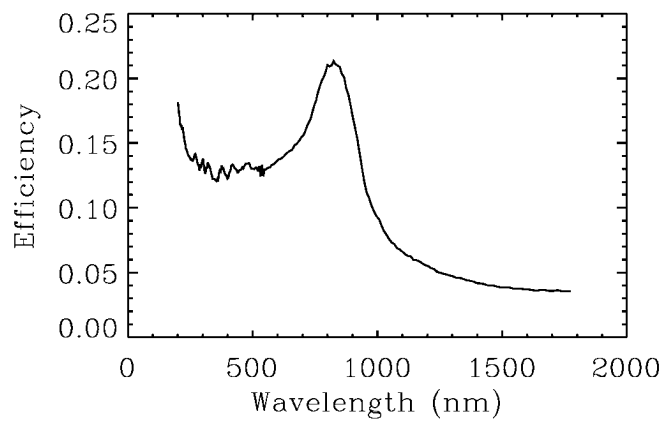

**Figure 5.18:** Photon absorption efficiency for a 100 nm thick Al film on top of a 500 μm thick sapphire substrate as a function of the photon wavelength. The incidence angle of the incoming photons is normal to the surface.

Figure 5.19 shows a typical optical spectrum as acquired with the Al junctions. The spectrum shown in the figure was taken with the 30 μm side length junction, which was illuminated with 2.48 eV photons (λ=500 nm). The bias voltage applied to the junction was 80 μV and the parallel magnetic field applied in order to suppress the Josephson current was equal to 32 Gauss. The peak in the middle of the figure represents the events absorbed in the Al junction. The peak on the right hand side of the figure is the electronic pulser peak, which measures the resolution broadening arising from the noise created by the charge sensitive pre-amplifier. The width of this peak can be quadratically subtracted from the photo peak width in order to yield the intrinsic resolution of the detector, cleared from any broadening effects by the electronics. Whereas the final measured resolution of the detector of course includes the broadening due to the electronics, the knowledge of the intrinsic resolution is still interesting, because it allows the comparison of the actual performance of the junction to theoretical predictions. On the left hand side of the spectrum a small number of events can be detected that arise from the ~2 μm infrared background created by the laboratory environment at 25°C. These thermal photons reach the detector through the optical fibre. In between the photo-peak and the microwave background peak an additional, very small number of events can be seen, which are events absorbed in the base lead and the plug of the detector.

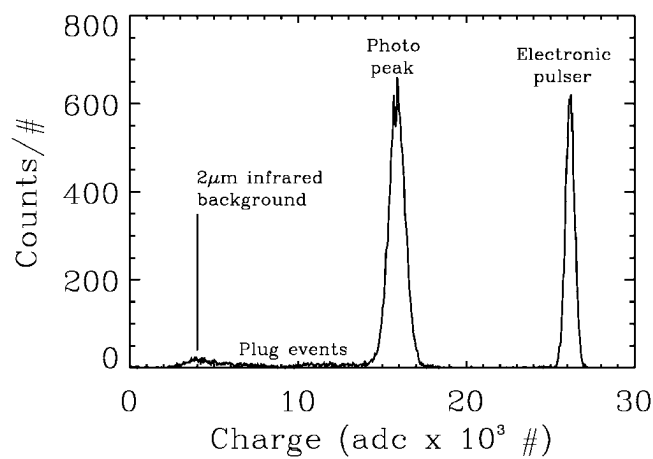

**Figure 5.19:** Spectrum acquired with the 30 μm side length Al junction under illumination with 500 nm photons (E = 2.48 eV). The applied bias voltage is 80 μV and the parallel magnetic field is equal to 32 Gauss.

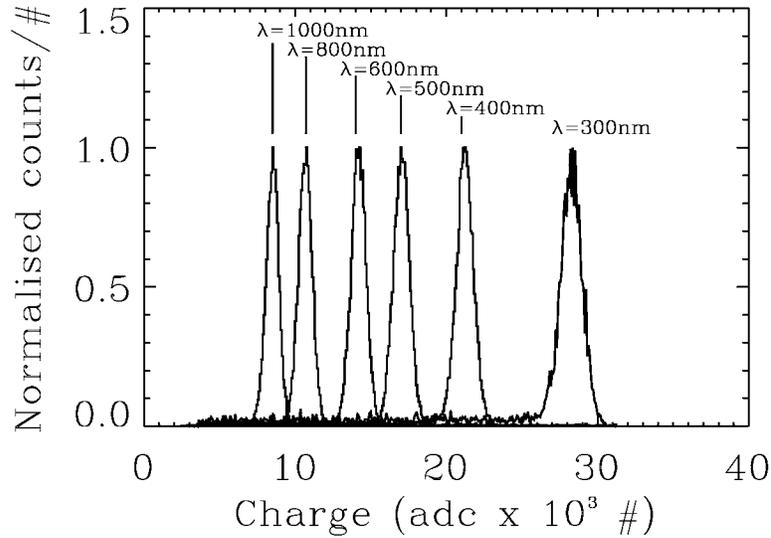

**Figure 5.20:** Collage of six different spectra acquired with the 30 μm side-length device with an applied bias voltage of 80 μV. The applied parallel magnetic field is 32 Gauss. The six different spectra correspond to six different photon energies varying between 1.24 eV (λ = 1000 nm) and 4.13 eV (λ = 300 nm). For every spectrum only the photo-peak is shown. The peak corresponding to the electronic noise is not plotted.

A collage of six different spectra acquired with the 30 μm junction is shown in Fig. 5.20. The same as for the spectrum of Fig. 5.19, the applied bias voltage was 80 μV and the magnetic field 32 Gauss. The energy of the incoming photons was varied between 1.24 and 4.13 eV. Only the photo peaks can be seen in the figure, as the electronic pulser peaks of the respective spectra are not shown.

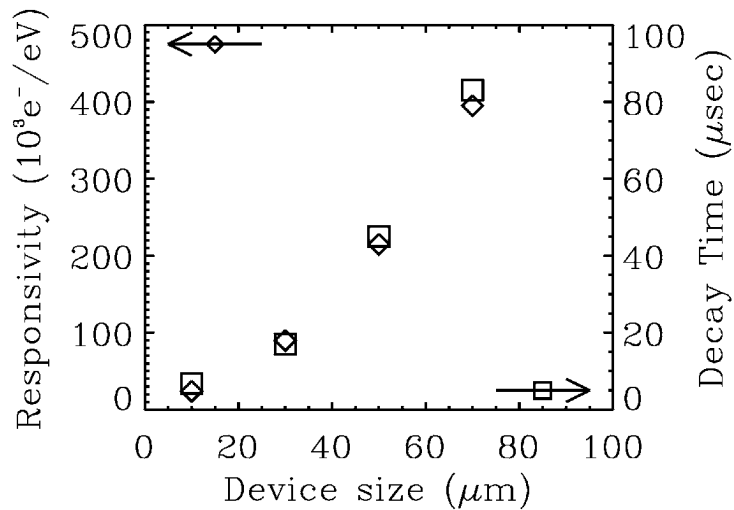

**Figure 5.21:** Responsivity and pulse decay time as a function of the device size of the different Al junctions. The applied bias voltage is equal to 100 μV and the parallel magnetic field of the order of 30 Gauss.

### 5.2.2.1    Responsivity and pulse decay time

From the analysis of the slow and fast channel spectra (see section 3.2.2) one can determine the responsivity and the decay time of the signal pulses. In the following it will be analysed how the responsivity and the decay time of the measured pulses depend on junction characteristics such as the side length and other characteristics such as the applied bias voltage, temperature and photon energy.

#### *5.2.2.1.1    Size dependence*

The responsivity and decay time of the junctions were determined as a function of the different junction side lengths with an applied bias voltage of 100 µV. The result is shown in Fig. 5.21. Both the responsivity and the decay time of the detectors increase approximately proportional to the area of the detector, indicating that the major loss source is not in the bulk of the Al, but rather very localised, probably located at the contacts of the detector. One eV of photon energy creates approximately 3300 free charge carriers in the Al electrode, whereas the measured charge at the output of the detectors is much higher. The ratio of the charge measured at the output of the junction Q to the number of charge carriers initially created in the junction $Q_0$ is defined as the charge amplification factor $\bar{n} = Q/Q_0$ of the junction. This amplification factor $\bar{n}$ is rather large for our Al junctions.  It is equal to 7 for the smallest junction (10 µm) and increases to 105 for the largest available junction size (70 µm). These large amplification factors increase the signal to noise ratio and are very helpful in reducing the electronic noise and increasing the resolving power of the detector. Note that the charge amplification factor $\bar{n}$ is not equivalent to the total number of tunnel events a quasiparticle undergoes on average during its lifetime <n>. The difference between the two comes from the fact that a quasiparticle can undergo a direct tunnel event, which adds a charge to the output signal, or a cancellation tunnel event, which removes a charge from the output signal. If $<n_{tun}>$ is the average number of tunnel events per quasiparticle and $<n_{can}>$ is the average number of cancellation events, one has $<n>=<n_{tun}>+<n_{can}>$ and $\bar{n}=<n_{tun}>-<n_{can}>$. The average total number of tunnel events per quasiparticle <n> is related to the charge amplification factor $\bar{n}$ via:

$$< n > = \frac{\sigma + 1}{\sigma - 1}\bar{n} \, , \qquad\qquad (5.12)$$

where σ is the tunnel to cancellation ratio $\sigma=<n_{tun}>/<n_{can}>$.
On the other hand the signal pulses are rather long. The decay time of the signal varies between 7 µsec for the smallest junction and 80 µsec for the largest junction. This of course limits the maximum count rate of the detector severely.

#### *5.2.2.1.2    Energy  dependence*

A series of spectra was acquired with the junctions of 30 to 70 µm side length with the wavelength of the incoming photons varying between 300 and 1000 nm. For all spectra the bias voltage was equal to 50 µV and the applied magnetic field was of the order of 50 Gauss. The responsivity and the decay time of all the spectra was determined from the slow and fast channel spectra and the calibration data. The result is shown in Fig. 5.22 For the 30 and the 50 μm side length junctions, spectra for 6 keV X-rays could also be acquired (see also section 5.2.3). For completeness of the energy-dependent data set, these points were added to the figure as well. In all the Al devices, even in the near-IR to soft-UV domain, the responsivity and the decay time are not constant with respect to the photon energy. This non-linearity in the optical domain is a very uncommon feature, which cannot be observed in Ta or Nb based junctions [Verhoeve 02b]. The non-linearity arises from the fact that the quasiparticle traps in the junctions gradually saturate as the number of quasiparticles in the electrodes increases [Poelaert 99a]. As a consequence the number of available traps in the electrodes cannot be much larger than the number of quasiparticles created by an optical photon. In Al, 3300 quasiparticles are created per eV of photon energy. Therefore, the number of available trapping states must be of the order of several thousand states. In order to get more information about the trap characteristics, simulations with the energy-dependent kinetic equations model for all the three available junction sizes were made. The five input parameters of the model were varied in order to find a fit to the energy dependent data in Fig 5.22. The results of the simulations are shown in the figure as well. Table 5.7 shows the values of the parameters, which were found to give the best fit to the experimental data for the three different device sizes.

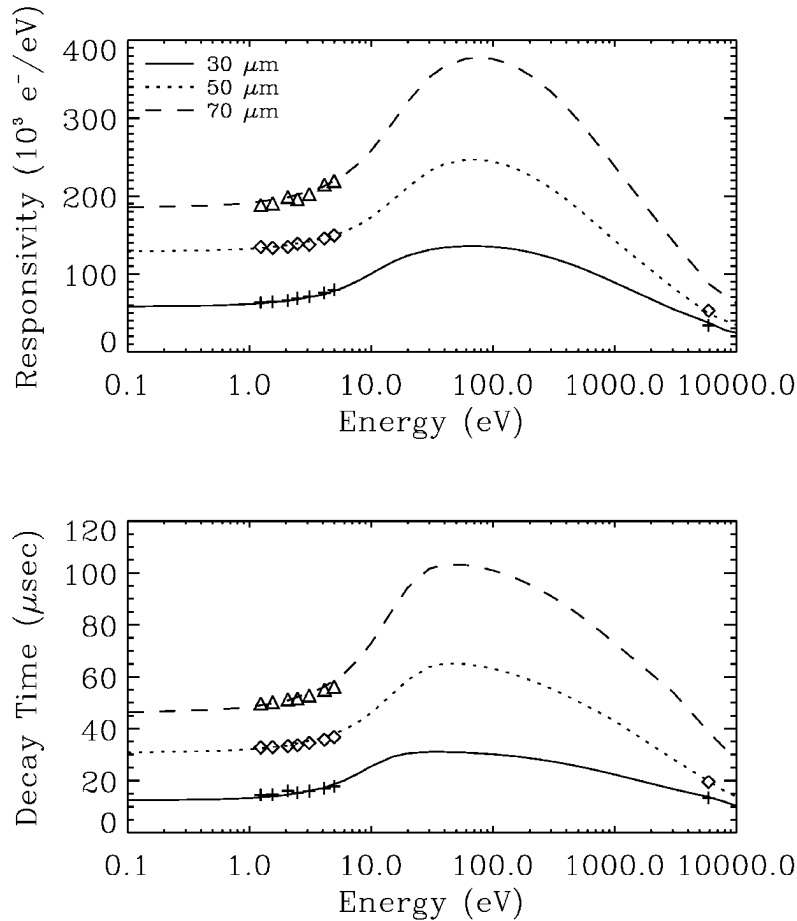

**Figure 5.22:** Measured responsivity and decay time of the 30 (crosses), 50 (diamonds) and 70 (triangles) μm Al junctions as a function of incoming photon energy for an applied bias voltage of 50 μV. The results of the simulations are shown as well.

**Table 5.7:** Fitting parameters of the model for three Al based junctions with side-lengths of respectively 30, 50 and 70 μm.

| Symbol | Name | Unit | 30 μm | 50 μm | 70 μm |
|---|---|---|---|---|---|
| $\tau_{loss}=1/\Gamma_{loss}$ | Quasiparticle loss time | μsec | 32 | 69 | 108 |
| $\Gamma_{esc}$ | Phonon escape rate | Hz | 1.4 10⁹ | 1.4 10⁹ | 1.4 10⁹ |
| $C_{trap}$ | Trapping probability | / | 0.055 | 0.02 | 0.014 |
| $n^{traps}$ | Number of traps in electrode | / | 5300 | 7800 | 7700 |
| $d_{trap}$ | Trap depth | μeV | 81 | 81 | 81 |

Indeed, the number of available traps is very low of the order of 5000-8000 traps per electrode. In addition, the number of traps is roughly independent of the device size, which shows that the traps are not in the bulk or at the perimeter of the Al junctions, as these trap locations would imply an increase of the number of trapping states with the size of the device. The most probable location of the trapping states is at the positions where the Nb of the top contact or the base plug is in contact with the Al from the tunnel junction. From the simulations in chapter 2 it is known that Nb forms in general a lot of trapping states, which makes the scenario of quasiparticle trapping in the Nb of the contacts and plugs even more likely. The trap depth, which is equal to 81 μeV, is lower than the corresponding trap depths in Nb or Ta. Note nevertheless that the best experiment to perform in order to determine the trap depth is the variation of the responsivity as a function of temperature. This data is not available for our Al junctions and therefore the uncertainty on the trap depth and the trapping probability is the largest. It could well be that the real trap depth is larger, whereas the trapping probability $C_{trap}$ is lower than the derived values. In order to get a better error margin for the trap depth one will have to analyse the responsivity data as a function of temperature. The trapping probability on the other hand decreases with increasing device size. For the absolute value of the probability the same argument holds as for the trap depth. The uncertainty is rather high, because no temperature dependent data is available. Nevertheless, the relative values of the probabilities do have a real meaning. The increasing probability of being trapped in smaller devices reflects the very localised position of the traps in the junction, namely in the Nb making contact with the Al electrodes. As the size of the junction decreases the quasiparticles have a higher chance to diffuse into the corners of the devices, where the traps are located.

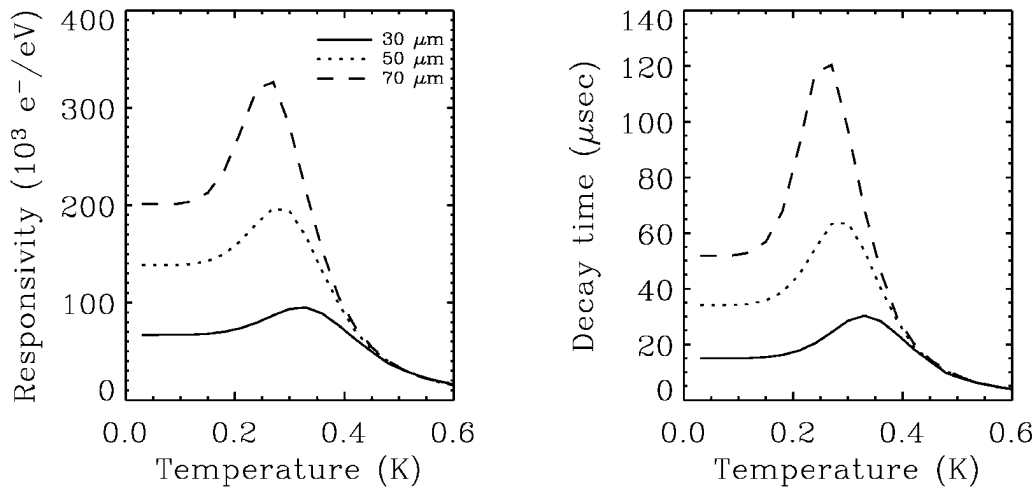

**Figure 5.23:** Simulated responsivity and decay time as a function of temperature for the 30, 50 and 70 μm Al devices. The applied bias voltage is equal to 50 μV.

### *5.2.2.1.3 Temperature dependence*

As already mentioned in the previous section, no temperature dependent data of the responsivity and the charge output is available for our Al junctions. The reason for this is our current inability to control the temperature of the ADR without creating excess noise in the read-out electronics of the junctions. In principle the temperature of the ADR can be very well adjusted to a required setting by applying a small magnetic field to the paramagnetic salts and by removing it gradually as the heat flow into the salt pill increases the entropy of the cold stage. In this way the temperature of an ADR can be controlled to within a fraction of a mK over the timescale of several hours. Unfortunately, the power supply we use to create the currents in the magnetic coil creates excess noise in the read-out electronics, which spoils the energy resolution in the optical regime severely. We therefore switched off the power supply completely and let the temperature of the system naturally drift up from base temperature (~35 mK) to approximately 100 mK. In this temperature domain the output of the Al based detector is not sensitive to temperature variations. The temperature dependence only sets in at higher temperatures, when the thermal phonon energy is sufficient to lift quasiparticles out of the trap. In the future the usage of a less noisy power supply should enable us to control the temperature of the cryostat without loosing too much in electronic noise. For the time being only simulations of the temperature dependence made with the parameters derived in the previous section can be presented. Figure 5.23 shows the results of simulations made for the 30, 50 and 70 μm side length junctions with an applied bias voltage of 50 μV. The model predicts that de-trapping by thermal phonons sets in at temperatures of about 150 mK, which as a consequence increases the responsivity and decay time. The exact value depends strongly on the trap depth. Therefore this kind of experiment is a good way to determine the depth of the traps. For temperatures of the order of 250 mK recombination with the very numerous thermal quasiparticles sets in and the responsivity and decay time start decreasing again.

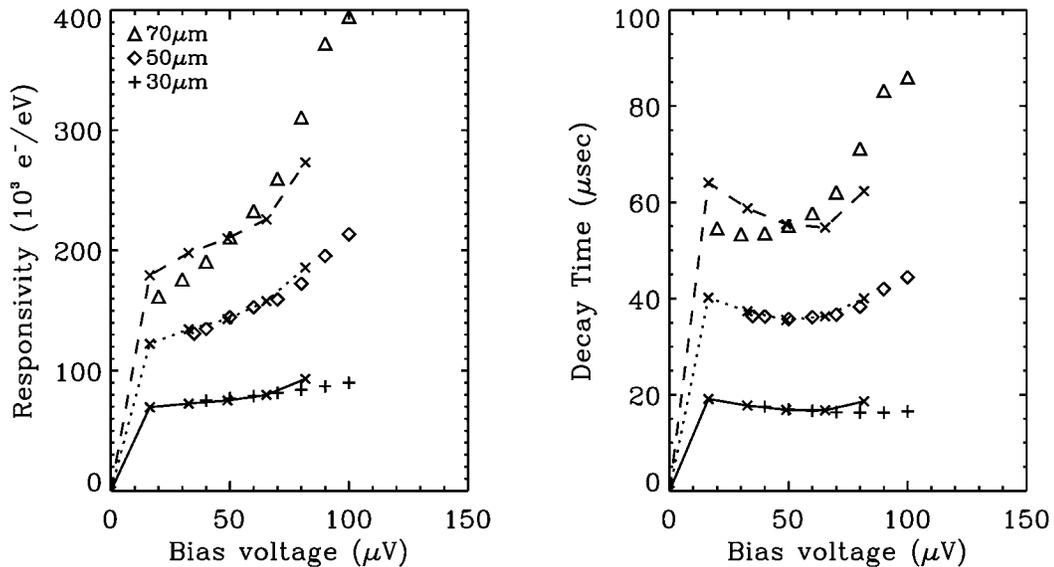

**Figure 5.24:** Responsivity and decay time as a function of applied bias voltage for the absorption of 300 nm (4.13 eV) photons in Al junctions. The experimental data is shown for the 30 (crosses), 50 (diamonds) and the 70 μm (triangles) side length junctions. The simulated data is shown as well for the 30 (solid line), 50 (dotted line) and the 70 μm (dashed line) device.

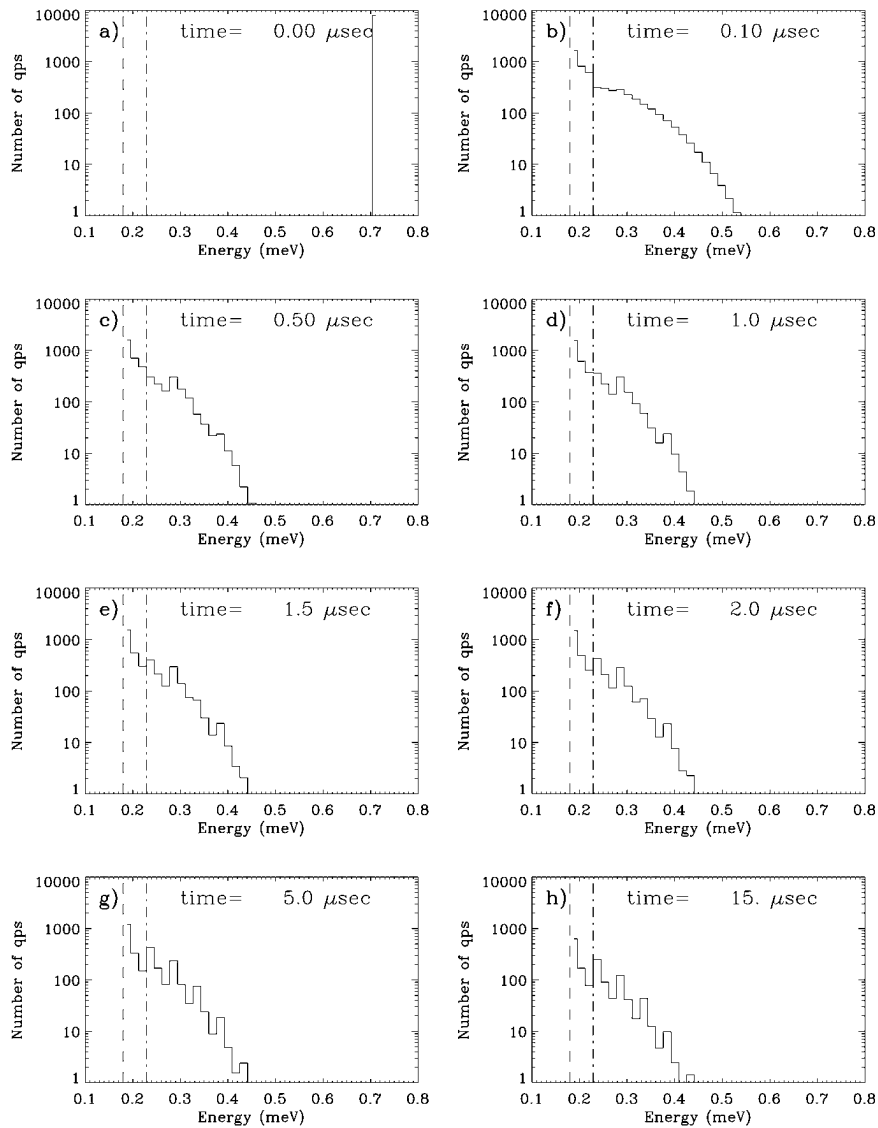

**Figure 5.25:** Time evolution of the quasiparticle energy distribution for the absorption of a 300 nm photon in the 30 μm side length Al junction with an applied bias voltage of 50 μV. The vertical dashed-dotted line shows the bias energy above the gap.

### 5.2.2.1.4  Bias voltage dependence

Figure 5.24 shows the variation of the responsivity and the decay time for the absorption of 300 nm photons as a function of the applied bias voltage for the Al devices with side lengths of respectively 30, 50 and 70 μm. The magnetic field applied is of the order of 50 Gauss. The experimental data points are shown as well as the results of simulations made with the parameters from table 5.7. For the 30 μm device the dependence of the responsivity and decay time on the bias voltage is rather flat. This is the traditional dependence, which is observed for most STJs [Hettl 97, Katagiri 97, Verhoeve 02b]. The 50 and 70 μm side length junctions on the other hand show a curious twofold effect. The first remarkable effect is the increase of the decay time of the pulse as the bias voltage increases. It looks like the quasiparticle losses decrease with increasing bias voltage. As a consequence the responsivity of the corresponding junctions increases as well, because the quasiparticles have on average more time available to tunnel. Now, the second curious effect is that the responsivity increases proportionally faster than the decay time. It looks like the tunnel rate of the quasiparticles increases with the applied bias voltage. In order to explain theses two effects, which are typical for low T$_C$ and low loss junctions, one will have to take a closer look at the evolution of the quasiparticle energy distribution during the current pulse.

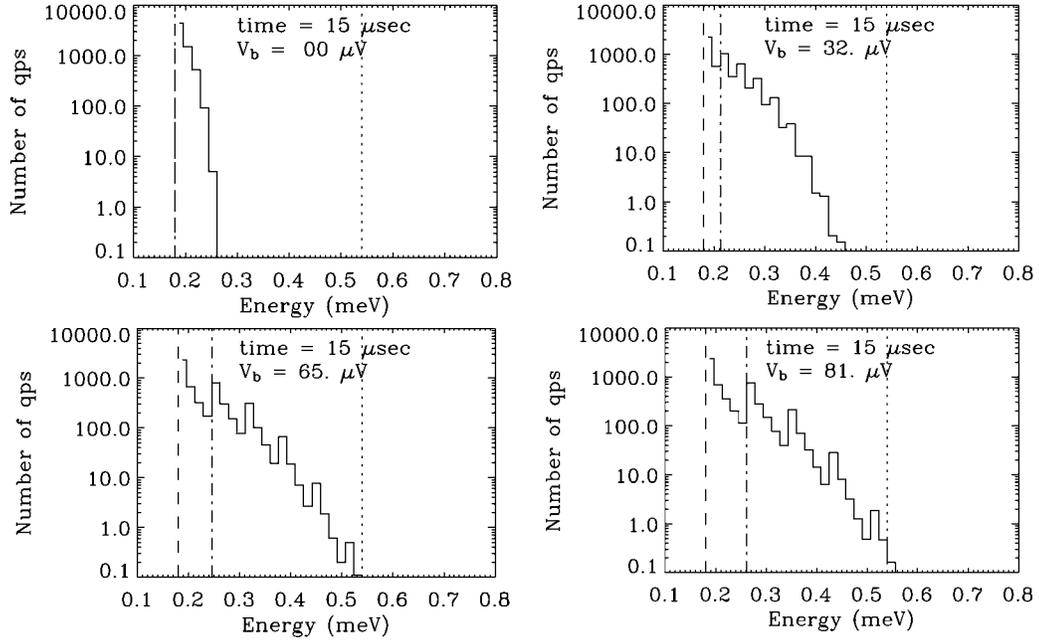

**Figure 5.26:** Variation of the quasiparticle energy distribution 15 μsec after the absorption of a 500 nm photon in a 30 μm side length Al junction as a function of the applied bias voltage. The quasi-equilibrium distributions of four different bias voltage settings are shown, which are respectively 0, 32, 65 and 81 μV. The vertical dashed line indicates the gap energy, the vertical dashed-dotted line indicates the bias energy above the gap and the vertical dotted line indicates the $3\Delta_g$ energy level, which is defined as the minimum energy of the active region.

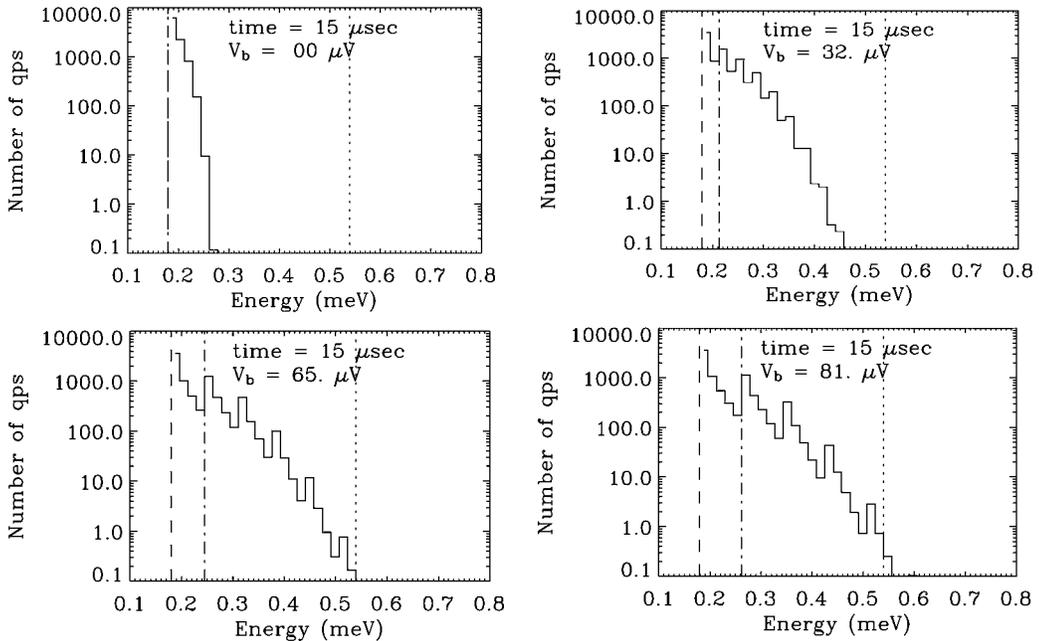

**Figure 5.27:** Variation of the quasiparticle energy distribution 15 μsec after the absorption of a 500 nm photon in a 50 μm side length Al junction as a function of the applied bias voltage. The quasi-equilibrium distributions of four different bias voltage settings are shown, which are respectively 0, 32, 65 and 81 μV. The vertical dashed line indicates the gap energy, the vertical dashed-dotted line indicates the bias energy above the gap and the vertical dotted line indicates the $3\Delta_g$ energy level, which is defined as the minimum energy of the active region.

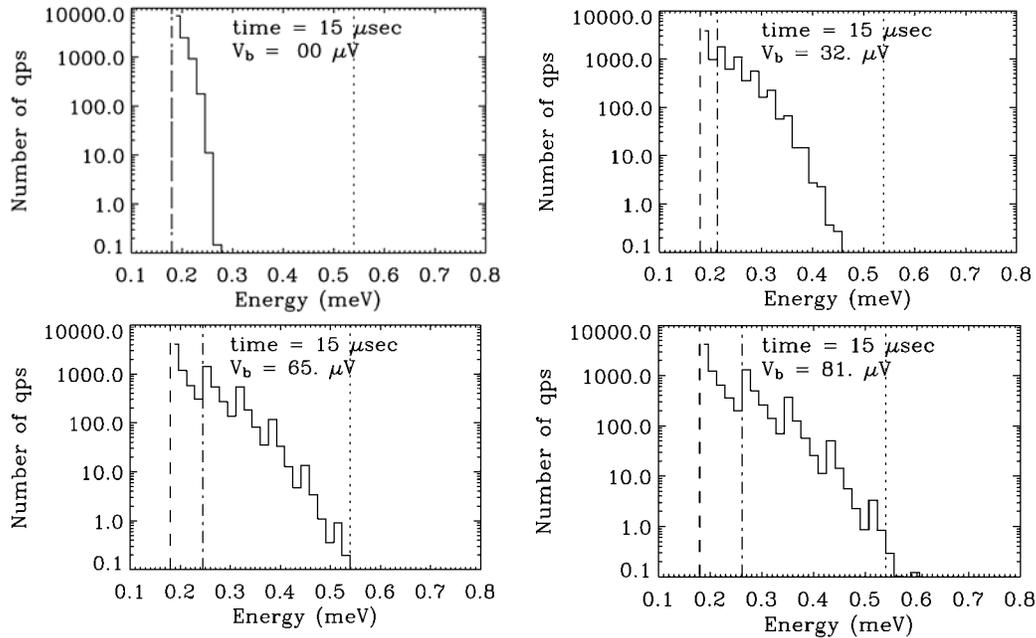

**Figure 5.28:** Variation of the quasiparticle energy distribution 15 μsec after the absorption of a 500 nm photon in a 70 μm side length Al junction as a function of the applied bias voltage. The quasi-equilibrium distributions of four different bias voltage settings are shown, which are respectively 0, 32, 65 and 81 μV. The vertical dashed line indicates the gap energy, the vertical dashed-dotted line indicates the bias energy above the gap and the vertical dotted line indicates the 3Δ$_g$ energy level, which is defined as the minimum energy of the active region.

Figure 5.25 shows the evolution of the quasiparticle energy distribution in the 30 μm side-length junction with an applied bias voltage of 50 μV. The quasi-equilibrium distribution (see section 2.3.3.1) is reached within one to two microseconds. It shows the typical step structure created via energy accumulation due to multiple tunnelling with steps at multiples of the bias energy. From the moment the quasi-equilibrium distribution is reached the energy distribution of the quasiparticles stays constant. Only the absolute number of quasiparticles in the electrodes decreases because of the different loss channels. On the other hand, the decay times of all the Al junctions are rather long, in excess of 15 μsec. Therefore, the quasiparticles spend most of their lifetime spread over the energy domain according to the quasi-equilibrium distribution. Figs 5.26, 5.27 and 5.28 show these quasi-equilibrium distributions as a function of the applied bias voltage for the 30, 50 and 70 μm side-length junctions respectively. For all the junctions it can be clearly seen that the quasiparticle distribution is lifted higher in energy as the bias voltage increases. Figure 5.29(a) shows the average quasiparticle energy as a function of bias voltage, which increases from approximately 200 μeV at zero bias voltage to 240 μeV at an applied bias voltage of 100 μV. It gets interesting when some of the quasiparticles reach the "active region", which is the energy region that starts $2\Delta_g$ above the energy gap. This energy level is indicated by the vertical dotted line in the figures 5.26–5.28. Quasiparticles that reach this energy level have a very large probability of scattering down to the gap energy and thereby releasing a phonon of energy $2\Delta_g$ or larger. This phonon can break a Cooper pair and create two new quasiparticles. The rate for this mechanisms, called quasiparticle multiplication, is given in section 2.3.1.8. This mechanism is therefore a constant source of new quasiparticles and effectively slows down the absolute loss of quasiparticles. The decay of the pulse decreases accordingly, which explains the rise of the decay time with bias voltage. If the generation of quasiparticles is larger than the loss, the quasiparticle population starts growing until losses by recombination outnumber the generation gains. This happens for bias voltages larger than ~100 μV and is the origin of the current steps observed in the IV-curves of the junctions (section 4.3.2.1). Because of the different loss times in the junctions there exist very slight differences in the distributions of the three junctions. The faster losses in the smaller devices cause a somewhat suppressed energy distribution in the high energy domain. For the same bias voltage more quasiparticles in the low loss junctions are able to reach the higher lying energy states. This explains why the effect is larger for the large junctions with the lower

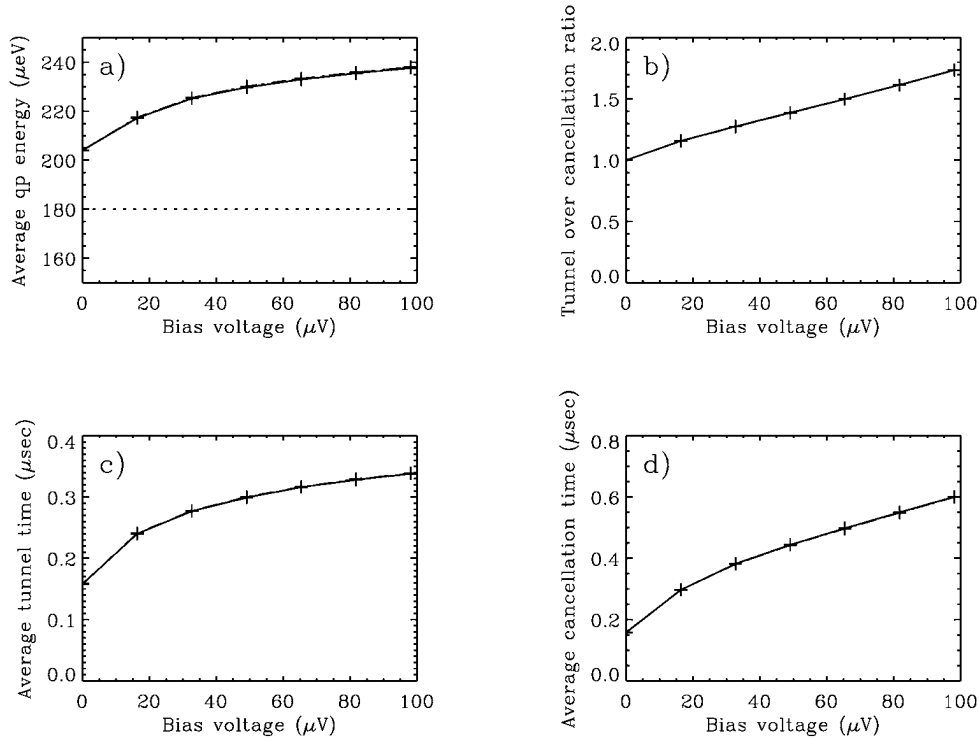

**Figure 5.29:** Calculated average quantities as a function of bias voltage for the absorption of a 500 nm photon in the 30, 50 and 70 μm Al junctions. The curves for the three device sizes are superimposed. The crosses indicate the bias voltages for which calculations were made. The lines in between the crosses are a guide to the eye. **(a)** Average quasiparticle energy, **(b)** tunnel to cancellation ratio σ (The dotted line indicates the gap energy), **(c)** average tunnel time and **(d)** average cancellation time.

loss times.

A second effect caused by the bias voltage is the increase of both tunnel and cancellation times with increasing bias voltage. Figures 5.29c) and d) show the average tunnel and cancellation times as a function of bias voltage. It can be observed that the cancellation time rises faster than the tunnel time. This can be seen in Fig. 5.29b), which shows the tunnel to cancellation ratio as a function of bias voltage. This ratio increases from one at zero bias to approximately 1.75 at 100 µV. As a consequence the fraction of charge lost due to cancellation currents decreases with increasing bias and the responsivity increases accordingly, explaining the faster than proportional rise of the responsivity as compared to the decay time in Fig. 5.24.

As can be seen in Fig. 5.24 the twofold effect can be very satisfactorily simulated with the model including the multiplication term for all three device sizes.

### 5.2.2.2   Energy resolution

The same as for the V based junctions the FWHM energy resolution of an Al STJ is given by:

$$\Delta E = 2.355 \sqrt{\sigma_{el}^2 + 1.7 E \Delta_g (F + G + H) + \alpha E^2} \ . \qquad (5.13)$$

The different resolution broadening terms are the same as in the V case and are explained in more detail in section 5.1.3.2.

- The Fano factor is expected to be equal to 0.2, the same value as in Nb and Sn.
- The mean number of tunnels <n> is very large for all Al junctions, in excess of 40. One therefore expects a tunnelling factor G equal to one. Effects due to finite charge integration times are neglected here.
- The cancellation noise factor depends strongly on the average number of tunnels $\bar{n}$ and the tunnel to cancellation ratio $\sigma$ (Equ. (5.5)). The tunnel to cancellation ratio $\sigma$ was calculated as a function of bias voltage (Fig. 5.29b). Knowing $\sigma$, the average number of tunnels <n> can be derived from the charge amplification factor $\bar{n}$ using (5.12), which in turn is known from measurements (Fig. 5.24). One can therefore calculate the expected resolution broadening factor H as a function of bias voltage for the 30, 50 and 70 μm side length junctions. The result is shown in Fig. 5.30. The increase of the tunnel to cancellation ration with increasing bias voltage causes the decrease of the cancellation broadening factor H with increasing bias. In addition, because of the lower responsivity of the smaller devices the cancellation broadening is larger for the smaller junctions.

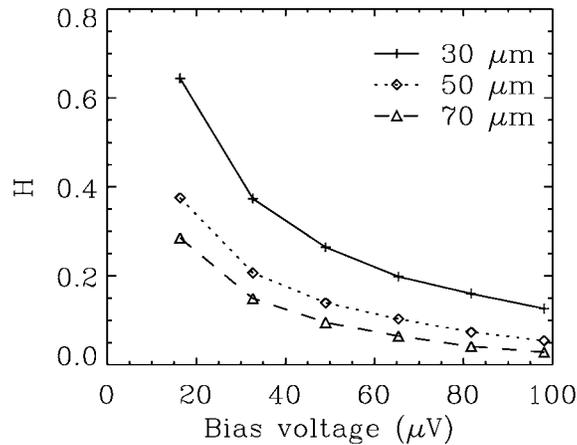

**Figure 5.30:** Cancellation noise factor H as a function of bias voltage for the Al devices with side lengths of 30, 50 and 70 µm.

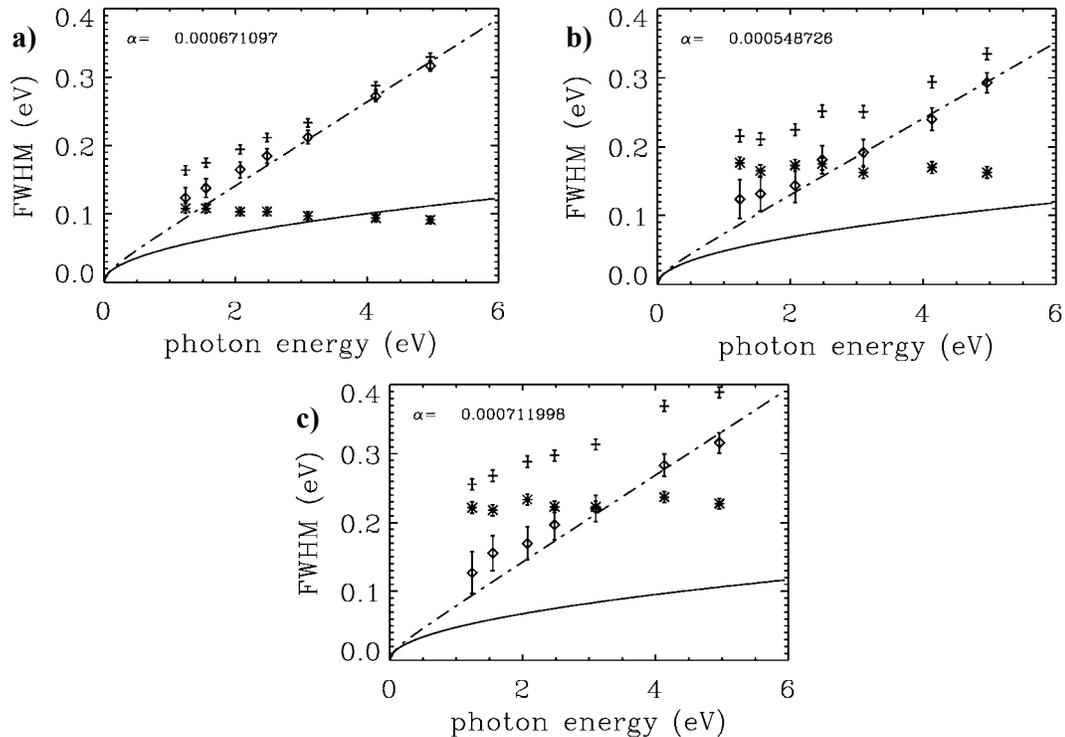

**Figure 5.31:** Measured energy resolution (crosses), measured electronic noise contribution (stars) and the derived intrinsic resolution (diamonds) as a function of incoming photon energy for the devices with 30 **(a)**, 50 **(b)** and 70 **(c)** μm side lengths. The bias voltage applied is equal to 50 μV. The solid line is the expected ideal energy resolution achievable. The dashed dotted line is a fit to the intrinsic resolution with the spatial broadening factor α used as fitting parameter.

Figure 5.31 shows the measured energy resolution, the measured electronic noise contribution and the derived intrinsic resolution as a function of incoming photon energy for the devices with 30, 50 and 70 μm side length. The bias voltage applied is equal to 50 μV. The intrinsic energy resolution is obtained by quadratic subtraction of the electronic noise contribution from the measured energy resolution. By varying the spatial broadening factor α and using the expected values for F, G and H, a fit to the intrinsic resolution was made. For all three device sizes the derived spatial broadening factor is of the order of 7 $10^{-4}$, which is a very large value compared to the best values obtained in Ta based junctions ($\alpha \sim 10^{-6}$). The fit to the intrinsic resolution is the worst in the low energy region (1 - 2.5 eV), where the intrinsic resolutions lie generally above the fitted curve. The reason for this bad fit could be because of the rather large uncertainty in the intrinsic resolution value caused by the quadratic subtraction of the electronic noise, which is the largest broadening contribution in that energy range. On the other hand it cannot be excluded that a currently unknown additional resolution broadening factor with a square-root dependent photon energy dependence adds additional broadening to the measured photo peaks. Nevertheless, it is absolutely clear that the largest contribution to the intrinsic photo peak width comes from the spatial broadening, which has a linear dependence on the photon energy. Table 5.8 summarises the different contributions to the measured photo peak width for the absorption of 500 nm photons in the Al devices.

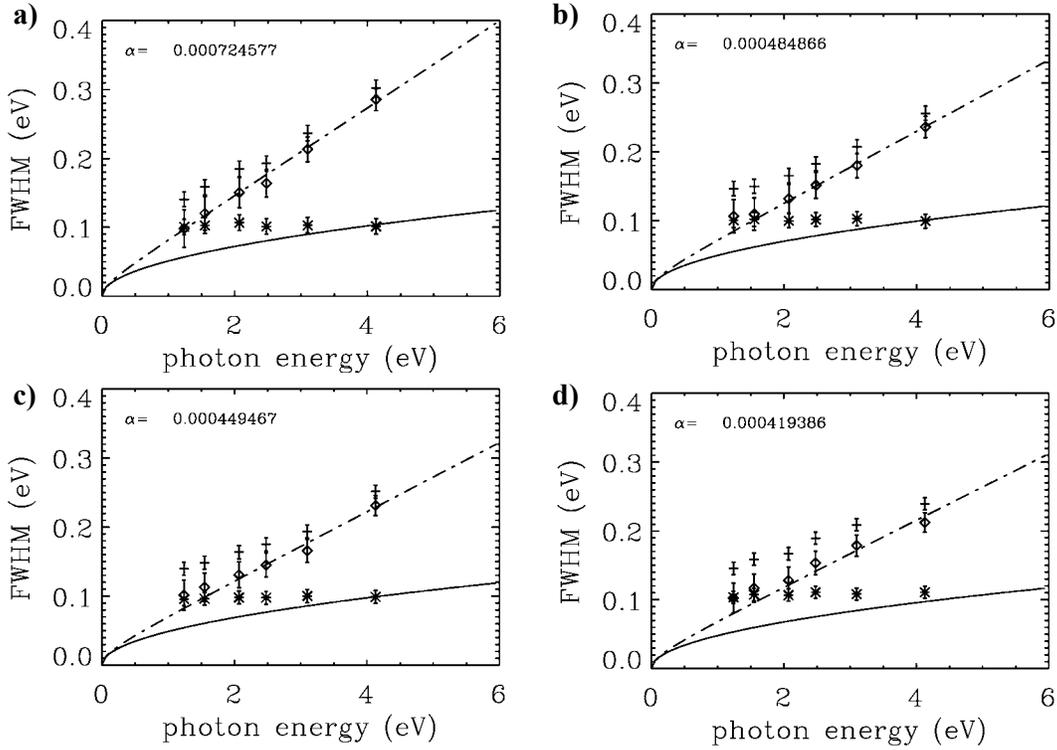

**Figure 5.32:** Measured energy resolution (crosses), measured electronic noise contribution (stars) and the derived intrinsic resolution (diamonds) as a function of incoming photon energy for the device with 30 μm side length. The bias voltage applied is respectively 40 **(a)**, 60 **(b)**, 80 **(c)** and 95 μV **(d)**. The solid line is the expected ideal energy resolution achievable. The dashed dotted line is a fit to the intrinsic resolution with the spatial broadening factor α used as fitting parameter.

**Table 5.8:** Different calculated and measured contributions to the resolution broadening of Al STJs as a function of the device size. The photon energy is 2.48 eV and the applied bias voltage is equal to 50 μV.

| Device size | Measured Total FWHM | Measured electronic noise | Calculated Fano noise | Calculated tunnel noise | Calculated cancellation noise | Derived spatial broadening | Derived α |
|---|---|---|---|---|---|---|---|
| (μm) | (meV) | (meV) | (meV) | (meV) | (meV) | (meV) | / |
| 30 | 211 | 98 | 29 | 65 | 34 | 151 | $6.7\ 10^{-4}$ |
| 50 | 252 | 175 | 29 | 65 | 25 | 137 | $5.5\ 10^{-4}$ |
| 70 | 298 | 222 | 29 | 65 | 22 | 156 | $7.1\ 10^{-4}$ |

The electronic noise contribution is proportional to the device size because of the dependence of the leakage currents on the area of the devices. The broadening term due to spatial inhomogeneities is by far the largest contribution to the intrinsic peak width. The cause for this large contribution probably arises from the localised nature of the sources of quasiparticle loss in the detector as already discussed in section 5.2.2.1.1. In order to decrease the spatial inhomogeneities we have tried to replace the Nb contacts with Ta, which is known to form less quasiparticle trapping sites. As a consequence the responsivity should become more homogeneous over the area of the detector. Unfortunately the deposited Ta used for the top contacts turned out to be the wrong phase. The body centered cubic α-Ta phase has a critical temperature of 4.5 K, whereas the tetragonal β-Ta phase has a critical temperature of about 0.5 K. The Ta film we deposited for plugs and top contacts turned out to be β-phase with a $T_C$ of 0.5 K, which made the Ta plugs and top contacts act as very strong trapping centres. This of course reduced the charge output very strongly and made the junctions unusable as optical photon detectors. In the near future we will try to determine deposition conditions for the Ta film, which will allow for the deposition of the high $T_C$ phase material. Another possibility to homogenise the responsivity would be to homogeneously introduce loss centres over the area of the detector, for example by introducing a very thin Nb or Ta layer on top or below the Al films forming the electrode.

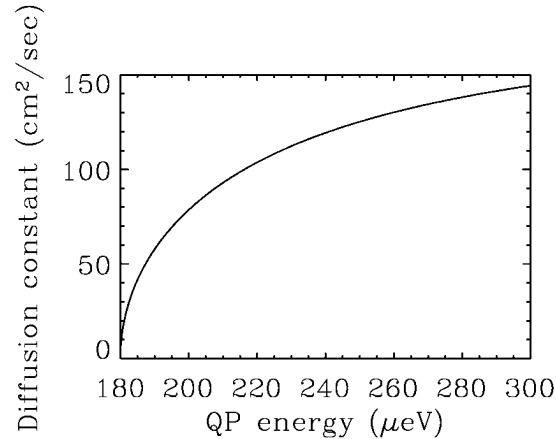

**Figure 5.33:** Calculated quasiparticle diffusion constant in superconducting Al as a function of quasiparticle energy.

The variation of the spatial broadening factor α with bias voltage is shown in Fig. 5.32. The spatial broadening factor α is larger for the lower bias voltages, which can be related to slower diffusion in the Al for lower bias voltages. Figure 5.33 shows the variation of the diffusion in superconducting Al, calculated according to (5.7), as a function of the quasiparticle energy. It was calculated that the average quasiparticle energy in the electrodes approximately varies between 210 and 240 μeV for applied bias voltages varying between 40 and 100 μV (Fig. 5.29a). As a consequence the diffusion in Al for a bias voltage setting of 40 μV must be some 30% slower than for a bias voltage equal to 100 μV, increasing the variations in charge output for photons absorbed close to the loss centres and photons absorbed far away from the loss centres accordingly. As a consequence the spectra with large bias voltage settings present the best observed energy resolution. Even though the spatial variations of the charge output are very large, the measured resolving power for the 30 μm Al device with an applied bias voltage of 100 μV is equal to 13 for 500 nm photons.

### 5.2.3   X-ray detection experiments in Al junctions

The 30 and 50 μm side length junctions were also tested as x-ray detectors in the ADR. The [55]Fe radioactive source was located 3-5 mm away from the sample in order to have a count-rate of approximately 100 photon absorption events per second. The 70 μm junction could unfortunately not be measured because of magnetic flux trapping in the junction, caused by the small inherent magnetic field of the $^{55}$Fe source. A parallel magnetic field of the order of 50 Gauss was applied in order to suppress the Josephson currents. The radioactive $^{55}$Fe source emits x-rays from the Mn-K$_\alpha$ (5895 eV) and the Mn-K$_\beta$ (6490 eV) emission lines with probabilities of respectively 90 % and 10 %.

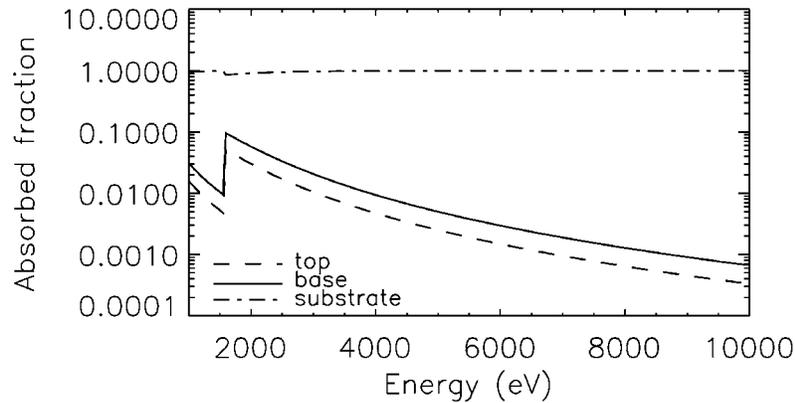

**Figure 5.34:** Fraction of x-ray photons absorbed in the 50 nm thick top and 100 nm thick base electrodes of an Al based junction and in the sapphire substrate.

Due to the very low absorption efficiency of Al in the x-ray domain, Al junctions are not very well suited as x-ray detectors with absorption of the photons in the electrodes of the junctions. Figure 5.34 shows the absorption efficiencies of the 50 nm thick top and 100 nm thick base Al films for 1 to 10 keV x-rays [Henke 93, LBNL]. At 6 keV the absorption probability in the base and top film is approximately 0.3 % and 0.15 % respectively, which shows that almost all (~99.55 %) of the x-rays emitted by the radioactive source pass through the detector and are absorbed in the sapphire substrate. In the 350 nm thick Nb forming the top contact and the base lead plug the absorption efficiency is equal to about 10 %.

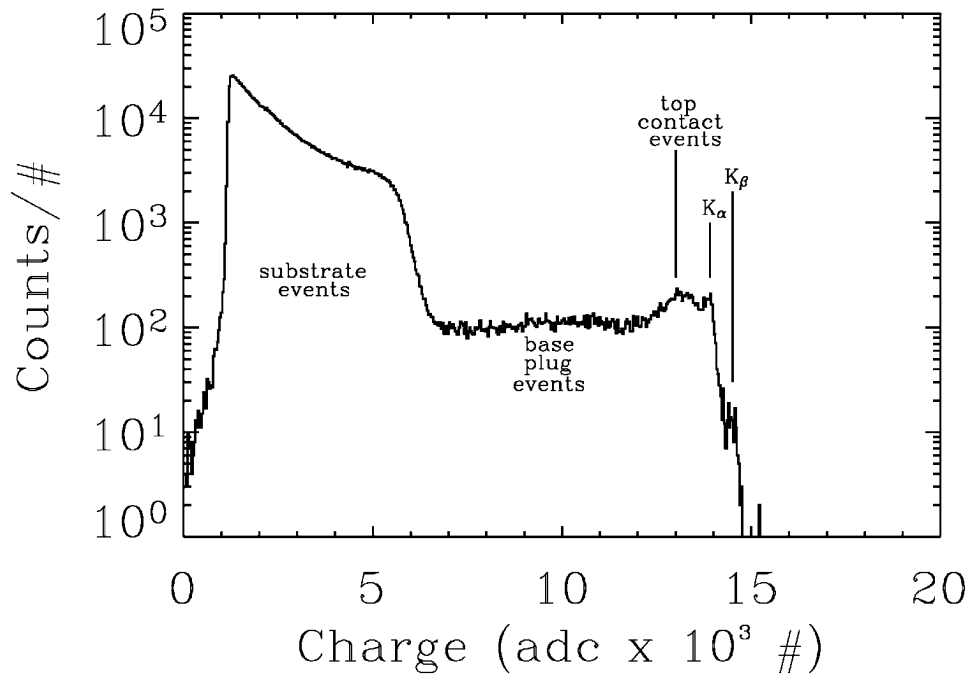

**Figure 5.35:** $^{55}$Fe spectrum acquired with the 30μm side length Al based junction. The applied bias voltage is equal to 100 μV, whereas the applied parallel magnetic field is 40 Gauss.

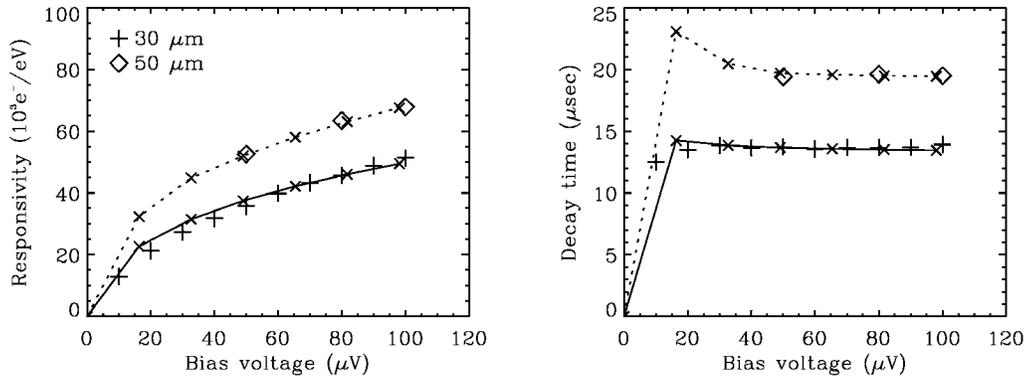

**Figure 5.36:** Experimental responsivity and decay time of the 30 (crosses) and 50 (diamonds) μm side length Al junctions as a function of bias voltage for the absorption of 6 keV x-rays. The applied parallel magnetic field is equal to 40 Gauss. The results of the simulations using the parameters from table 5.7 are shown as well for the 30 (solid line) and the 50 (dotted line) μm junctions. The bias voltages for which simulations were made are indicated by an x.

A typical [55]Fe spectrum acquired with the 30 μm junction biased at 100 μV is shown in Fig. 5.35. As expected, the spectrum is completely dominated by events absorbed in the substrate and in the Nb contact and base plug material. The events absorbed in the Al are concentrated in a small peak on the high charge output part of the spectrum. Of course, these [55]Fe spectra are not very useful, as even the determination of the energy resolution is practically impossible, because of the very numerous top contact events. More useful Al soft x-ray detectors should include an absorber of a different material coupled to the Al junction. Two designs have already been realised by other groups. The first design is a lead absorber deposited on top of the junction, electrically separated from the top Al electrode by the natural Al oxide that forms by contact with the normal atmosphere. The coupling to the detector is made through phonons passing the Al oxide layer separating the absorber and the junction. Such a lay-up gives energy resolutions of 12 eV for 6 keV x-rays [Angloher 01]. A second design consists of a strip of superconducting absorber material with a higher $T_C$ than Al, which is coupled to two Al read-out junctions on both ends of the absorber. Quasiparticles created in the superconducting absorber diffuse to the Al junctions, where they get trapped in the lower $T_C$ material and create a current pulse. In addition to the energy resolution information, the knowledge of the relative intensities of the current pulses in the two junctions gives additional information about the absorption position in the absorber. A design involving Ta as the absorber material gives an energy resolution of 13 eV at 6 keV for events absorbed in a small sub-section of the absorber [Li 01].

Even though the spectra are not very useful for practical applications, one can determine the responsivity and the decay times of the current pulses created by the events absorbed in the Al. Figure 5.36 shows the variation of the responsivity and decay time with bias voltage for the 30 and 50 μm side length junctions. For these large photon energies strong recombination losses set in, which prevent the quasiparticles from reaching the active region, where quasiparticle multiplication starts. As a consequence the decay time of the pulse is shorter than in the optical and roughly independent of the bias voltage, as opposed to the strong dependence seen in the optical regime. The increase of the responsivity with increasing bias voltage is due to the variation of the quasiparticle energy distribution with bias, which causes the variation of the tunnel to cancellation ratio (Fig. 5.29b). For larger bias voltages the fraction of cancellation tunnel events decreases, which increases the charge output accordingly. As can be observed in Fig. 5.36, the experimental results can be very satisfactorily simulated with the model and the parameters from table 5.7.

## *5.3  Mo-Al based junctions*

Unfortunately, because of the very large leakage currents of the Mo-Al based junctions stable biasing for photon detection experiments was not possible. The leakage currents in these devices first have to be further reduced.

## *5.4  Conclusions*

Photon detection experiments with V-Al and Al based STJs were presented.

6 keV spectra could be acquired with V-Al junctions with side lengths varying between 7 and 30 μm. The responsivity and rise time of all the junctions turned out to be low of the order of 600 e⁻/eV and 0.7 μsec respectively. No dependence on the side length of the junctions could be observed, showing that the quasiparticle loss channels are distributed homogeneously over the area of the detector. Simulations with the energy dependent kinetic equations model revealed that quasiparticle trapping in V is very strong. The number of local trapping states in the V devices is of the order of 900 000, the trap depth equal to 330 μeV and the trapping probability equal to 25 %. These numbers are large compared to typical numbers for Ta-Al (20 000, 160 μeV and 7.2 % respectively) and Nb-Al (185 000, 240 μeV and 22 % respectively) junctions. The local trapping states are believed to be located at the top surface and edges of the junctions, where metallic V oxides form, due to reaction with the normal atmosphere when the junctions are removed from the UHV system. Because of the proximity effect, metallic impurities in a superconductor form islands with a local energy gap that is lower than the energy gap of the surrounding superconductor and quasiparticles relaxing to theses lower energy states are prevented from diffusing.

The bias voltage dependence of the responsivity and decay time could not be simulated with the kinetic equation model in its current form. Both the responsivity and decay time increase almost linearly with the applied bias voltage, whereas the simulated behaviour is rather flat over the whole bias domain, with a small maximum occurring at low bias voltages. This flat dependence is also generally observed for Ta-Al and Nb-Al junctions with a similar lay-up. Two scenarios able to cause such a dependence were proposed, based on the quasiparticle energy distribution as calculated with the model. Because of the fast tunnelling for quasiparticles at energies below the bulk energy gap of V and slow tunnelling for quasiparticles at the bulk energy gap of V, an inversion of the quasiparticle population is observed. Most quasiparticles actually reside at the bulk energy gap of V and the population inversion increases with increasing bias voltage. Two possible causes for the linear increase of the responsivity with bias are strong Andreev reflections at the V-Al interface and stimulated relaxation of the quasiparticles residing at the bulk V energy gap by non-equilibrium phonons created during the photon absorption process. Both effects are not yet included in the model and further developments are necessary in order to find the exact mechanism that causes this almost linear bias dependence of the responsivity.

The measured energy resolution for 6 keV photons absorbed in the 7 μm side length detectors is equal to 80 eV and increases with devices size to 900 eV for the 30 μm side length junctions. This is much larger than the expected theoretical resolution for V junctions equal to 7.5 eV. The broadening is caused by spatial inhomogeneities in the detectors response, which are due to the larger loss channels at the edges of the detector and the slow diffusion in V.

In summary, V-based junctions are not well suited as photon detectors because of the strong reactivity of V with oxygen and the metallic nature of some of the oxides, creating a large amount of localised quasiparticle trapping states. Therefore, work on V-based devices was discontinued. Nevertheless, the main goal of the V-based junctions was achieved, as it was in the first instance intended as a process route development vehicle for the fabrication process of the lower $T_C$ junctions based on pure Al and Mo-Al.

Single pixel pure Al junctions with side lengths varying between 10 and 70 μm were for the first time successfully operated as near-IR to soft-UV photon detectors. The responsivity and decay time of the junctions are very large and increase proportional to the area of the junction, showing the localised nature of the loss sources, probably located at the Nb plugs and top electrode contacts. The responsivity and decay time of the smallest 10 μm junction are approximately equal to $2 \cdot 10^4$ e⁻/eV and 5 μsec respectively, increasing to $4 \cdot 10^5$ e⁻/eV and 80 μsec for the largest 70 μm junction. The charge amplification factor increases from 7 for the smallest to 105 for the largest junction.

The responsivity and decay time show a strong non-linearity in the optical domain caused by the very low amount of localised trapping states in the junction. Simulations with the kinetic equation model revealed a number of approximately 7000 trapping states independent of the devices size of the junction. The trapping probability on the other hand decreases with increasing device size and is equal to 5.5% and 1.4% for the 30 and 70 μm side length junctions respectively. This again reflects the localised nature of the quasiparticle traps. The depth of the traps was determined to be approximately equal to 80 μeV.

The bias dependence of the responsivity and the decay time show a curious increase with increasing bias voltage, the effect being stronger for the larger junctions. This bias dependence of the responsivity and the decay time could be simulated with the kinetic equations model including a quasiparticle multiplication term. This term accounts for additional quasiparticle generation, caused by pair breaking by $2\Delta_g$ phonons released by quasiparticles with energy larger than $3\Delta_g$ relaxing down to the energy gap. This quasiparticle generation process increases with increasing bias voltage as the energy gain via tunnelling is stronger for large bias voltages and therefore more quasiparticles can reach energy levels as large as $3\Delta_g$. An additional increase of the responsivity with bias voltage is due to the increase of the average quasiparticle energy with bias voltage, which decreases the proportion of charge lost due to cancellation tunnel events. This bias dependence of the responsivity and decay time can only be explained with the model including the full energy dependence of the quasiparticles, showing the increasing importance of the knowledge of the full quasiparticle energy distribution in the lower energy gap junctions.

The energy resolution of the Al detectors contains a spatial non-uniformities contribution caused by the localised nature of the loss channels. This spatial broadening contribution limits the intrinsic resolving power $\lambda/\Delta\lambda$ for 500 nm photons to approximately 17, instead of the theoretical value which is equal to 30. The spatial broadening contribution increases with decreasing bias voltage because of the lower average quasiparticle energy for lower bias voltages, which slows down the diffusion of the quasiparticles. The measured resolving power for 500 nm photons is equal to 13 for the 30 μm side length junction, which includes a 0.1 eV electronic noise contribution. Nevertheless, the capability of lower $T_C$ junctions used as optical photon detectors was demonstrated, which show already a resolving power comparable to the best Ta-based junctions despite the spatial broadening factor. Replacing the Nb top contacts and plugs by Ta should reduce the losses at these positions, because of the fewer trapping states in Ta. This should homogenise the responsivity over the area of the detector and accordingly increase the resolving power. The replacement of the Nb plugs and top contacts by Ta is foreseen in the near future. Another possibility to homogenise the responsivity over the area of the detector is to introduce a homogeneous loss source over the area of the detector. This could for example be achieved by adding a very thin Ta or Nb film on top of the detector.

The Al junctions were also irradiated by 6 keV x-rays, but because of the very low stopping power of Al at these energies 99.6% of the x-rays were absorbed in the substrate and in the Nb contacts and plugs. In order to make low energy gap Al junctions useful as x-ray detectors these have to be coupled to an absorber with larger stopping power. This is also foreseen to be done in the near future.

In summary, the capabilities of pure Al based junctions used as optical photon detectors was demonstrated and some more work on the homogeneity of the response over the area of the detector is required in order to achieve the better predicted energy resolutions of these detectors compared to Ta-based junctions.

# Summary

This thesis describes the development of low-energy gap superconducting tunnel junctions (STJs) for use as photon detectors, with as a main goal the improvement of the energy resolution in both the optical and the x-ray energy domain.

A new model for the photon detection process with STJs is presented, which includes the full energy dependence of all the quasiparticle processes occurring in the junctions. This model allows for the calculation of the time- and energy-dependent quasiparticle distribution from the moment of generation of the quasiparticles by the photon absorption process until the end of the current pulse, when all the quasiparticles have disappeared. The exact knowledge of the quasiparticle energy distribution in the junctions is of increasing importance for the lower energy gap junctions, as the quasiparticle relaxation rate is approximately proportional to the cube of the energy gap of the superconductor. As a consequence, energy down-conversion of quasiparticles in lower-$T_C$ superconductors becomes much slower and the bias energy gained by the quasiparticles due to successive tunnel and back-tunnel events leads to a very broad energy distribution of the non-equilibrium quasiparticle population. Two effects related to the broad quasiparticle energy distribution, which cannot be explained with an energy-independent Rothwarf-Taylor approach, are on one hand the proportion of charge lost due to cancellation tunnel events and on the other hand quasiparticle multiplication. When a quasiparticle has energy above the bias energy level, it can undergo a tunnel event against the bias and annihilate a charge from the current pulse. In this way a certain percentage of the charge output is lost, which can be as large as 80 % for lower energy gap junctions with fast tunnelling. When a quasiparticle has energy larger than $3\Delta_g$, it will release a phonon of energy larger than $2\Delta_g$, when relaxing down to the gap energy $\Delta_g$. This released phonon can in turn break a Cooper pair and create two additional quasiparticles. In this way the quasiparticle population increases drastically, which has a strong effect on the measured tunnel currents as well. This mechanism, called quasiparticle multiplication, is typical for lower energy gap, low loss junctions.

Illustrations of the energy-dependent kinetic equations model simulating the response of Ta- and Nb-based junctions show that the quasiparticle energy distribution converges quickly to a "quasi-equilibrium" distribution. The distribution is called to be in quasi-equilibrium in the sense that the normalised distribution is invariable and only the total number of quasiparticles diminishes because of the different quasiparticle loss channels. This quasi-equilibrium distribution shows a step-like structure, with local maxima occurring at multiples of the bias energy because of the energy gain due to subsequent tunnel and back-tunnel events. Even in the relatively larger energy gap junctions based on Ta and Nb the average quasiparticle energy lies well above the energy gap of the superconductor and the main condition of the Rothwarf-Taylor approach is therefore not justified.

The fabrication processes for three different types of STJs are presented in this thesis, based on Vanadium-Aluminium, Aluminium, and Molybdenum-Aluminium electrodes respectively. The V-based junctions were intended as a process route development vehicle for the fabrication of the lower $T_C$ junctions based on Al and Mo. The reason for this is the possibility to operate the V-based junctions at 300 mK in a $^3$He sorption cooler with a very fast turn-around time. The Al and Mo based junctions on the other hand have to be operated in an Adiabatic Demagnetisation Refrigerator at temperatures below 100 mK, which has a much longer turn-around time. The progress made with V-based junctions, while varying certain parameters of the processing steps common to all three junctions, can then be directly transferred to the lower energy gap junctions.

The fabricated V-Al based junctions are of good quality with a normal resistivity $\rho_{nn}$ approximately equal to 1.2 $\mu\Omega$ cm$^2$ and a dynamical resistivity $\rho_d$ in the bias range approximately equal to 1.1 $\Omega$ cm$^2$, which corresponds to a quality factor Q = $\rho_d/\rho_{nn}$ of ~10$^6$. The Josephson current suppression pattern is very regular, indicating the good homogeneity of the insulating barrier separating the two electrodes of the junctions. 6 keV photon detection experiments could be performed with V-based junctions having side lengths varying between 7 and 30 $\mu$m. The responsivity was shown to be very low of the order of 600 e$^-$/eV and independent of the device size of the junctions. Simulations with the energy dependent kinetic equations model show that the number of localised trapping states is very large, about one and two orders of magnitude larger than in similar Nb and Ta based junctions respectively. This large number of quasiparticle trapping states is believed to be related to the strong reactivity of V with oxygen and the metallic nature of some of the oxides forming small islands in the superconductor with a locally suppressed energy gap. The 6 keV energy resolution of the junctions is 80 eV full width at half maximum (FWHM) for the smallest device sizes and increases to approximately 900 eV FWHM for the 30 $\mu$m side length devices. The reason for this poor energy resolution is variation of the responsivity as a function of absorption position over the area of the detector. The work on V-based junctions was discontinued, because they are not well suited as photon detectors. Nevertheless, the main goal for these junctions was achieved, as they were mainly intended as a process route development vehicle for the lower energy gap junctions based on Al and Mo.

High quality single pixel Al STJs were fabricated with side lengths varying between 10 and 70 $\mu$m. The normal resistivity of these junctions is ~7 $\mu\Omega$ cm$^2$ and the dynamical resistance in the bias domain is 1.9 $\Omega$ cm$^2$, corresponding to a quality factor of approximately 2.7 10$^5$. The Josephson current suppression pattern is very regular with a pronounced minimum for an applied magnetic field equal to 50 Gauss, allowing the successful suppression of the zero bias Josephson currents. Optical photon detection experiments could be performed with the Al based junctions. The responsivity of the devices is very large of the order of 10$^5$ e$^-$/eV and proportional to the area of the detector. Responsivity and pulse decay time of the Al detectors show a strong photon energy non-linearity in the optical domain, indicating the small number of localised quasiparticle trapping states. Simulations with the kinetic equation model reveal that the number of traps in the Al junctions is only of the order of 7000 states, which is a factor 3 and 30 lower than in comparable Ta-Al and Nb-Al junctions respectively. The number of trapping states does not depend on the device size, which is a very strong indication that the trapping states are located in the Nb that forms the contacts to the top and base electrodes. The bias voltage dependence of the responsivity of the junctions shows an increase with increasing bias voltage, the effect being stronger for the larger junctions with the longer quasiparticle loss times. This dependency is related to the broad quasiparticle energy distribution. The lower cancellation currents and stronger quasiparticle multiplication at higher bias voltage result in a higher responsivity. These effects could be successfully simulated with the energy dependent model. The energy resolution of the Al junctions includes a spatial broadening contribution, probably because all the loss sites are located at the positions of contact of the Nb leads to the top and base electrodes. This spatial broadening contribution limits the intrinsic resolving power $\lambda/\Delta\lambda$ for 500 nm optical photons to approximately 17, well below the theoretical value of approximately 30. Nevertheless, the capabilities of Al STJs as optical photon detectors were demonstrated. Further work on Al based junctions will include the replacement of the Nb leads by Ta, which should homogenise the responsivity over the area of the detector and as a consequence increase the resolving power of the detector. In addition, coupling of the Al junctions to x-ray absorbers is planned in order to take advantage of the good theoretical energy resolution of Al STJs in the x-ray energy domain as well.

For the fabricated Mo-Al based junctions there still exists a problem with the edge structure created by the base etch step. The top Mo film is etched 600 nm further than the base Mo film, resulting in a step like structure at the edges. The Al film in between the two Mo layers does not form a vertical edge structure, which damages the aluminium oxide insulating layer. As a consequence the junctions show large perimeter related leakage currents of 1.25 µA per µm of perimeter length. No photon detection experiments could be performed with these junctions as the large leakage currents prevent stable biasing of the STJ. Further work on Mo based junctions, in particular on the base etch procedure, will have to be performed in order to reduce the leakage currents in these devices.

# Samenvatting

Dit proefschrift beschrijft de ontwikkeling van supergeleidende tunnel junkties (STJ's) met een lage energiekloof voor gebruik als foton detektoren, met als belangrijkste doel het verbeteren van het energie-oplossend vermogen in zowel het zichtbare als het röntgen energiebereik.

Er wordt een nieuw model gepresenteerd voor het detektiemechanisme van fotonen met STJ's, dat de volledige energie-afhankelijkheid van alle processen met quasideeltjes in de junkties omvat. Dit model maakt de berekening mogelijk van de tijd- en energie-afhankelijke verdeling van quasideeltjes vanaf het moment van ontstaan van de quasideeltjes door de absorptie van een foton, tot het einde van de stroompuls als alle quasideeltjes weer verdwenen zijn. Een nauwkeurige beschrijving van de energieverdeling van de quasideeltjes is van toenemend belang naarmate de energiekloof van de junkties kleiner is, omdat de snelheid waarmee de quasideeltjes naar een evenwichtstoestand evolueren evenredig is met de derde macht van die energiekloof. Dit heeft tot gevolg dat de herverdeling van energie van de quasideeltjes in supergeleiders met een lage kritische temperatuur $T_C$ veel langzamer gaat, zodat energie die de quasideeltjes winnen door opeenvolgende tunnel- en terugtunnelprocessen in de aanwezigheid van een aangelegde spanning, leidt tot een erg brede energieverdeling van de quasideeltjes populatie in niet-evenwichtstoestand. Er zijn twee effecten die gerelateerd zijn aan deze brede energie verdeling van de quasideeltjes, en die niet met een Rothwarf-Taylor beschrijving zonder energie-afhankelijkheid verklaard kunnen worden. Dit zijn, ten eerste, het gemeten tekort aan tunnelsignaal tengevolge van tegengestelde tunnel processen, en, ten tweede, vermenigvuldiging van quasideeltjes. Zodra een quasideeltje een energieniveau bereikt boven de aangelegde spanning, kan er een tunnel proces in tegengestelde richting optreden, waardoor de stroompuls effectief verlaagd wordt. Op deze manier gaat een deel van het gemeten ladingssignaal verloren, wat kan oplopen tot 80% voor junkties met een kleine energiekloof en een korte tunneltijd. Als een quasideeltje een energie boven driemaal de energiekloof bereikt, kan het in het relaxatieproces naar de energiekloof een fonon uitzenden met een energie groter dan tweemaal de energiekloof. Zo'n fonon kan op zijn beurt weer een Cooperpaar opsplitsen in twee extra quasideeltjes. Hierdoor zal de quasideeltjespopulatie snel groeien, met een dienovereenkomstig effect voor de gemeten tunnelstroom. Dit mechanisme wordt quasideeltjes vermenigvuldiging genoemd, en is typisch voor junkties met een kleine energiekloof en langzame quasideeltjes-verliesprocessen.

Voorbeelden van het energie-afhankelijke kinetische model waarin junkties gebaseerd op Ta en Nb worden gesimuleerd, laten zien dat de energieverdeling van de quasideeltjes snel convergeert naar een 'quasi-evenwichtsverdeling'. De verdeling wordt in quasi-evenwicht genoemd omdat de genormaliseerde verdeling onveranderlijk is, terwijl het totale aantal quasideeltjes wel vermindert tengevolge van de verschillende verlieskanalen voor quasideeltjes. Deze 'quasi-evenwichtsverdeling' vertoont een stapstructuur, met lokale maxima bij veelvouden van de biasenergie, veroorzaakt door de energietoename door opeenvolgende tunnel- en terugtunnelprocessen. Zelfs in junkties gebaseerd op Ta en Nb, met relatief grote energiekloven, ligt de gemiddelde energie van de quasideeltjes duidelijk boven de energiekloof van deze materialen, zodat aan de belangrijkste voorwaarde voor de Rothwarf-Taylor benadering niet voldaan is.

De fabricageprocessen voor drie soorten STJ's, gebaseerd op vanadium-aluminium, aluminium en molybdeen-aluminium electroden, worden in dit proefschrift gepresenteerd. De junkties gebaseerd op V waren bedoeld als object om de processtappen te ontwikkelen voor de fabricage de junkties met lagere $T_C$, gebaseerd op Al en Mo. De junkties gebaseerd op V hebben als voordeel dat ze gebruikt kunnen worden bij een temperatuur van 300 mK in een $^3$He adsorptie koeler, waardoor een snelle testcyclus mogelijk is. Junkties gebaseerd op Al en Mo daarentegen, moeten gebruikt worden bij temperaturen van 100 mK of lager, waarvoor een Adiabatisch Demagnetisatie koeler (ADR) vereist is, die typisch een veel langere koelcyclus heeft. De optimalisatie van processtappen die gemeenschappelijk zijn voor alle drie typen junkties, gevonden met behulp van de vanadium junkties, kan dan direct toegepast worden voor de junkties met lagere energiekloof.

De gefabriceerde V-Al junkties zijn van hoge kwaliteit met een resistiviteit in de normale toestand $\rho_{nn}$ gelijk aan ~1.2 $\mu\Omega$ cm$^2$ en een dynamische resistiviteit van ~1.1 $\Omega$ cm$^2$, wat overeenkomt met een kwaliteitsfactor Q = $\rho_d/\rho_{nn}$ ~10$^6$. De variatie van de Josephsonstroom met toenemend magnetisch veld vertoont een zeer regelmatig patroon, kenmerkend voor een hoge mate van uniformiteit van de isolerende barriere tussen de twee electroden van de junkties. Röntgenfotonen met een energie van 6 keV konden gedetekteerd worden met V junkties met afmetingen tussen 7 en 30 $\mu$m. Hun responsiviteit bleek met een waarde van ongeveer 600 e$^-$/eV echter erg laag te zijn, en onafhankelijk van de grootte van de detektoren. Simulaties met het energie-afhankelijke kinetische model toonden aan dat het aantal gelokaliseerde 'trap'-toestanden, waarin quasideeltjes kunnen worden ingevangen, erg groot was: respectievelijk één en twee orden van grootte meer dan in soortgelijke junkties van Nb en Ta. Dit grote aantal van dit soort toestanden wordt in verband gebracht met de hoge reactiviteit van vanadium met zuurstof, en de metallische eigenschappen van sommige van de oxiden, die eilandjes met lagere energiekloof vormen in de supergeleider. Het energieoplossend vermogen bij een fotonenergie van 6 keV was 80 eV volle breedte op halve hoogte (FWHM) voor de kleinste junkties, oplopend tot ongeveer 900 eV FWHM voor de 30 $\mu$m grote junkties. De oorzaak van dit slechte oplossend vermogen is een variatie van de responsiviteit met de laterale positie van fotonabsorptie in de detector. Het werk aan junkties uit vanadium is stopgezet vanwege hun ongeschiktheid als fotondetectoren. Desalniettemin is het belangrijkste doel voor deze junkties wel bereikt, namelijk het ontwikkelen van de processtappen voor de junkties met een kleinere energiekloof gebaseerd op Al en Mo.

 Afzonderlijke Al STJ's van hoge kwaliteit zijn gefabriceerd in afmetingen variërend van 10 tot 70 $\mu$m. De resistiviteit in de normale toestand $\rho_{nn}$ van deze junkties is ~7 $\mu\Omega$ cm$^2$ en hun dynamische resistiviteit ~1.9 $\Omega$ cm$^2$, wat overeenkomt met een kwaliteitsfactor Q ~2.7 10$^5$. De variatie van de Josephsonstroom met toenemend magnetisch veld vertoont een zeer regelmatig patroon, met een geprononceerd minimum bij een aangelegd veld van 50 Gauss, waardoor de Josephsonstroom effectief kan worden onderdrukt. Optische fotonen met een energie van 1-6 eV konden met deze STJ's gedetekteerd worden. De responsiviteit van de Al junkties is erg hoog met een typische waarde van 10$^5$ e$^-$/eV, en evenredig met de oppervlakte van de detektoren. Zowel de responsiviteit als de afvaltijd van de pulsen van de Al STJ's zijn een sterk niet-lineaire functie van de fotonenergie, wat duidt op een relatief klein aantal 'trap'toestanden waarin quasideeltjes kunnen worden ingevangen. Simulaties met het energie-afhankelijke kinetische model tonen aan dat het aantal gelokaliseerde 'trap'-toestanden in de Al junkties slechts ongeveer 7000 bedraagt, respectievelijk 3 en 30 keer minder dan in soortgelijke junkties van Nb-Al en Ta-Al. Het aantal 'trap'-toestanden is onafhankelijk van de grootte van de detektoren, wat er sterk op duidt dat deze toestanden gelokaliseerd zijn in het Nb waaruit de contacten naar de bovenste en onderste electroden zijn gemaakt. De responsiviteit van de junkties neemt toe met toenemende aangelegde spanning, en dit effect is sterker voor de grotere junkties met langere verliestijden voor quasideeltjes. Deze relatie houdt verband met de brede energieverdeling van de quasideeltjes. Doordat de tunnelstromen in tegengestelde richting lager zijn, en de quasideeltjes vermenigvuldiging sterker is bij hogere aangelegde spanning, neemt de responsiviteit toe met toenemende spanning. Deze effecten konden met succes gesimuleerd worden met het energie-afhankelijke kinetische model. Het energie-oplossend vermogen van de Al junkties wordt beperkt door een positie-afhankelijke component, waarschijnlijk doordat de verliezen van quasideeltjes voornamelijk optreden ter plaatse van de Nb contacten aan de bovenste en onderste electroden. Deze positie-afhankelijke component beperkt het intrinsieke oplossend vermogen $\lambda/\Delta\lambda$ voor optische fotonen met een golflengte van 500 nm tot ongeveer 17, ver onder de theoretische waarde van ongeveer 30. Desalniettemin zijn de mogelijkheden van Al STJ's als detektoren van optische fotonen aangetoond. In toekomstig werk aan Al STJ's zullen de Nb contacten vervangen worden door Ta, waardoor de positie-afhankelijkheid van de responsiviteit verminderd moet worden, en daarmee het energie-oplossend vermogen van de detector verhoogd. Daarnaast is voorzien om de Al junkties te koppelen aan röntgenabsorbers en zo het goede theoretische energie-oplossende vermogen van Al STJ's ook in het röntgengebied te exploiteren.

De gefabriceerde junkties uit Mo-Al vertonen nog een probleem met de struktuur van hun randen, zoals die gevormd worden in de etsprocedure waarin de vorm van de junkties bepaald wordt. De bovenste Mo laag wordt 600 nm meer zijwaarts wegge-etst dan de onderste Mo laag, waardoor een stapprofiel aan de randen ontstaat. De Al laag tussen de twee Mo lagen vormt geen vertikale wand, waardoor de dunne isolerende laag van aluminiumoxide beschadigd is. Dientengevolge hebben deze junkties hoge lekstromen die schalen met de omtrek van de detektoren, en waarden hebben van ongeveer 1.25 µA per µm van de omtrek. Dergelijke hoge lekstromen maken de detektie van fotonen onmogelijk. Een eerste vereiste voor bruikbare STJ's uit Mo is daarom een verbetering van de bovengenoemde etsprocedure, zodat de lekstromen drastisch worden verlaagd.

# Acknowledgments

Having finished the final manuscript, it becomes clear how much help and support was provided throughout the years I have worked on this thesis and even before.

A first group of people who have very considerably contributed are my colleagues at the European Space Agency, where all of the work was performed.

First, I would like to thank Tone Peacock and Abel Poelaert for making it all possible. Their flexibility and dedication gave me the opportunity to write my graduation thesis at ESTEC, which finally led to this work. Tone has always pointed me towards the right directions, but leaving me enough room to make developments of my own. I want to thank Abel for all the great work done before I arrived, as this thesis is largely based on his previous findings. I am also very grateful to him for answering my email, where I was looking for a subject for my graduation thesis, and for teaching me the basics of STJ operation.

Then, I would like to thank all my colleagues from Aurora and in the Science Payload and Advanced Concepts Office (or whatever other Division/Office they might be in) for a very competent and interesting working environment and a lot of very useful tips and discussions: Peter Verhoeve, Nicola Rando, Roland den Hartog, Didier Martin, Alan Owens, Jason Page, Marcos Bavdaz, Alex Jeanes, Diane Barton, Jacques Verveer, John van den Biezen, Solve Andersson, Axel van Dordrecht, Hans Smit, Birgit Schröder, Mylene Riemens, Georges Sirbi, Franz Moser, Gunther Thorner, Robert Strade, Stefan Kraft, Thierry Beaufort, Ludovic Duvet, Christian Erd, Richard Hijmering and everybody else I possibly forgot, I learned a lot from you.

A second group of people I would like to thank is the team of the Low Temperature Division of the University of Twente. I would especially like to thank Prof. Horst Rogalla for accepting me as an external PhD student at the University of Twente and my co-promotor Alexander Golubov, with whom I had a lot of very fruitful discussions and collaborations, which led to the proximity effect results of the thesis.

Then, I would like to thank Rob Venn from Cambridge MicroFab for the fabrication of the very numerous junctions and for a very good cooperation throughout the years.

I am also very grateful to Alex Kozorezov and Keith Wigmore from the School of Physics and Chemistry of the University of Lancaster. The theoretical part of this work has gained a lot from their insight into the physics of phonons and quasiparticles. The endless discussions and meetings we had have greatly influenced my understanding of STJs.

Finally, and most importantly, I would like to thank my parents, my sister Kim and all my friends, whose enormous support dates back to times where I wasn't even thinking about writing a thesis.

# List of publications

## First author

1. *Modelling the energy gap in transition metal/aluminium bilayers*
   G. Brammertz, A. Golubov, A. Peacock, P. Verhoeve, D. J. Goldie, R. Venn, Physica C
   **350**, 227 (2001).

2. *Development of Practical Soft X-ray Spectrometers*
   G. Brammertz, P. Verhoeve, A. Peacock, D. Martin, N. Rando, R. den Hartog, D. J.
   Goldie, IEEE Transactions on Applied Superconductivity **11**, 828 (2001).

3. *Generalized proximity effect model in superconducting bi- and trilayer films*
   G. Brammertz, A. Poelaert, A. A. Golubov, P. Verhoeve, A. Peacock, H. Rogalla, , J.
   Appl. Phys. **90**, 355 (2001).

4. *Strong quasiparticle trapping in a 6x6 array of Vanadium-Aluminium Superconducting
   Tunnel Junctions*
   G. Brammertz, A. Peacock, P. Verhoeve, A. Kozorezov, R. den Hartog, N. Rando, R.
   Venn, , AIP Conference Proceedings **605**, Low Temperature Detectors, Edited by F.S.
   Porter, D. McCammon, M. Galeazzi, C. K. Stahle, 59 (2002).

5. *Critical temperature of superconducing bilayers: theory and experiment*
   G. Brammertz, A. A. Golubov, P. Verhoeve, R. den Hartog, A. Peacock, H. Rogalla, ,
   Appl. Phys. Lett. **80**, 2955 (2002).

6. *Energy-dependent model of photon absorption by superconducting tunnel junctions*
   G. Brammertz, A. G. Kozorezov, J. K. Wigmore, R. den Hartog, P. Verhoeve, D. Martin,
   A. Peacock, A. A. Golubov, H. Rogalla, accepted for publication in J. Appl. Phys. **94**, 5854
   (2003).

7. *Optical photon detection in Al superconducting tunnel junctions*
   G. Brammertz, A. Peacock, P. Verhoeve, D. Martin, R. Venn, to be published in the
   proceedings of the 10[th] international conference on low temperature detectors, 7-11 July
   2003, Genoa, Italy (2003).

8. *Enhancement of photo-responsivity of small-gap, multiple tunnelling superconducting
   tunnel junctions due to quasiparticle multiplication*
   G. Brammertz, A. G. Kozorezov, R. den Hartog, P. Verhoeve, A. Peacock, J. K. Wigmore,
   D. Martin, R. Venn, to be published in the proceedings of the european conference of
   applied superconductivity 2003, 14-18 September 2003, Sorrento, Italy (2003).

9. *Optical photon detection with high quality Al superconducting tunnel junctions*
   G. Brammertz, A. Peacock, P. Verhoeve, D. Martin, R. den Hartog, R. Venn, submitted to
   Rev. Scient. Instr. (2003).

# Co-author

1. *Distributed Read-Out Imaging Devices for X-ray Imaging Spectroscopy*
   R. den Hartog, D. Martin, A. Kozorezov, P. Verhoeve, N. Rando, A. Peacock, G. Brammertz, M. Krumrey, D. J. Goldie, R. Venn, Proceedings of SPIE 4012, 237 (2000).

2. *Large-format Distributed Read-Out Imaging Devices for Optical and X-ray Imaging Spectroscopy*
   R. den Hartog, D. Martin, A. Kozorezov, G. Brammertz, P. Verhoeve, A. Peacock, F. Scholze, D.J Goldie, AIP Conference Proceedings 605 Low Temperature Detectors, Edited by F.S. Porter, D. McCammon, M. Galeazzi, C. K. Stahle, 11 (2002).

3. *Quasiparticle diffusion and the energy resolution of superconducting tunnelling junctions as photon detectors. II. Experiment*
   R. den Hartog, A. G. Kozorezov, J. K. Wigmore, D. Martin, P. Verhoeve, A. Peacock, A. Poelaert, G. Brammertz, Phys. Rev. B 66, 094511 (2002).

4. *Large Format Distributed Read-Out Imaging Devices for X-Ray Imaging Spectroscopy*
   R. den Hartog, A. Kozorezov, D. Martin, G. Brammertz, P. Verhoeve, A. Peacock, F. Scholze, D. J. Goldie, Proceedings of SPIE 4497, 50 (2002).

5. *Hard X-ray test and evaluation of a prototype 32x32 pixel Gallium-Arsenide Array*
   C. Erd, A. Owens, G. Brammertz, M. Bavdaz, A. Peacock, V. Laemsae, S. Nenonen, H. Andersson, N. Haack, Nucl. Instr. and Meth. A 487, 78 (2002).

6. *New measurements of quantum efficiency and depletion depth in gallium-arsenide detectors*
   C. Erd, A. Owens, G. Brammertz, M. Bavdaz, A. Peacock, S. Nenonen, H. Andersson, Proceedings of SPIE, Seattle, July 2002.

7. *Self-heating phenomena in superconducting tunnelling junctions*
   A. G. Kozorezov, J. K. Wigmore, A. Peacock, R. den Hartog, G. Brammertz, D. Martin, P. Verhoeve, N. Rando., AIP Conference Proceedings 605 Low Temperature Detectors, Edited by F.S. Porter, D. McCammon, M. Galeazzi, C. K. Stahle, 51(2002).

8. *Local trap spectroscopy in superconducting tunnel junctions*
   A. G. Kozorezov, J. K. Wigmore, A. Peacock, A. Poelaert, P. Verhoeve, R. den Hartog, G. Brammertz, Appl. Phys. Lett. 78, 3651 (2001).

9. *The singularities of quasiparticle current in superconducting multiple tunnelling junctions*
   A. G. Kozorezov, J. K. Wigmore, A. Peacock, R. den Hartog, G. Brammertz, D. Martin, P. Verhoeve, N. Rando, Physica B 316-317, 586 (2002).

10. *Observation of a new non-equilibrium state in superconductors caused by sequential tunnelling*
    A. G. Kozorezov, J. K. Wigmore, A. Peacock, R. den Hartog, D. Martin, G. Brammertz, P. Verhoeve, N. Rando, submitted to Phys. Rev. Lett. (2003).